\documentclass[12pt]{article}
\pdfoutput=1
\usepackage{epsfig}
\usepackage{psfrag}
\usepackage{latexsym}
\usepackage{indentfirst}
\usepackage{fancyhdr}
\usepackage{amssymb}
\usepackage{amsfonts}
\usepackage{cite}
\usepackage{bbold}
\usepackage[footnotesize]{caption}
\usepackage[center,footnotesize,hang]{subfigure}
\usepackage{url}

\usepackage{indentfirst}
\RequirePackage{mathrsfs}
\usepackage{subfigure}
\usepackage{multicol}
\usepackage[usenames,dvipsnames]{color}
\usepackage{cite}
\RequirePackage[colorlinks=true
,urlcolor=blue
,anchorcolor=blue
,citecolor=blue
,filecolor=blue
,linkcolor=blue
,menucolor=blue
,pagecolor=blue
,linktocpage=true
,pdfproducer=medialab
]{hyperref}
\usepackage{cancel}

\newcommand{\be}{\beta}

\newcommand{\beq}{\begin{equation}}
\newcommand{\eeq}{\end{equation}}
\newcommand{\bac}{\beq\begin{array}}
\newcommand{\eac}{\end{array}\eeq}
\newcommand{\ba}{\begin{array}}
\newcommand{\ea}{\end{array}}
\newcommand{\bea}{\begin{eqnarray}}
\newcommand{\eea}{\end{eqnarray}}
\newcommand{\beaa}{\begin{eqnarray*}}
\newcommand{\eeaa}{\end{eqnarray*}}

\def\beq{\begin{equation}}
\def\eeq{\end{equation}}
\def\bea{\begin{eqnarray}}
\def\eea{\end{eqnarray}}
\def\bet{\begin{tabular}}
\def\eet{\end{tabular}}
\def\bes{\begin{subequations}\bea}
\def\ees{\eea\end{subequations}}

\def\be{\begin{equation}}
\def\ee{\end{equation}}
\def\bc{\begin{center}}
\def\ec{\end{center}}
\def\bea{\begin{eqnarray}}
\def\eea{\end{eqnarray}}

\catcode`@=11
\def\marginnote#1{}
\newcount\hour
\newcount\minute
\newtoks\amorpm
\hour=\time\divide\hour by60
\minute=\time{\multiply\hour by60 \global\advance\minute by-\hour}
\edef\standardtime{{\ifnum\hour<12 \global\amorpm={am}%
        \else\global\amorpm={pm}\advance\hour by-12 \fi
        \ifnum\hour=0 \hour=12 \fi
        \number\hour:\ifnum\minute<10 0\fi\number\minute\the\amorpm}}
\edef\militarytime{\number\hour:\ifnum\minute<10 0\fi\number\minute}
\def\draftlabel#1{{\@bsphack\if@filesw {\let\thepage\relax
   \xdef\@gtempa{\write\@auxout{\string
      \newlabel{#1}{{\@currentlabel}{\thepage}}}}}\@gtempa
   \if@nobreak \ifvmode\nobreak\fi\fi\fi\@esphack}
        \gdef\@eqnlabel{#1}}
\def\@eqnlabel{}
\def\@vacuum{}
\def\draftmarginnote#1{\marginpar{\raggedright\scriptsize\tt#1}}
\def\draft{\oddsidemargin 0.0truein
        \def\@oddfoot{\sl preliminary draft \hfil
        \rm\thepage\hfil\sl\today\quad\militarytime}
        \let\@evenfoot\@oddfoot \overfullrule 3pt
        \let\label=\draftlabel
        \let\marginnote=\draftmarginnote
   \def\@eqnnum{(\theequation)\rlap{\kern\marginparsep\tt\@eqnlabel}%
\global\let\@eqnlabel\@vacuum}  }
\catcode`@=12
\begin{document}
\begin{titlepage}
\vspace*{-1cm}
\phantom{hep-ph/***}

\hfill{RM3-TH/13-1}
\hfill{CERN-PH-TH/2013-020}

\vskip 2.5cm
\begin{center}
{\Large\bf Collider Physics within the Standard Model: a Primer}
\end{center}
\vskip 0.2  cm
\vskip 0.5  cm
\begin{center}
{\large Guido Altarelli}~\footnote{e-mail address: guido.altarelli@cern.ch}
\\
\vskip .1cm
Dipartimento di Matematica e Fisica, Universit\`a di Roma Tre 
\\
INFN, Sezione di Roma Tre, I-00146 Rome, Italy
\\
\vskip .1cm
and
\\
CERN, Department of Physics, Theory Unit
\\
CH-1211 Geneva 23, Switzerland
\\
\end{center}
\vskip 0.7cm
\begin{abstract}
\noindent
The first LHC results at 7-8 TeV, with the discovery of a candidate Higgs boson and the non observation of new particles or exotic phenomena, have made a big step towards completing the experimental confirmation of the Standard Model (SM) of fundamental particle interactions. It is thus a good moment for me to collect, update and improve my graduate lecture notes on Quantum Chromodynamics (QCD) and the theory of Electroweak (EW) Interactions, with main focus on Collider Physics. I hope that these lectures can provide an introduction to the subject for the interested reader, assumed to be already familiar with quantum field theory and some basic facts in elementary particle physics as taught in undergraduate courses.
\end{abstract}
\end{titlepage}
\setcounter{footnote}{0}
\vskip2truecm 

\pdfbookmark[1]{Table of Contents}{tableofcontents}
\tableofcontents
\newpage

%
\section{\Large\bf{Gauge Theories and the Standard Model}}
\label{Chap:1}
\vskip 0.7cm

\subsection{An Overview of the Fundamental Interactions}
\label{sec:1}

A possible goal of fundamental physics is to reduce all natural phenomena to a set of basic laws and theories that, at least in principle, can quantitatively reproduce and predict the experimental observations. At microscopic level all the phenomenology of matter and radiation, including molecular, atomic, nuclear and subnuclear physics, can be understood in terms of three classes of fundamental interactions: strong, electromagnetic and weak interactions. For all material bodies on the Earth and in all geological, astrophysical and cosmological phenomena a fourth interaction, the gravitational force, plays a dominant role, while it is instead negligible in atomic and nuclear physics. In atoms the electrons are bound to nuclei by electromagnetic forces and the properties of electron clouds explain the complex phenomenology of atoms and molecules. Light is a particular vibration of electric and magnetic fields (an electromagnetic wave). Strong interactions bind the protons and neutrons together in nuclei, being so intensively attractive at short distances that they prevail over the electric repulsion due to the equal sign charges of protons. Protons and neutrons, in turn, are composites of three quarks held together by strong interactions to which quarks and gluons are subject (hence these particles are called "hadrons" from the Greek word for "strong"). To the weak interactions  are due the beta radioactivity that makes some nuclei unstable as well as the nuclear reactions that produce the enormous energy radiated by the stars and by our Sun in particular. The weak interactions also cause the disintegration of the neutron, the charged pions, the lightest hadronic particles with strangeness, charm, and beauty (which are "flavour" quantum numbers) as well as the decay of the quark top and of the heavy charged leptons (the muon $\mu^-$ and the tau $\tau^-$). In addition all observed neutrino interactions are due to these weak forces. 

All these interactions (with the possible exception of gravity) are described within the framework of quantum mechanics and relativity, more precisely by a local relativistic quantum field theory. To each particle, described as pointlike, is associated a field with suitable (depending on the particle spin) transformation properties under the Lorentz group (the relativistic space-time coordinate transformations). It is remarkable that the description of all these particle interactions is based on a common principle: "gauge" invariance. A "gauge" symmetry is invariance under transformations that rotate the basic internal degrees of freedom  but with rotation angles that depend on the space-time point. At the classical level gauge invariance is a property of the Maxwell equations of electrodynamics and it is in this context that the notion and the name of gauge invariance were introduced. The prototype of all quantum gauge field theories, with a single gauged charge, is QED, Quantum Electro- Dynamics, developed in the years from 1926 until about 1950, which indeed is the quantum version of Maxwell theory. Theories with gauge symmetry in 4 space-time dimensions are renormalizable and are completely determined given the symmetry group and the representations of the interacting fields.  The whole set of strong, electromagnetic and weak interactions is described by a gauge theory, with 12 gauged non-commuting charges, which is called "the Standard Model" of particle interactions (SM). Actually only a subgroup of the SM symmetry is directly reflected in the spectrum of physical states. A part of the electroweak symmetry is hidden by the Higgs mechanism for spontaneous symmetry breaking of the gauge symmetry.

The theory of general relativity is a classic description (in the sense of non quantum mechanical)  of gravity that goes beyond the static approximation described by Newton law and includes dynamical phenomena like, for example, gravitational waves. The problem of  formulating a quantum theory of gravitational interactions is one of the central problems of contemporary theoretical physics. But quantum effects in gravity become only important for energy concentrations in space-time which are not in practice accessible to experimentation in the laboratory. Thus the search for the correct theory can only be done by a purely speculative approach. All attempts at a description of quantum gravity in terms of a well defined and computable local field theory along similar lines as for the SM have so far failed to lead to a satisfactory framework. Rather, at present the most complete and plausible description of quantum gravity is a theory formulated in terms of non pointlike basic objects, the so called "strings", extended over distances much shorter than those experimentally accessible, that live in a space-time with 10 or 11 dimensions. The additional dimensions beyond the familiar 4 are, typically, compactified which means that they are curled up with a curvature radius of the order of the string dimensions. Present string theory is an all-comprehensive framework that suggests a unified description of all interactions together with gravity of which the SM would be only a low energy or large distance approximation.

A fundamental principle of quantum mechanics, the Heisenberg indetermination principle, implies that, for studying particles with spatial dimensions of order $\Delta x$ or interactions taking place at distances of order $\Delta x$, one needs as a probe a beam of particles  (typically produced by an accelerator) with impulse $p\gtrsim \hbar/\Delta x$, where $\hbar$ is the reduced Planck constant ($\hbar=h/2\pi$). Accelerators presently in operation, like the Large Hadron Collider (LHC) at CERN near Geneva, allow to study collisions between two particles with total center of mass energy up to $2E\sim2pc\lesssim 7 -14~TeV$. These machines, in principle, can allow to study physics down to distances $\Delta x\gtrsim 10^{-18} cm$. Thus, on the basis of results from experiments at existing accelerators, we can confirm that, down to distances of that order of magnitude, indeed electrons, quarks and all the fundamental SM particles do not show an appreciable internal structure and look elementary and pointlike. We expect that quantum effects in gravity will certainly become important at distances $\Delta x \leq 10^{-33} cm$ corresponding to energies up to $E\sim M_{Planck}c^2\sim 10^{19}~GeV$, where $M_{Planck}$ is the Planck mass, related to Newton constant by $G_N = \hbar c/M_{Planck}^2$. 
At such short distances the particles that so far appeared as pointlike could well reveal an extended structure, like for strings, and be described by a more detailed theoretical framework of which the local quantum field theory description of the SM would be just a low energy/large distance limit. 

From the first few moments of the Universe, after the Big Bang, the temperature of the cosmic background went down gradually, starting from $kT\sim M_{Planck}c^2$, where $k=8.617...10^{-5}~eV~^0K^{-1}$ is the Boltzmann constant, down to the present situation where $T\sim 2.725 ^0K$. Then all stages of high energy physics from string theory, which is a purely speculative framework, down to the SM phenomenology, which is directly accessible to experiment and well tested, are essential for the reconstruction of the evolution of the Universe starting from the Big Bang. This is the basis for the ever increasing relation between high energy physics and cosmology.

\subsection{The Architecture of the Standard Model}
\label{sec:2}

The SM is a  gauge field theory based on the symmetry group $SU(3) \bigotimes SU(2) \bigotimes U(1)$. The transformations of the group act on the basic fields. This group has 8+3+1= 12 generators with a non trivial commutator algebra (if all generators  commute the gauge theory is said to be "abelian", while the SM is a "non abelian" gauge theory).   $SU(2) \bigotimes U(1)$ describes the electroweak (EW) interactions \cite{SM1}- \cite{SM3} and the electric charge Q, the generator of the QED gauge group $U(1)_Q$, is the sum of $T_3$, one of the $SU(2)$ generators and of $Y/2$, where $Y$ is the $U(1)$ generator: $Q=T_3+Y/2$. $SU(3)$ is the "colour" group of the theory of strong interactions (QCD: Quantum Chromo-Dynamics \cite{SM4}- \cite{SM6}).

In a gauge theory \footnote{Much of the material in this Chapter is a revision and update of ref. \cite{spri1}} to each generator $T$ is associated a vector boson (also said gauge boson) with the same quantum numbers as $T$, and, if the gauge symmetry is unbroken, this boson is of vanishing mass. These vector (i.e. of spin 1) bosons act as mediators of the corresponding interactions. For example, in QED the vector boson associated to the generator $Q$ is the photon $\gamma$. The interaction between two charged particles in QED, for example two electrons, is mediated by the exchange of one (or seldom more than one) photon emitted by one electron and reabsorbed by the other one. Similarly in the SM there are 8 gluons associated to the $SU(3)$ colour generators, while for $SU(2) \bigotimes U(1)$ there are 4 gauge bosons $W^+$, $W^-$, $Z^0$ and  $\gamma$. Of these, only  the gluons and the photon $\gamma$ are massless because the symmetry induced by the other 3 generators is actually spontaneously broken. The masses of $W^+$, $W^-$ and $Z^0$ are quite large indeed on the scale of elementary particles: $m_W \sim 80.4~GeV$, $m_Z \sim 91.2~GeV$ are as heavy as atoms of intermediate size like rubidium and molibdenum, respectively. In the electroweak theory the breaking of the symmetry is of a particular type, denoted as spontaneous symmetry breaking. In this case charges and currents are as dictated by the symmetry but the fundamental state of minimum energy, the vacuum, is not unique and there is a continuum of degenerate states that all together respect the symmetry (in the sense that the whole vacuum orbit is spanned by applying the symmetry transformations). The symmetry breaking is due to the fact that the system (with infinite volume and infinite number of degrees of freedom) is found in one particular vacuum state, and this choice, which for the SM occurred in the first instants of the Universe life, makes the symmetry violated in the spectrum of states. In a gauge theory like the SM the spontaneous symmetry breaking is realized by the Higgs mechanism \cite{ebh,ghk,hig,kib} (described in detail in Sect. (\ref{sec:7})): there are a number of scalar (i.e. of zero spin) Higgs bosons with a potential that produces an orbit of degenerate vacuum states. One or more of these scalar Higgs particles must necessarily be present in the spectrum of physical states with masses very close to the range so far explored. The Higgs particle has now been found at the LHC with $m_H \sim 126$ GeV \cite{ATLH,CMSH} thus making a big step towards completing the experimental verification of the SM. The Higgs boson acts as the mediator of a new class of interactions that, at the tree level, are coupled in proportion to the particle masses and thus have a very different strength for, say, an electron and a top quark. 

The fermionic (all of spin 1/2) matter fields of the SM are quarks and leptons. Each type of quark is a colour triplet (i.e. each quark flavour comes in three colours) and also carries electroweak charges, in particular electric charges +2/3 for up-type quarks and -1/3 for down-type quarks. So quarks are subject to all SM interactions. Leptons are colourless and thus do not interact strongly (they are not hadrons) but have electroweak charges, in particular electric charges -1 for charged leptons ($e^-$, $\mu^-$ and $\tau^-$) and charge 0 for neutrinos ($\nu_e$, $\nu_\mu$ and $\nu_\tau$). Quarks and leptons are grouped in 3 "families" or "generations" with equal quantum numbers but different masses. At present we do not have an explanation for this triple repetition of fermion families: 
 \beq
\left[\matrix{
u&u&u&\nu_e\cr
d&d&d&e\cr
} 
\right],~~~~~
\left[\matrix{
c&c&c&\nu_{\mu}\cr
s&s&s&\mu\cr
} 
\right],~~~~~\left[\matrix{
t&t&t&\nu_{\tau}\cr
b&b&b&\tau\cr
} 
\right]. 
\label{123f}
\eeq
 
The QCD sector of the SM (see Chapter 2) has a simple structure but a very rich dynamical content, including the observed complex spectroscopy with a large number of hadrons. The most prominent properties of QCD are asymptotic freedom
and confinement. In field theory the effective coupling of a given interaction vertex is modified by the
interaction. As a result, the measured intensity of the force depends on the transferred (four)momentum squared, 
$Q^2$, among the participants. In QCD the relevant coupling parameter that appears in physical processes is
$\alpha_s=e_s^2/4\pi$ where $e_s$ is the coupling constant of the basic interaction vertices of quark and gluons: $qqg$ or $ggg$ (see Eqs.(\ref{LagQCD},\ref{der},\ref{alfa},\ref{F})). Asymptotic freedom means that the effective
coupling becomes a function of $Q^2$: $\alpha_s(Q^2)$ decreases for increasing $Q^2$ and vanishes asymptotically. Thus, the QCD
interaction becomes very weak in processes with large $Q^2$, called hard processes or deep inelastic processes
(i.e. with a final state distribution of momenta and a particle content very different than those in the initial
state). One can prove that in 4 space-time dimensions all pure-gauge theories based on a non commuting group of
symmetry are asymptotically free and conversely. The effective
coupling decreases very slowly at large momenta with the inverse logarithm of $Q^2$:
$\alpha_s(Q^2)=1/b\log{Q^2/\Lambda^2}$ where b is a known constant and $\Lambda$ is an energy of order a few
hundred MeV. Since in quantum mechanics large momenta imply short wavelenghts, the result is that at short
distances (or $Q > \Lambda$) the potential between two colour charges is similar to the Coulomb potential, i.e. proportional to
$\alpha_s(r)/r$, with an effective colour charge which is small at short distances. On the contrary the
interaction strenght becomes large at large distances or small transferred momenta, of order $Q <
\Lambda$. In fact all observed hadrons are tightly bound composite states of quarks (baryons are made of $qqq$ and mesons of $q\bar q$), with compensating colour
charges so that they are overall neutral in colour. In fact, the property of confinement is the impossibility of
separating colour charges, like individual quarks and gluons or any other coloured state. This is because in QCD the interaction
potential between colour charges increases at long distances linearly in r. When we try to separate the quark
and the antiquark that form a colour neutral meson the interaction energy grows until pairs of quarks and
antiquarks are created from the vacuum and new neutral mesons are coalesced and observed in the final state instead of free quarks. For
example, consider the process $e^+e^- \rightarrow q\bar{q}$ at large center of mass energies. The final
state quark and antiquark have large energies, so they separate in opposite directions very fast. But the
colour confinement forces create new pairs in between them. What is observed is two back-to-back jets of
colourless hadrons with a number of slow pions that make the exact separation of the two jets impossible. In
some cases a third well separated jet of hadrons is also observed: these events correspond to the radiation
of an energetic gluon from the parent quark-antiquark pair.

In the EW sector the SM (see Chapter 3) inherits the phenomenological successes of the old $(V-A) \otimes (V-A)$ four-fermion
low-energy description of weak interactions, and provides a well-defined and consistent theoretical framework
including weak interactions and quantum electrodynamics in a unified picture. The weak
interactions derive their name from their intensity. At low energy the strength of the effective four-fermion
interaction of charged currents is determined by the Fermi coupling constant $G_F$. For example, the effective
interaction for muon decay is given by
\beq {\cal L}_{\rm eff} = (G_F/\sqrt 2) \left[ \bar
\nu_{\mu}\gamma_{\alpha}(1-\gamma_5)\mu \right]
\left[ \bar e\gamma^{\alpha}(1-\gamma_5)\nu_e \right]~,
\label{1}
\eeq with \cite{pdg12} 
\beq G_F = 1.1663787(6) \times 10^{-5}~{\rm GeV}^{-2}~.
\label{2}
\eeq In natural units $ \hbar = c = 1$, $G_F$ (which we most often use in this work) has dimensions of (mass)$^{-2}$. As a result, the intensity of weak
interactions at low energy is characterized by
$G_FE^2$, where $E$ is the energy scale for a given process ($E \approx m_{\mu}$  for muon decay). Since
\begin{equation} G_FE^2 = G_Fm^2_p(E/m_p)^2 \simeq 10^{-5}(E/m_p)^2~,
\label{3}
\end{equation} where $m_p$ is the proton mass, the weak interactions are indeed weak at low energies (up to energies of order
a few ten's of GeV). Effective four
fermion couplings for neutral current interactions have comparable intensity and energy behaviour. The quadratic increase with
energy cannot continue for ever, because it would lead to a violation of unitarity. In fact, at large energies the propagator
effects can no longer be neglected, and the current--current interaction is resolved into current--$W$ gauge boson vertices
connected by a $W$ propagator. The strength of the weak interactions at high energies is then measured by $g_W$, the
$W-\mu$-$\nu_{\mu}$ coupling, or, even better, by
$\alpha_W = g^2_W/4\pi$ analogous to the fine-structure constant $\alpha$ of QED (in Chapter \ref{Chap:3}, $g_W$ is simply denoted by $g$ or $g_2$). In the standard EW theory, we have
\begin{equation}
\alpha_W = \sqrt 2~G_F~m^2_W/\pi \cong 1/30~.
\label{4}
\end{equation} That is, at high energies the weak interactions are no longer so weak.

 The range $r_W$ of weak interactions is very short: it is only with the experimental discovery of the $W$ and $Z$
gauge bosons that it could be demonstrated that $r_W$ is non-vanishing. Now we know that
\begin{equation} r_W = \frac{\hbar}{m_Wc} \simeq 2.5 \times 10^{-16}~{\rm cm}~,
\label{5}
\end{equation} corresponding to $m_W \simeq 80.4$~GeV. This very large value for the $W$ (or the
$Z$) mass makes a drastic difference, compared with the massless photon and the infinite range of the QED force. The
direct experimental limit on the photon mass is \cite{pdg12}
$m_{\gamma} <1~10^{-18}~eV$. Thus, on the one hand, there is very good evidence that the photon is massless. On the
other hand, the weak bosons are very heavy. A unified theory of EW interactions has to face this striking
difference.

Another apparent obstacle in the way of EW unification is the chiral structure of weak interactions: in the
massless limit for fermions, only left-handed quarks and leptons (and right-handed antiquarks and antileptons) are
coupled to $W$'s. This clearly implies parity and charge-conjugation violation in weak interactions.

The universality of weak interactions and the algebraic properties of the electromagnetic and weak currents [the
conservation of vector currents (CVC), the partial conservation of axial currents (PCAC), the algebra of currents,
etc.] have been crucial in pointing to a symmetric role of electromagnetism and weak interactions at a more fundamental
level. The old Cabibbo universality \cite{cab} for the weak charged current:
\begin{eqnarray} J^{\rm weak}_{\alpha} &=&
\bar \nu_{\mu}\gamma_{\alpha} (1-\gamma_5)\mu +
\bar \nu_e\gamma_{\alpha}(1-\gamma_5) e +
\cos\theta_c~\bar u \gamma_{\alpha}(1-\gamma_5)d + \nonumber \\ 
 &+&\sin \theta_c~\bar u \gamma_{\alpha}(1-\gamma_5)s +
...~,
\label{6}
\end{eqnarray} suitably extended, is naturally implied by the standard EW theory. In this theory the weak
gauge bosons couple to all particles with couplings that are proportional to their weak charges, in the same way as the
photon couples to all particles in proportion to their electric charges [in Eq.~(\ref{6}), $d' =
\cos\theta_c~d + \sin \theta_c~s$ is the weak-isospin partner of $u$ in a doublet.  The $(u,d')$ doublet has the same
couplings as the $(\nu_e,\ell)$ and 
$(\nu_{\mu},\mu)$ doublets].

Another crucial feature is that the charged weak interactions are the only known interactions that can change flavour:
charged leptons into neutrinos or up-type quarks into down-type quarks. On the contrary, there are no flavour-changing
neutral currents at tree level. This is a remarkable property of the weak neutral current, which is explained by the
introduction of the Glashow-Iliopoulos-Maiani (GIM) mechanism \cite{GIM} and has led to the successful prediction of charm.

The natural suppression of flavour-changing neutral currents, the separate conservation of $e, \mu$  and $\tau$
leptonic flavours that is only broken by the small neutrino masses, the mechanism of CP violation through the phase in the quark-mixing matrix \cite{KM}, are all crucial
features of the SM. Many examples of new physics tend to break the selection rules of the standard theory.
Thus the experimental study of rare flavour-changing transitions is an important window on possible new physics.
 
The SM is a renormalizable field theory which means that the ultra-violet divergences that appear in loop diagrams can be eliminated by a suitable redefinition of the parameters already appearing in the bare lagrangian: masses, couplings and field normalizations. As it will be discussed later, a necessary condition for a theory to be renormalizable is that only operator vertices of dimension not larger than 4 (that is $m^4$ where $m$ is some mass scale) appear in the lagrangian density $\cal L$ (itself of dimension 4, because the action $S$ is given by the integral of $\cal L$ over $d^4x$ and is dimensionless in natural units: $\hbar = c = 1$). Once this condition is added to the specification of a gauge group and of the matter field content the gauge theory lagrangian density is completely specified. We shall see the precise rules to write down the lagrangian of a gauge theory in the next Section.

\vspace{-6pt}
\subsection{The Formalism of Gauge Theories}
\label{sec:3}

In this Section we summarize the definition and the structure of a gauge Yang--Mills theory \cite{YM}. We
will list here the general rules for constructing such a theory. Then these results will be applied
to the SM.

Consider a lagrangian density ${\cal L}[\phi,\partial_{\mu}\phi]$ which is invariant under a $D$ dimensional continuous
group $\Gamma$ of transformations:
\begin{equation}
\phi'(x) = U(\theta^A)\phi(x)\quad\quad (A = 1, 2, ..., D)~.
\label{7}
\end{equation} 
with:
\begin{equation}
U(\theta^A) = \exp{[ig \sum_A~\theta^AT^A]} \sim~1 + ig \sum_A~\theta^AT^A~+\dots, 
\label{7a}
\end{equation}
The quantities $\theta^A$ are numerical parameters, like angles in the particular case of a rotation group in some internal space.  The approximate expression on the right is valid for $\theta^A$ infinitesimal.
Then, $g$ is the coupling constant and
$T^A$ are the generators of the group $\Gamma$ of transformations (\ref{7}) in the (in general reducible)
representation of the fields $\phi$. Here we restrict ourselves to the case of internal symmetries, so that $T^A$ are
matrices that are independent of the space-time coordinates and the arguments of the fields $\phi$ and $\phi'$ in Eq.(\ref{7}) are the same. If $U$ is unitary, then the generators $T^A$ are hermitian, but this need not be the case in general (though it is true for the SM). Similarly if $U$ is a group of matrices with unit determinant, then the traces of $T^A$ vanish: ${\rm tr}(T^A)=0$. In general, the generators satisfy the commutation relations
\begin{equation} [T^A,T^B] = iC_{ABC}T^C~.
\label{9}
\end{equation}
For $A,B,C....$ up or down indices make no difference: $T^A=T_A$ etc.
The structure constants $C_{ABC}$ are completely antisymmetric in their indices, as can be easily seen. 
Recall that if all generators  commute the gauge theory is said to be "abelian" (in this case all the structure constants $C_{ABC}$ vanish), while the SM is a "non abelian" gauge theory. 
We choose to normalize the generators $T^A$  in such a way that
for the lowest dimensional non-trivial representation of the group $\Gamma$ (we use $t^A$ to denote the generators in
this particular representation) we have
\begin{equation} {\rm tr}(t^At^B) = \frac{1}{2} \delta^{AB}~.
\label{8}
\end{equation}
A normalization convention is needed to fix the  normalization of the coupling $g$ and of the structure constants $C_{ABC}$.
In the following, for each quantity $f^A$ we define
\begin{equation} {\bf f} = \sum_A~T^Af^A~.
\label{10}
\end{equation}
For example, we can rewrite Eq. (\ref{7a}) in the form:
\begin{equation}
U(\theta^A) = \exp{[ig\mbox{\boldmath $\theta$}]} \sim~1 + ig\mbox{\boldmath $\theta$}~+\dots, 
\label{10a}
\end{equation} 
If we now make the parameters $\theta^A$ depend on the space--time coordinates
$\theta^A = \theta^A(x_{\mu}),$ ${\cal L}[\phi,\partial_{\mu}\phi]$ is in general no longer invariant under the gauge
transformations $U[\theta^A(x_{\mu})]$, because of the derivative terms: indeed $\partial_{\mu}\phi'=\partial_{\mu}(U\phi)\ne U\partial_{\mu}\phi$. Gauge invariance is recovered if the ordinary
derivative is replaced by the covariant derivative:
\begin{equation} D_{\mu} = \partial_{\mu} + ig{\bf V}_{\mu}~,
\label{11}
\end{equation} where $V^A_{\mu}$ are a set of $D$ gauge vector fields (in one-to-one correspondence with the group generators)
with the transformation law
\begin{equation} {\bf V}'_{\mu} = U{\bf V}_{\mu}U^{-1} - (1/ig)(\partial_{\mu}U)U^{-1}~.
\label{12}
\end{equation} For constant $\theta^A$, {\bf V} reduces to a tensor of the adjoint (or regular) representation of the
group:
\begin{equation} {\bf V}'_{\mu} = U{\bf V}_{\mu}U^{-1} \simeq {\bf V}_{\mu} + ig[\mbox{\boldmath $\theta$}, {\bf V}_{\mu}]~\dots,
\label{13}
\end{equation} which implies that
\begin{equation} V'^C_{\mu} = V^C_{\mu} - gC_{ABC}\theta^AV^B_{\mu}~\dots,
\label{14}
\end{equation} where repeated indices are summed up.

As a consequence of Eqs. (\ref{11}) and (\ref{12}), $D_{\mu}\phi$  has the same transformation properties as $\phi$:
\begin{equation} (D_{\mu}\phi)' = U(D_{\mu}\phi)~.
\label{15}
\end{equation}
In fact
\begin{eqnarray} (D_{\mu}\phi)' &=& (\partial_{\mu} + ig{\bf V'}_{\mu})\phi' =
(\partial_{\mu}U) \phi+U\partial_{\mu}\phi+igU{\bf V}_{\mu} \phi-(\partial_{\mu}U) \phi =U(D_{\mu}\phi)~.
\label{15a}
\end{eqnarray}

Thus ${\cal L}[\phi,D_{\mu}\phi]$ is indeed invariant under gauge transformations. But, at this stage, the gauge fields $V_{\mu}^A$ appear as external fields that do not propagate. In order to construct a
gauge-invariant kinetic energy term for the gauge fields $V_{\mu}^A$, we consider
\begin{equation} [D_{\mu},D_{\nu}] \phi =  ig\{\partial_{\mu}{\bf V}_{\nu} - \partial_{\nu}{\bf V}_{\mu} + ig[{\bf
V}_{\mu},{\bf V}_{\nu}]\}\phi \equiv ig {\bf F}_{\mu\nu}\phi~,
\label{16}
\end{equation} which is equivalent to
\begin{equation} F^A_{\mu\nu} = \partial_{\mu}V^A_{\nu} - \partial_{\nu}V^A_{\mu} - gC_{ABC}V^B_{\mu}V^C_{\nu}~.
\label{17}
\end{equation} From Eqs. (\ref{7}), (\ref{15}) and (\ref{16}) it follows that the transformation properties of
$F^A_{\mu\nu}$ are those of a tensor of the adjoint representation
\begin{equation} {\bf F}'_{\mu\nu} = U{\bf F}_{\mu\nu}U^{-1}~.
\label{18}
\end{equation} The complete Yang--Mills lagrangian, which is invariant under gauge transformations, can be written in
the form
\begin{equation} {\cal L}_{\rm YM} = - \frac{1}{2}Tr{\bf F}_{\mu\nu}{\bf F}^{\mu\nu}+ {\cal L} [\phi,D_{\mu}\phi]~ = - \frac{1}{4} \sum_A F^A_{\mu\nu}F^{A\mu\nu} + {\cal L} [\phi,D_{\mu}\phi]~.
\label{19}
\end{equation}
Note that the kinetic energy term is an operator of dimension 4. Thus if $\cal L$ is renormalizable, also ${\cal L}_{\rm YM}$ is renormalizable. If we give up renormalizability then more gauge invariant higher dimension terms could be added. It is already clear at this stage that no mass term for gauge bosons of the form $m^2V_{\mu}V^{\mu}$ is allowed by gauge invariance.

\subsection{Application to QED and QCD}
\label{sec:4}

For an abelian theory, as for example QED, the gauge transformation reduces to
$U[\theta(x)] = {\rm exp} [ieQ\theta(x)]$, where $Q$ is the charge generator (for more commuting generators one simply has a product of similar factors). The associated gauge field (the photon),
according to Eq. (\ref{12}), transforms as
\begin{equation} V'_{\mu} = V_{\mu} - \partial_{\mu}\theta(x)~.
\label{20}
\end{equation} and the familiar gauge transformation by addition of a 4--gradient of a scalar function is recovered.
The QED lagrangian density is given
by:
\beq
{\cal L}~=~-\frac{1}{4}F^{\mu\nu}F_{\mu\nu}~+~\sum_{\psi}\bar {\psi}(iD\llap{$/$}-m_{\psi})\psi~~.\label{LQED}\\
\eeq  
Here $D\llap{$/$}=D_{\mu}\gamma^{\mu}$, where
$\gamma^{\mu}$ are the Dirac matrices and the covariant derivative is given in terms of the photon field $A_{\mu}$ and the charge operator Q by: 
\beq
D_{\mu}=\partial_{\mu}+ieA_{\mu}Q\label{derQED}\\
\eeq
and 
\beq
F_{\mu\nu}~=~\partial_{\mu} A_{\nu}-\partial_{\nu} A_{\mu}\label{FQED}\\
\eeq
Note that in QED one usually takes the $e^-$ to be the particle, so that $Q=-1$ and the covariant derivative is $D_\mu=\partial_\mu-ieA_\mu$ when acting on the electron field. In the abelian case, the $F_{\mu\nu}$ tensor is linear in the gauge field $V_{\mu}$ so that in the absence of
matter fields the theory is free. On the other hand, in the non abelian case the $F^A_{\mu\nu}$ tensor contains both
linear and quadratic terms in $V^A_{\mu}$, so that the theory is non-trivial even in the absence of matter fields.

According to the formalism of the previous section, the
statement that QCD is a renormalizable gauge theory based on the group $SU(3)$ with colour triplet quark matter fields fixes the QCD
lagrangian density to be 
\beq
{\cal L}~=~-\frac{1}{4}\sum_{A=1}^8F^{A\mu\nu}F^A_{\mu\nu}~+~\sum_{j=1}^{n_f}\bar q_j(iD\llap{$/$}-m_j)q_j\label{LagQCD}\\
\eeq
Here $q_j$ are the quark fields (of $n_f$ different flavours) with mass $m_j$ and $D_{\mu}$ is the covariant derivative: 
\beq
D_{\mu}=\partial_{\mu}+ie_s{\bf g_{\mu}};\label{der}\\
\eeq $e_s$ is the gauge coupling and later we will mostly use, in analogy with QED
\beq
\alpha_s=\frac{e_s^2}{4\pi}.\label{alfa}\\
\eeq
Also, ${\bf g_{\mu}}= \sum_A~t^Ag_{\mu}^A~$
where
$g_{\mu}^A$, $A=1,8$, are the gluon fields and
$t^A$ are the
$SU(3)$ group generators in the triplet representation of quarks (i.e. $t_A$ are 3x3 matrices acting on $q$); the
generators obey the commutation relations  $[t^A,t^B]=iC_{ABC}t^C$ where $C_{ABC}$ are the complete antisymmetric
structure constants of $SU(3)$ (the normalization of $C_{ABC}$ and of $e_s$ is specified by that of the generators $t^A$:
$Tr[t^At^B]=\delta^{AB}/2$, see Eq.(\ref{8})). Finally we have:
\beq
F^A_{\mu\nu}~=~\partial_{\mu} g^A_{\nu}-\partial_{\nu} g^A_{\mu}~-~e_sC_{ABC}g^B_{\mu}g^C_{\nu}\label{F}
\eeq

Chapter \ref{Chap:2} is devoted to a detailed description of QCD as the theory of strong interactions. The physical vertices in QCD include the gluon-quark-antiquark vertex,
analogous to the QED photon-fermion-antifermion coupling, but also the 3-gluon and 4-gluon vertices, of order $e_s$ and
$e_s^2$ respectively, which have no analogue in an abelian theory like QED. In QED the photon is coupled to all electrically charged particles but itself is neutral. In QCD the gluons are coloured
hence self-coupled. This is reflected in the fact that in QED $F_{\mu\nu}$ is linear in the gauge field, so that the term
$F_{\mu\nu}^2$ in the lagrangian is a pure kinetic term, while in QCD $F^A_{\mu\nu}$ is quadratic in the gauge field so
that in
$F^{A2}_{\mu\nu}$ we find cubic and quartic vertices beyond the kinetic term. Also instructive is to consider the case of
scalar QED: 
\beq
{\cal L}~=~-\frac{1}{4}F^{\mu\nu}F_{\mu\nu}~+~(D_{\mu}\phi)^\dagger (D^{\mu}\phi)-m^2(\phi^\dagger \phi)\label{LSQED}\\
\eeq
For $Q=1$ we have:
\beq
(D_{\mu}\phi)^\dagger (D^{\mu}\phi)~=~(\partial_{\mu}\phi)^\dagger (\partial^{\mu}\phi)~+~ieA_{\mu}[(\partial^{\mu}\phi)^\dagger \phi
~-~\phi^\dagger (\partial^{\mu}\phi)]~+~e^2A_{\mu}A^{\mu}\phi^\dagger \phi
\label{seagull}\\
\eeq
We see that for a charged boson in QED, given that the kinetic term for bosons is quadratic in the derivative, there is a
gauge-gauge-scalar-scalar vertex of order
$e^2$. We understand that in QCD the 3-gluon vertex is there because the gluon is coloured and the 4-gluon vertex because the gluon is
a boson.

\subsection{Chirality}
\label{sec:5}

We recall here the notion of chirality and related issues which is crucial for the formulation of the EW Theory. The fermion fields can be described through their Right Handed (RH) (chirality +1)  and Left Handed (LH) (chirality -1) components:
\begin{equation}
\psi_{L,R} = [(1 \mp \gamma_5)/2]\psi, \quad
\bar \psi_{L,R} = \bar \psi[(1 \pm \gamma_5)/2]~,
\label{24}
\end{equation} with $\gamma_5$ and other Dirac matrices defined as in the book by Bjorken--Drell \cite{qedtb}. In particular, $\gamma^2_5
= 1, \gamma_5^{\dag} = \gamma_5$. Note that, as follows from Eq. (\ref{24}), one has:
\begin{eqnarray*}
\bar\psi_L = 
\psi^{\dag}_L\gamma_0 = \psi^{\dag}[(1-\gamma_5)/2]\gamma_0 =
\bar\psi \gamma_0[(1-\gamma_5)/2]\gamma_0 = \bar \psi[(1 + \gamma_5)/2]~.
\end{eqnarray*}
The matrices $P_{\pm} = (1 \pm \gamma_5)/2$ are projectors. They satisfy the relations $P_{\pm}P_{\pm} = P_{\pm}$, 
$P_{\pm}P_{\mp} = 0$, $P_+ + P_- = 1$. The $P$ projectors project on fermions of definite chirality. For massless particles, chirality coincides with helicity. For massive particles, a chirality +1 state only coincides with a +1 helicity state up to terms suppressed by powers of $m/E$.

The sixteen linearly independent Dirac matrices ($\Gamma$) can be divided into
$\gamma_5$-even  ($\Gamma_E$) and $\gamma_5$-odd ($\Gamma_O$) according to whether they commute or anticommute with $\gamma_5$. For the
$\gamma_5$-even, we have
\begin{equation}
\bar \psi \Gamma_E \psi = \bar \psi_L \Gamma_E \psi_R + \bar \psi_R \Gamma_E \psi_L
\quad\quad (\Gamma_E \equiv 1, i\gamma_5, \sigma_{\mu\nu})~,
\label{25}
\end{equation} whilst for the $\gamma_5$-odd,
\begin{equation}
\bar \psi\Gamma_O \psi = \bar \psi_L\Gamma_O\psi_L + \bar \psi_R\Gamma_O\psi_R
\quad\quad (\Gamma_O \equiv \gamma_{\mu}, \gamma_{\mu}\gamma_5)~.
\label{26}
\end{equation}

We see that in a gauge lagrangian fermion kinetic terms and  interactions of gauge bosons with vector and axial vector fermion currents all conserve chirality while fermion mass terms flip chirality. For example, in QED if an electron emits a photon, the electron chirality is unchanged. In the ultrarelativistic limit, when the electron mass can be neglected, chirality and helicity are approximately the same and we can state that the helicity of the electron is unchanged by the photon emission. In a massless gauge theory the LH and the RH fermion components are uncoupled and can be transformed separately. If in a gauge theory the LH and RH components transform as different representations of the gauge group one speaks of a chiral gauge theory, while if they have the same gauge transformations one has a vector gauge theory. Thus, QED and QCD are vector gauge theories because, for each given fermion, $\psi_L$ and $\psi_R$ have the same electric charge and the same colour.
Instead, the standard EW theory is a chiral theory, in the sense that $\psi_L$ and $\psi_R$ behave
differently under the gauge group (so that parity and charge conjugation non conservation are made possible in principle). Thus, mass terms for fermions (of the form
$\bar\psi_L\psi_R$ + h.c.) are forbidden in the EW gauge-symmetric limit. In particular, in the
Minimal Standard Model (MSM: i.e. the model that only includes all observed particles plus a single Higgs doublet), all $\psi_L$ are $SU(2)$ doublets while all $\psi_R$ are singlets.

\subsection{Quantization of a Gauge Theory}
\label{sec:6}
The lagrangian density ${\cal L}_{YM}$ in Eq.(\ref{19}) fully describes the theory at the classical level. The formulation of the theory at the quantum level requires that procedures of quantization, of regularization and, finally, of renormalization are also specified. To start with, the formulation of Feynman rules is not straightforward. A first problem, common to all gauge theories, including the abelian case of QED, can be realized by observing that the free equations of motion for $V_{\mu}^A$, as obtained from Eqs.(\ref{17},\ref{19}), are given by
\beq
[\partial^2 g_{\mu\nu}-\partial_\mu \partial_\nu]V^{A\nu}=0  \label{Q1}\\
\eeq
Normally the propagator of the gauge field should be determined by the inverse of the operator $[\partial^2 g_{\mu\nu}-\partial_\mu \partial_\nu]$ which, however, has no inverse, being a projector over the transverse gauge vector states. This difficulty is removed by fixing a particular gauge. If one chooses a covariant gauge condition $\partial^\mu V_{\mu}^A=0$ then a gauge fixing term of the form
\beq
\Delta {\cal L}_{GF}=-\frac{1}{2\lambda}\sum_A |\partial^\mu V_{\mu}^A|^2 \label{Q2}\\
\eeq
has to be added to the lagrangian ($1/\lambda$ acts as a lagrangian multiplier). The free equations of motion are then modified as follows:
\beq
[\partial^2 g_{\mu\nu}-(1-1/\lambda)\partial_\mu \partial_\nu]V^{A\nu}=0.  \label{Q3}\\
\eeq
This operator now has an inverse whose Fourier transform is given by:
\beq
D_{\mu\nu}^{AB}(q)= \frac{i}{q^2+i\epsilon}~[- g_{\mu\nu}+(1-\lambda)\frac{q_\mu q_\nu}{q^2+i\epsilon}]~\delta^{AB}  \label{Q4}\\
\eeq
which is the propagator in this class of gauges. The parameter $\lambda$ can take any value and it disappears from the final expression of any gauge invariant, physical quantity. Commonly used particular cases are $\lambda=1$ (Feynman gauge) and $\lambda=0$ (Landau gauge).

While in an abelian theory the gauge fixing term is all that is needed for a correct quantization, in a non abelian theory the formulation of complete Feynman rules involves a further subtlety. This is formally taken into account by introducing a set of D fictitious ghost fields that must be included as internal lines in closed loops (Faddeev-Popov ghosts \cite{FP}). Given that gauge fields connected by a gauge transformation describe the same physics, clearly there are less physical degrees of freedom than gauge field components. Ghosts appear, in the form of a transformation Jacobian in the functional integral, in the process of elimination of the redundant variables associated with fields on the same gauge orbit \cite{AL}. By performing some path integral acrobatics the correct ghost contributions can be translated into an additional term in the lagrangian density. For each choice of the gauge fixing term the ghost langrangian is obtained by considering the effect of an infinitesimal gauge transformation $V_{\mu}^{'C}=V_{\mu}^C-gC_{ABC}\theta^AV_{\mu}^B-\partial_\mu \theta^C$ on the gauge fixing condition. For $\partial^\mu V_{\mu}^C=0$ one obtains: 
\beq
\partial^\mu V_{\mu}^{'C} =\partial^\mu V_{\mu}^C-gC_{ABC}\partial^\mu (\theta^A V_{\mu}^B)-\partial^2 \theta^C~=~ -[\partial^2 \delta_{AC} +g C_{ABC} V_\mu^B\partial^\mu]\theta^A~\label{Q5}\\
\eeq
where the gauge condition $\partial^\mu V_{\mu}^C=0$ has been taken into account in the last step. The ghost lagrangian is then given by:
\beq
\Delta {\cal L}_{Ghost}= \bar \eta^C[\partial^2 \delta_{AC}+gC_{ABC}V_\mu^B\partial^\mu ]\eta^A~\label{Q6}\\
\eeq
where $\eta^A$ is the ghost field (one for each index $A$) which has to be treated as a scalar field except that a factor $(-1)$ for each closed loop has to be included as for fermion fields.

Starting from non covariant gauges one can construct ghost-free gauges. An example, also important in other respects, is provided by the set of "axial" gauges: $n^\mu V_\mu^A=0$ where $n_\mu$ is a fixed reference 4-vector (actually for  $n_\mu$ spacelike one has an axial gauge proper, for $n^2=0$ one speaks of a light-like gauge and for $n_\mu$ timelike one has a Coulomb or temporal gauge). The gauge fixing term is of the form:
\beq
\Delta {\cal L}_{GF}=-\frac{1}{2\lambda}\sum_A |n^\mu V_{\mu}^A|^2 \label{Q7}\\
\eeq
With a procedure that can be found in QED textbooks \cite{qedtb} the corresponding propagator, in Fourier space, is found to be:
\beq
D_{\mu\nu}^{AB}(q)= \frac{i}{q^2+i\epsilon}~[- g_{\mu\nu}+\frac{n_\mu q_+n_\nu q_\mu}{(nq)}-\frac{n^2 q_\mu q_\nu}{(nq)^2}]~\delta^{AB}  \label{Q8}\\
\eeq
In this case there are no ghost interactions because $n^\mu V_\mu^{'A}$, obtained by a gauge transformation from $n^\mu V_\mu^A$, contains no gauge fields, once the the gauge condition  $n^\mu V_\mu^A=0$ has been taken into account. Thus the ghosts are decoupled and can be ignored.

The introduction of a suitable regularization method that preserves gauge invariance is essential for the definition and the calculation of loop diagrams and for the renormalization programme of the theory. The method that is by now currently adopted is dimensional regularization \cite{dr} which consists in the formulation of the theory in $n$ dimensions. All loop integrals have an analytic expression that is actually valid also for non integer values of $n$. Writing the results for $n=4-\epsilon$ the loops are ultraviolet finite for $\epsilon>0$ and the divergences reappear in the form of poles at $\epsilon=0$.

\subsection{Spontaneous Symmetry Breaking in Gauge Theories}
\label{sec:7}
The gauge symmetry of the SM was difficult to discover because it is well hidden in nature. The only
observed gauge boson that is massless is the photon. The gluons are presumed massless but cannot be directly observed because of
confinement, and the $W$ and $Z$ weak bosons carry a heavy mass. Indeed a major difficulty in unifying the weak and
electromagnetic interactions was the fact that e.m. interactions have infinite range $(m_{\gamma} = 0)$, whilst the weak
forces have a very short range, owing to
$m_{W,Z} \not= 0$. The solution of this problem is in the concept of spontaneous symmetry breaking, which was borrowed from condensed matter physics. 

\begin{figure}
\noindent
\includegraphics*[width=7.2cm]{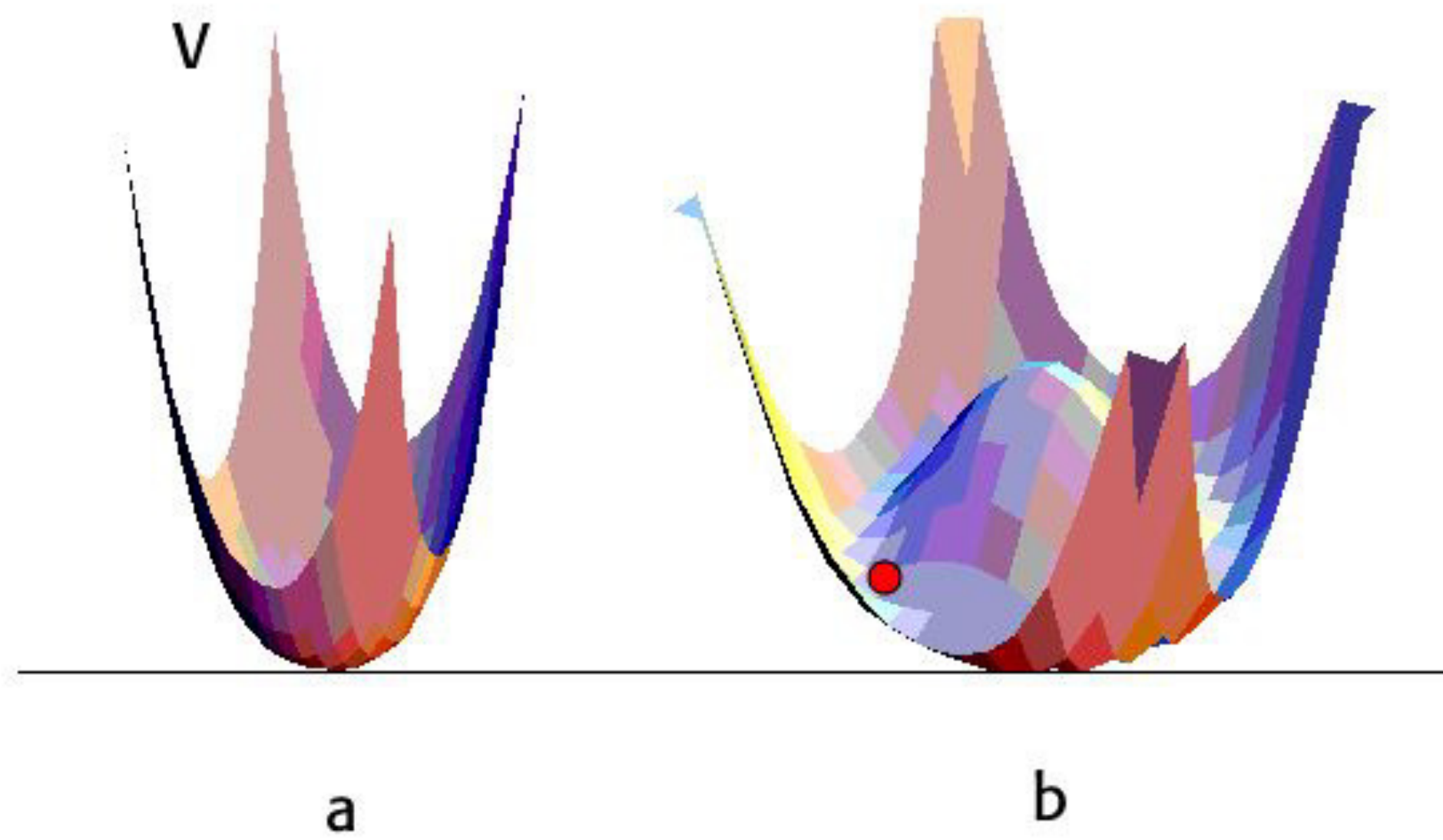} 
\hfill
\begin{minipage}[b]{7.0cm}
\caption{The potential $V=1/2~\mu^2{\bf M}^2 + 1/4~\lambda({\bf M}^2)^2$ for positive (a) or negative $\mu^2$ (b) (for simplicity, $\bf M$ is a 2-dimensional vector). The small sphere indicates a possible choice for the direction of $\bf M$.}
\label{Hpot}
\vspace{12pt}
\end{minipage}
\end{figure}

Consider a ferromagnet at zero magnetic field in the Landau--Ginzburg approximation. The free energy in terms of the
temperature $T$ and the magnetization {\bf M} can be written as
\begin{equation} F({\bf M}, T) \simeq F_0(T) + 1/2~\mu^2(T){\bf M}^2 + 1/4~\lambda(T)({\bf M}^2)^2 + ...~.
\label{64}
\end{equation} This is an expansion which is valid at small magnetization.  The neglect of terms of higher order in
$\vec M^2$ is the analogue in this context of the renormalizability criterion. Also, $\lambda(T) > 0$ is assumed for
stability; $F$ is invariant under rotations, i.e. all directions of {\bf M} in space are equivalent. The minimum
condition for $F$ reads
\begin{equation}
\partial F/\partial M_i = 0, \quad [\mu^2(T) + \lambda(T){\bf M}^2]{\bf M} = 0~.
\label{65}
\end{equation} There are two cases, shown in Fig. \ref {Hpot}. If $\mu^2 \gtrsim 0$, 
then the only solution is ${\bf M} = 0$, there is no magnetization,
and the rotation symmetry is respected. In this case the lowest energy state (in a quantum theory the vacuum) is unique and invariant under rotations. If $\mu^2 < 0$, then another solution appears, which is
\begin{equation} |{\bf M}_0|^2 = -\mu^2/\lambda~.
\label{66}
\end{equation} In this case there is a continuous orbit of lowest energy states, all with the same value of $|\bf M|$ but different orientations. A particular direction chosen by the vector ${\bf M}_0$ leads to a breaking of the rotation symmetry. 

For a piece of iron we can imagine to bring it to high temperature and to let it melt in an external magnetic field $\bf B$. The presence of $\bf B$ is an explicit breaking of the rotational symmetry and it induces a non zero magnetization $\bf M$ along its direction. Now we lower the temperature while keeping $\bf B$ fixed. Both $\lambda$ and $\mu^2$ depend on the temperature. With lowering $T$, $\mu^2$ goes from positive to negative values. The critical
temperature $T_{\rm crit}$ (Curie temperature) is where $\mu^2(T)$ changes sign:
$\mu^2(T_{\rm crit}) = 0$. For pure iron $T_{\rm crit}$ is below the melting temperature. So at $T=T_{\rm crit}$ iron is a solid. Below $T_{\rm crit}$ we remove the magnetic field. In a solid the mobility of the magnetic domains is limited and a non vanishing $M_0$ remains. The form of the free energy is again rotationally invariant as in Eq.(\ref{64}). But now the system allows a minimum energy state with non vanishing $\bf M$ in the direction where $\bf B$ was. As a consequence the symmetry is broken by this choice of one particular vacuum state out of a continuum of them.

We now prove the Goldstone theorem \cite{gold}. It states that when spontaneous symmetry
breaking takes place, there is always a zero-mass mode in the spectrum. In a classical context this can be proven as
follows. Consider a lagrangian
\begin{equation} {\cal L} =\frac{1}{2} |\partial_{\mu}\phi|^2 - V(\phi).
\label{68}
\end{equation}
The potential $V(\phi)$ can be kept generic at this stage but, in the following, we will be mostly interested in a renormalizable potential of the form (with no more than quartic terms):
\begin{equation}
V(\phi)=-\frac{1}{2} \mu^2~ \phi^2+\frac{1}{4}\lambda ~\phi^4.
\label{68a}
\end{equation}
Here by $\phi$ we mean a column vector with real components $\phi_i$ (1=1,2...N) (complex fields can always be decomposed into a pair of real fields), so that, for example, $\phi^2=\sum_i\phi_i^2$. This particular potential is symmetric under a NxN orthogonal matrix rotation $\phi'=O\phi$, where $O$ is a SO(N) transformation.
For simplicity, we have omitted odd powers of $\phi$, which means that we assumed an extra discrete  symmetry under $\phi \leftrightarrow  -\phi$. Note that, for positive $\mu^2$, the mass term in the potential has the "wrong" sign: according to the previous discussion this is the condition for the existence of a non unique lowest energy state.
More in general, we only assume here that the potential is symmetric under the infinitesimal transformations
\begin{equation}
\phi \rightarrow  \phi' = \phi + \delta \phi, \quad
\delta \phi_i = i \delta\theta^A  t_{ij}^A\phi_j~.
\label{69}
\end{equation} where $\delta \theta^A$ are infinitesimal parameters and $t_{ij}^A$ are the matrices that represent the symmetry group on the representation of the fields $\phi_i$ (a sum over A is understood). The minimum condition on $V$ that identifies the equilibrium position (or the vacuum state in quantum field theory
language) is
\begin{equation} (\partial V/\partial \phi_i)(\phi_i = \phi^0_i) = 0~.
\label{70}
\end{equation} The symmetry of $V$ implies that
\begin{equation}
\delta V = (\partial V/\partial \phi_i)\delta \phi_i = i \delta \theta^A(\partial V/\partial \phi_i)t_{ij}^A\phi_j = 0~.
\label{71}
\end{equation} By taking a second derivative at the minimum $\phi_i = \phi^0_i$ of both sides of the previous equation, we obtain that, for each A: 
\begin{equation}
\frac{\partial^2V}{\partial \phi_k\partial \phi_i} (\phi_i =
\phi^0_i)t_{ij}^A\phi^0_j + \frac{\partial V}{\partial \phi_i} (\phi_i =
\phi^0_i)t_{ik}^A = 0~.
\label{72}
\end{equation} The second term vanishes owing to the minimum condition, Eq. (\ref{70}). We then find
\begin{equation}
\frac{\partial^2V}{\partial \phi_k\partial \phi_i} (\phi_i = \phi^0_i)t_{ij}^A\phi^0_j = 0~.
\label{73}
\end{equation} The second derivatives $M^2_{ki} = (\partial^2V/\partial \phi_k \partial
\phi_i)(\phi_i = \phi^0_i)$ define the squared mass matrix. Thus the above equation in matrix notation can be written as
\begin{equation} M^2 t^A \phi^0 = 0~.
\label{74}
\end{equation} In the case of no spontaneous symmetry breaking the ground state is unique, all symmetry transformations leave it invariant, so that, for all $A$, $t^A\phi^0=0$. On the contrary, if, for some values of $A$, the vectors $(t^A \phi^0)$ are non-vanishing, i.e. there is some generator that shifts
the ground state into some other state with the same energy (hence the vacuum is not unique), then each $t^A \phi^0 \ne 0$ is an eigenstate of the squared mass
matrix with zero eigenvalue. Therefore, a massless mode is associated with each broken generator. The charges of the massless modes (their quantum numbers in quantum language) differ from those of the vacuum (usually taken as all zero) by the values of the $t^A$ charges: one says that the massless modes have the same quantum numbers of the broken generators, i.e. those that do not annihilate the vacuum. 

The previous proof of the Goldstone theorem has been given in the classical case. In the quantum case the classical potential corresponds to the tree level approximation of the quantum potential. Higher order diagrams with loops introduce quantum corrections. The functional integral formulation of quantum field theory \cite{AL}, \cite{ftbooks} is the most appropriate framework to define and compute, in a loop expansion, the quantum potential which specifies, exactly as described above, the vacuum properties of the quantum theory. If the theory is weakly coupled, e.g. if $\lambda$ is small, the tree level expression for the potential is not too far from the truth, and the classical situation is a good approximation. We shall see that this is the situation that occurs in the electroweak theory with a moderately light Higgs (see Chapter \ref{Chap:3}, Sec.  \ref{sec:24}).

\begin{figure}
\noindent
\includegraphics{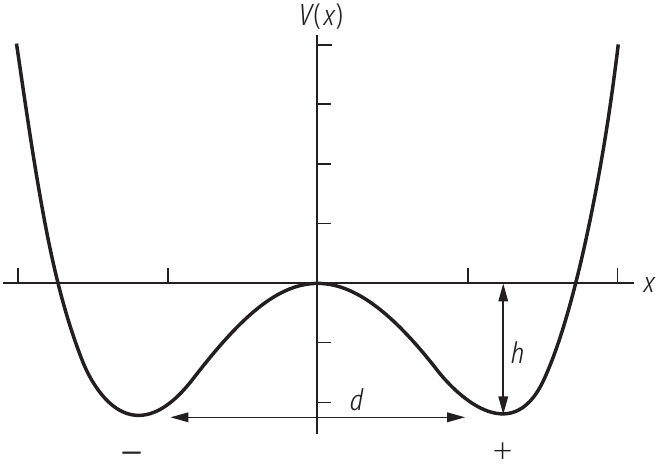} 
\hfill
\begin{minipage}[b]{7.0cm}
\caption{A Schroedinger potential $V(x)$ analogous to the Higgs potential.}
\label{Schro}
\vspace{12pt}       
\end{minipage}
\end{figure}

We note that for a quantum system with a finite number of degrees of freedom, for example one described by the Schroedinger equation, there are no degenerate vacua: the vacuum is always unique.  For example, in the one dimensional Schroedinger problem with a potential:
\beq
V(x)=- \mu^2/2~ x^2+\lambda ~x^4/4~,
\label{nr1}
\eeq
there are two degenerate minima at $x=\pm x_0= (\mu^2/\lambda)^{1/2}$ which we denote by $|+\rangle$ and $|-\rangle$. But the potential is not diagonal in this basis: the off diagonal matrix elements: 
 \beq
 \langle+|V|-\rangle=\langle-|V|+\rangle \sim \exp{(-khd)}= \delta
 \label{nr2}
\eeq
are different from zero due to the non vanishing amplitude for a tunnel effect between the two vacua given in Eq.(\ref{nr2}), proportional to the exponential of minus the product of the distance d between the vacua and the height h of the barrier with k a constant (see Fig. \ref{Schro}). After diagonalization the eigenvectors are $(|+\rangle+|-\rangle)/\sqrt{2}$ and $(|+\rangle-|-\rangle)/\sqrt{2}$, with different energies (the difference being proportional to $\delta$). Suppose now that you have a sum of n equal terms in the potential, $V=\sum_i V(x_i)$. Then the transition amplitude would be proportional to $\delta^n$ and would vanish for infinite n: the probability that all degrees of freedom together jump over the barrier vanishes. In this example there is a discrete number of minimum points. The case of a continuum of minima is obtained, always in the Schroedinger context, if we take
\beq
V=1/2~\mu^2{\bf r}^2 + 1/4~\lambda({\bf r}^2)^2~,
\label{nr3}
\eeq 
with ${\bf r}=(x,y,z)$. Also in this case the ground state is unique: it is given by a state with total orbital angular momentum zero, an s-wave state, whose wave function only depends on $|\bf r|$, independent of all angles. This is a superposition of all directions with the same weight, analogous to what happened in the discrete case.  But again, if we replace a single vector $\bf r$, with a vector field $\bf M(x)$, that is a different vector at each point in space, the amplitude to go from a minimum state in one direction to another in a different direction goes to zero in the limit of infinite volume. In simple words, the vectors at all points in space have a vanishing small amplitude to make a common rotation, all  together at the same time. In the infinite volume limit all vacua along each direction have the same energy and spontaneous symmetry breaking can occur.

A massless Goldstone boson correspond to a long range force. Unless the massless particles are confined, as for the gluons in QCD, these long range forces would be easily detectable. Thus, in the construction of the EW theory we cannot accept massless physical scalar particles. Fortunately, when spontaneous symmetry breaking takes place in a gauge theory, the massless Goldstone modes exist, but they are 
unphysical and disappear from the spectrum. Each of them becomes, in fact, the third helicity state of a gauge boson that takes
mass. This is the Higgs mechanism \cite{ebh,ghk,hig,kib} (it should be called Englert-Brout-Higgs mechanism, because of the simultaneous paper by Englert and Brout). Consider, for example, the simplest Higgs model described by the lagrangian \cite{hig,kib}
\begin{equation} {\cal L} = -\frac{1}{4}~F^2_{\mu\nu} + |(\partial_{\mu} + ieA_{\mu}Q)\phi|^2 +
\mu^2 \phi^*\phi - \frac{\lambda}{2}(\phi^*\phi)^2~.
\label{75}
\end{equation} Note the `wrong' sign in front of the mass term for the scalar field $\phi$, which is necessary for the
spontaneous symmetry breaking to take place. The above lagrangian is invariant under the $U(1)$ gauge symmetry
\begin{equation} A_{\mu} \rightarrow A'_{\mu} = A_{\mu} - \partial_{\mu}\theta(x), \quad
\phi \rightarrow \phi' = {\rm exp}[ieQ\theta(x)]~\phi .
\label{76}
\end{equation} For the U(1) charge Q we take $Q\phi=-\phi$, like in QED, where the particle is $e^-$. Let $\phi^0 = v \not= 0$, with $v$ real, be the ground state that minimizes the potential and induces
the spontaneous symmetry breaking. In our case $v$ is given  by $v^2=\mu^2/\lambda$. 
Making use of gauge invariance, we can do the change of variables
\begin{eqnarray} &&\phi(x) \rightarrow[v+\frac{h(x)}{\sqrt 2}]~{\rm exp}[-i \frac{\zeta(x)}{v\sqrt 2}]~, \nonumber \\ &&A_{\mu}(x)
\rightarrow A_{\mu} - \partial_{\mu} \frac{\zeta(x)}{ev\sqrt 2}.
\label{77}
\end{eqnarray} Then the position of the minimum at $\phi^0 = v$ corresponds to $h = 0$, and the lagrangian becomes
\begin{equation} {\cal L} = -\frac{1}{4}F^2_{\mu\nu} +e^2v^2A^2_{\mu} + \frac{1}{2} e^2h^2A^2_{\mu} +
\sqrt 2e^2h vA^2_{\mu} + {\cal L}(h)~.
\label{78}
\end{equation} The field $\zeta(x)$ is the would-be Goldstone boson, as can be seen by considering only the $\phi$ terms in the lagrangian, i.e. setting $A_\mu=0$ in Eq.(\ref{75}). In fact in this limit the kinetic term $\partial_\mu \zeta \partial^\mu \zeta$ remains but with no $\zeta^2$ mass term. Instead, in the gauge case of Eq.(\ref{75}), after changing variables in the lagrangian, the field $\zeta(x)$ completely disappears (not even the kinetic term remains), whilst the mass
term $e^2v^2A^2_{\mu}$ for $A_{\mu}$ is now present: the gauge boson mass is $M=\sqrt{2}ev$. The field $h$ describes the massive Higgs particle. Leaving a constant term aside, the last term in Eq.(\ref{78}) is given by:
\begin{equation}  {\cal L}(h)=\frac{1}{2} \partial_\mu h \partial^\mu h -h^2 \mu^2+....
\label{78a}
\end{equation} 
where the dots stand for cubic and quartic terms in $h$. We see that the $h$ mass term has the "right" sign, due to the combination of the quadratic tems in $h$ that, after the shift, arise from the quadratic and quartic terms in $\phi$. The $h$ mass is given by $m^2_h=2\mu^2$.

The Higgs mechanism is realized in well-known physical situations. It was actually discovered in condensed matter physics by Anderson \cite{and}. For a superconductor in the Landau--Ginzburg
approximation the free energy can be written as
\begin{equation} F = F_0 + \frac{1}{2}{\bf B}^2 + |({\bf \nabla} - 2ie{\bf A})\phi|^2/4m -
\alpha|\phi|^2 + \beta|\phi|^4~.
\label{79}
\end{equation} 
Here {\bf B} is the magnetic field, $|\phi|^2$ is the Cooper pair $(e^-e^-)$ density, 2$e$ and 2$m$ are the charge and
mass of the Cooper pair. The 'wrong' sign of $\alpha$ leads to $\phi \not= 0$ at the minimum. This is precisely the
non-relativistic analogue of the Higgs model of the previous example. The Higgs mechanism implies the absence of
propagation of massless phonons (states with dispersion relation ~$\omega = kv$ with constant $v$). Also the mass term
for {\bf A} is manifested by the exponential decrease of {\bf B} inside the superconductor (Meissner effect). But in condensed matter examples the Higgs field is not elementary but, rather, a condensate of elementary fields (like for the Cooper pairs).

\subsection{Quantization of Spontaneously Broken Gauge Theories: $R_\xi$ Gauges}
\label{sec:8}

We have discussed in Sect. (\ref{sec:6}) the problems arising in the quantization of a gauge theory and in the formulation of the correct Feynman rules (gauge fixing terms, ghosts etc). Here we give a concise account of the corresponding results for spontaneously broken gauge theories. In particular we describe the $R_\xi$ gauge formalism  \cite{AL}, \cite{ftbooks},\cite{fls}: in this formalism the interplay of transverse and longitudinal gauge boson degrees of freedom is made explicit and their combination leads to the cancellation from physical quantities of the gauge parameter $\xi$. We work out in detail an abelian example that later will be easy to generalize to the non abelian case.

We restart from the abelian model of Eq.(\ref{75}) (with $Q=-1$). In the treatment presented there the would-be Goldstone boson $\zeta(x)$ was completely eliminated from the lagrangian by a non linear field transformation formally identical to a gauge transformation corresponding to the $U(1)$ symmetry of the lagrangian. In that description, in the new variables we eventually obtain a theory with only physical fields: a massive gauge boson $A_\mu$ with mass $M=\sqrt{2} e v$ and a Higgs particle $h$ with mass $m_h=\sqrt{2}\mu$. This is called a "unitary" gauge, because only physical fields appear. But if we work out the propagator of the massive gauge boson :
\beq
iD_{\mu\nu}(k)=-i\frac{g_{\mu\nu}-k_\mu k_\nu/M^2}{k^2-M^2+i\epsilon}~,
\label{xi1}
\eeq
we find that it has a bad ultraviolet behaviour due to the second term in the numerator. This choice does not prove to be the most convenient for a discussion of the ultraviolet behaviour of the theory. Alternatively one can go to an alternative formulation where the would-be Goldstone boson remains in the lagrangian but the complication of keeping spurious degrees of freedom is compensated by having all propagators with good ultraviolet behaviour ("renormalizable" gauges). To this end we replace the non linear transformation for $\phi$ in Eq.(\ref{77}) with its linear equivalent (after all perturbation theory deals with the small oscillations around the minimum):
\beq
\phi(x) \rightarrow[v+\frac{h(x)}{\sqrt 2}]~{\rm exp}[-i \frac{\zeta(x)}{v\sqrt 2}]~\sim~[v+\frac{h(x)}{\sqrt 2}-i \frac{\zeta(x)}{\sqrt 2}]~.
\label{xi2}
\eeq
Here we have only applied a shift by the amount $v$ and separated the real and imaginary components of the resulting field with vanishing vacuum expectation value. If we leave $A_\mu$ as it is and simply replace the linearized expression for $\phi$, we obtain the following quadratic terms (those important for propagators):
\bea
 {\cal L}_{\rm quad} &=& - \frac{1}{4} \sum_A F^A_{\mu\nu}F^{A\mu\nu} + \frac{1}{2} M^2A_\mu A^\mu+\nonumber \\ 
 &+& \frac{1}{2}(\partial_\mu \zeta)^2 + MA_\mu \partial^\mu\zeta+\frac{1}{2}(\partial_\mu h)^2 -h^2 \mu^2
 \label{xi3}
 \eea
 The mixing term between $A_\mu$ and $\partial_\mu \zeta$ does not allow to directly write diagonal mass matrices. But this mixing term can be eliminated by an appropriate modification of the covariant gauge fixing term given in Eq.(\ref{Q2}) for the unbroken theory. We now take:
\beq
\Delta {\cal L}_{GF}=-\frac{1}{2\xi}(\partial^\mu A_{\mu}-\xi M\zeta)^2~.
 \label{xi4}
\eeq
By adding $\Delta {\cal L}_{GF}$ to the quadratic terms in Eq.(\ref{xi3}) the mixing term cancels (apart from a total derivative that can be omitted) and we have:
\bea
 {\cal L}_{\rm quad} &=& - \frac{1}{4} \sum_A F^A_{\mu\nu}F^{A\mu\nu} + \frac{1}{2} M^2A_\mu A^\mu-\frac{1}{2\xi}(\partial^\mu A_{\mu})^2+\nonumber \\ 
 &+& \frac{1}{2}(\partial_\mu \zeta)^2 -\frac{\xi}{2}M^2\zeta^2+ \frac{1}{2}(\partial_\mu h)^2 -h^2 \mu^2
 \label{xi5}
 \eea
 We see that the $\zeta$ field appears with a mass $\sqrt{\xi}M$ and its propagator is:
 \beq
 iD_\zeta=\frac{i}{k^2-\xi M^2+i\epsilon}.
 \label{xi6}
 \eeq
 The propagators of the Higgs field $h$ and of gauge field $A_\mu$ are:
 \beq
 iD_h=\frac{i}{k^2-2\mu^2+i\epsilon}~,
 \label{xi7}
 \eeq\beq
iD_{\mu\nu}(k)=\frac{-i}{k^2-M^2+i\epsilon}(g_{\mu\nu}-(1-\xi)\frac{k_\mu k_\nu}{k^2-\xi M^2})~.
\label{xi8}
\eeq
As anticipated, all propagators have a good behaviour at large $k^2$. This class of gauges are called "$R_\xi$ gauges" \cite {fls}. Note that for $\xi=1$ we have a sort of generalization of the Feynman gauge with a  Goldstone boson of mass $M$ and a gauge propagator:
\beq
iD_{\mu\nu}(k)=\frac{-ig_{\mu\nu}}{k^2-M^2+i\epsilon}~.
\label{xi8a}
\eeq
Also for $\xi \rightarrow \infty$ the unitary gauge description is recovered in that the would-be Goldstone propagator vanishes and the gauge propagator reproduces that of the unitary gauge in Eq.(\ref{xi1}). All $\xi$ dependence, including the unphysical singularities of the $\zeta$ and $A_\mu$ propagators at $k^2=\xi M^2$, present in individual Feynman diagrams, must cancel in the sum of all contributions to any physical quantity.

An additional complication is that a Faddeev-Popov ghost is also present in $R_\xi$ gauges (while it is absent in an unbroken abelian gauge theory). In fact under an infinitesimal gauge transformation with parameter $\theta(x)$ we have the transformations:
\bea
A_\mu &\rightarrow& A_\mu-\partial_\mu \theta \nonumber \\
\phi &\rightarrow& (1-ie\theta)[v+\frac{h(x)}{\sqrt{2}}-i \frac{\zeta(x)}{\sqrt{2}}]~,
\label{xi9}
\eea
so that:
\beq
\delta A_\mu=-\partial_\mu \theta,~~~~\delta h=-e\zeta \theta,~~~~\delta \zeta= e \theta \sqrt{2}(v+\frac{h}{\sqrt{2}})~.
\label{xi10}
\eeq
The gauge fixing condition $\partial_\mu A^\mu-\xi M\zeta=0$ undergoes the variation:
\beq
\partial_\mu A^\mu-\xi M\zeta \rightarrow \partial_\mu A^\mu-\xi M\zeta -[\partial^2 +\xi M^2(1+\frac{h}{v\sqrt{2}})]\theta~,
\label{xi11}
\eeq
where we used $M=\sqrt{2}ev$. From this, recalling the discussion in Sect.(\ref{sec:6}), we see that the ghost is not coupled to the gauge boson (as usual for an abelian gauge theory) but has a coupling to the Higgs field $h$. The ghost lagrangian is:
\beq
\Delta {\cal L}_{Ghost}= \bar \eta[\partial^2 +\xi M^2(1+\frac{h}{v\sqrt{2}})]\eta~.
\label{xi12}
\eeq
The ghost mass is seen to be $m_{gh} =\sqrt{\xi} M$ and its propagator is:
\beq
 iD_{gh}=\frac{i}{k^2-\xi M^2+i\epsilon}.
 \label{xi13}
 \eeq
The detailed Feynman rules follow for all the basic vertices involving the gauge boson, the Higgs, the would-be Goldstone boson and the ghost and can be easily derived, with some algebra, from the total lagrangian including the gauge fixing and ghost additions.  
The generalization to the non abelian case is in principle straightforward, with some formal complications involving the projectors over the space of the would-be Goldstone bosons and over the orthogonal space of the Higgs particles. But for each gauge boson that takes mass $M_a$ we still have a corresponding would-be Goldstone boson and a ghost with mass $\sqrt{\xi}M_a$. The Feynman diagrams, both for the abelian and the non abelian case, are listed explicitly, for example,  in the Cheng and Li textbook in ref.\cite{ftbooks}.

We conclude that the renormalizability of non abelian gauge theories, also in presence of spontaneous symmetry breaking, was proven in the fundamental works of t'Hooft and Veltman \cite{HV} and discussed in detail in \cite{LZJ}.

\newpage

\section{\Large\bf{QCD: The Theory of Strong Interactions}}
\label{Chap:2}

\subsection{Introduction}
\label{sec:9}
This Chapter is devoted to a concise introduction to Quantum Chromo-Dynamics (QCD), the theory of strong
interactions \cite{SM4,SM5,SM6} (for a number of dedicated books on QCD, see \cite{Book}, see also \cite{spri2}). The main emphasis will be on ideas without too many technicalities.
As an introduction we present here a broad overview of the strong interactions (for reviews of the subject, see, for example, \cite{PR,AR}). Then some methods of non perturbative QCD will be briefly described including both analytic approaches and simulations of the theory on a discrete space-time lattice. Then we will proceed to the main focus of the Chapter which is on the principles and
the applications of perturbative QCD that will be discussed in detail.  

\begin{figure}
\noindent
\includegraphics [width=17cm]{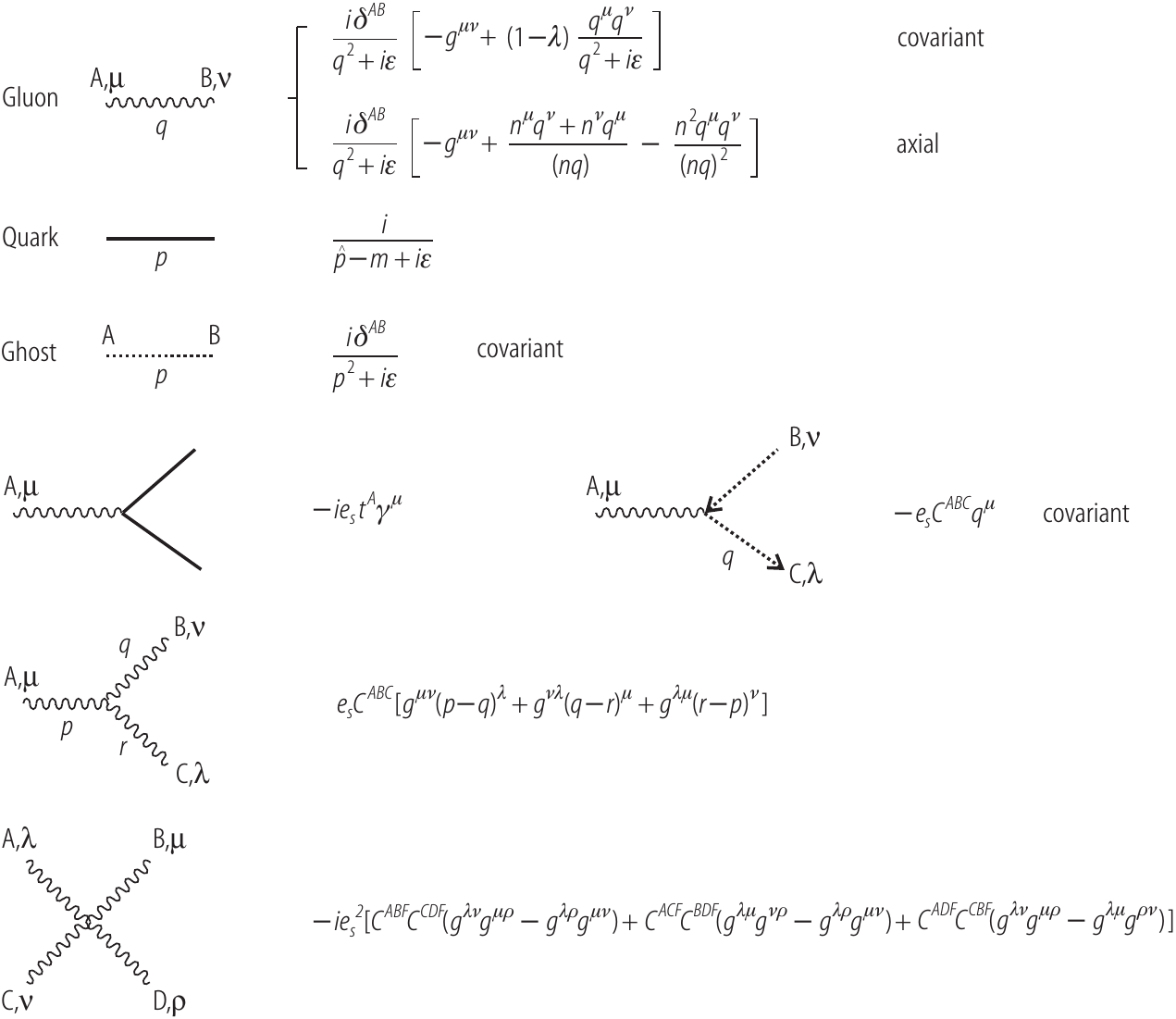} 
\caption{Feynman rules for QCD. Solid lines
 represent the quarks,  curly lines the gluons, and dotted
 lines the ghosts (see Chapter \ref{Chap:1}). The gauge parameter is denoted by $\lambda$. The 3-gluon vertex is written as if all gluon lines are outgoing. }
 \label{QCDFeyn}
\end{figure}

As discussed in Chapter \ref{Chap:1}  the QCD theory of strong interactions is an unbroken gauge theory based on the group $SU(3)$ of colour. The eight massless gauge bosons are the gluons $g^A_\mu$ and the matter fields are colour triplets of quarks $q^a_i$ (in different flavours $i$). Quarks and gluons are the only fundamental fields of the Standard Model (SM) with  strong interactions (hadrons). The QCD Lagrangian was introduced in Sect. \ref{sec:4}, Eqs. \ref{LagQCD}- \ref{F}. For quantization the classical Lagrangian in Eq.~(\ref{LagQCD}) must be enlarged to contain gauge fixing and ghost terms, as described in Chapter \ref{Chap:1}. The Feynman rules of QCD are listed in Fig.~\ref{QCDFeyn}. The physical vertices in QCD include the gluon-quark-antiquark vertex,
analogous to the QED photon-fermion-antifermion coupling, but also the 3-gluon and 4-gluon vertices, of order $e_s$ and
$e_s^2$ respectively, which have no analogue in an abelian theory like QED. 

Why $SU(N_C=3)_{colour}$? The selection of $SU(3)$ as colour gauge group is unique in view of a number of constraints.
(a) The group must admit complex representations because it must be able to distinguish a quark from an antiquark \cite{quarks}. In
fact there are meson states made up of  $q\bar q$ but not analogous $qq$ bound states. Among simple groups this
restricts the choice to $SU(N)$ with $N\geq 3$, $SO(4N+2)$ with $N\geq 2$ (taking into account that $SO(6)$ has the
same algebra as $SU(4)$) and $E(6)$. (b) The group must admit a completely antisymmetric colour singlet baryon made up of 3
quarks:
$qqq$. In fact, from the study of hadron spectroscopy we know that the low lying baryons, completing an octet and a decuplet of
(flavour)
$SU(3)$ (the approximate symmetry that rotates the 3 light quarks u, d and s), are made up of three quarks and are colour
singlets. The
$qqq$ wave function must be completely antisymmetric in colour in order to agree with Fermi statistics. Indeed if we
consider, for example, a
$N^{*++}$ with spin z-component +3/2, this is made up of $(u\Uparrow u\Uparrow u\Uparrow)$ in an s-state. Thus its wave
function is totally symmetric in space, spin and flavour so that complete antisymmetry in colour is required by Fermi
statistics. In QCD this requirement is very simply satisfied by $\epsilon_{abc}q^aq^bq^c$ where a, b, c are
$SU(3)_{colour}$ indices. (c) The choice of
$SU(N_C=3)_{colour}$ is confirmed by many processes that directly measure $N_C$. Some examples are listed here. The total
rate for hadronic production in
$e^+e^-$ annihilation is linear in $N_C$. Precisely if we consider $R=R_{e^+e^-}=\sigma(e^+e^-\rightarrow
hadrons)/\sigma_{point}(e^+e^-\rightarrow \mu^+\mu^-)$ above the $b \bar b$ threshold and below $m_Z$ and we neglect small
computable radiative corrections (that will be discussed later in Sect.~\ref{sec:15}) we have a sum of individual contributions (proportional
to $Q^2$, where $Q$ is the electric charge in units of the proton charge) from
$q\bar q$ final states with $q=u,~c,~d,~s,~b$:
\beq
R~\approx~N_C [ 2 \cdot \frac{4}{9}~+~3\cdot \frac{1}{9}]~\approx~N_C \frac{11}{9}\label{R}\\  
\eeq 
The data neatly indicate $N_C=3$ as seen from Fig.~\ref{Ree} \cite{pdg06}.  The slight excess of the data with respect to the value 11/3 is due to the QCD radiative corrections ( Sect.~\ref{sec:15}). Similarly we can consider the branching ratio
$B(W^-\rightarrow e^-\bar{\nu})$, again in Born approximation. The possible fermion-antifermion ($f\bar f$) final states are
for $f= e^-,~\mu^-,~\tau^-, d, s$ (there is no $f=b$ because the top quark is too heavy for $b\bar t$ to occur). Each channel
gives the same contribution, except that for quarks we have $N_C$ colours:

\begin{figure}[t]
\noindent
\includegraphics[width=16cm]{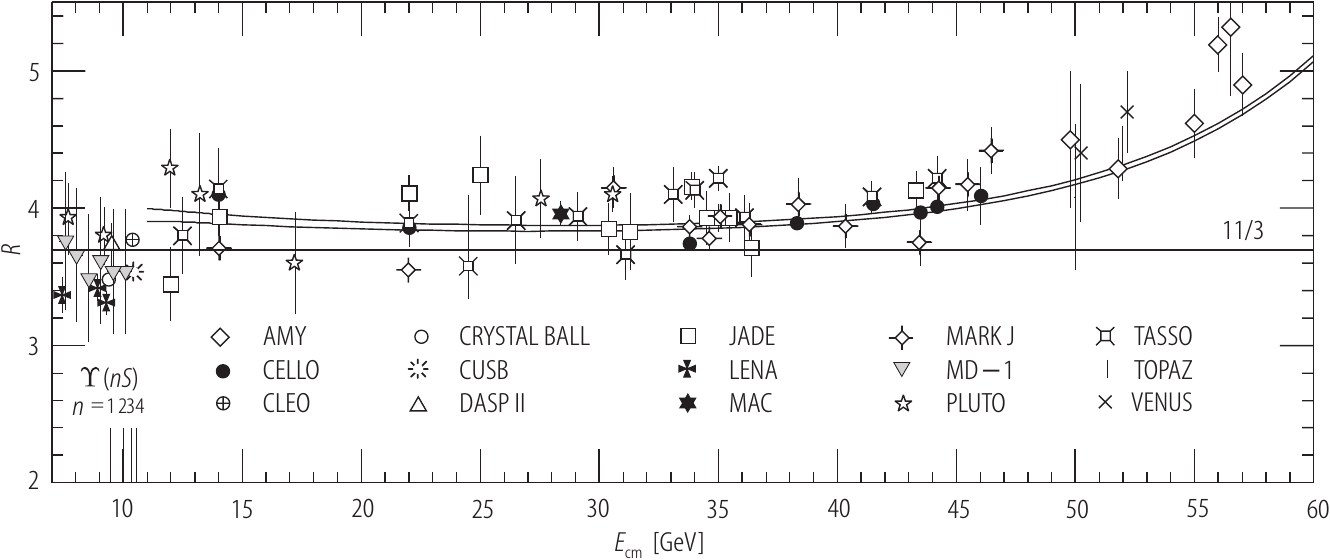} 
\caption{Comparison of the data on $R=\sigma(e^+e^-\rightarrow
hadrons)/\sigma_{point}(e^+e^-\rightarrow \mu^+\mu^-)$ with the QCD prediction \cite{pdg06}. $N_C=3$ is indicated by the data points above ~ 10 GeV (the $b \bar b$ threshold) and ~ 40 GeV where the rise due to the $Z_0$ resonance become appreciable.}
\label{Ree}
\end{figure}

\beq
B(W^-\rightarrow e^-\bar{\nu})~\approx~\frac{1}{3+2N_C}\label{BW}\\
\eeq
For $N_C=3$ we obtain $B=11\%$ and the experimental number is $B=10.7\%$. Another analogous example is the branching ratio
$B(\tau^-\rightarrow
e^-\bar{\nu_e}\nu_{\tau})$. From the final state channels with $f=
e^-,~\mu^-,~d$ we find
\beq
B(\tau^-\rightarrow e^-\bar{\nu_e}\nu_{\tau})~\approx~\frac{1}{2+N_C}\label{Btau}\\
\eeq
For $N_C=3$ we obtain $B=20\%$ and the experimental number is $B=18\%$ (the less accuracy in this case is explained by
the larger radiative and phase-space corrections because the mass of $\tau^-$ is much smaller than $m_W$). An important
process that is quadratic in $N_C$ is the rate $\Gamma(\pi^0\rightarrow 2\gamma)$. This rate can be reliably calculated
from a solid theorem in field theory which has to do with the chiral anomaly:
\beq 
\Gamma(\pi^0\rightarrow 2\gamma)\approx~(\frac{N_C}{3})^2\frac{\alpha^2m_{\pi^0}^3}{32\pi^3f_{\pi}^2}~=~
(7.73\pm0.04)(\frac{N_C}{3})^2~{\rm eV}\label{pi0}\\
\eeq
where the prediction is obtained for $f_{\pi}=(130.7\pm0.37)$~MeV. The experimental result is $\Gamma~=~(7.7\pm0.5)$~eV in
remarkable agreement with $N_C=3$. There are many more experimental confirmations that $N_C=3$: for example the rate for
Drell-Yan processes (see Sect.~\ref{sec:17} ) is inversely proportional to $N_C$.

\subsection{Non Perturbative QCD}
\label{sec:110}

The QCD lagrangian in Eq.~(\ref{LagQCD}) has a simple structure but a very rich dynamical content. It gives rise to a
complex spectrum of hadrons, implies the striking properties of confinement and asymptotic freedom, is endowed with
an approximate chiral symmetry which is spontaneously broken, has a highly non trivial topological vacuum structure
(instantons, $U(1)_A$ symmetry breaking, strong CP violation (which is a problematic item in QCD possibly connected with new physics, like axions), ...), an intriguing phase transition diagram (colour
deconfinement, quark-gluon plasma, chiral symmetry restoration, colour superconductivity, ...). 

How do we get testable predictions from QCD? On the one hand there are non perturbative methods. The most important
at present is the technique of lattice simulations (for a recent review, see ref. \cite{kron}): it is based on first principles, it has produced very valuable
results on confinement, phase transitions, bound states, hadronic matrix elements and so on, and it is by now an
established basic tool. The main limitation is from computing power and therefore there is continuous progress and a lot
of good perspectives for the future. Another class of approaches is based on effective lagrangians which provide
simpler approximations than the full theory, valid in some definite domain of physical conditions. Typically at energies below a given scale $L$ particles with mass larger than $L$ cannot be produced and thus only contribute short distance effects as virtual states in loops. Under suitable conditions one can write down a simplified effective lagrangian where the heavy fields have been eliminated (one says "integrated out"). Virtual heavy particle short distance effects are absorbed into the coefficients of the various operators in the effective Lagrangian. These coefficients are determined in a matching procedure, by requiring that the effective theory reproduces the matrix elements of the full theory up to power corrections. Chiral
lagrangians are based on soft pion theorems \cite{chir} and are valid for suitable processes at energies below 1~GeV (for a recent, concise review see ref. \cite{gas} and references therein). Heavy
quark effective theories \cite{HQ} are obtained from expanding in inverse powers of the heavy quark mass and are
mainly important for the study of  b and, to less accuracy, c decays (for reviews, see, for example, ref. \cite{neu}). Soft-collinear effective theories (SCET) \cite{SCET}, are valid for processes where quarks with energy much larger than their mass appear.  Light energetic quarks not only emit soft gluons, but  also collinear gluons (a gluon in the same direction as the original quark), without changing their virtuality. In SCET the logs associated with these soft and collinear gluons are resummed. The approach of QCD sum rules \cite{shi,na} has led to
interesting results but now appears not to offer  much potential for further development. On the other hand, the perturbative approach, based on
asymptotic freedom, still remains the main quantitative connection to experiment, due to its wide range of
applicability to all sorts of "hard" processes.

\subsubsection{Progress in Lattice QCD}
\label{sec:11}

A main approach to non perturbative problems in QCD is by simulations of the theory on the lattice, a technique started by K. Wilson in 1974 \cite{wils} which has shown continuous progress over the last decades. In this approach the QCD theory is reformulated on a discrete space time, an hypercubic lattice of sites (in the simplest realizations) with spacing $a$ and 4-volume $L^4$; on each side there are $N$ sites with $L=Na$. Over the years one has learned how to efficiently describe a field theory on a discrete space time and how to implement gauge symmetry, chiral symmetry and so on (for a recent review see, for example, ref. \cite{kron}). Gauge and matter fields are specified on the lattice sites and the path integral is computed numerically as a sum over the field configurations. Much more powerful computers than in the past now allow for a number of essential improvements. As eventually one is interested in the continuum limit, $a \rightarrow 0$, it is important to work with as fine lattice spacing $a$ as possible. Methods have been developed for ''improving'' the lagrangian in such a way that the discretization errors vanish faster than linearly in $a$. Larger lattice volume (i. e. large $L$ or $N$) is also useful as the dimensions of the lattice should be as large as possible in comparison with the dimensions of the hadrons to be studied. In many cases the volume corrections are exponentially damped, but this is not always the case. Lattice simulation is limited to large enough masses of light quarks: in fact, heavier quarks have shorter wavelenghts and can be accommodated in a smaller volume. In general computations are done for  quark and pion masses  heavier than in reality and then extrapolated to the physical values, but at present one can work with smaller quark masses than in the past.  One can also take advantage of the chiral effective theory in order to control the chiral logs: $\log(m_q/4\pi f_\pi)$ and guide the extrapolation. A recent big step, made possible by the availability of more powerful dedicated computers, is the evolution from quenched (i.e. with no dynamical fermions) to unquenched calculations. In doing so an evident improvement in the agreement of predictions with the data is obtained. For example \cite{kron}, modern unquenched simulations reproduce the hadron spectrum quite well. Calculations with dynamical fermions (which take into account the effects of virtual quark loops) imply the evaluation of the quark determinant which is a difficult task. How difficult depends on the particular calculation method. There are several approaches (Wilson, twisted mass,  Kogut-Susskind staggered, Ginsparg-Wilson fermions), each with its own advantages and disadvantages (including the time it takes to run the simulation on a computer): a compromise between efficiency and theoretical purity is needed. The most reliable lattice calculations are today for 2 + 1 light quarks (degenerate up and down quarks and a heavier strange quark s). The first calculations for 2 + 1 +1 including charm quarks are starting to appear.

Lattice QCD is becoming increasingly predictive and plays a crucial role in different domains. For example, in flavour physics it is essential for computing the relevant hadronic matrix elements. In high temperature QCD the most illuminating studies of the phase diagram, the critical temperature and the nature of the phase transitions are obtained by lattice QCD: as we now discuss  the best arguments to prove that QCD implies confinement come from the lattice.

\subsubsection{Confinement}
\label{sec:11a}

\begin{figure}[t]
\noindent
\centerline{\includegraphics[width=10cm]{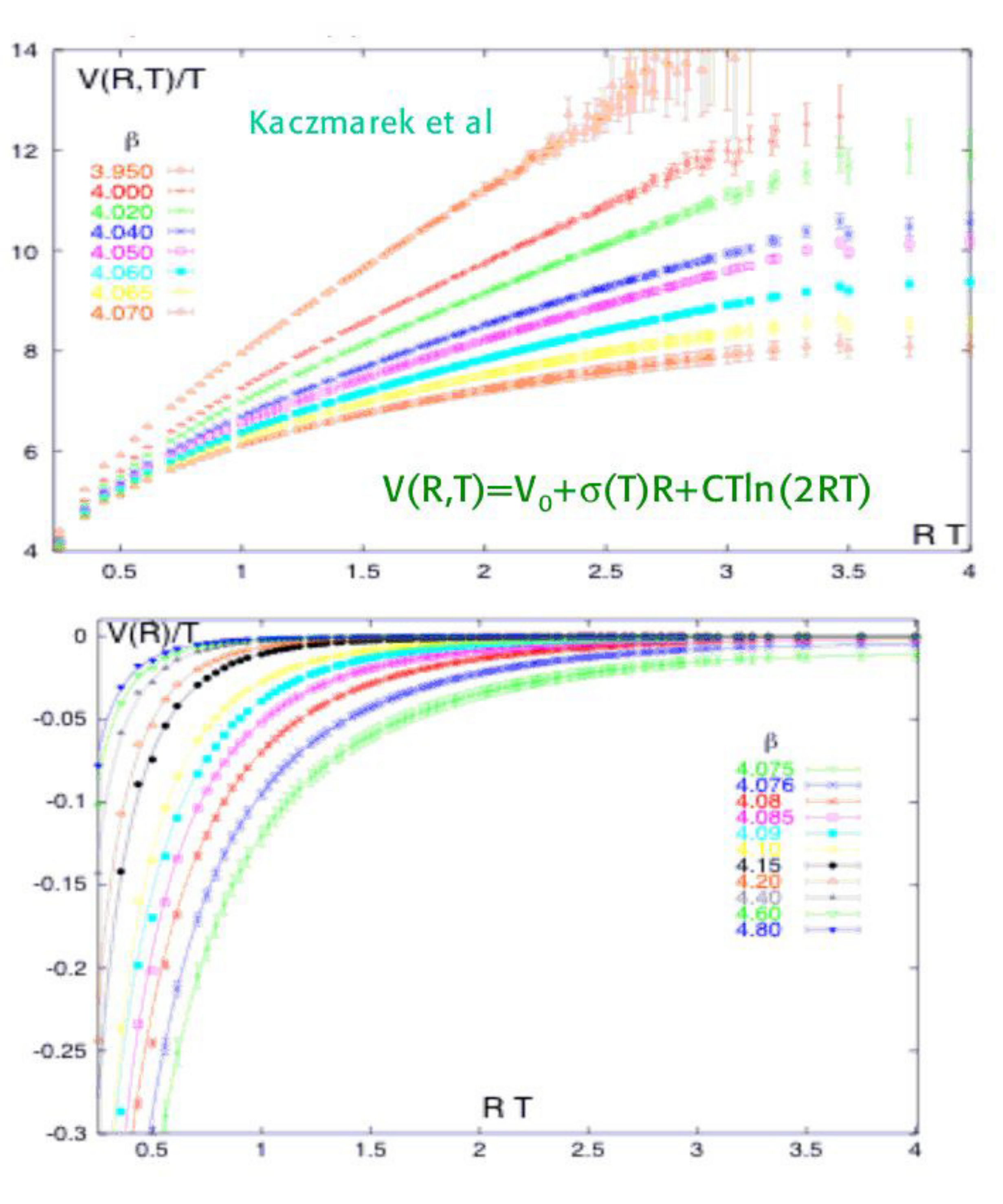}}
\caption{The potential between a quark and an antiquark computed on the lattice in the quenched approximation \cite{kaczma}. The upper panel shows that the slope of the linearly rising term decreases with temperature and, at the critical temperature $T_C$, it vanishes. At $T \ge T_C$ the slope remains put at zero (lower panel).}
\label{kacz}
\end{figure}

Confinement is the property that no isolated coloured charge can exist but only colour singlet particles. Our understanding of the confinement mechanism has much improved thanks to lattice simulations of QCD at finite temperatures and densities (for reviews, see, for example ref.\cite{detar,fodo,baza}). For example,
the potential between a quark and an antiquark has been studied on the lattice \cite{kaczma}. It has a Coulomb part at short
distances and a linearly rising term at long distances:
\beq
V_{q\bar q}~\approx~C_F[\frac{\alpha_s(r)}{r}~+....+\sigma r]\label{V}\\
\eeq
where
\beq
C_F~=~\frac{1}{N_C}\sum_At^At^A~=~\frac{N_C^2-1}{2N_C}\label{CF}
\eeq
with $N_C$ the number of colours ($N_C=3$ in QCD). The scale dependence of $\alpha_s$ (the distance r is
Fourier-conjugate to momentum transfer) will be explained in detail later. 
The slope decreases with increasing temperature until it vanishes at a critical temperature $T_C$; then above $T_C$ the slope remains zero, as shown in Fig. \ref{kacz}. The value of the critical temperature is estimated around $T_C \sim$ 175 MeV.

The linearly rising term in the potential makes it
energetically impossible to separate a $q-\bar q$ pair. If the pair is created at one space-time point, for example in
$e^+e^-$ annihilation, and then the quark and the antiquark start moving away from each other in the center of mass frame,
it soon becomes energetically favourable to create additional pairs, smoothly distributed in rapidity between the two leading
charges, which neutralize colour and allow the final state to be reorganized into two jets of colourless hadrons, that
communicate in the central region by a number of "wee" hadrons with small energy. It is just like the familiar example of
the broken magnet: if you try to isolate a magnetic pole by stretching a dipole, the magnet breaks down and two new poles
appear at the breaking point. 

Confinement is essential to explain why nuclear forces have very short range while massless gluon exchange would be long
range. Nucleons are colour singlets and they cannot exchange colour octet gluons but only colourless states. The lightest
colour singlet hadronic particles are pions. So the range of nuclear forces is fixed by the pion mass $r\simeq
m_{\pi}^{-1} \simeq 10^{-13}~cm$ : $V\approx \exp(-m_{\pi}r)/r$.

\begin{figure}[h]
\noindent
\centerline{\includegraphics[width=10cm]{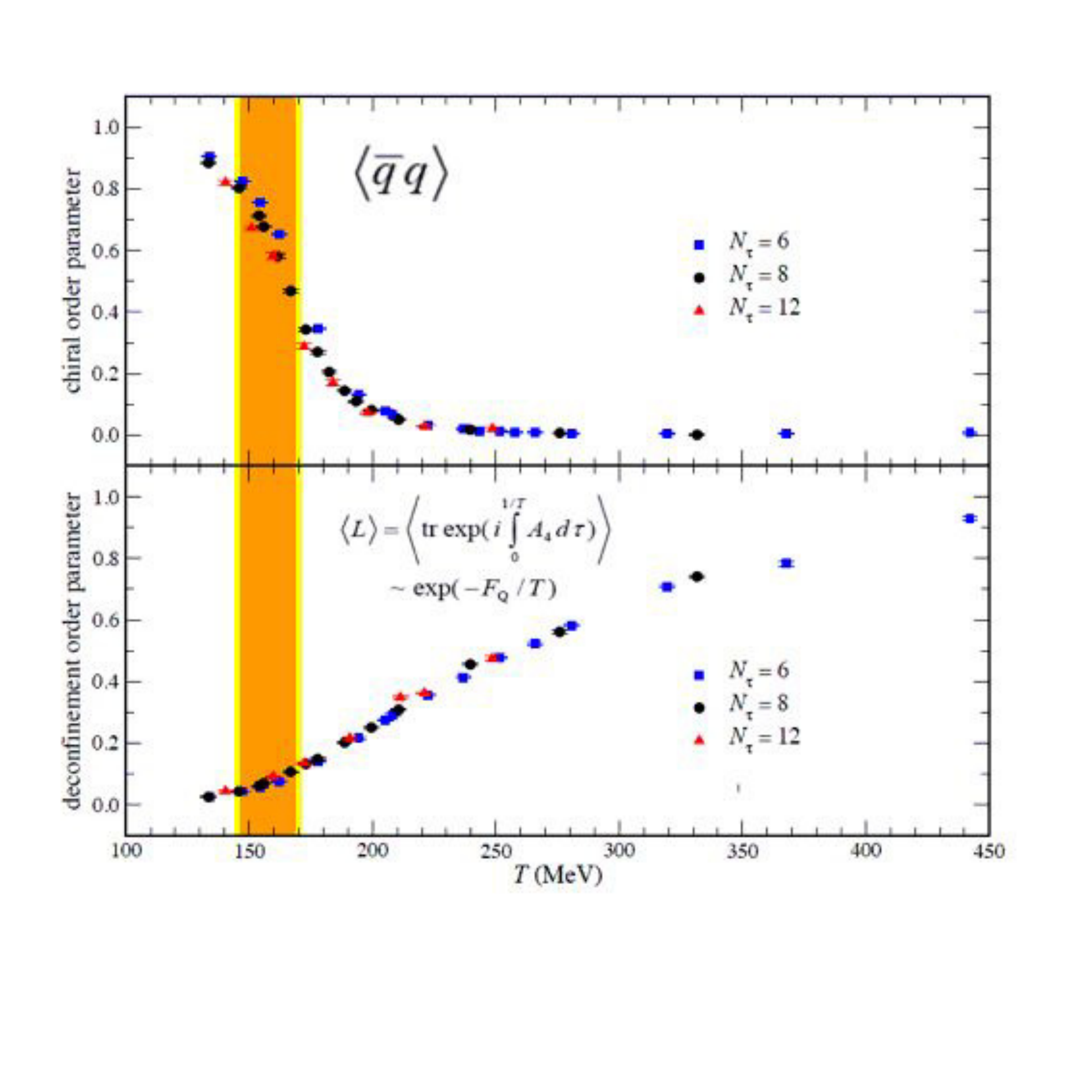}}
\caption{Order parameters for deconfinement (bottom) and chiral symmetry
restoration (top), as a function of temperature \cite{baza,kron}. On a finite lattice the singularities associated to phase transitions are not present but their development is indicated by a rapid rate of change. The vacuum expectation value of the quark-antiquark condensate, with increasing temperature, goes from the finite value that breaks chiral symmetry down to zero where chiral symmetry is restaured. In a comparable temperature range, the Wilson plaquette, the order parameter for deconfinement, goes from zero to a finite value.}
\label{deconf}
\end{figure}

\begin{figure}[t]
\noindent
\centerline{\includegraphics[width=10cm]{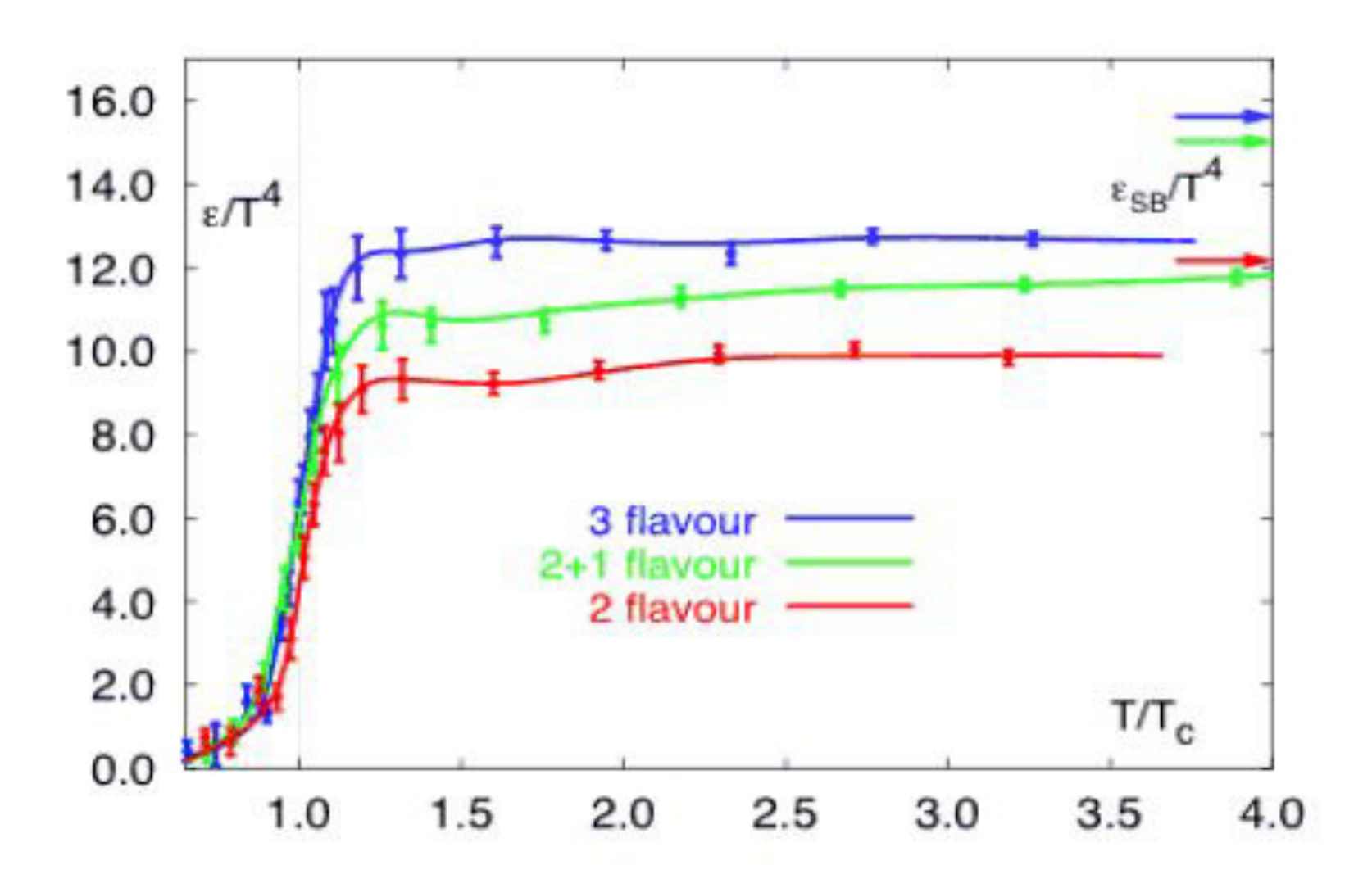}}
\caption{The energy density divided by the 4rth power of the temperature, computed on the lattice with different number of sea flavours, shows a marked rise near the critical temperature \cite{baza,kron}. The arrows on top show the limit for a perfect Bose gas (while the hot dense hadronic fluid is not expected to be a perfect gas).}
\label{endens}
\end{figure}

\begin{figure}[h]
\noindent
\centerline{\includegraphics[width=15cm]{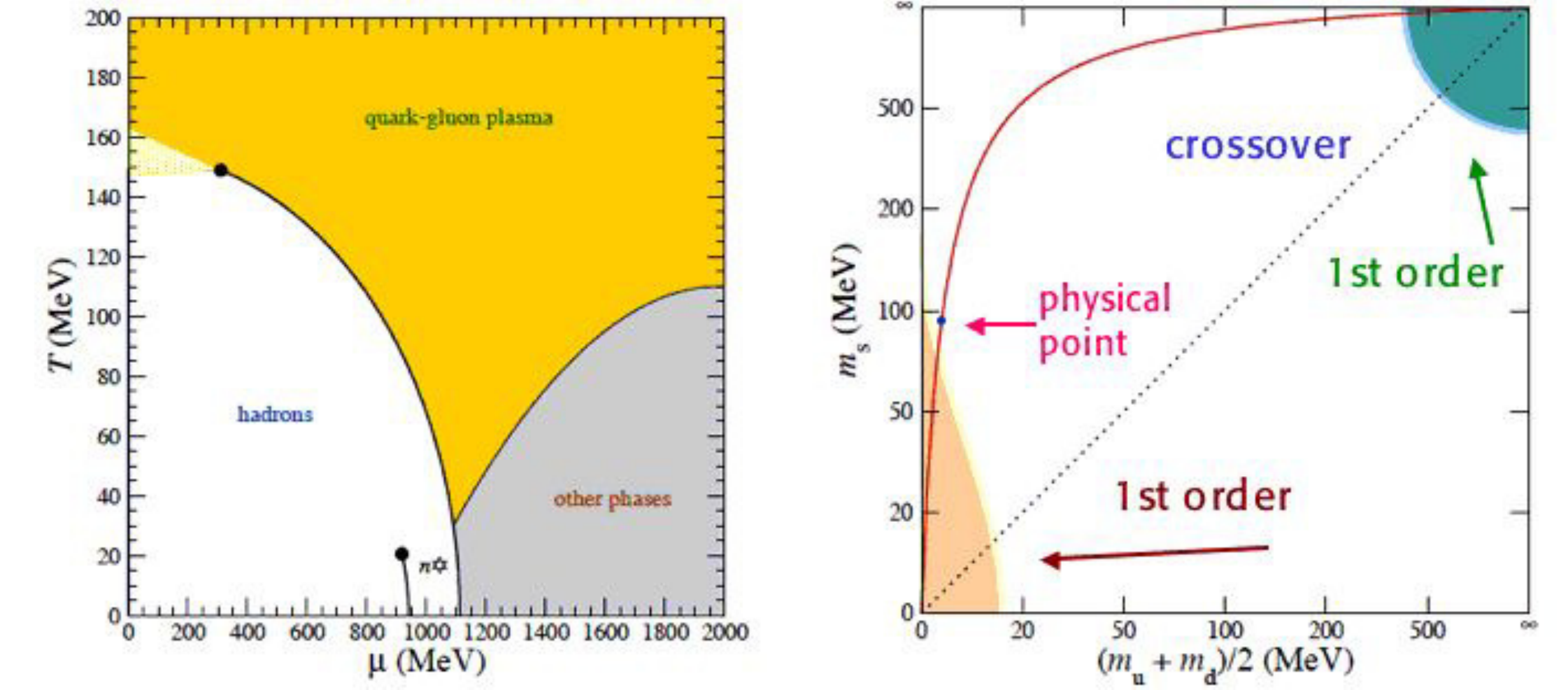}}
\caption{Left: A schematic view of the QCD phase diagram. Right: On the lattice the nature of the phase transition depends on the number of quark flavours and their masses as indicated \cite{kron}}
\label{phasdiagr}
\end{figure}

The phase transitions of colour deconfinement and of chiral restauration appear to happen together on the lattice \cite{kron,detar,fodo,baza} (see Fig.\ref{deconf}). A rapid transition is observed in lattice simulations where the energy density $\epsilon(T)$ is seen to sharply increase near the critical temperature for deconfinement and chiral restauration (see Fig.\ref{endens}). The critical parameters and the nature of the phase transition depend on the number of quark flavours $n_f$ and on their masses (see Fig.\ref{phasdiagr}). For example, for  $n_f$ = 2 or 2+1 (i.e. 2 light u and d quarks and 1 heavier s quark), $T_C \sim 175$~MeV  and $\epsilon(T_C) \sim 0.5-1.0$~GeV/fm$^3$. For realistic values of the masses $m_s$ and $m_{u,d}$ the two phases are connected by a smooth crossover, while the phase transition becomes first order for very small or very large $m_{u,d,s}$. Accordingly the hadronic phase and the deconfined phase are separated by a crossover region at small densities and by a critical line at high densities that ends with a critical point. Determining the exact location of the critical point in T and $\mu_B$ is an important challenge for theory and is also important for the interpretation of heavy ion collision experiments. At high densities the colour superconducting phase is also present with bosonic diquarks acting as Cooper pairs. 

A large investment is being done in experiments of heavy ion collisions with the aim of finding some evidence of the quark gluon plasma phase. Many exciting results have been found at the CERN SPS in the past  years, more recently at RHIC and now at the LHC in dedicated heavy ion runs \cite{schu} (the ALICE detector is especially designed for the study of  heavy ion collisions).

\subsubsection{Chiral Symmetry in QCD and the Strong CP Problem}
\label{sec:10}

In the QCD lagrangian, Eq.~(\ref{LagQCD}), the quark mass terms are of the general form [$m \bar \psi_L \psi_R$ +h.c.] (recall the definition of $\psi_{L,R}$ in Sect. \ref{sec:5} and the related discussion). These terms are the only ones that show a chirality flip. In the absence of these terms, i.e. for $m=0$, the QCD lagrangian would be invariant under independent unitary tranformations separately on $\psi_L$ and on $\psi_R$. Thus, if the masses of the $N_f$ lightest quarks are neglected the QCD lagrangian is invariant under a global $U(N_f)_L\bigotimes U(N_f)_R$ chiral group. Consider $N_f=2$: $SU(2)_V$ corresponds to the observed approximate isospin symmetry and $U(1)_V$ to the portion of  baryon number associated with u and d quarks. Since no approximate parity doubling of light quark bound states is observed the $U(2)_A$ symmetry must be spontaneously broken (for example, no opposite parity analogues of protons and neutrons exist with a few tens of MeV separation in mass from the ordinary nucleons). The breaking of chiral symmetry is induced by the VEV of a quark condensate: for $N_f=2$ this is  [$\bar u_L u_R+\bar d_L d_R$+h.c.]. A recent lattice calculation \cite{fuk} has given for this condensate the value $[234\pm18~ MeV]^3$ (in $\bar {MS}$, $N_f=2+1$, with the physical $m_s$ value, at the scale of 2 GeV). This scalar operator is an isospin singlet, so that it preserves $U(2)_V$ but breaks $U(2)_A$ (it transforms like (1/2,1/2) under $U(2)_L\bigotimes U(2)_R$ but is a singlet under the diagonal group $U(2)_V$). The pseudoscalar mesons are obvious candidates for the would-be Goldstone bosons associated with the breakdown of the axial group in that they have the quantum number of the broken generators: the three pions are the approximately massless Goldstone bosons (exactly massless in the limit of vanishing u and d quark masses) associated with the breaking of three generators of $U(2)_L\bigotimes U(2)_R$ down to $SU(2)_V\bigotimes U(1)_V\bigotimes U(1)_A$. The couplings of Goldstone bosons are very special: in particular only derivative couplings are allowed. The pions as pseudo-Goldstone bosons have couplings that satisfy strong constraints. An effective chiral lagrangian formalism \cite{chir} allows to systematically reproduce the low energy theorems implied by the approximate status of Goldstone particles for the pion and successfully describes QCD at energies at scales below $\sim 1$ GeV.

The breaking mechanism for the remaining $U(1)_A$ arises from an even subtler mechanism. A state in the $\eta-\eta'$ space cannot be the associated Goldstone particle because the masses are too large \cite{wei75} and the $\eta'$ mass does not vanish in the chiral limit \cite{wive}. Rather the conservation of the singlet axial current $ j_5^\mu=\Sigma \bar q_i\gamma^\mu\gamma_5q_i$  is broken by the Adler-Bell-Jackiw anomaly \cite{ABJ}:
\beq
\partial_\mu j_5^\mu \equiv I(x) = N_f \frac{\alpha_s}{4\pi}\sum_A F^A_{\mu\nu}\tilde F^{A\mu\nu}=N_f \frac{\alpha_s}{2\pi}Tr({\bf F_{\mu\nu}\tilde F^{\mu\nu}})
\label{anom}
\eeq
(recall that ${\bf F_{\mu\nu}}=\Sigma F^A_{\mu\nu}t^A$ and the normalization is $Tr(t^At^B)=1/2\delta^{AB}$)
with $F^A_{\mu\nu}$ given in Eq.\ref{F} and $j_5^\mu$ being the u+d singlet axial current (the factor of $N_f$, in this case $N_f=2$, in front of the right hand side takes into account that $N_f$ flavours are involved) and
\beq
\tilde F^A_{\mu\nu}=\frac{1}{2}\epsilon_{\mu\nu\rho\sigma}F^{A\rho\sigma}
\label{Ftwid}
\eeq

An important point is that the pseudoscalar quantity $I(x)$ is a four divergence. Precisely one can check that:
\beq
Tr({\bf F_{\mu\nu}\tilde F^{\mu\nu}})=\partial^\mu k_\mu
\label{delk}
\eeq
with
\beq
 k_\mu= \epsilon_{\mu\nu\lambda\sigma}Tr[{\bf A^\nu(F^{\lambda\sigma}}-\frac{2}{3}ie_s{\bf A^\lambda A^\sigma)}]
\label{defk}
\eeq
As a consequence the modified current $\tilde j_5^\mu$ and its associated charge $\tilde Q_5$ appear to be still conserved:
\beq
\partial_\mu \tilde j_5^\mu=\partial_\mu  ( j_5^\mu -N_f \frac{\alpha_s}{2\pi}k^\mu)=0
\label{newcurr}
\eeq
and could act as modified chiral current and charge with an additional gluonic component. But actually this charge is not conserved due to the topological structure of the QCD vacuum (instantons) as discussed in the following (for an introduction, see ref. \cite{peccei}).

The configuration where all gauge fields are zero $A^A_\mu=0$ can be called ''the vacuum''. However all configurations that are connected to $A^A_\mu=0$ by a gauge transformation do also correspond to the same physical vacuum. For example, in a abelian theory all gauge fields that can be written as the gradient of a scalar, $A^A_\mu=\partial_\mu \chi(x)$  are equivalent to $A^A_\mu=0$. In non abelian gauge theories there are some ''large'' gauge transformations that are topologically non trivial and correspond to non vanishing integer values of a topological charge, the ''winding number''. Taking $SU(2)$ for simplicity (in QCD it could be any such subgroup of colour $SU(3)$) we can consider the following, time independent gauge transformation:
\beq
\Omega_1(\vec{x})=\frac{\vec{x}^2-d^2+2id \vec{\tau}^. \vec{x}}{\vec{x}^2+d^2}
\label{omeg}
\eeq
where $d$ is a positive constant. Note that $\Omega_1^{-1}=\Omega_1^*$. Starting from ${\bf A_\mu}=(A_0,A_i)=(0,0)$ (i=1,2,3) with ${\bf A_\mu}=\Sigma A_\mu^a \tau^a/2$ the gauge transformed potential by $\Omega_1$ is (recall the general expression of a gauge transformation in Eq. \ref{12}):
\beq
{\bf A^{(1)}_j}=-\frac{i}{e_s}[\bigtriangledown_j \Omega_1(\vec{x})]\Omega_1^{-1}(\vec{x})
\label{a1}
\eeq
For the vector potential ${\bf A^{(1)}}$, which, being a pure gauge, is part of the ''vacuum'',  the winding number $n$, defined in general by
\beq
n=\frac{ie_s^3}{24\pi^2} \int d^3 x Tr[{\bf A_i(x)A_j(x)A_k(x)}]\epsilon^{ijk}
\eeq
is equal to 1: $n=1$. Similarly, for ${\bf A^{(m)}}$ obtained from $\Omega_m$ = [$\Omega_1]^m$ one has $n=m$. Given Eq. \ref{delk} we could expect
that the integrated four-divergence would vanish but instead one finds:
\beq
\frac{\alpha_s}{4\pi}\int d^4x~Tr({\bf F_{\mu\nu}\tilde F^{\mu\nu}})=\frac{\alpha_s}{4\pi}\int d^4x~\partial_\mu k^\mu=\frac{\alpha_s}{4\pi}[\int d^3x~
k_0]^{+\infty}_{-\infty}=n_+-n_-
\label{inst}
\eeq
for a configuration of gauge fields that vanish fast enough on the space sphere at infinity and the winding numbers are $n_{-,+}$ at time $t=-,+\infty$  (''instantons'').

From the above discussion it follows that in QCD all gauge fields can be classified in sectors with different $n$: there is a vacuum for each $n$, $\vert n \rangle$, and $\Omega_1\vert n \rangle = \vert n +1\rangle$ (not gauge invariant!). The true vacuum must be gauge invariant (up to a phase) and is obtained as a superposition of all $\vert n \rangle$:
\beq
\vert \theta \rangle=\Sigma_{-\infty}^{+\infty}e^{-in\theta}\vert n \rangle
\label{defth}
\eeq
In fact:
\beq
\Omega_1 \vert \theta \rangle=\Sigma e^{-in\theta}\vert n+1 \rangle=e^{i\theta}\vert \theta \rangle
\label{GTth}
\eeq
If we compute the expectation value of any operator $O$ in the $\theta$ vacuum we find:
\beq
\langle \theta \vert O \vert \theta \rangle= \Sigma_{m,n} e^{i(m-n)\theta} \langle m \vert O \vert n \rangle
\label{vacO}
\eeq
The path integral describing the $O$ vacuum matrix element at $\theta=0$ must be modified to reproduce the extra phase, taking Eq. \ref{inst} into account:
\beq
\langle \theta \vert O \vert \theta \rangle= \int dA d\bar \psi d\psi ~O ~exp[iS_{QCD} +i\theta\frac{\alpha_s}{4\pi}\int d^4x~Tr({\bf F_{\mu\nu}\tilde F^{\mu\nu}})]
\eeq
This is equivalent to adding a $\theta$ term to the QCD lagrangian:
\beq
{\cal L_{QCD}}=\theta \frac{\alpha_s}{4\pi}\int d^4x~Tr({\bf F_{\mu\nu}\tilde F^{\mu\nu}})
\label{CPterm}
\eeq

The $\theta$ term is parity P odd and charge conjugation C even, so that it introduces CP violation in the theory (and also time reversal T violation). A priori one would expect $\tilde \theta$ to be O(1). But it would contribute to the neutron electric dipole moment, according to $d_n(e^.cm) \sim 3~10^{-16} \tilde \theta$. The strong experimental bounds on $d_n$ ( $d_n(e^.cm) \leq 3~10^{-26}$ \cite{pdg12}) imply that $\tilde \theta$ must be very small: $\tilde \theta \leq 10^{-10}$. The so-called ''Strong CP-problem'' or ''$\theta$-problem'' is to find an explanation for such a small value \cite{peccei,kimcp}. An important point that is relevant for a possible solution is that a chiral transformation translates $\theta$ by a fixed amount. By recalling Eq. \ref{newcurr} we have:
\beq
e^{i\delta \tilde Q_5}\vert \theta \rangle =\vert \theta -2N_f \delta \rangle
\label{q5act}
\eeq
To prove this relation we first observe that $\tilde Q_5$ is not gauge invariant under $\Omega_1$ because it involves $k_0$:
\beq
\Omega_1\tilde Q_5 \Omega_1^{-1}=Q_5-\Omega_1 2 N_f \frac{\alpha_s}{4\pi}[\int d^3x~k_0 \Omega_1^{-1}=\tilde Q_5-2N_f
\eeq
It then follows that 
\beq
\Omega_1 e^{i\delta \tilde Q_5}\vert \theta \rangle=\Omega_1 e^{i\delta \tilde Q_5}\Omega_1^{-1}\Omega_1\vert \theta \rangle=e^{i(\theta-2N_f \delta)}
e^{i\delta \tilde Q_5}\vert \theta \rangle
\eeq 
which implies Eq. \ref{q5act}. Thus in a chiral invariant theory one could dispose of $\theta$. For this it would be
sufficient that a single quark mass is zero and the obvious candidate would be $m_u = 0$. But apparently this possibility has been excluded \cite{kimcp}. 
For non vanishing quark masses the transformation $m \rightarrow U_L^\dagger m U_R$ needed to make the mass matrix hermitian (which implies $\gamma_5$ - free) and diagonal involves a chiral transformation that affects $\theta$. Considering that $U(N)=U(1)\bigotimes SU(N)$, that for hermitian $m$ the argument of the determinant vanishes: $Arg Det m =0$ the transformation from a generic $m'$ to a real and diagonal $m$ gives:
\beq 
Arg Det~m = 0  =Arg Det ~U_L^* +Arg Det ~m' + Arg Det ~U_R= -2 N_f(\delta_L - \delta_R) +Arg Det~ m' 
\eeq
From this equation one derives the phase $(\delta_R - \delta_L)$ of the chiral transformation and then, by Eq. \ref{q5act}, the important result for the effective $\theta$ value:
\beq
\theta_{eff} = \theta +Arg Det~ m'
\label{tetaeff}
\eeq

As we have seen the small empirical value of $\theta_{eff}$ poses a serious naturalness problem for the SM. Among the possible solutions perhaps the most interesting option is a mechanism proposed by Peccei-Quinn \cite{pequi}. One assumes that the SM or an enlarged theory is invariant under an additional chiral symmetry $U(1)_{PQ}$ acting on the fields of the theory. This symmetry is spontaneously broken by the vacuum expectation value $v_{PQ}$ of a scalar field. The associated Goldstone boson, the axion, is actually not massless because of the chiral anomaly. The parameter $\theta$ is canceled by the vacuum expectation value of the axion field due to the properties of the associated potential, also determined by the anomaly.  Axions could contribute to the Dark Matter in the Universe, if their mass falls in a suitable narrow range (for a recent review, see, for example, \cite{kim}. Alternative solutions to the $\theta$-problem have also been suggested. Some of them can probably be discarded (for example, that the up quark is exactly massless), while other ones are still possible: for example, in supersymmetric theories, if the smallness of $\theta$ could be guaranteed at the Planck scale by some feature of the more fundamental theory valid there, then the non rinormalization theorems of supersymmetry would preserve its small value throughout the running down to low energy.

\subsection{Massless QCD and Scale Invariance}
\label{sec:12}

As discussed in Chapter 2, the QCD lagrangian in Eq.~(\ref{LagQCD}) only specifies the theory at the classical level. The procedure for quantization of
gauge theories involves a number of complications that arise from the fact that not all degrees of freedom of gauge
fields are physical because of the constraints from gauge invariance which can be used to eliminate the dependent
variables. This is already true for abelian theories and one is familiar with the QED case. One introduces a gauge fixing
term (an additional term in the lagrangian density that acts as a Lagrange multiplier in the action extremization).
One can choose to preserve manifest Lorentz invariance. In this case, one adopts a covariant gauge, like the Lorentz
gauge, and  in QED one proceeds according to the formalism of Gupta-Bleuler \cite{qedtb}. Or one can give up explicit formal covariance
and work in a non covariant gauge, like the Coulomb or the axial gauges, and only quantize the physical degrees of freedom
(in QED the transverse components of the photon field). While this is all for an abelian gauge theory, in the non-abelian case
some additional complications arise, in particular the necessity to introduce ghosts for the formulation of Feynman rules.
As we have seen, there are in general as many ghost fields as gauge bosons and they appear in the form of a transformation Jacobian in the
Feynman functional integral. Ghosts only propagate in closed loops and their vertices with gluons can be included
as additional terms in the lagrangian density which are fixed once the gauge fixing terms and their infinitesimal gauge
transformations are specified. Finally the complete Feynman rules in either the covariant or the axial gauges can be obtained and they
appear in Fig.~\ref{QCDFeyn}. 

Once the Feynman rules are derived we have a formal perturbative expansion but loop diagrams generate infinities. First a
regularization must be introduced, compatible with gauge symmetry and Lorentz invariance. This is possible in QCD. In
principle one can introduce a cut-off $K$ (with dimensions of energy), for example, a' la Pauli-Villars \cite{qedtb}. But at
present the universally adopted regularization procedure is dimensional regularization that we will briefly describe later
on. After regularization the next step is renormalization. In a renormalizable theory (which is the case for all gauge theories in 4
spacetime dimensions and for QCD in particular) the dependence on the cutoff can be completely reabsorbed in a
redefinition of particle masses, of gauge coupling(s) and of wave function normalizations. After renormalization is
achieved the perturbative definition of the quantum theory that corresponds to a classical lagrangian
like in Eq.~(\ref{LagQCD}) is completed. 
In the QCD Lagrangian of Eq.~(\ref{LagQCD}) quark masses are the only parameters with physical dimensions (we work in the
natural system of units $\hbar=c=1$). Naively we would expect that massless QCD is scale invariant. This is actually true
at the classical level. Scale invariance implies that dimensionless observables should not depend on the absolute scale
of energy but only on ratios of energy-dimensional variables. The massless limit should be relevant for the asymptotic
large energy limit of processes which are non singular for $m\rightarrow 0$.

The naive expectation that massless QCD should be scale invariant is false in the quantum theory. The scale symmetry of
the classical theory is unavoidably destroyed by the regularization and renormalization procedure which introduce a
dimensional parameter in the quantum version of the theory. When a symmetry of the classical theory is necessarily
destroyed by quantization, regularization and renormalization one talks of an "anomaly". So, in this sense, scale
invariance in massless QCD is anomalous.

While massless QCD is finally not scale invariant, the departures from scaling are asymptotically small, logarithmic and
computable. In massive QCD there are additional mass corrections suppressed by powers of m/E, where E is the energy scale (for processes that are non singular in the limit $m\rightarrow 0$). At the parton level (q and g) we can conceive to apply the asymptotic predictions
of massless QCD to processes and observables (we use the word "processes" for both) with the following properties ("hard
processes"). (a) All relevant energy variables must be large:
\beq
E_i~=~z_iQ,~~~~~~~~~Q>>m_j;~~~~~~~~~~z_i\rm{:scaling~variables~O(1)}
\label{hp}
\eeq
(b) There should be no infrared singularities (one talks of "infrared safe" processes). (c) The processes
concerned must be finite for $m\rightarrow 0$ (no mass singularities). To possibly satisfy these criteria processes must be
as "inclusive" as possible: one should include all final states with massless gluon emission and add all mass degenerate
final states (given that quarks are massless also $q-\bar q$ pairs can be massless if "collinear", that is moving together
in the same direction at the common speed of light).

In perturbative  QCD one computes inclusive rates for partons (the fields in the lagrangian, that
is, in QCD, quarks and gluons) and takes them as equal to rates for hadrons. Partons and hadrons are considered as two
equivalent sets of complete states. This is called "global duality" and it is rather safe in the rare instance of a totally
inclusive final state. It is less so for distributions, like distributions in the invariant mass M ("local duality") where
it can be reliable only if smeared over a sufficiently wide bin in M.

\begin{figure}[h]
\noindent
\includegraphics{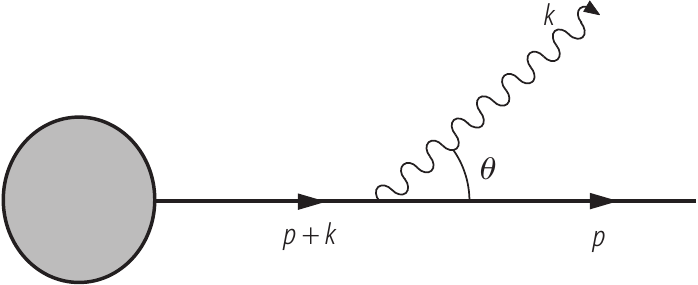} 
\caption[]{The splitting of a virtual quark into a quark and a gluon.}
\label{fig3}
\end{figure}

Let us discuss more in detail infrared and collinear safety. Consider, for example, a quark virtual line that ends up into
a real quark plus a real gluon (Fig.~\ref{fig3}).

For the propagator we have:
\beq
\rm{propagator}~=~\frac{1}{(p+k)^2-m^2}~=~\frac{1}{2(p\cdot k)}~=~\frac{1}{2E_kE_p}\cdot\frac{1}{1-\beta_p\cos{\theta}}
\label{prop}\\
\eeq 
Since the gluon is massless, $E_k$ can vanish and this corresponds to an infrared singularity. Remember that we have to
take the square of the amplitude and integrate it over the final state phase space, or, in this case, all together, $dE_k/E_k$. Indeed we get $1/E_k^2$ from the squared
amplitude and $d^3k/E_k\sim E_kdE_k$ from the phase space. Also, for
$m\rightarrow 0$,
$\beta_p=\sqrt{1-m^2/E_p^2}\rightarrow 1$ and $(1-\beta_p\cos{\theta})$ vanishes at $\cos{\theta}=1$. This leads to a
collinear mass singularity.

\begin{figure}[h]
\noindent
\includegraphics[width=12cm]{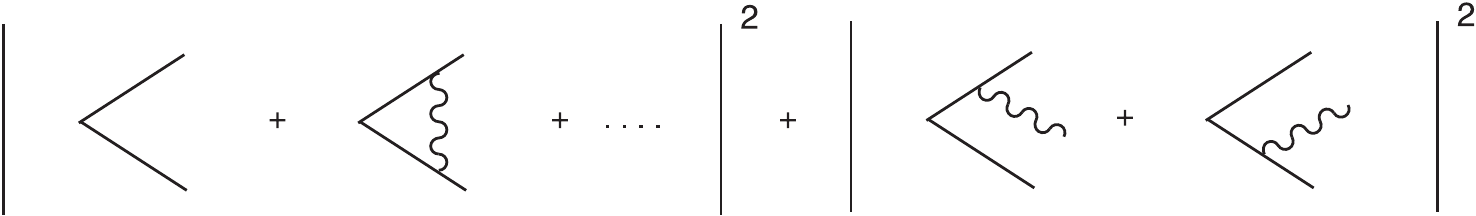} 
\caption[]{The diagrams contributing to the total cross-section $e^+e^-\rightarrow \rm{hadrons}$ at order $\alpha_s$. For simplicity, only the final state quarks and (virtual or real) gluons are drawn.}
\label{fig4}
\end{figure}

There are two very important theorems on infrared and mass singularities. The first one is the Bloch-Nordsieck theorem \cite{BlNo}:
infrared singularities cancel between real and virtual diagrams (see Fig.~\ref{fig4}) when all resolution indistinguishable final
states are added up. For example, for each real detector there is a minimum energy of gluon radiation that can be
detected. For the cancellation of infrared divergences, one should add all possible gluon emission with a total energy
below the detectable minimum. The second one is the Kinoshita-Lee, Nauenberg theorem \cite{KLN}: mass singularities connected with an
external particle of mass $m$ are canceled if all degenerate states (that is with the same mass) are summed up. That is for
a final state particle of mass $m$ we should add all final states that in the limit
$m\rightarrow 0$ have the same mass, also including gluons and massless pairs. If a completely inclusive final state is
taken, only the mass singularities from the initial state particles remain (we shall see that they will be absorbed inside
the non perturbative parton densities, which are probability densities of finding the given parton in the initial hadron).

\begin{figure}[h]
\noindent
\includegraphics[width=6cm]{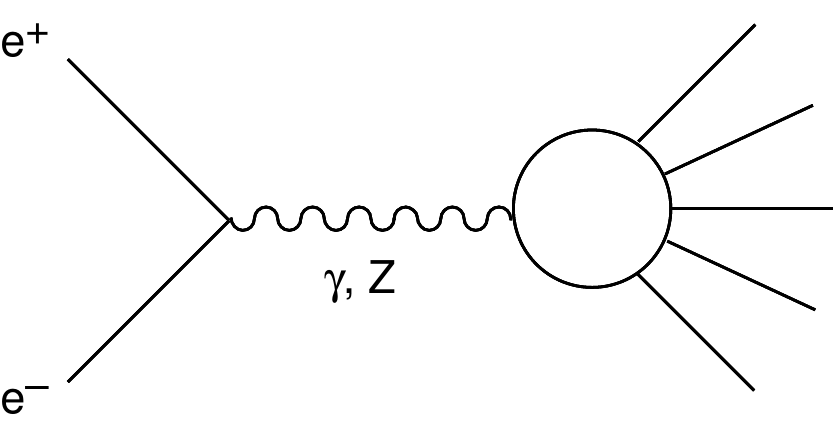} 
\caption[]{The total cross-section $e^+e^-\rightarrow \rm{hadrons}$.}
\label{fig5}
\end{figure}

Hard processes to which the massless QCD asymptotics can possibly apply must be infrared and collinear safe, that is they must
satisfy the requirements from the Bloch-Nordsieck and the Kinoshita-Lee-Nauenberg theorems. We give now some examples of
important hard processes. One of the simplest hard processes is the totally inclusive cross section for hadron production in
$e^+e^-$ annihilation, Fig.~\ref{fig5}, parameterized in terms of the already mentioned dimensionless observable
$R=\sigma(e^+e^-\rightarrow hadrons)/\sigma_{point}(e^+e^-\rightarrow \mu^+\mu^-)$. The pointlike cross section in the
denominator is given by
$\sigma_{point} = 4\pi\alpha^2/3s$, where $s=Q^2=4E^2$ is the squared total center of mass energy and $Q$ is the mass of
the exchanged virtual gauge boson. At parton level the final state is $(q\bar q~+~n~g~+~n'~q'\bar q')$ and n and n' are
limited at each order of perturbation theory. It is assumed that the conversion of partons into hadrons does not affect
the rate (it happens with probability 1). We have already mentioned that in order for this to be true within a given
accuracy an averaging over a sufficiently large bin of $Q$ must be understood. The binning width is larger in the
vicinity of thresholds: for example when one goes across the charm $c\bar c$ threshold the physical cross-section shows
resonance bumps which are absent in the smooth partonic counterpart which however gives an average of the cross-section.

\begin{figure}[h]
\noindent
\includegraphics[width=4cm]{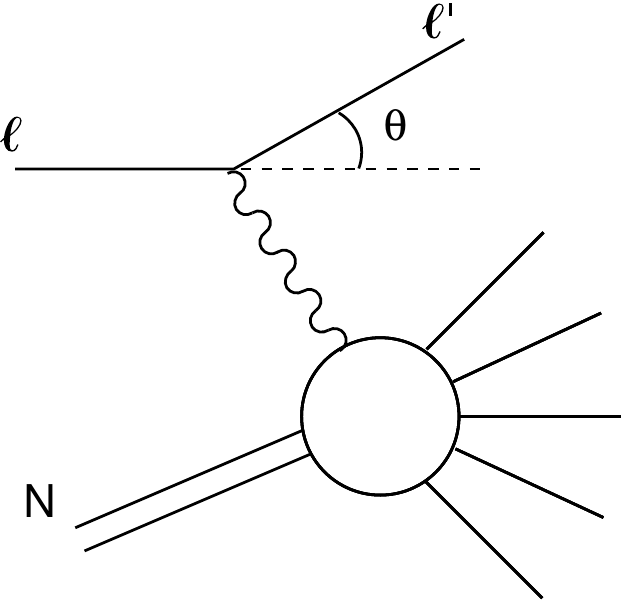} 
\caption[]{Deep inelastic lepto-production.}
\label{fig6}
\end{figure}

A very important class of hard processes is Deep Inelastic Scattering (DIS)
\beq 
l~+~N\rightarrow l'~+~X~~~~~~~~~~~~l=e^{\pm}, \mu^{\pm}, \nu, \bar{\nu}\label{DIS}
\eeq 
which has played and still plays a very important role for our understanding of QCD and of nucleon structure. 
For the processes in Eq.~(\ref{DIS}), shown in Fig.~\ref{fig6}, we have, in the lab system where the nucleon of mass $m$ is
at rest:
\beq Q^2~=~-q^2~=~-(k-k')^2~=~4EE'\sin^2{\theta/2};~~~~~~~~m\nu~=~(p.q);~~~~~~~~x~=~\frac{Q^2}{2m\nu}\label{kin}  
\eeq
In this case the virtual momentum $q$ of the gauge boson is spacelike. $x$ is the familiar Bjorken variable. The DIS processes in QCD will be extensively discussed in Sect. \ref{sec:16}

\subsection{The Renormalization Group and Asymptotic Freedom}
\label{sec:13}
In this section we aim at providing a reasonably detailed introduction to the renormalization group formalism and the
concept of running coupling which leads to the result that QCD has the property of asymptotic freedom. We start with a
summary on how renormalization works. 

In the simplest conceptual situation imagine that we implement regularization of divergent integrals by introducing a
dimensional cut-off $K$ that respects gauge and Lorentz invariance. The dependence of renormalized quantities on
$K$ is eliminated by absorbing it into a redefinition of $m$ (the quark mass: for simplicity we assume a single
flavour here), the gauge coupling $e$ (can be $e$ in QED or $e_s$ in QCD) and the wave function renormalization factors
$Z^{1/2}_{q,g}$ for $q$ and $g$, using suitable renormalization conditions (that is precise definitions of $m$, $g$ and $Z$ that can be implemented order by order in perturbation theory). For
example we can define the renormalized mass $m$ as the position of the pole in the quark propagator and, similarly, the
normalization $Z_q$ as the residue at the pole:
\beq
\rm{Propagator}~=~\frac{Z_q}{p^2-m^2}~+~\rm{no-pole~terms}\label{mZ}\\
\eeq
The renormalized coupling $e$ can be defined in terms of a renormalized 3-point vertex at some specified values of the
external momenta. Precisely, we consider a one particle irreducible vertex (1PI). We recall that a connected Green function is the sum of all connected diagrams, while 1PI Green functions are the sum of all diagrams that cannot be separated into two disconnected parts by cutting only one line.

We now become more specific by concentrating on the case of massless QCD. If we start from a vanishing
mass at the classical (or "bare") level, $m_0=0$, the mass is not renormalized because it is protected by a symmetry,
chiral symmetry. The conserved currents of chiral symmetry are axial currents: $\bar q\gamma_{\mu}\gamma_5q$. The
divergence of the axial current gives, by using the Dirac equation, $\partial^{\mu}(\bar q\gamma_{\mu}\gamma_5q)~=~2m\bar
q\gamma_5q$. So the axial current and the corresponding axial charge are conserved in the massless limit. Actually the singlet axial current is not conserved due to the anomaly, but, since QCD 
is a vector theory we have not to worry about chiral anomalies in the present context. As there are no $\gamma_5$ around the chosen regularization 
preserves chiral symmetry besides gauge and Lorentz symmetry and the renormalized mass remains zero. The renormalized
propagator has the form in Eq.~(\ref{mZ}) with $m=0$. 

\begin{figure}[h]
\noindent
\includegraphics[width=9cm]{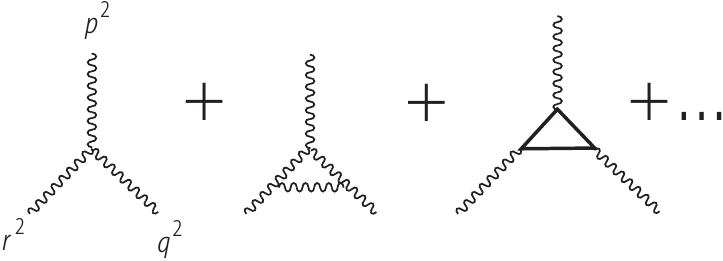} 
\caption[]{Diagrams contributing to the 1PI 3-gluon vertex at the one-loop approximation level.}
\label{fig7}
\end{figure}

The renormalized coupling $e_s$ can be defined from the
renormalized 1PI 3-gluon vertex at a scale $-\mu^2$ (Fig.~\ref{fig7}):
\beq
V_{bare}(p^2,q^2,r^2)~=~ZV_{ren}(p^2,q^2,r^2),~~~~Z=Z_g^{-3/2},~~~~V_{ren}(-\mu^2,-\mu^2,-\mu^2) \rightarrow e_s~\label{e}\\
\eeq
We could as well use the quark-gluon vertex or any other vertex which coincides with $e_{s0}$ in lowest order (even the ghost-gluon vertex, if we want). With a regularization and renormalization that preserves gauge invariance we are guaranteed that all these different definitions are equivalent.

Here $V_{bare}$ is what is obtained from computing the Feynman diagrams including, for example, the 1-loop corrections at 
the lowest non
trivial order. $V_{bare}$ is defined as the scalar function multiplying the 3-gluon vertex tensor (given in Fig. \ref{QCDFeyn}), normalized in such a way that it coincides with $e_{s0}$ in lowest order. $V_{bare}$ contains the
cut-off
$K$ but does not know about
$\mu$. $Z$ is a factor that depends both on the cut-off and on $\mu$ but not on momenta. Because of infrared singularities the defining scale
$\mu$ cannot vanish. The negative value
$-\mu^2<0$ is chosen to stay away from physical cuts (a gluon with negative virtual mass cannot decay). Similarly, 
in the massless theory, we can define $Z_g^{-1}$ as the inverse gluon propagator (the 1PI 2-point function) at the same scale $-\mu^2$ (the
vanishing mass of the gluon is guaranteed by gauge invariance). 

After computing
all 1-loop diagrams indicated in Fig.~\ref{fig7} we have:
\bea
V_{bare}(p^2,p^2,p^2)~&=&~e_{s0}[1+c\alpha_{s0}\cdot \log{\frac{K^2}{p^2}~+...]}~=~\nonumber\\
&=&~[1+c\alpha_{s}\cdot
\log{\frac{K^2}{-\mu^2}~+~...]}e_{s0}[1+c\alpha_{s0}\cdot
\log{\frac{-\mu^2}{p^2}]}\cr
&=&~Z_V^{-1}e_{s0}[1+c\alpha_{s}\cdot
\log{\frac{-\mu^2}{p^2}]}\cr
&=&~[1+d\alpha_{s}\cdot
\log{\frac{K^2}{-\mu^2}~+~...]}e_s[1+c\alpha_{s}\cdot
\log{\frac{-\mu^2}{p^2}]}\cr
&=&~Z_g^{-3/2}V_{ren}
\label{Vren}
\eea
Note the replacement of $\alpha_{s0}$ with $\alpha_s$ in the second step, as we work at 1-loop accuracy. Then we change $e_{s0}$ into $e_s$ given by $e_0=Z_g^{-3/2}Z_Ve$ and this implies changing $c$ into $d$ in the first bracket. The definition of $e_s$ demands that one precisely
specifies what is included in $Z$. For this, in a given renormalization scheme, a prescription is
fixed to specify the finite terms that go into Z (i.e. the terms of order $\alpha_s$ that accompany $\log{K^2}$).
Then
$V_{ren}$ is specified and the renormalized coupling is defined from it according to Eq.~(\ref{e}). For example, in the
momentum subtraction scheme we define $V_{ren}(p^2,p^2,p^2)=e_s~+~V_{bare}(p^2,p^2,p^2)-V_{bare}(-\mu^2,-\mu^2,-\mu^2)$,
which is equivalent to say, at 1-loop, that all finite terms that do not vanish at $p^2=-\mu^2$ are included in Z.

A crucial observation is that $V_{bare}$ depends on $K$ but not on $\mu$, which is only introduced when Z,
$V_{ren}$ and hence $\alpha_{s}$ are defined. (From here on, for shorthand, we write $\alpha$ to indicate either the QED
coupling or the QCD coupling $\alpha_{s}$). More in general for a generic Green function G, we similarly have:
\beq
G_{bare}(K^2,\alpha_0,p_i^2)~=~Z_G G_{ren}(\mu^2,\alpha,p_i^2)\label{G}\\
\eeq
so that we have:
\beq
\frac{dG_{bare}}{d\log{\mu^2}}~=~\frac{d}{d\log{\mu^2}}[Z_G G_{ren}]~=~0\label{dG}\\
\eeq
or
\beq
Z_G [\frac{\partial}{\partial\log{\mu^2}}~+~\frac{\partial
\alpha}{\partial\log{\mu^2}}\frac{\partial}{\partial \alpha}~+~
\frac{1}{Z_G }\frac{\partial
Z_G }{\partial\log{\mu^2}}]G_{ren}~=~0\label{RGE1}\\
\eeq
Finally the renormalization group equation (RGE) can be written as:
\beq
[\frac{\partial}{\partial\log{\mu^2}}~+~\beta(\alpha)\frac{\partial}{\partial \alpha}~+~
\gamma_G(\alpha)]G_{ren}~=~0\label{RGE2}\\
\eeq
where
\beq
\beta(\alpha)~=~\frac{\partial\alpha}{\partial\log{\mu^2}}\label{beta}\\
\eeq
and
\beq
\gamma_G(\alpha)~=~\frac{\partial\log{Z_G }}{\partial\log{\mu^2}}\label{gamma}\\
\eeq
Note that $\beta(\alpha)$ does not depend on which Green function $G$ we are considering; actually it is a property of the
theory and of the renormalization scheme adopted, while $\gamma_G(\alpha)$ also depends on $G$. Strictly speaking the RGE as written above is only valid in the Landau gauge ($\lambda=0$). In other gauges an additional term that takes the variation of the gauge fixing parameter $\lambda$ should also be included. We omit this term, for simplicity, as it is not relevant at the 1-loop level.

Assume that we want to apply the RGE to some hard process at a large scale $Q$, related to a Green function G that we can
always take as dimensionless (by multiplication by a suitable power of $Q$). Since the interesting dependence on $Q$ 
will be logarithmic we introduce the variable $t$ as :
\beq
t~=~\log{\frac{Q^2}{\mu^2}}\label{t}\\
\eeq
Then we can write $G_{ren}\equiv F(t,\alpha,x_i)$ where $x_i$ are scaling variables (we often omit to write them in the
following). In the naive scaling limit
$F$ should be independent of $t$, according to the classical intuition that massless QCD is scale invariant. To find the actual dependence on $t$, we want to solve the RGE
\beq
[-\frac{\partial}{\partial t}~+~\beta(\alpha)\frac{\partial}{\partial \alpha}~+~
\gamma_G(\alpha)]G_{ren}~=~0\label{RGE3}\\
\eeq
with a given boundary condition at $t=0$ (or $Q^2=\mu^2$): $F(0,\alpha)$. 

We first solve the RGE in the simplest case that $\gamma_G(\alpha)=0$. This is not an unphysical case: for example, it
applies to $R=R_{e^+e^-}=\sigma(e^+e^-\rightarrow
hadrons)/\sigma_{point}(e^+e^-\rightarrow \mu^+\mu^-)$ where the vanishing of $\gamma$ is related to the non renormalization of the electric charge in
QCD (otherwise the proton and the electron charge would not exactly compensate each other: this will explained in Sect.~\ref{sec:15}).
So we consider the equation:
\beq
[-\frac{\partial}{\partial t}~+~\beta(\alpha)\frac{\partial}{\partial \alpha}]G_{ren}~=~0\label{RGE4}\\
\eeq
The solution is simply
\beq
F(t,\alpha)~=~F[0,\alpha(t)]\label{Fsol1}\\
\eeq
where the "running coupling" $\alpha(t)$ is defined by:
\beq
t~=~\int_{\alpha}^{\alpha(t)}\frac{1}{\beta(\alpha')}d\alpha'\label{run}\\
\eeq
Note that from this definition it follows that $\alpha(0)=\alpha$, so that the boundary condition is also satisfied.
To prove that $F[0,\alpha(t)]$ is indeed the solution, we first take derivatives with respect of $t$ and $\alpha$ (the two
independent variables) of both sides of Eq.~(\ref{run}). By taking $d/dt$ we obtain
\beq
1~=~\frac{1}{\beta(\alpha(t))}\frac{\partial\alpha(t)}{\partial t}\label{ddt}\\
\eeq
We then take $d/d\alpha$ and obtain
\beq
0~=~-\frac{1}{\beta(\alpha)}~+~\frac{1}{\beta(\alpha(t))}\frac{\partial\alpha(t)}{\partial \alpha}\label{dda}\\
\eeq
These two relations make explicit the dependence of the running coupling on $t$ and $\alpha$:
\bea
\frac{\partial\alpha(t)}{\partial t}~=~\beta(\alpha(t))\label{runt}\\
\frac{\partial\alpha(t)}{\partial \alpha}~=~\frac{\beta(\alpha(t))}{\beta(\alpha)}\label{runa}\\
\nonumber
\eea
Using these two equations one immediately checks that $F[0,\alpha(t)]$ is indeed the solution.

Similarly, one finds that the solution of the more general equation with $\gamma\not=0$, Eq.~(\ref{RGE3}), is given by:
\beq
F(t,\alpha)~=~F[0,\alpha(t)]\exp{\int_{\alpha}^{\alpha(t)}\frac{\gamma(\alpha')}{\beta(\alpha')}d\alpha'}\label{Fsol2}\\
\eeq
In fact the sum of the two derivatives acting on the factor $F[0,\alpha(t)]$ vanishes (as we have just seen) and the exponential is by itself a
solution of the complete equation. Note that the boundary condition is also satisfied.

The important point is the appearance of the running coupling that determines the asymptotic departures from scaling. The
next step is to study the functional form of the running coupling. From Eq.~(\ref{runt}) we see that the rate of change
with $t$ of the running coupling is determined by the $\beta$ function. In turn $\beta(\alpha)$ is determined by the $\mu$
dependence of the renormalized coupling through Eq.~(\ref{beta}). Clearly there is no dependence on $\mu$ of the basic
3-gluon vertex in lowest order (order $e$). The dependence starts at 1-loop, that is at order $e^3$ (one extra gluon has
to be emitted and reabsorbed). Thus we obtain that in perturbation theory:
\beq
\frac{\partial e}{\partial \log{\mu^2}}~\propto~e^3\label{de}\\
\eeq
Recalling that $\alpha~=~e^2/4\pi$, we have:
\beq
\frac{\partial \alpha}{\partial \log{\mu^2}}~\propto~2e\frac{\partial e}{\partial \log{\mu^2}}~\propto~e^4
~\propto \alpha^2\label{da}\\
\eeq
Thus the behaviour of $\beta(\alpha)$ in perturbation theory is as follows:
\beq
\beta(\alpha)~=~\pm b\alpha^2[1~+~b'\alpha~+...]\label{betapert}\\
\eeq
Since the sign of the leading term is crucial in the following discussion, we stipulate that always $b>0$ and we
make the sign explicit in front. 

Let us make the procedure more precise for computing the 1-loop beta function in QCD (or, similarly, in QED). The result of the 1loop 1PI diagrams for $V_{ren}$ can be written down as:
\beq
V_{ren}~=~e[1+\alpha B_{3g}\log{ \frac {\mu^2}{-p^2}}] \label{1loopV} \\
\eeq
$V_{ren}$ satisfies the RGE:
\beq
[\frac{\partial}{\partial\log{\mu^2}}~+~\beta(\alpha)\frac{\partial e}{\partial \alpha}\frac{\partial }{\partial e}~-~
\frac{3}{2}\gamma_g(\alpha)]V_{ren}~=~0\label{RGEV}\\
\eeq
With respect to Eq.~(\ref{RGE2}) the beta function term has been rewritten taking into account that $V_{ren}$ starts with $e$ and the anomalous dimension term arises from a factor $Z_g^{-1/2}$ for each gluon leg. In general for a n-leg 1PI Green function $V_{n,bare}=Z_g^{-n/2}V_{n,ren}$, if all external legs are gluons.  Note that in the particular case of $V=V_3$ that is used to define $e$ other Z factors are absorbed in the replacement $Z_V^{-1}Z_g^{3/2}e_0=e$. At 1-loop accuracy we replace $\beta(\alpha)=-b\alpha^2$ and $\gamma_g(\alpha)=\gamma_g^{(1)} \alpha$. All together one obtains:
\beq
b=2(B_{3g}-\frac{3}{2}\gamma_g^{(1)})\label{bvsB}\\
\eeq
Similarly we can write the diagrammatic expression and the RGE for the 1PI 2-gluon Green function which is the inverse gluon propagator $\Pi$ (a scalar function after removing the gauge invariant tensor):
\beq
\Pi_{ren}~=~[1+\alpha B_{2g}\log{\frac{\mu^2}{-p^2}}+\dots] \label{1lV2}\\
\eeq
and
\beq
[\frac{\partial}{\partial\log{\mu^2}}~+~\beta(\alpha)\frac{\partial}{\partial \alpha}~-~
\gamma_g(\alpha)]\Pi_{ren}~=~0\label{RGEPi}\\
\eeq
Notice that the normalization and the phase of $\Pi$ are specified by the lowest order term being 1. In this case the $\beta$ function term is negligible being of order $\alpha^2$ (because $\Pi$ is a function of $e$ only through $\alpha$) and we obtain:
\beq
\gamma_g^{(1)}=B_{2g} \label{B2g}\\
\eeq
Thus, finally:
\beq
b=2(B_{3g}-\frac{3}{2}B_{2g})\label{bfin}\\
\eeq

By direct calculation at 1-loop one finds:
\beq
\rm{QED:}~~~~~~~~\beta(\alpha)~\sim~+b\alpha^2~+.....~~~~~~~~~~~b~=~\sum_i\frac{N_CQ^2_i}{3\pi}
\label{beQED}\\
\eeq
where $N_C = 3$ for quarks and $N_C = 1$ for leptons and the sum runs over all fermions of charge $Q_ie$ that are coupled. Also, one finds:
\beq
\rm{QCD:}~~~~~~~~\beta(\alpha)~\sim~-b\alpha^2~+.....~~~~~~~~~~~b~=~\frac{11N_C-2n_f}{12\pi}\label{beQCD}\\
\eeq
where, as usual, $n_f$ is the number of coupled (see below) flavours of quarks (we assume here that $n_f~\le~16$ so that $b>0$ in QCD).
If
$\alpha(t)$ is small we can compute
$\beta(\alpha(t))$ in perturbation theory. The sign in front of $b$ then decides the slope of the coupling: $\alpha(t)$
increases with t (or
$Q^2$) if $\beta$ is positive at small $\alpha$ (QED), or $\alpha(t)$ decreases with t (or
$Q^2$) if $\beta$ is negative at small $\alpha$ (QCD). A theory like QCD where the running coupling vanishes
asymptotically at large $Q^2$ is called (ultraviolet) "asymptotically free". An important result that has been proven \cite{cogro} is
that in 4 spacetime dimensions all and only non-abelian gauge theories are asymptotically free.

Going back to Eq.~(\ref{run}) we replace $\beta(\alpha)~\sim~\pm b\alpha^2$, do the integral and perform a simple algebra.
We find 
\beq
\rm{QED:}~~~~~~~~\alpha(t)~\sim~\frac{\alpha}{1-b\alpha t}\label{beQED1}\\
\eeq
and
\beq
\rm{QCD:}~~~~~~~~\alpha(t)~\sim~\frac{\alpha}{1+b\alpha t}\label{beQCD1}\\
\eeq
A slightly different form is often used in QCD. Defining $1/\alpha~=~b\log{\mu^2/\Lambda_{QCD}^2}$ we can write:
\beq
\alpha(t)~\sim~\frac{1}{\frac{1}{\alpha}~+~bt}~=~\frac{1}{b\log{\frac{\mu^2}{\Lambda_{QCD}^2}}~+~b\log{\frac{Q^2}{\mu^2}}}
~=~\frac{1}{b\log{\frac{Q^2}{\Lambda_{QCD}^2}}}\label{alfaQCD}\\
\eeq
The parameter $\mu$ has been traded for the parameter $\Lambda_{QCD}$. We see that $\alpha(t)$ decreases logarithmically with $Q^2$ and that one can introduce a dimensional parameter
$\Lambda_{QCD}$ that replaces $\mu$. Often in the following we will simply write $\Lambda$ for $\Lambda_{QCD}$. Note that it is clear that $\Lambda$
depends on the particular definition of $\alpha$, not only on the defining scale $\mu$ but also on the renormalization
scheme (see, for example, the discussion in the next section). Through the parameter $b$, and in general through the
$\beta$ function, it also depends on the number $n_f$ of coupled flavours. It is very important to note that QED and QCD
are theories with "decoupling":  up to the scale $Q$ only quarks with masses $m<<Q$
contribute to the running of $\alpha$. This is clearly very important, given that
all applications of perturbative QCD so far apply to energies below the top quark mass $m_t$. For the validity of the
decoupling theorem \cite{ApCa} it is necessary that the theory where all the heavy particle internal lines are eliminated is still
renormalizable and that the coupling constants do not vary with the mass. These requirements are true for the mass of heavy
quarks in QED and QCD, but are not true in the electroweak theory where the elimination of the top would violate $SU(2)$
symmetry (because the t and b left-handed quarks are in a doublet) and the quark couplings to the Higgs multiplet (hence to the
longitudinal gauge bosons) are proportional to the mass. In conclusion, in QED and QCD, quarks with $m>>Q$ do not contribute
to $n_f$ in the coefficients of the relevant $\beta$ function. The effects of heavy quarks are power suppressed and can be
taken separately into account. For example, in $e^+e^-$ annihilation for
$2m_c<Q<2m_b$ the relevant asymptotics is for $n_f=4$, while for $2m_b<Q<2m_t$ $n_f=5$. Going accross the $b$ threshold
the $\beta$ function coefficients change, so the $\alpha(t)$ slope changes. But $\alpha(t)$ is continuous, so that
$\Lambda$ changes so as to keep $\alpha(t)$ constant at the matching point at $Q\sim O(2m_b)$. The effect on $\Lambda$ is
large: approximately $\Lambda_5~\sim~0.65\Lambda_4$ where $\Lambda_{4,5}$ are for $n_f=4,5$. 

Note the presence of a pole in Eqs.(\ref{beQED1},\ref{beQCD1}) at $\pm b\alpha t~=~1$, called the Landau pole, who realised
its existence in QED already in the '50's. For $\mu~\sim m_e$ (in QED) the pole occurs beyond the Planck mass. In QCD the Landau
pole is located for negative $t$ or at $Q<\mu$ in the region of light hadron masses. Clearly the issue of the definition
and the behaviour of the physical coupling (which is always finite, when defined in terms of some physical process) in the region around the perturbative Landau pole is a problem that lies
outside the domain of perturbative QCD.

The non leading terms in the asymptotic behaviour of the running coupling can in principle be evaluated going back to
Eq.~(\ref{betapert}) and computing $b'$ at 2-loops and so on. But in general the perturbative coefficients of
$\beta(\alpha)$ depend on the definition of the renormalized coupling $\alpha$ (the renormalization scheme), so one
wonders whether it is worthwhile to do a complicated calculation to get $b'$ if then it must be repeated for a
different definition or scheme. In this respect it is interesting to remark that actually both $b$ and $b'$ are independent
of the definition of $\alpha$, while higher order coefficients do depend on that. Here is the simple proof. Two different
perturbative definitions of $\alpha$ are related by $\alpha'~\sim~\alpha(1~+~c_1\alpha~+~...)$. Then we have:
\bea
\beta(\alpha')~=~\frac{d\alpha'}{d\log{\mu^2}}~&=&~\frac{d\alpha}{d\log{\mu^2}}(1~+~2c_1\alpha~+~...)\nonumber\\
&~=~&\beta(\alpha)(1~+~2c_1\alpha~+~...)\nonumber\\
&~=~&\pm b\alpha^2(1~+~b'\alpha~+~...)(1~+~2c_1\alpha~+~...)\nonumber\\
&~=~&\pm b\alpha'^2(1~+~b'\alpha'~+~...)
\label{proof}
\eea
which shows that, up to the first subleading order, $\beta(\alpha')$ has the same form as $\beta(\alpha)$. 

In QCD ($N_C=3$)
one has calculated \cite{bpr}:
\beq
b'~=~\frac{153-19n_f}{2\pi(33-2n_f)}\label{b'}\\
\eeq
By taking $b'$ into account one can write the expression of the running coupling at next to the leading order (NLO):
\beq
\alpha(Q^2)~=~\alpha_{LO}(Q^2)[1~-~b'\alpha_{LO}(Q^2)\log{\log{\frac{Q^2}{\Lambda^2}}}~+~...]\label{NLOa}\\
\eeq
where $\alpha_{LO}^{-1}~=~b\log{Q^2/\Lambda^2}$ is the LO result (actually at NLO the definition of $\Lambda$ is modified according to $b\log{\mu^2/\Lambda^2}=1/\alpha+b'\log{b\alpha}$).

Summarizing, we started from massless classical QCD which is scale invariant. But we have seen that the procedure of
quantization, regularization and renormalization necessarily breaks scale invariance. In the quantum QCD theory there is a
scale of energy, $\Lambda$, which from experiment is of the order of a few hundred MeV, its precise value depending
on the definition, as we shall see in detail. Dimensionless quantities depend on the energy scale through the running
coupling which is a logarithmic function of
$Q^2/\Lambda^2$. In QCD the running coupling decreases logarithmically at large $Q^2$ (asymptotic freedom), while in QED
the coupling has the opposite behaviour.

\subsection{More on the Running Coupling}
\label{sec:14}
 
In the previous section we have introduced the renormalized coupling $\alpha$ in terms of the 3-gluon vertex at
$p^2=-\mu^2$ (momentum subtraction). The Ward identities of QCD then ensure that the coupling defined from other vertices
like the $\bar q qg$ vertex  are renormalized in the same way and the finite radiative corrections are related. But at
present the universally adopted definition of $\alpha_s$ is in terms of dimensional regularization \cite{dimreg}, because of computational
simplicity, which is essential given the great complexity of present day calculations. So we now briefly review the
principles of dimensional regularization and the definition of Minimal Subtraction ($MS$) \cite{MS} and Modified Minimal Subtraction
($\overline{MS} $) \cite{MSbar}. The $\overline{MS}$ definition of $\alpha_s$ is the one most commonly adopted in the literature and a
value quoted for it is normally referring to this definition.

Dimensional Regularization (DR) is a gauge and Lorentz invariant regularization that consists in formulating the theory in 
$D<4$ spacetime dimensions in order to make loop integrals ultraviolet finite. In DR one rewrites the theory in D
dimensions (D is integer at the beginning, but then one realizes that the calculated expression of  diagrams makes sense at all D except
for isolated singularities). The metric tensor is extended into a $D\times D$ matrix $g_{\mu\nu}~=~diag(1,-1,-1,....,-1)$ and
4-vectors are given by $k^{\mu}~=~(k^0,k^1,...,k^{D-1})$. The Dirac $\gamma^{\mu}$ are $f(D) \times f(D)$ matrices and the precise form of the function $f(D)$ is not important. It is sufficient to extend the usual algebra in a
straightforward way like $\{ \gamma_\mu,\gamma_\nu \}=2g_{\mu,\nu}I$, with $I$ the D-dimensional identity matrix, $\gamma^{\mu}\gamma^{\nu}\gamma_{\mu}~=~-(D-2)\gamma^{\nu}$ or
$Tr(\gamma^{\mu}\gamma^{\nu})~=~f(D)g_{\mu\nu}$. 

The physical dimensions of fields change in D dimensions and, as a
consequence, the gauge couplings become dimensional $e_D~=~\mu^\epsilon e$, where $e$ is dimensionless, $D~=~4-2\epsilon$
and $\mu$ is a scale of mass (this is how a scale of mass is introduced in the DR of massless QCD!). In fact, the dimension of
fields is determined by requiring that the action $S~=~\int d^Dx {\cal L}$ is dimensionless. By inserting for ${\cal L}$ terms
like
$m\bar\Psi \Psi$ or $m^2\phi^{\dagger} \phi$ or $e\bar\Psi \gamma^{\mu}\Psi A_{\mu}$ the dimensions of the fields and
coupling are determined as: $m, \Psi, \phi, A_\mu, e~=~1, (D-1)/2, (D-2)/2, (D-2)/2, (4-D)/2$, respectively. The formal
expression of loop integrals can be written for any D. For example:
\beq
\int\frac{d^Dk}{(2\pi)^D}\frac{1}{(k^2-m^2)^2}~=~\frac{\Gamma(2-D/2)(-m^2)^{D/2-2}}{(4\pi)^{D/2}}\label{int}\\
\eeq
For $D~=~4-2\epsilon$ one can expand using:
\beq
\Gamma(\epsilon)~=~\frac{1}{\epsilon}~-~\gamma_E~+~O(\epsilon),~~~~~~~~~~\gamma_E~=~0.5772.....\label{gamexp}\\
\eeq
For some Green function G, normalized to 1 in lowest order, (like $V/e$ with $V$ the 3-$g$ vertex function at the symmetric
point $p^2=q^2=r^2$, considered in the previous section) we typically find at 1-loop:
\beq
G_{bare}~=~1~+~\alpha_0(\frac{-\mu^2}{p^2})^\epsilon~[B(\frac{1}{\epsilon}+\log{4\pi}-\gamma_E)~+~A~+~O(\epsilon)]\label{Gexp}\\
\eeq
In $\overline {MS}$  one rewrites this as (diagram by diagram: this is a virtue of the method): 
\bea
G_{bare}&=&ZG_{ren}\nonumber\\
Z&=&1~+~\alpha~[B(\frac{1}{\epsilon}+\log{4\pi}-\gamma_E)]\nonumber\\
G_{ren}&=&1~+~\alpha~[B\log{\frac{-\mu^2}{p^2}}+A]\label{MSbar}
\eea
Here $Z$ stands for the relevant product of renormalization factors. In the original $MS$ prescription only $1/\epsilon$ was subtracted (that clearly plays the role of a cutoff) and not also
$\log{4\pi}$ and
$\gamma_E$. Later, since these constants  always appear from the expansion of $\Gamma$ functions it was decided to modify
$MS$ into
$\overline {MS}$. Note that the $\overline {MS}$ definition of $\alpha$ is different than that in the momentum subtraction
scheme because the finite terms (those beyond logs) are different. In particular here the order $\alpha$ correction to $G_{ren}$ does not vanish at
$p^2=-\mu^2$. 

The third \cite{tar} and fourth \cite{ver} coefficients of the QCD $\beta$ function are also known in the $\overline{MS}$ prescription (recall that only the
first two coefficients are scheme independent). The calculation of the last term involved the evaluation of some 50,000 4-loop diagrams. Translated in numbers, for $n_f=5$ one obtains :
\beq
\beta(\alpha)~=~-0.610\alpha^2[1~+~1.261...\frac{\alpha}{\pi}~+~1.475...(\frac{\alpha}{\pi})^2~+~9.836...(\frac{\alpha}{\pi})^3...]\label{beta4}\\
\eeq 
It is interesting to remark that the expansion coefficients are of order 1 or 10 (only for the last one), so that the $\overline {MS}$ expansion looks
reasonably well behaved. 

\subsection{On the Non-convergence of Perturbative Expansions}
\label{borel}

It is important to keep in mind that the QED and QCD perturbative series, after renormalization, have all their
coefficients finite, but the expansion does not converge. Actually the perturbative series is not
even Borel summable (for reviews, see, for example refs. \cite{sixteen}). After the Borel resummation, for a given process one is left with a result which is ambiguous by terms
typically down by $\exp{-n/(b\alpha)}$, with $n$ an integer and $b$ is the absolute value of the first $\beta$ function coefficient. In QED these
corrective terms are extremely small and not very important in practice. On the contrary in QCD $\alpha=\alpha_s(Q^2)\sim
1/(b\log{Q^2/\Lambda^2})$ and the ambiguous terms are of order $(1/Q^2)^n$, that is are power suppressed. It is interesting that, through this mechanism, the perturbative version of the theory is able to somehow take into account the power suppressed corrections. A sequence of diagrams with factorial growth at large order $n$ is made up by dressing  gluon propagators by any number of quark bubbles together with their gauge completions (renormalons). The problem of the precise relation between the ambiguities of the perturbative expansion and the power suppressed corrections has been discussed in recent years, also in processes without light cone operator expansion \cite{sixteen,shifren}.

\clearpage

\subsection{$e^+e^-$ Annihilation and Related Processes}
\label{sec:15}

\subsubsection{$R_{e^+e^-}$ }
\label{sec:15.1}

\begin{figure}[h]
\noindent
\includegraphics[width=11cm]{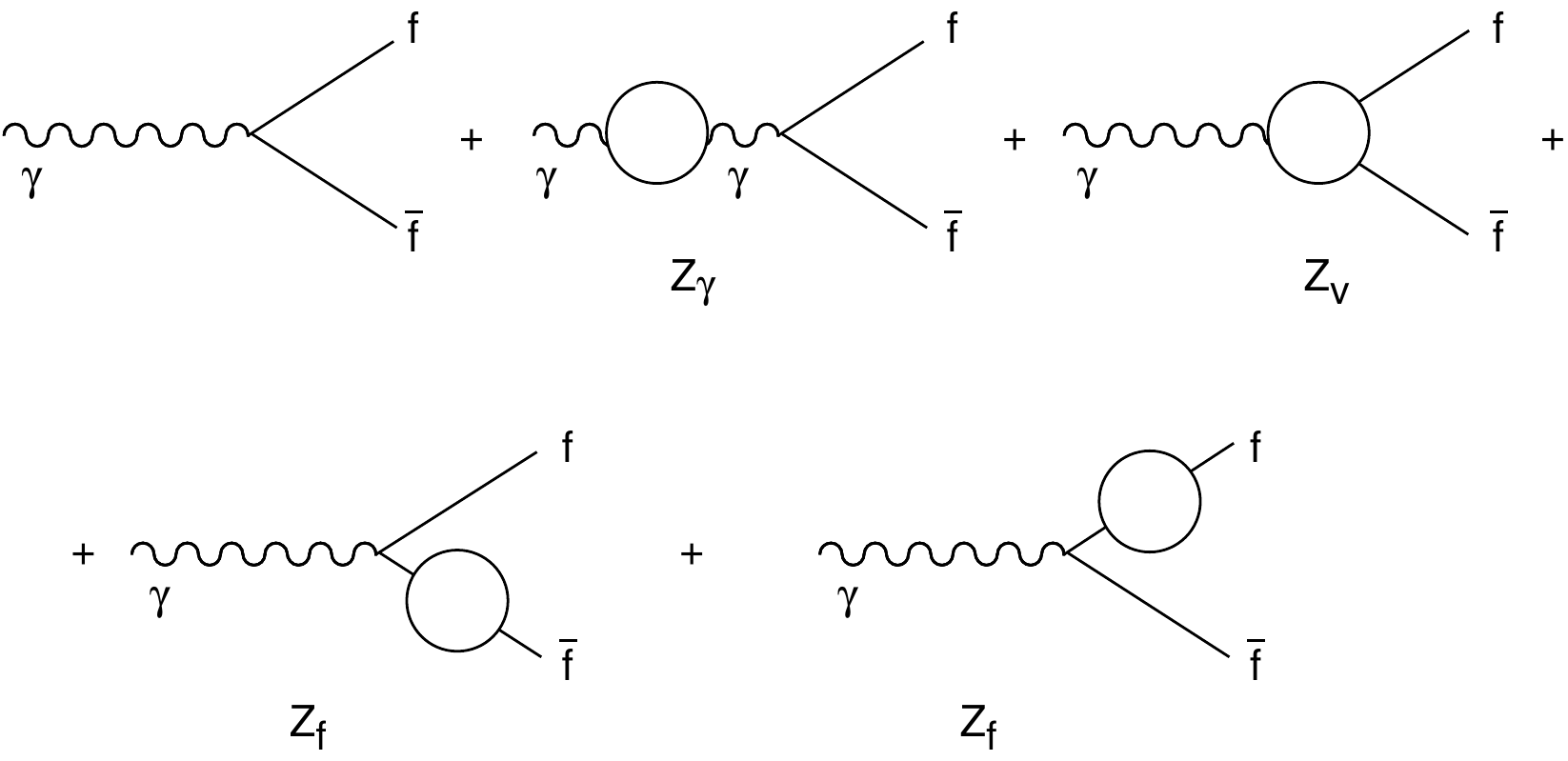} 
\caption[]{Diagrams for charge renormalization in QED at 1-loop (the blob, in each diagram, represents the loop).}
\label{fig8}
\end{figure}

The simplest hard process is $R=R_{e^+e^-}=\sigma(e^+e^-\rightarrow
hadrons)/\sigma_{point}(e^+e^-\rightarrow \mu^+\mu^-)$ that we have already introduced. $R$ is dimensionless and in perturbation
theory is given by
$R~=~N_C\sum_i Q^2_i F(t,\alpha_s)$, where $F~=~1+O(\alpha_s)$. \footnote{ Actually starting from the order $\alpha_s^2$ there are some "singlet" terms proportional to $[\sum_i Q_i]^2$. These small terms are included in $F$ by dividing and multiplying by $\sum_i Q^2_i$.}  We have already mentioned that for this process the
"anomalous dimension" function vanishes: $\gamma(\alpha_s)=0$ because of electric charge non renormalization by strong
interactions. Let us recall how this happens in detail. The diagrams that are relevant for charge renormalization in QED
at 1-loop are shown in Fig.~\ref{fig8}. The Ward identity that follows from gauge invariance in QED imposes that the vertex
($Z_V$) and the self-energy
($Z_f$) renormalization factors cancel and the only divergence remains in $Z_\gamma$, the vacuum polarization of the
photon. So the charge is only renormalized by the photon vacuum polarization blob, hence it is universal (the same factor for all fermions,
independent of their charge) and is not affected by QCD at 1-loop. It is true that at higher orders the photon vacuum
polarization diagram is affected by QCD (for example, at 2-loops we can exchange a gluon between the quarks in the loop) 
but the renormalization induced by the divergent logs from the vacuum polarization diagram remain independent of the nature of the fermion to
which the photon line is attached. The gluon contributions to the vertex
($Z_V$) and to the self-energy
($Z_f$) cancel because they have exactly the same structure as in QED, and there is no gluon contribution to the photon blob at 1-loop, so that $\gamma(\alpha_s)=0$. 

\begin{figure}[h]
\noindent
\includegraphics[width=11cm]{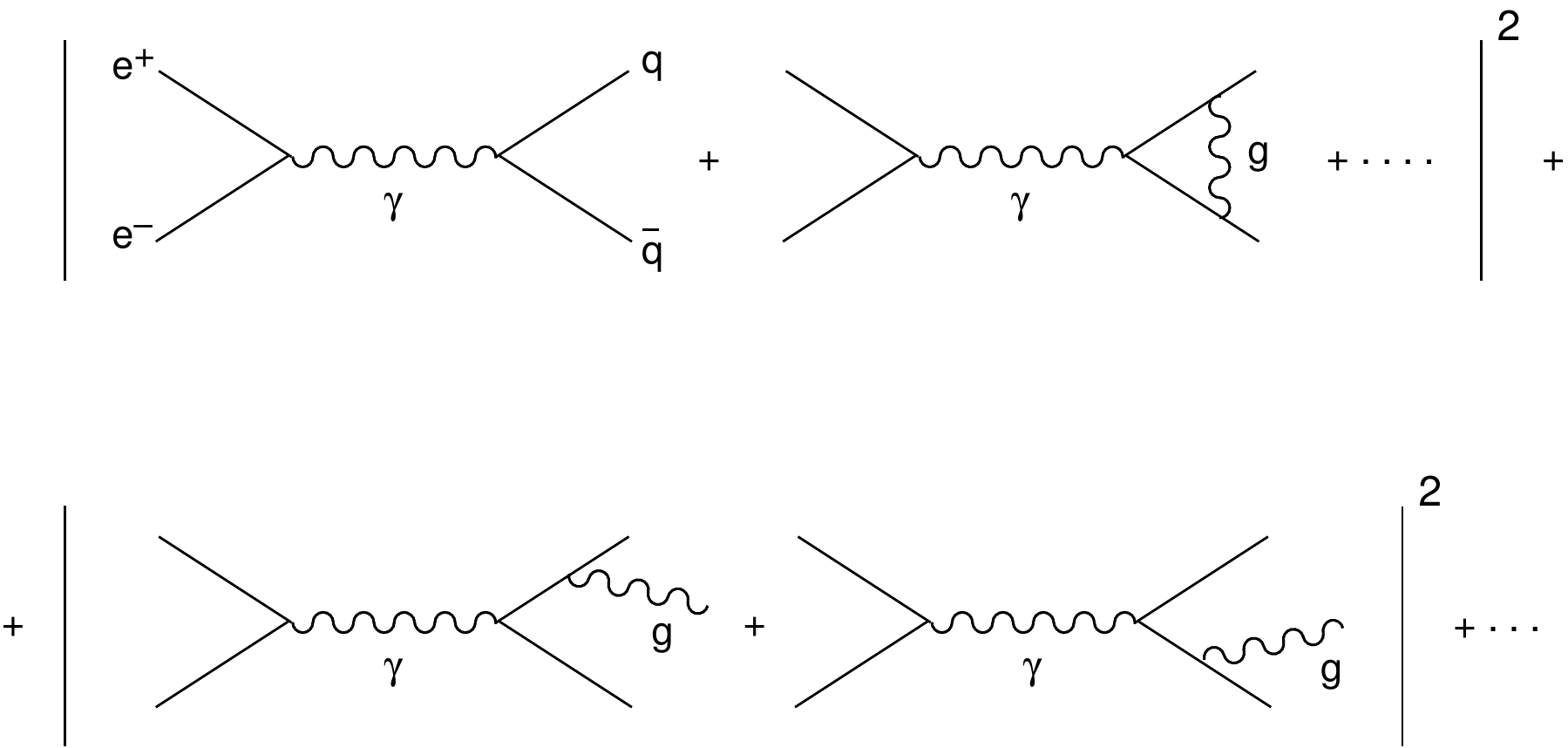} 
\caption[]{Real and virtual diagrams relevant for the computation of R at 1-loop accuracy (the initial $e^+e^-$ has been omitted to make the drawing simpler).}
\label{fig9}
\end{figure}

At 1-loop the diagrams relevant for the computation of R are shown in Fig.~\ref{fig9}. There are virtual diagrams and also real
diagrams with one additional gluon in the final state. Infrared divergences cancel  between the interference term of
the virtual diagrams and the absolute square of the real diagrams, according to the Bloch-Nordsieck theorem. Similarly
there are no mass singularities, in agreement with the Kinoshita-Lee-Nauenberg theorem, because the initial state is
purely leptonic and all degenerate states that can appear at the given order are included in the final state. Given that
$\gamma(\alpha_s)=0$ the RGE prediction is simply given, as we have already seen, by $F(t,\alpha_s)=F[0,\alpha_s(t)]$.
This means that if we do, for example, a 2-loop calculation, we must obtain a result of the form:
\beq
F(t,\alpha_s)~=~1~+~c_1\alpha_s(1-b\alpha_st)~+~c_2\alpha_s^2~+O(\alpha_s^3)\label{Fexp2}\\
\eeq  
In fact, taking into account the expression of the running coupling in Eq.~(\ref{beQCD1}):
\beq
\alpha_s(t)~\sim~\frac{\alpha_s}{1+b\alpha_s t}\sim\alpha_s(1~-~b\alpha_s t~+~....)\label{xx}\\
\eeq
Eq. \ref{Fexp2} can be rewritten as
\beq
F(t,\alpha_s)~=~1~+~c_1\alpha_s(t)~+~c_2\alpha_s^2(t)~+O(\alpha_s^3(t))~=~F[0,\alpha_s(t)]\label{firstFexp3}\\
\eeq  
The content of the RGE prediction is, at this order, that there are no $\alpha_s t$ and $(\alpha_s t)^2$ terms (the leading
log sequence must be absent) and the term of order $\alpha_s^2 t$ has the appropriate coefficient to be reabsorbed in the
transformation of $\alpha_s$ into $\alpha_s(t)$.

At present the first 4 coefficients $c_1,...,c_4$ have been computed in the $\overline {MS}$ scheme (the references are: for $c_2$ \cite{c2}, for $c_3$ \cite{gor} and for $c_4$ \cite{baik4}). Clearly $c_1=1/\pi$ does not
depend on the definition of $\alpha_s$ but $c_n$ with $n \geq 2$ do. The subleading coefficients also depend on the scale choice:
if instead of expanding in $\alpha_s(Q)$ we decide to choose $\alpha_s(Q/2)$ the coefficients $c_n$ $n \geq 2$ change. In
the  
$\overline {MS}$ scheme, for $\gamma$-exchange and $n_f=5$, which are good approximations for $2m_b<<Q<<m_Z$, one has:
\beq
F[0,\alpha_s(t)]~=~1~+~\frac{\alpha_s(t)}{\pi}~+~1.409...(\frac{\alpha_s(t)}{\pi})^2~-~12.8....(\frac{\alpha_s(t)}{\pi})^3~-
~80.0...(\frac{\alpha_s(t)}{\pi})^4+...\label{Fexp3}\\
\eeq

Similar perturbative results at 3-loop accuracy also exist for $R_Z=\Gamma(Z\rightarrow hadrons)/\Gamma(Z\rightarrow
leptons)$, $R_{\tau}=\Gamma(\tau\rightarrow \nu_{\tau}+hadrons)/\Gamma(\tau\rightarrow \nu_{\tau}+ leptons)$, etc. We
will discuss these results in Sect. \ref{sec:18} where we deal with measurements of $\alpha_s$.

The perturbative expansion in powers of $\alpha_s(t)$ takes into account all contributions that are suppressed by powers
of logarithms of the large scale $Q^2$ ("leading twist" terms). In addition there are corrections suppressed by powers of
the large scale $Q^2$ ("higher twist" terms).  The pattern of power corrections is controlled by the light-cone Operator
Product Expansion (OPE) \cite{willc,BrPr} which (schematically) leads to:
\beq
F~=~\rm{pert.}~+~r_2\frac{m^2}{Q^2}~+~r_4\frac{<0|Tr[\bf{F_{\mu\nu}}\bf{F^{\mu\nu}}]|0>}{Q^4}~+~...~+~
r_6\frac{<0|O_6|0>}{Q^6}~+~...\label{lce+e-}\\
\eeq
Here $m^2$ generically indicates mass corrections, for example from b quarks, beyond the b threshold, while top quark mass corrections only arise from loops, vanish in the limit $m_t\rightarrow \infty$ and are included in
the coefficients as those in Eq.~(\ref{Fexp3}) and the analogous ones for higher twist terms;
${\bf F_{\mu\nu}} =\sum_A F_{\mu\nu}^At^A$, $O_6$ is typically a 4-fermion operator, etc. For each possible gauge invariant
operator the corresponding negative power of $Q^2$ is fixed by dimensions. 

We now consider the light-cone OPE in some more detail. $R_{e^+e^-}\sim \Pi(Q^2)$ where $\Pi(Q^2)$ is the scalar spectral
function related to the hadronic contribution to the imaginary part of the photon vacuum polarization $T_{\mu\nu}$:
\bea
T_{\mu\nu}~&=&~(-g_{\mu\nu}Q^2~+~q_{\mu}q_{\nu})\Pi(Q^2)~=~\int d^4x \exp{i(q\cdot x)} <0|J_{\mu}^\dagger (x)J_{\nu}(0)|0> ~=~\nonumber\\
&=&
\sum_n<0|J_{\mu}^\dagger (0)|n><n|J_{\nu}(0)|0>(2\pi)^4\delta^4(q-p_n)\label{Tmunu}\\
\nonumber\eea
For $Q^2\rightarrow \infty$ the $x^2\rightarrow 0$ region is dominant. The light cone OPE is valid to all orders in perturbation theory. Schematically, dropping Lorentz indices, for simplicity, near $x^2\sim0$ we have:
\bea
J^\dagger (x)J(0)&=& I(x^2)~+~E(x^2)\sum_{n=0}^\infty c_n(x^2)x^{\mu_1}...x^{\mu_n}\cdot
O^n_{\mu_1...\mu_n}(0)+\rm{less~sing.~terms}~~\label{OPEJJ}
\eea
Here $I(x^2)$, $E(x^2)$,..., $c_n(x^2)$ are c-number singular functions, $O^n$ is a string of local operators. $E(x^2)$ is
the singularity of free field theory, $I(x^2)$ and $c_n(x^2)$ in the interacting theory contain powers of $\log{(\mu^2 x^2)}$. Some $O^n$
are already present in free field theory, other ones appear when interactions are switched on. Given that
$\Pi(Q^2)$ is related to the Fourier transform of the vacuum expectation value of the product of currents, less singular terms in $x^2$ lead to power suppressed terms in $1/Q^2$.   
The perturbative terms, like those in Eq. \ref{firstFexp3}, come from $I(x^2)$ which is the leading twist
term and the dominant logarithmic scaling violations induced by the running coupling are the logs in $I(x^2)$.

\subsubsection{The Final State in $e^+e^-$ Annihilation}
\label{sec:15.2}

Experiments on $e^+e^-$ annihilation at high energy provide a remarkable possibility of systematically testing the distinct
signatures predicted by QCD for the structure of the final state averaged over a large number of events. Typical of
asymptotic freedom is the hierarchy of configurations emerging as a consequence of the smallness of $\alpha_s(Q^2)$. When all
corrections of order $\alpha_s(Q^2)$ are neglected one recovers the naive parton model prediction for the final state:
almost collinear events with two back-to-back jets with limited transverse momentum and an angular distribution as
$(1+\cos^2{\theta})$ with respect to the beam axis (typical of spin 1/2 parton quarks: scalar quarks would lead to a
$\sin^2{\theta}$ distribution). At order $\alpha_s(Q^2)$ a tail of events is predicted to appear with large transverse
momentum $p_T\sim Q/2$ with respect to a suitably defined jet axis (for example the thrust axis, see below). This small fraction of events with large $p_T$ mostly consists of
three-jet events with an almost planar topology. The skeleton of a three-jet event, at leading order in $\alpha_s(Q^2)$, is
formed by three hard partons $q\bar qg$, the third being a gluon emitted by a quark or antiquark line. 
At order
$\alpha_s^2(Q^2)$ a hard perturbative non planar component starts to build up and a small fraction of four-jet events $q\bar
qgg$ or $q\bar q q\bar q$ appear, and so on.

Event shape variables defined from the set of 4-momenta of final state particles are introduced to quantitatively describe the topological structure of the final state energy flow \cite{evsh}. The most well known event shape variable is thrust (T) \cite{far} defined as:
\beq
T=max \frac{
\sum_i |\vec{p_i}\cdot \vec{n_T}|}{\sum_i |\vec{p_i}|}\label{thr}\\
\eeq
where the maximization is in terms of the axis defined by the unit vector $n_T$: the thrust axis is the axis that maximizes the sum of the absolute values of the longitudinal momenta of the final state particles. The thrust $T$ varies between 1/2, for a spherical event, to 1 for a collinear (2-jet) event. Event shape variables are important for QCD tests and measurements of $\alpha_s$ and also for more practical purposes like a laboratory for assessing the reliability of event simulation programmes and a tool for the separation of signals and background.

A quantitatively specified definition of jets and of the number of jets in one event (jet counting) must be introduced for precise QCD tests and for measuring $\alpha_s$,
which must be infrared safe (i.e. not altered by soft particle emission or collinear splittings of massless particles) in
order to be computable at parton level and as much as possible insensitive to the transformation of partons into hadrons (see, for example, ref. \cite{mor}). For $e^+e^-$ physics one
has used a jet algorithm based on a resolution parameter $y_{cut}$ and a suitable pair variable; for example \cite{dur}:
\beq
y_{ij}~=~\frac{2min(E_i^2,E_j^2)(1-\cos{\theta_{ij}})}{s}\label{yij}\\
\eeq
Note that $1-\cos{\theta_{ij}} \sim  \theta_{ij}^2/2$ so that the relative transverse momentum $k_T^2$ is involved (hence the name $k_T$ algorithm). The particles i,j belong to different jets for $y_{ij}>y_{cut}$. Clearly the number of jets becomes a function of $y_{cut}$:
there are more jets for smaller $y_{cut}$. 

Recently, motivated by the LHC experiments there has been a flurry of  improved jet algorithm studies: it is essential that a correct jet finding is implemented by LHC experiments for an optimal matching of theory and experiment \cite{sal,sell}.  In particular the existing sequential recombination algorithms like $k_T$ \cite {dur}, \cite{kT} and Cambridge/Aachen \cite{camb}  have been generalized. In this recursive definitions one introduces distances $d_{ij}$ between particles or clusters of particles $i$ and $j$ and $d_{iB}$ between $i$ and the beam (B). The inclusive clustering proceeds by identifying the smallest of the distances and, if it is a $d_{ij}$, by recombining particles $i$ and $j$, while, if it is $d_{iB}$,  calling $i$ a jet and removing it from the list. The distances are recalculated and the procedure repeated until no $i$ and $j$ are left.
The extension relative to the $k_T$ \cite{kT} and Cambridge/Aachen \cite{camb} algorithms lies in the definition of the distance measures:
\beq
d_{ij}~=~min(k_{Ti}^{2p},k_{Tj}^{2p})\frac{\Delta_{ij}^2}{R^2},          \label{yij2}\\
\eeq
where $\Delta_{ij}^2=(y_i-y_j)^2 + (\phi_i-\phi_j)^2$ and $k_{Ti}$, $y_i$ and $\phi_i$ are respectively the transverse momentum, rapidity and azimuth of particle $i$. $R$ is the radius of the jet, i.e. the radius of a cone that, by definition, contains the jet. The exponent $p$ fixes the relative power of the energy versus geometrical ($\Delta_{ij}$) scales.
For $p = 1$ one has the inclusive $k_{T}$ algorithm. It can be shown in general that for $p \geq 0$ the behaviour of the jet algorithm with respect to soft radiation is rather similar to that observed for the $k_{T}$ algorithm. The case $p = 0$ is special and it corresponds to the inclusive Cambridge/Aachen algorithm \cite{camb}. Surprisingly (at first sight ), taking $p$ to be negative also yields an algorithm that is infrared and collinear safe and has sensible phenomenological behaviour. For $p=-1$ one obtains the recently introduced  Òanti-$k_{T}$Ó  jet-clustering algorithm \cite{cs} which has particularly stable jet boundaries with respect to soft radiation and is suitable for practical use by the experiments.

\subsection{Deep Inelastic Scattering}
\label{sec:16}

Deep Inelastic Scattering (DIS) processes
have played and still play a very important role for our understanding of QCD and of nucleon structure. This set of
processes actually provides us with a rich laboratory for theory and experiment. There are several structure functions that
can be studied, $F_i(x,Q^2)$, each a function of two variables. This is true separately for different beams and targets and
different polarizations. Depending on the charges of $\ell$ and $\ell$' (see Eq.~(\ref{DIS})) we can have neutral currents ($\gamma$,Z) or charged currents in
the $\ell$-$\ell$' channel (Fig.~\ref{fig6}). In the past DIS processes were crucial for establishing QCD as
the theory of strong interactions and quarks and gluons as the QCD partons. At present DIS remains very important for quantitative studies and tests of QCD. The theory
of scaling violations for totally inclusive DIS structure functions, based on operator expansion or diagrammatic techniques and renormalization group
methods, is crystal clear and the predicted $Q^2$ dependence can be tested at each value of $x$. The measurement of quark and
gluon densities in the nucleon, as functions of x at some reference value of $Q^2$, which is an essential starting point for
the calculation of all relevant hadronic hard processes, is performed in DIS processes. At the same time one measures
$\alpha_s(Q^2)$ and the DIS values of the running coupling can be compared with those obtained from other processes. At all times new theoretical
challenges arise from the study of DIS processes. Recent examples (see the following) are the so-called "spin crisis" in polarized DIS and the
behaviour of singlet structure functions at small $x$ as revealed by HERA data. In the following we will review the past
successes and the present open problems in the physics of DIS.

The cross-section $\sigma\sim L^{\mu \nu}W_{\mu \nu}$ is given in terms of the product of a leptonic ($L^{\mu \nu}$) and a
hadronic ($W_{\mu
\nu}$) tensor. While $L^{\mu \nu}$ is simple and easily obtained from the lowest order electroweak (EW) vertex plus QED
radiative corrections, the complicated strong interaction dynamics is contained in $W_{\mu \nu}$. The latter is proportional
to the Fourier transform of the forward matrix element between the nucleon target states of the product of two EW
currents:
\beq W_{\mu \nu}~=~\int{~d^4y~\exp{i(q\cdot y)}~<p|J^{\dagger}_{\mu}(y)J_{\nu}(0)|p>}\label{FTx}
\eeq Structure functions are defined starting from the general form of $W_{\mu \nu}$ given Lorentz invariance and current
conservation. For example, for EW currents between unpolarized nucleons we have (for the definition of variables recall Eqs. \ref{DIS}, \ref{kin}):
\bea
W_{\mu \nu}~&=&~(-g_{\mu \nu}~+~\frac{q_{\mu}q_{\nu}}{q^2})~W_1(\nu,Q^2)~+~(p_{\mu}~-~\frac{m
\nu}{q^2}q_{\mu})(p_{\nu}~-~\frac{m
\nu}{q^2}q_{\nu})~\frac{W_2(\nu,Q^2)}{m^2}~-~\nonumber\\
&&~-~\frac{i}{2m^2}\epsilon_{\mu \nu \lambda
\rho}p^{\lambda}q^{\rho}~W_3(\nu,Q^2)\nonumber\\
\label{sf}
\nonumber\eea
$W_3$ arises from VA interference and is absent for pure vector currents. In the limit $Q^2>>m^2$, with the Bjorken variable $x$ fixed, the structure functions obey approximate Bjorken
scaling which in reality is broken by logarithmic corrections that can be computed in QCD:
\bea 
mW_1(\nu,Q^2)&\rightarrow& F_1(x)\nonumber \\
\nu W_{2,3}(\nu,Q^2)&\rightarrow& F_{2,3}(x) \label{Bj}
\eea 
The $\gamma-N$ cross-section is given by ($W_i~=~W_i(Q^2,\nu)$):
\beq
\frac{d\sigma^{\gamma}}{dQ^2d\nu}~=~\frac{4\pi\alpha^2E'}{Q^4E}\cdot
[2\sin^2{\frac{\theta}{2}}W_1~+~\cos^2{\frac{\theta}{2}}W_2]\label{gN}\\
\eeq
while for the $\nu-N$ or $\bar{\nu}-N$ cross-section one has:
\beq
\frac{d\sigma^{\nu,\bar{\nu}}}{dQ^2d\nu}~=~\frac{G_F^2E'}{2\pi E}(\frac{m_W^2}{Q^2+m_W^2})^2\cdot
[2\sin^2{\frac{\theta}{2}}W_1~+~\cos^2{\frac{\theta}{2}}W_2\pm\frac{E+E'}{m}\sin^2{\frac{\theta}{2}}W_3]\label{nuN}\\
\eeq 
($W_i$ for photons, $\nu$ and $\bar{\nu}$ are all different, as we shall see in a moment). 

In the scaling limit the longitudinal and transverse cross sections are given by:
\bea
\sigma_L&\sim&\frac{1}{s}[\frac{F_2(x)}{2x}~-~F_1(x)]\nonumber \\
\sigma_{RH,LH}&\sim& \frac{1}{s}[F_1(x)~\pm~F_3(x)]\nonumber \\
\sigma_T&=&\sigma_{RH}~+~\sigma_{LH} \label{sig}
\eea 
where L, RH, LH refer to the helicity 0, 1, -1, respectively, of the exchanged gauge vector boson. For the photon case $F_3=0$ and  $\sigma_{RH}=\sigma_{LH}$.

In the '60's the demise of hadrons from the  status of fundamental particles to that of bound states of constituent quarks was
the breakthrough that made possible the construction of a renormalizable field theory for strong interactions. The presence of
an unlimited number of hadrons species, many of them with large spin values, presented an obvious dead-end for a manageable field
theory. The evidence for constituent quarks emerged clearly from the systematics of hadron spectroscopy. The complications of
the hadron spectrum could be explained in terms of the quantum numbers of spin 1/2, fractionally charged, u, d and s quarks.
The notion of colour was introduced to reconcile the observed spectrum with Fermi statistics. But confinement that forbids 
the
observation of free quarks was a clear obstacle towards the acceptance of quarks as real constituents and not just as
fictitious entities describing some mathematical pattern (a doubt expressed even by Gell-Mann at the time). The early
measurements at SLAC of DIS dissipated all doubts: the observation of Bjorken scaling and the success of the "naive" (not so
much after all) parton model of Feynman imposed quarks as the basic fields for describing the nucleon structure (parton
quarks). 

\begin{figure}[h]
\noindent
\includegraphics[width=5cm]{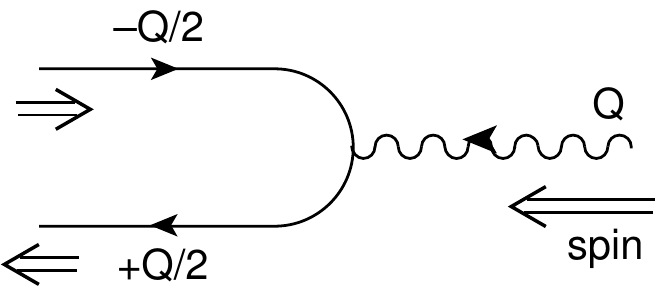}
\caption[]{Schematic diagram for the interaction of the virtual photon with a parton quark in the Breit frame.}
\label{fig10}
\end{figure}

In the language of Bjorken and Feynman the virtual $\gamma$ (or, in general, any gauge boson) sees the quark partons
inside the nucleon target as quasi-free, because their (Lorentz dilated) QCD interaction time is much longer than
$\tau_{\gamma}\sim 1/Q$, the duration of the virtual photon interaction. Since the virtual photon 4-momentum is spacelike, we can go to a Lorentz frame where $E_{\gamma}=0$
(Breit frame). In this frame $q=(E_{\gamma}=0;0,0,Q)$ and the nucleon momentum, neglecting the mass $m<<Q$, is
$p=(Q/2x;0,0,-Q/2x)$ (note that this correctly gives $q^2=-Q^2$ and $x=Q^2/2(p\cdot q)$). Consider (Fig.~\ref{fig10}) the interaction of the photon
with a quark carrying a fraction y of the nucleon 4-momentum: $p_q=yp$ (we are neglecting the transverse components of $p_q$
which are of order $m$). The incoming parton with $p_q=yp$ absorbs the photon and the final parton has 4-momentum
$p'_q$. Since in the Breit frame the photon carries no energy but only a longitudinal momentum $Q$, the photon can only be 
absorbed by those partons with $y=x$: then the longitudinal component of $p_q=yp$ is $-yQ/2x=-Q/2$ and can be flipped into
$+Q/2$ by the photon. As a result, the photon longitudinal momentum $+Q$ disappears, the parton quark momentum changes of
sign from
$-Q/2$ into $+Q/2$ and the energy is not changed. So the structure functions are proportional to the density of partons
with fraction $x$ of the nucleon momentum, weighted with the squared charge. Also, recall that the helicity of a massless quark
is conserved in a vector (or axial vector) interaction (recall Sect. \ref {sec:5}). So when the momentum is reversed also the spin must flip. Since the
process is collinear there is no orbital contribution and only a photon with helicity
$\pm 1$ (transverse photon) can be absorbed. Alternatively, if partons were spin zero only longitudinal photons would instead contribute. 

Using these results, which are maintained in QCD at leading order, the quantum numbers of the quarks were confirmed by
early experiments. The observation that
$R~=~\sigma_L/\sigma_T\rightarrow 0$ implies that the charged partons have spin 1/2. The quark charges were derived from the
data on the electron and neutrino structure functions:
\bea
F_{ep}&=&\frac{4}{9}u(x)~+~\frac{1}{9}d(x)~+~.....~;~~~~~~F_{en}~=~\frac{4}{9}d(x)~+~\frac{1}{9}u(x)~+~....\nonumber\\
F_{\nu p}&=&F_{\bar{\nu}n}~=~2d(x)~+~.....~;~~~~~~~~~~~~~~F_{\nu n}~=~F_{\bar{\nu}p}~=~2u(x)~+~.....\label{charges}
\eea
where $F\sim 2F_1\sim F_2/x$ and $u(x)$, $d(x)$ are the parton number densities in the proton (with fraction $x$ of the proton
longitudinal momentum), which, in the scaling limit, do not depend on $Q^2$. The normalization of the structure functions
and the parton densities are such that the charge relations hold:
\beq
\int_0^1[u(x)-\bar u(x)]dx=2,~~~\int_0^1[d(x)-\bar d(x)]dx=1,~~~\int_0^1[s(x)-\bar s(x)]dx=0\label{cha}\\
\eeq
Also it was proven by experiment that at values of
$Q^2$ of a few GeV$^2$, in the scaling region, about half of the nucleon momentum, given by the momentum sum rule:
\beq
\int_0^1[\sum_i(q_i(x)+\bar{q}_i(x))~+~g(x)]xdx~=~1\label{momsr}\\
\eeq
is carried by neutral partons (gluons).

\begin{figure}[h]
\noindent
\includegraphics[width=14cm]{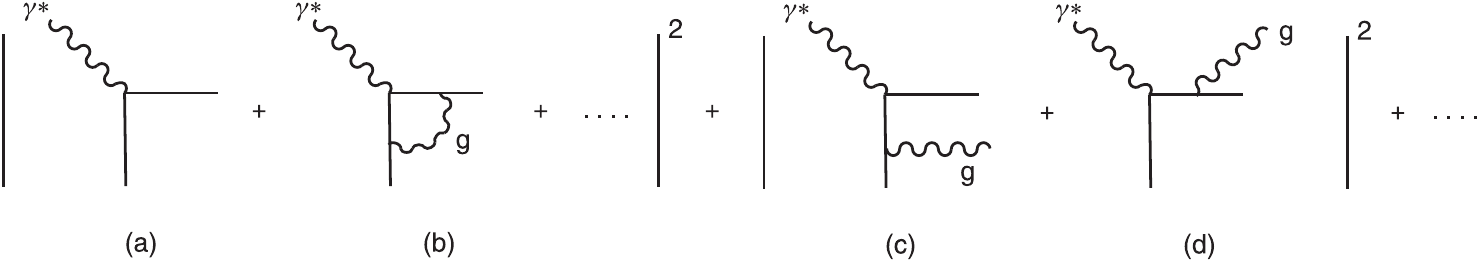} 
\caption{First order QCD corrections to the virtual photon-quark cross-section.}
\label{fig11}
\end{figure}

In QCD there are calculable log scaling violations induced by $\alpha_s(t)$. The parton rules in Eq.\ref{charges} can be
summarized in the schematic formula:
\beq
F(x,t)~=~\int_x^1dy\frac{q_0(y)}{y}\sigma_{point}(\frac{x}{y},\alpha_s(t))~+~O(\frac{1}{Q^2})\label{conv1}\\
\eeq
Before QCD corrections $\sigma_{point}=e^2\delta(x/y-1)$ and $F=e^2q_0(x)$ (here we denote by $e$ the charge of the quark in
units of the positron charge, i.e. $e=2/3$ for the $u$ quark). QCD modifies
$\sigma_{point}$ at order
$\alpha_s$ via the diagrams of Fig.~\ref{fig11}.  From a
direct computation of the diagrams one obtains a result of the following form:
\beq
\sigma_{point}(z,\alpha_s(t))~\simeq ~e^2[\delta (z-1)~+~\frac{\alpha_s}{2\pi}(t\cdot P(z)~+~f(z))]\label{sigalf}\\
\eeq

Note that the $y$ integral in Eq. \ref{conv1} is from $x$ to 1, because the energy can only be lost by
radiation before interacting with the photon (which eventually wants to find a fraction $x$, as we have explained). For $y>x$ the correction arises from diagrams with real gluon emission. Only the sum of the two real-gluon diagrams in Fig.~\ref{fig11} is
gauge invariant, so that the contribution of one given diagram is gauge dependent. But in an axial gauge, which for this reason is
some times also called the "physical gauge", the diagram of Fig.~\ref{fig11}(c), among real diagrams, gives the whole $t$-proportional term at $0 < x < 1$. It is
obviously not essential to go to this gauge, but this diagram has a direct physical interpretation: a quark in the proton
has a fraction
$y>x$ of the parent 4-momentum; it then radiates a gluon and looses energy down to a fraction $x$ before interacting with
the photon. The log arises from the virtual quark propagator, according to the discussion of collinear mass singularities in
Eq.~(\ref{prop}). In fact in the massless limit one has (k and h are the 4-momenta of the initial quark and the emitted gluon, respectively):
\bea
\rm{propagator}~&=&~\frac{1}{r^2}~=~\frac{1}{(k-h)^2}~=~\frac{-1}{2E_kE_h}\cdot\frac{1}{1-\cos{\theta}}\nonumber\\
&=&\frac{-1}{4E_kE_h}\cdot\frac{1}{\sin^2{\theta/2}}~\propto \frac{-1}{p_T^2}
\label{prop1}
\eea 
where $p_T$ is the transverse momentum of the virtual quark. So the square of the propagator goes like $1/p_T^4$. But there
is a $p_T^2$ factor in the numerator, because in the collinear limit, when $\theta=0$ and the initial and final quarks and
the emitted gluon are all aligned, the quark helicity cannot flip (vector interaction) so that the gluon should carry helicity zero while a real gluon can only have $\pm 1$ helicity. Thus the numerator vanishes as $p_T^2$ in the forward direction and the cross-section behaves as:
\beq
\sigma~\sim ~\int^{Q^2}\frac{1}{p_T^2}dp_T^2~\sim ~\log{Q^2}\label{log}\\
\eeq
Actually the log should be read as $\log{Q^2/m^2}$ because in the massless limit a genuine mass singularity appears. In fact
the mass singularity connected with the initial quark line is not cancelled because we do not have the sum of all degenerate
initial states \cite{KLN}, but only a single quark. But in correspondence to the initial quark we have the (bare) quark density
$q_0(y)$ that appears in the convolution integral. This is a non perturbative quantity  determined by the nucleon wave
function. So we can factorize the mass singularity in a redefinition of the quark density: we replace $q_0(y)\rightarrow
q(y,t)~=~q_0(y)~+~\Delta q(y,t)$ with:
\beq
\Delta q(x,t)~=~\frac{\alpha_s}{2\pi}t\int_x^1dy\frac{q_0(y)}{y}\cdot P(\frac{x}{y})\label{deltaq}\\
\eeq
Here the factor of $t$ is a bit symbolic: it stands for $\log{Q^2/m^2}$ but what we exactly put under $Q^2$ depends on the
definition of the renormalized quark density, which also fixes the exact form of the finite term $f(z)$ in
Eq.~(\ref{sigalf}).

The effective parton density $q(y,t)$ that we have defined is now scale dependent. In terms of this scale dependent density
we have the following relations, where we have also replaced the fixed coupling with the running coupling according to the 
prescription derived from the RGE:
\bea
F(x,t)&=&\int_x^1dy\frac{q(y,t)}{y}e^2[\delta
(\frac{x}{y}-1)~+~\frac{\alpha_s(t)}{2\pi}f(\frac{x}{y}))]~=~e^2q(x,t)~+~O(\alpha_s(t))\nonumber\\
\frac{d}{dt}q(x,t)&=&\frac{\alpha_s(t)}{2\pi}\int_x^1dy\frac{q(y,t)}{y}\cdot P(\frac{x}{y})~+~O(\alpha_s(t)^2)\label{APNS}\\
\nonumber\eea
We see that in lowest order we reproduce the naive parton model formulae for the structure functions in terms of effective
parton densities that are scale dependent. The evolution equations for the parton densities are written down in terms of
kernels (the "splitting functions" \cite{AP}) that can be expanded in powers of the running coupling. At leading order, we can
interpret the evolution equation by saying that the variation of the quark density at $x$ is given by the convolution of the
quark density at $y$ times the probability of emitting a gluon with fraction $x/y$ of the quark momentum.

It is interesting that the integro-differential QCD evolution equation for densities can be transformed into an infinite set of ordinary
differential equations for Mellin moments \cite{SM5}. The Mellin moment $f_n$ of a density $f(x)$ is defined as:
\beq
f_n~=~\int_0^1dxx^{n-1}f(x)\label{Mel}\\
\eeq
By taking moments of both sides of the second of Eqs.(\ref{APNS}) one finds, with a simple interchange of the integration
order, the simpler equation for the n-th moment:
\beq
\frac{d}{dt}q_n(t)~=~\frac{\alpha_s(t)}{2\pi}\cdot P_n \cdot q_n(t)\label{momev}\\
\eeq
To solve this equation we observe that it is equivalent to:
\beq
\log{\frac{q_n(t)}{q_n(0)}}~=~\frac{P_n}{2\pi}\int_0^t\alpha_s(t)dt~=~\frac{P_n}{2\pi}\int_{\alpha_s}^{\alpha_s(t)}
\frac{d\alpha'}{-b\alpha'}\label{nn}\\
\eeq
(to see the equivalence just take the t derivative of both sides) where we used Eq.~(\ref{runt}) to change the integration variable from $dt$ to $d\alpha(t)$ (denoted as $d\alpha'$) and
$\beta(\alpha)\simeq -b\alpha^2+...$. Finally the solution is:
\beq
q_n(t)~=~[\frac{\alpha_s}{\alpha_s(t)}]^{\frac{P_n}{2\pi b}}\cdot q_n(0)\label{solmom}\\
\eeq
 
The connection of these results with the RGE general formalism occurs via the light cone OPE (recall Eq.~(\ref{FTx}) for
$W_{\mu\nu}$ and Eq.~(\ref{OPEJJ}) for the OPE of two currents). In the case of DIS the c-number term $I(x^2)$ does not
contribute, because we are interested in the connected part of the matrix element $<p|...|p>-<0|...|0>$. The relevant terms are:
\beq
J^\dagger (x)J(0)~=~E(x^2)\sum_{n=0}^\infty c_n(x^2)x^{\mu_1}...x^{\mu_n}\cdot
O^n_{\mu_1...\mu_n}(0)~+~
\rm{less~sing.~terms}
\label{OPE}
\eeq
A formally intricate but conceptually simple argument based on the analiticity properties of the forward virtual Compton
amplitude shows that the Mellin moments $M_n$ of structure functions are related to the individual terms in the OPE,
precisely to the Fourier transform $c_n(Q^2)$ (we will write it as $c_n(t,\alpha)$)  of the coefficient $c_n(x^2)$ times a
reduced matrix element
$h_n$ from the operators
$O^n$:
$<p|O^n_{\mu_1...\mu_n}(0)|p>=h_n p_{\mu_1}...p_{\mu_n}$:
\beq
c_n<p|O^n|p>\rightarrow M_n=\int_0^1dxx^{n-1}F(x)\label{MomOPE}\\
\eeq
Since the matrix element of the products of currents satisfy the RGE so do the moments $M_n$. Hence the general form of the
$Q^2$ dependence is given by the RGE solution (see Eq.~(\ref{Fsol2})):
\beq
M_n(t,\alpha)~=~c_n[0,\alpha(t)]\exp{\int_{\alpha}^{\alpha(t)}\frac{\gamma_n(\alpha')}{\beta(\alpha')}d\alpha'}\cdot
h_n(\alpha)\label{Msol}\\
\eeq
In lowest order, identifying in the simplest case $M_n$ with $q_n$, we have:
\beq
\gamma_n(\alpha)~=~\frac{P_n}{2\pi} \alpha~+~...,~~~~~~~~~\beta(\alpha)~=~-b\alpha^2~+~...\label{Msol1}\\
\eeq
and  
\beq
q_n(t)=q_n(0)\exp{\int_{\alpha}^{\alpha(t)}\frac{\gamma_n(\alpha')}{\beta(\alpha')}d\alpha'}~=~
[\frac{\alpha_s}{\alpha_s(t)}]^{\frac{P_n}{2\pi b}}\cdot q_n(0)\label{solmom2}\\
\eeq
which exactly coincides with Eq.~(\ref{solmom}).

\begin{figure}[h]
\noindent
\includegraphics[width=5cm]{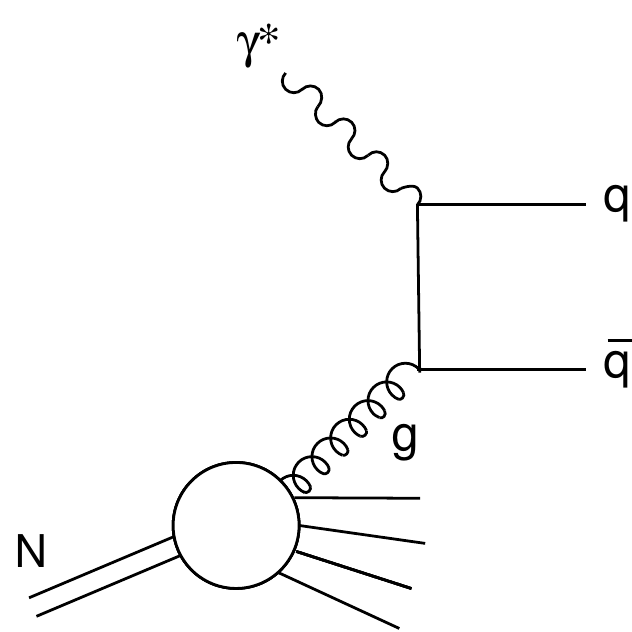} 
\caption[]{Lowest order diagram for the interaction of the virtual photon with a parton gluon.}
\label{fig12}
\end{figure}

Up to this point we have implicitly restricted our attention to non-singlet (under the flavour group) structure
functions. The $Q^2$ evolution equations become non diagonal as soon as we take into account the presence of gluons in the
target. In fact the quark which is seen by the photon can be generated by a gluon in the target (Fig.~\ref{fig12}). 

The quark evolution
equation becomes:
\beq
\frac{d}{dt}q_i(x,t)~=~\frac{\alpha_s(t)}{2\pi}[q_i\otimes P_{qq}]~+~\frac{\alpha_s(t)}{2\pi}[g\otimes
P_{qg}]\label{APqS}\\
\eeq
where we introduced the shorthand notation:
\beq
[q\otimes P]~=~[P\otimes q]~=~\int_x^1dy\frac{q(y,t)}{y}\cdot P(\frac{x}{y})\label{convdef}\\
\eeq
(it is easy to check that the so-defined convolution, like an ordinary product, is commutative). At leading order, the interpretation of
Eq.~(\ref{APqS}) is simply that the variation of the quark density is due to the convolution of the quark density at a higher
energy times the probability of finding a quark in a quark (with the right energy fraction) plus the gluon density  at a
higher energy times the probability of finding a quark (of the given flavour i) in a gluon. The evolution equation for the
gluon density, needed to close the system \footnote{The evolution equations are now often called DGLAP equations (Dokshitzer Gribov Lipatov Altarelli Parisi). The first article by Gribov and Lipatov was published in 72 \cite{GL} (even before the works by Gross and Wilczek and by Politzer!) and was followed in 74 by a paper by Lipatov \cite{Lip} (these dates correspond to the publication in russian). All these articles refer to an abelian vector theory (treated in parallel with a pseudoscalar theory). Seen from the point of view of the evolution equations, these papers, in the context of the abelian theory, ask the right question and extract the relevant logarithmic terms from the dominant class of diagrams. But from their formal presentation the relation to real physics is somewhat hidden (in this respect the 74 paper by Lipatov makes some progress and explicitly refers to the parton model). The article by Dokshitser \cite{Dok} was exactly contemporary to that by Altarelli Parisi \cite{AP}. It now refers to the non abelian theory (with running coupling) and the discussion is more complete and explicit than in the Gribov-Lipatov articles. But, for example, the connection to the parton model, the notion of the evolution as a branching process and the independence of the kernels from the process are not emphasized.}, can be obtained by suitably extending the same line of reasoning to a gedanken
probe sensitive to colour charges, for example a virtual gluon. The resulting equation is of the form:
\beq
\frac{d}{dt}g(x,t)~=~\frac{\alpha_s(t)}{2\pi}[\sum_i (q_i+\bar q_i)\otimes P_{gq}]~+~\frac{\alpha_s(t)}{2\pi}[g\otimes
P_{gg}]\label{APgS}\\
\eeq

\begin{figure}[htbp]
\noindent
\includegraphics[width=10cm]{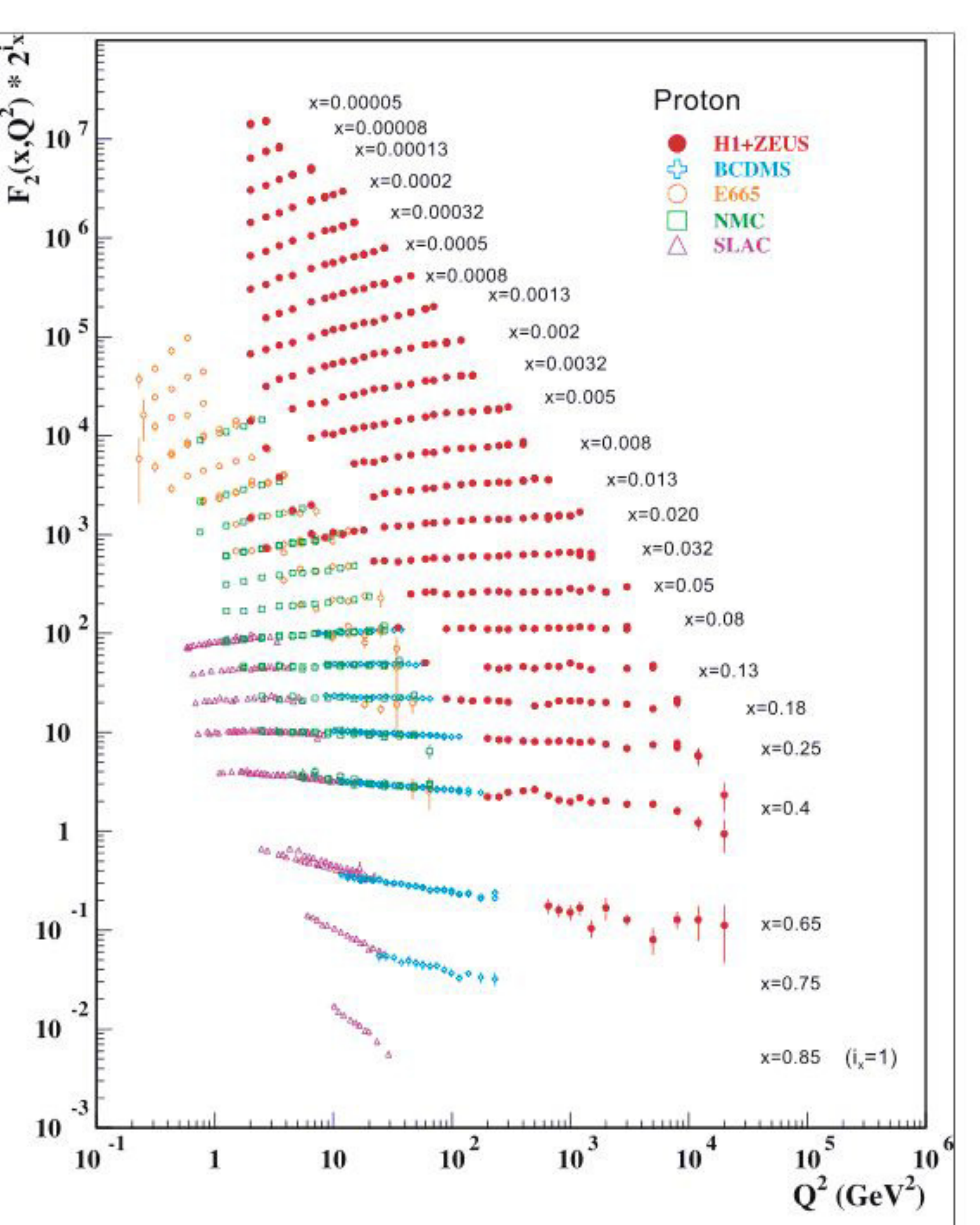}
\caption{A representative selection of data on the proton electromagnetic structure function $F_2^p$ from Collider (HERA) and fixed target experiments \cite{pdg12} that clearly shows the pattern of scaling violations.}
\label{fig13a}
\end{figure}

The explicit form of the splitting functions in lowest order \cite {GL,AP, Dok} can be directly derived from the QCD vertices \cite{AP}. They are a
property of the theory and do not depend on the particular process the parton density is taking part in. The results are :
\bea
P_{qq}&=&\frac{4}{3}[\frac{1+x^2}{(1-x)_+}~+~\frac{3}{2}\delta (1-x)]~+~O(\alpha_s)\nonumber\\
P_{gq}&=&\frac{4}{3}\frac{1+(1-x)^2}{x}~+~O(\alpha_s)\nonumber\\
P_{qg}&=&\frac{1}{2}[x^2+(1-x)^2]~+~O(\alpha_s)\nonumber\\
P_{gg}&=&6[\frac{x}{(1-x)_+}~+~\frac{1-x}{x}~+~x(1-x)]~+~\frac{33-2n_f}{6}\delta (1-x)~+~O(\alpha_s)
\label{Splfs}
\eea
For a generic non singular weight function $f(x)$, the "+" distribution is defined as:
\beq
\int_0^1\frac{f(x)}{(1-x)_+}dx~=~\int_0^1\frac{f(x)-f(1)}{1-x}dx\label{plus}\\
\eeq
The $\delta(1-x)$ terms arise from the virtual corrections to the lowest order tree diagrams. Their coefficient can be simply obtained by
imposing the validity of charge and momentum sum rules. In fact, from the request that the charge sum rules in
Eq.~(\ref{cha}) are not affected by the $Q^2$ dependence one derives that
\beq
\int_0^1P_{qq}(x)dx~=~0\label{Pqqsr}\\
\eeq
which can be used to fix the coefficient of the $\delta(1-x)$ terms of $P_{qq}$. Similarly, by taking the t-derivative of the
momentum sum rule in Eq.~(\ref{momsr}) and imposing its vanishing for generic $q_i$ and $g$, one obtains:
\beq
\int_0^1[P_{qq}(x)~+~P_{gq}(x)]xdx~=~0,~~~~~~\int_0^1[2n_fP_{qg}(x)~+~P_{gg}(x)]xdx~=~0.\label{Pmomsr}\\
\eeq

At higher orders the evolution equations are easily generalized but the calculation of the splitting functions rapidly
becomes very complicated. For many years the splitting functions were only completely known at NLO accuracy \cite{pet}: $\alpha_s P~\sim~\alpha_s
P_1~+~\alpha_s^2 P_2~+...$. Then in recent years the NNLO results
$P_3$ have been first derived in analytic form for the first few moments and, then the full NNLO analytic calculation, a really monumental work, was completed in 2004 by Moch, Vermaseren and Vogt \cite{verspl}. Beyond leading order a precise definition of parton densities should be specified. One can take a physical definition: for example, quark densities can be defined as to keep the LO expression for the structure function $F_2$ valid at all orders, the so called DIS definition \cite{F2}, and the gluon density could be defined starting from $F_L$, the longitudinal structure function. Alternatively one can adopt a more abstract specification as, for example, in terms of the $\overline{MS}$ prescription. Once the definition of parton densities is fixed, the coefficients that relate the different structure functions to the parton densities at each fixed order can be computed. Similarly the higher order splitting functions also depend, to some extent, from the definition of parton densities, and a consistent set of coefficients and splitting functions must be used at each order. 

\begin{figure}[htbp]
\noindent
\centerline{\includegraphics[width=10cm]{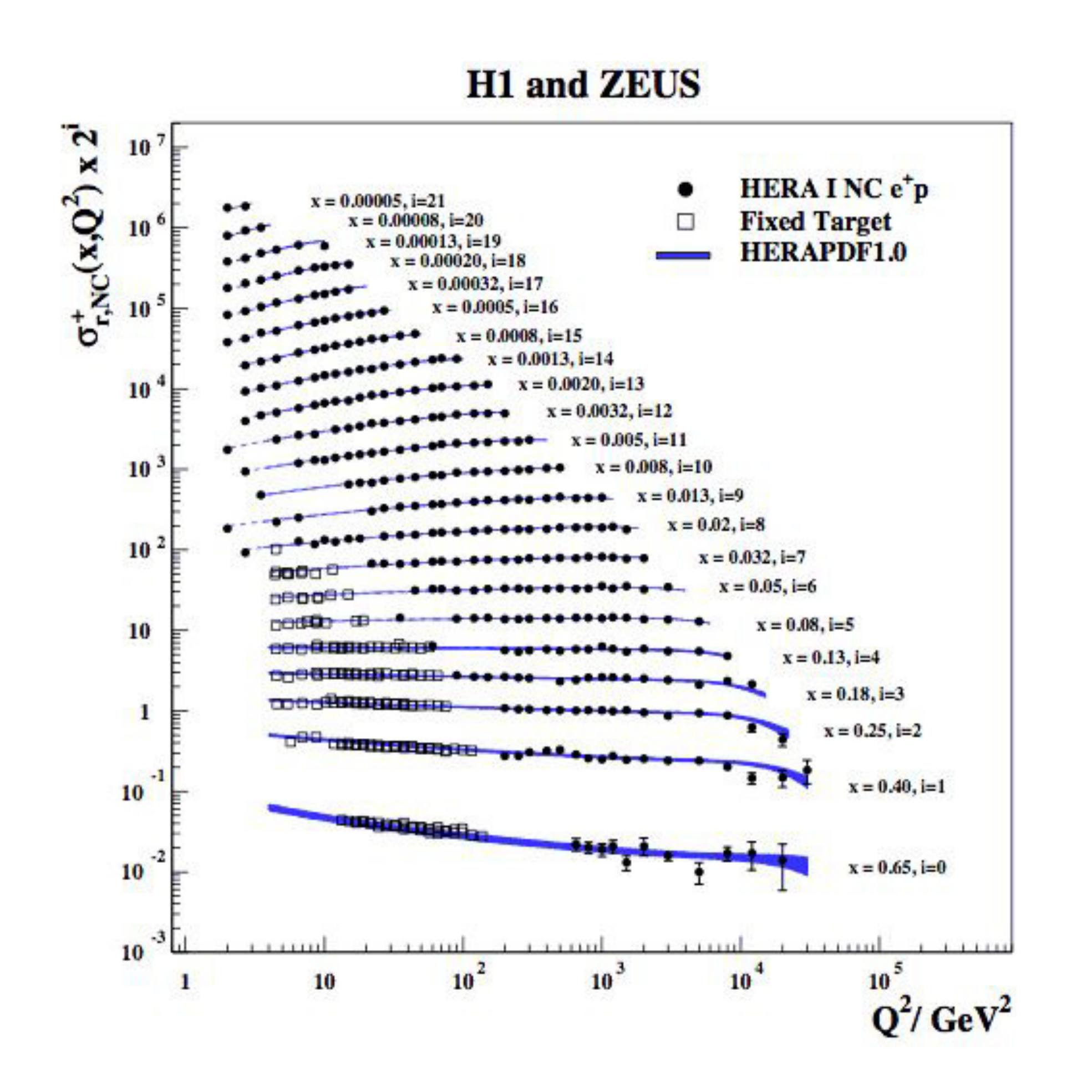}} 
\caption{A NLO QCD fit to the combined HERA
data with $Q^2 \geq $ 3.5 $GeV^2$): $\chi^2/dof$ = 574/582 \cite{HERA}.}
\label{fig13b}
\end{figure}

The scaling violations are clearly observed by experiment (Fig.~\ref{fig13a}) and their pattern is very well reproduced by QCD fits at
NLO Fig.~\ref{fig13b} \cite{HERA}. These fits provide an impressive confirmation of a quantitative QCD
prediction, a measurement of $q_i(x,Q_0^2)$ and $g(x,Q_0^2)$, at some reference value $Q_0^2$ of $Q^2$, and a precise
measurement of
$\alpha_s(Q^2)$.

\begin{figure}[htbp]
\noindent
\centerline{\includegraphics[width=10cm]{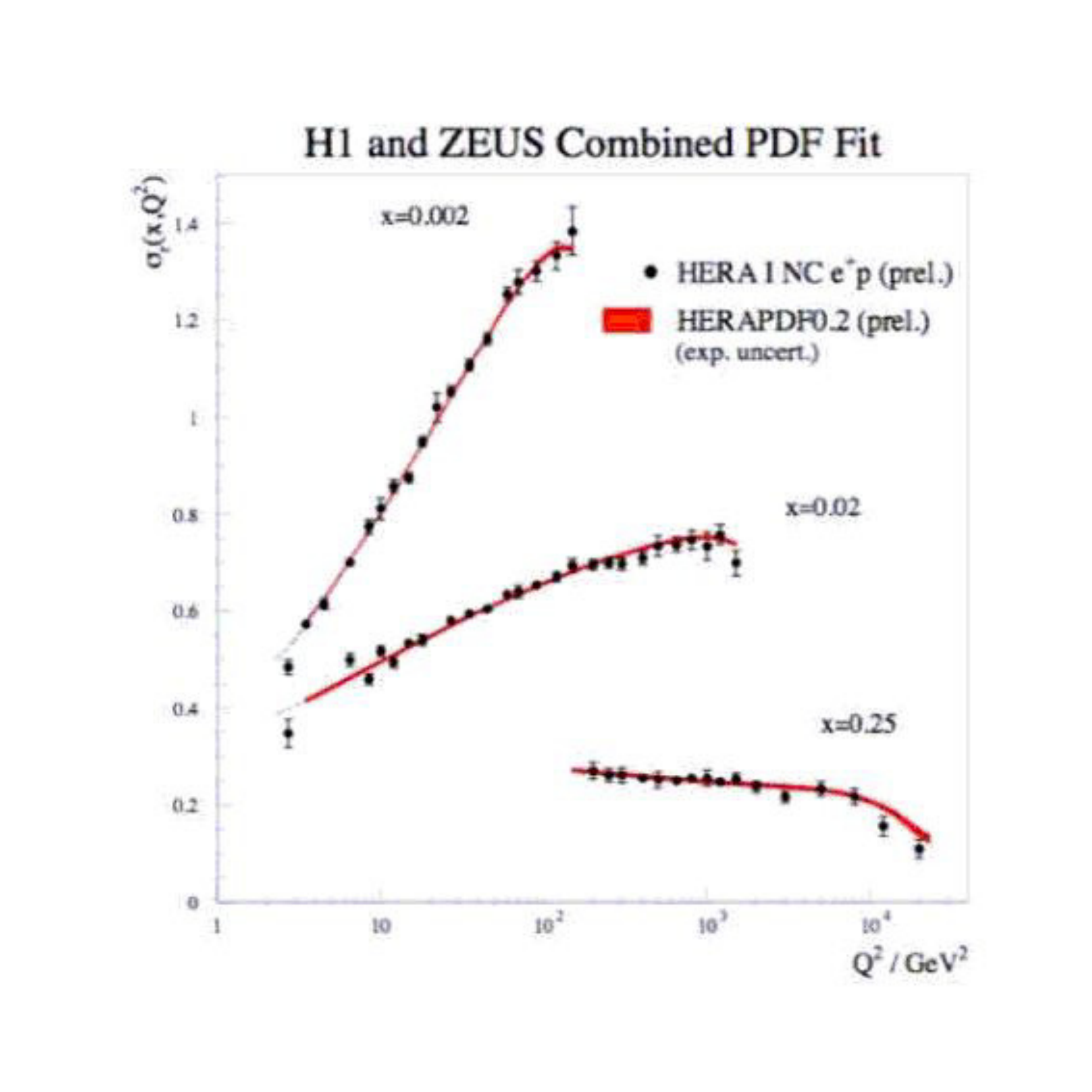}}
\caption{A more detailed view of the NLO QCD fit to a selection of the HERA data \cite{HERA}.}
\label{fig13c}
\end{figure}

\subsubsection{The Longitudinal Structure Function}
\label{sec:16.0}

\begin{figure}[!t]
\includegraphics[width=0.8\columnwidth]{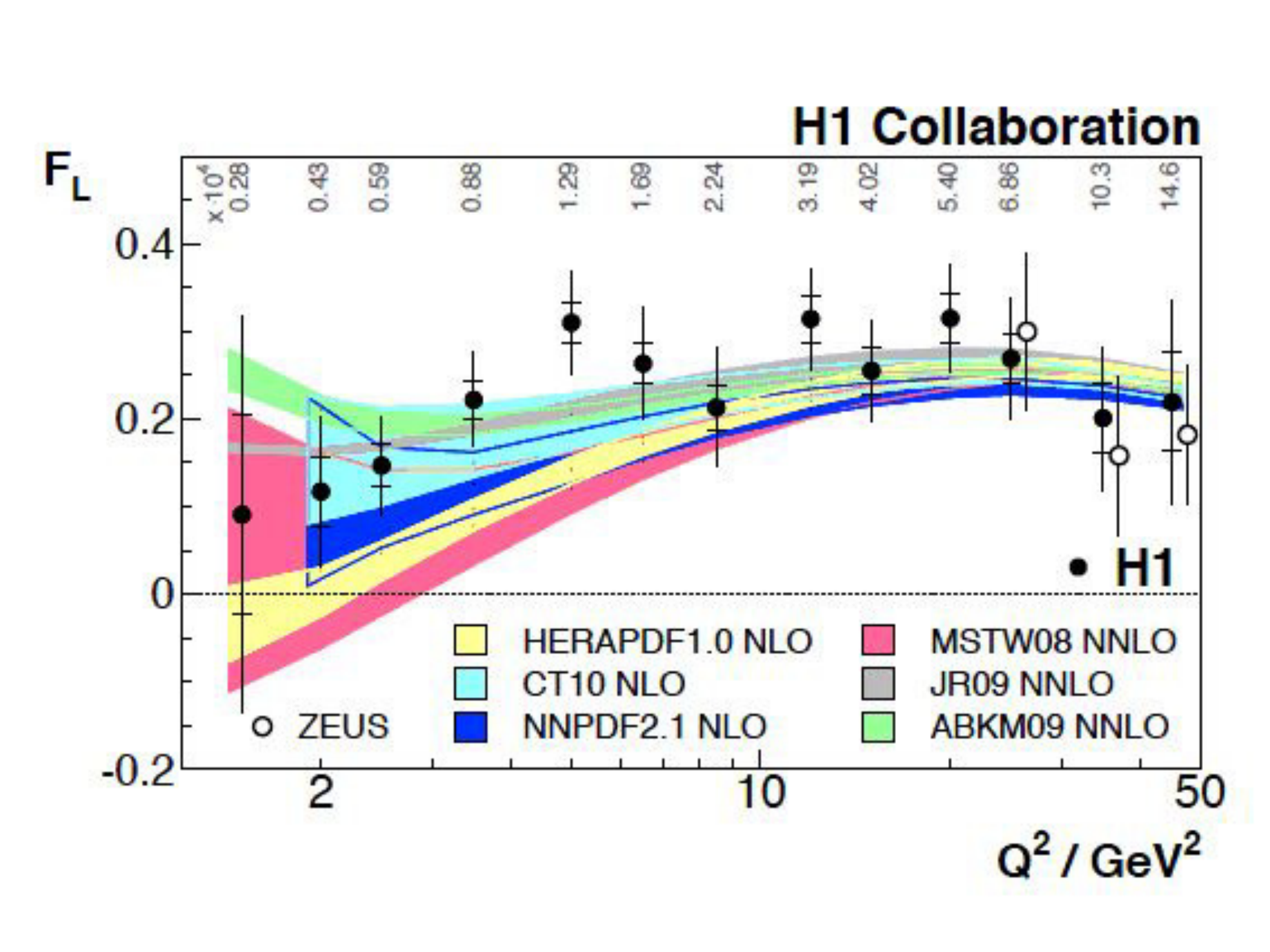} 
\caption[]{The longitudinal structure function $F_L$ measured by H1 at HERA, as function of $Q^2$ for different values of $x$. The theoretical curves are obtained from different sets of parton densities as indicated.}
\label{longi}
\end{figure}

After SLAC established the dominance of the transverse cross section it took ~40 years to get meaningful data on the longitudinal structure function $F_L$ (see Eq.(\ref{sig})! These data represent an
experimental highlight of recent years. They have been obtained by H1 at HERA \cite{H1}. The data are shown in Fig.(\ref{longi}). For spin 1/2 charged partons $F_L$ vanishes asymptotically. In QCD $F_L$ starts at order $\alpha_s(Q^2)$. At LO the simple, 30 years old, formula is valid (for $N_f=4$) \cite{am}:
\beq
F_L(x,Q^2)=\frac{\alpha_s(Q^2)}{2\pi}x^2\int_x^1\frac{dy}{y^3}\left[ \frac{8}{3}F_2(y,Q^2)+\frac{40}{9}yg(y,Q^2)(1- \frac{x}{y})\right]\\
\label{FL}
\eeq
The $O(\alpha_s^2)$ \cite{FL2} and $O(\alpha_s^3)$ \cite{FL3} corrections are at present also known. One would not have expected that it would take such a long time to have a meaningful test of this simple prediction! And in fact better data would be highly desirable. But how and when they will be obtained is at present not clear at all.

\subsubsection{Large and Small $x$ Resummations for Structure Functions}
\label{sec:16.1}

At values of $x$ either near 0 or near 1 (with $Q^2$ large) those terms of higher order in $\alpha_s$ in both the coefficients or the splitting functions which are multiplied by powers of $\log{1/x}$ or $\log{(1-x)}$ eventually become important and should be taken into account. Fortunately the sequences of leading and subleading logs can be evaluated at all orders by special techniques and resummed to all orders. 

For $x \sim 1$ resummation \cite{sterman} I refer to the recent papers \cite{largex}, \cite{largexHT}  (the latter also involving higher twist corrections, which are important at large $x$) where a list of references to previous work can be found. 

\begin{figure}[!t]
\centerline{\includegraphics[width=0.8\columnwidth]{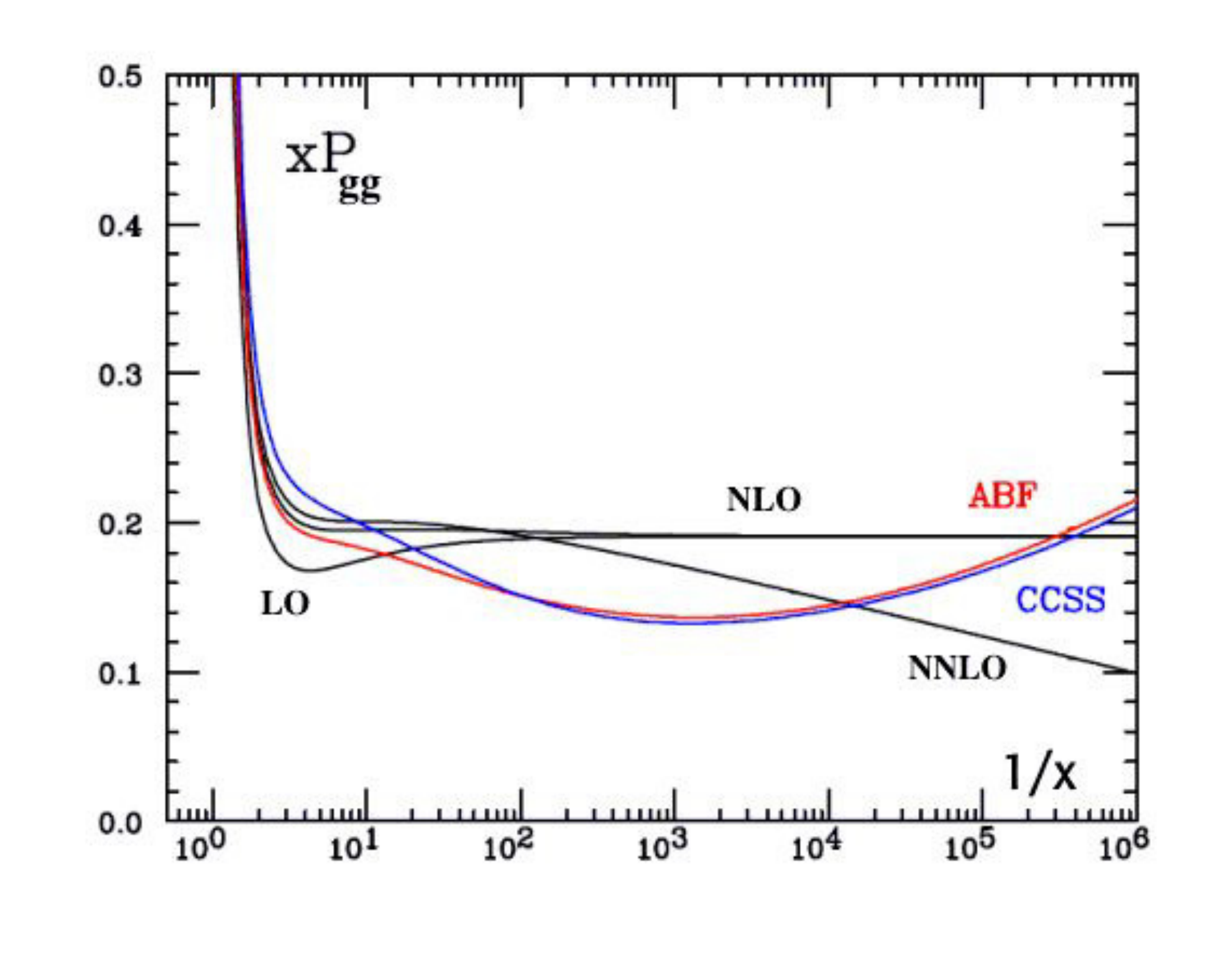}} 
\caption[]{The dominant singlet splitting function $xP_{gg}(x,\alpha_s(Q^2))$ for $\alpha_s(Q^2) \sim 0.2$. The resummed results from ref.\cite{xresus} (labeled ABF) and from ref.\cite{xres} (CCSS), which are in good agreement among them, are compared with the LO, NLO and NNLO perturbative results.}
\label{fig14}
\end{figure}

More important is the small $x$ resummation because the singlet structure functions are large in this domain of $x$ (while all structure functions vanish near $x = 1$). Here we will briefly summarize the small-$x$ case for the singlet structure function which is the dominant channel at HERA, dominated by the sharp rise of the gluon and sea parton densities at small $x$.
The small $x$ data collected
by HERA can be fitted reasonably well even at the smallest measured values of $x$ by the NLO QCD evolution equations, so that
there is no dramatic evidence in the data for departures. This is surprising also in view of the fact that the NNLO effects in the evolution have recently become available and are quite large \cite{verspl}. Resummation effects have been shown to resolve this apparent paradox. For the singlet splitting function the coefficients of all LO and NLO corrections of order $[\alpha_s(Q^2)\log{1/x}]^n$ and $\alpha_s(Q^2)[\alpha_s(Q^2)\log{1/x}]^n$, respectively, are explicitly known from the Balitski, Fadin, Kuraev, Lipatov (BFKL) analysis of virtual gluon-virtual gluon scattering \cite{BFKL}, \cite{BFKLNLO}. But the simple addition of these higher order terms to the perturbative result (with subtraction of all double counting) does not lead to a converging expansion (the NLO logs completely overrule the LO logs in the relevant domain of $x$ and $Q^2$). A sensible expansion is only obtained by a proper treatment of momentum conservation constraints, also using the underlying symmetry of the BFKL kernel under exchange of the two external gluons, and especially, of the running coupling effects (see the analysis in \cite{xresus, xres} and references therein). In Fig.~\ref{fig14} we present the results for the dominant singlet splitting function $xP_{gg}(x,\alpha_s(Q^2))$ for $\alpha_s(Q^2) \sim 0.2$. We see that while the NNLO perturbative splitting function sharply deviates from the NLO approximation at small $x$, the resummed result only shows a moderate dip with respect to the NLO perturbative splitting function in the region of HERA data, and the full effect of the true small $x$ asymptotics is only felt at much smaller values of $x$. The related effects are not very important for most processes at the LHC but could become relevant for next generation of hadron colliders.

\subsubsection{Polarized Deep Inelastic Scattering}
\label{sec:16.2}

Polarized DIS  is a subject where our knowledge is still far from satisfactory in spite of a great experimental effort  (for recent reviews, see, for example, \cite{bass}). One main question is how the proton helicity is distributed among quarks, gluons and orbital angular momentum: 
\beq
\frac{1}{2}\Delta \Sigma + \Delta g + L_z= \frac{1}{2}\\
\label{spin}
\eeq
Experiments with polarized leptons on polarized nucleons are sensitive to the polarized parton densities $\Delta q = q_+ - q_-$, the difference of quark densities with helicity plus and minus  in a proton with helicity plus. These differences are related to the quark matrix elements of the axial current. The polarized densities satisfy evolution equations analogous to Eqs. \ref{APqS},\ref{APgS}  but with modified splitting functions that were derived in ref. \cite{AP} (the corresponding anomalous dimensions were obtained in ref. \cite{ahmed}). The measurements have shown that the quark moment $\Delta \Sigma$ is small (the "spin crisis" started by ref.(\cite{EMC}): values from recent fits \cite{NNPol,deFlorian:2009vb,Blumlein:2010rn,Leader:2010rb,Hirai:2008aj,sis} are in the range $\Delta \Sigma \sim 0.2-0.3$: in any case, a less pronounced crisis than it used to be in the past. From the spin sum rule one obtains that either $\Delta g + L_z$ is relatively large or there are contributions to $\Delta \Sigma$ at very small $x$ outside of the measured region. Denoting, for short hand,  by $\Delta q$ the first moment of the net helicity carried by the sum  $q+\bar q$ we have the relations \cite{deFlorian:2009vb,Blumlein:2010rn}:
\beq
a_3= \Delta u -  \Delta d = (F+D)(1+\epsilon _2)=1.269\pm0.003\\
\label{a3}
\eeq
\beq
a_8= \Delta u +  \Delta d -2\Delta s= (3F-D)(1+\epsilon _3)=0.586\pm0.031\\
\label{a8}
\eeq
where the $F$ and $D$ couplings are defined in the SU(3) flavour symmetry limit and $\epsilon_2$ and $\epsilon_3$ describe the SU(2) and SU(3) breakings, respectively. From the measured first moment of the structure function $g_1$ one obtains the value of $a_0=\Delta \Sigma$:
\beq
\Gamma_1= \int dx g_1(x)= \frac{1}{12}\left[a_3+\frac{1}{3}(a_8+4a_0)\right]\\
\label{g1}
\eeq
with the result, at $Q^2\sim 4 \rm{GeV}^2$:
\beq
a_0= \Delta \Sigma=\Delta u +  \Delta d+\Delta s= a_8+3\Delta s \sim 0.25\\
\label{a0}
\eeq
In turn, in the SU(3) limit $\epsilon_2=\epsilon_3=0$, one then obtains:
\beq
\Delta u \sim 0.82,~~~~\Delta d \sim -0.45,~~~~\Delta s \sim -0.11\\
\label{uds}
\eeq
This is an important result! Given $F$, $D$ and $\Gamma_1$ we know $\Delta u$, $\Delta d$, $\Delta s$ and $\Delta \Sigma$ in the SU(3) limit which should be reasonably accurate.
The $x$ distribution of $g_1$ is known down to $x\sim 10^{-4}$ on proton and deuterium and the 1st moment of $g_1$ does not seem to get much from the unmeasured range at small $x$  (also theoretically $g_1$ should be smooth at small $x$ \cite{gr}). The value of  $\Delta s \sim -0.11$ from totally inclusive data and $SU(3)$ appears to be at variance with the value extracted from single particle inclusive DIS (SIDIS) where one obtains
a nearly vanishing result for $\Delta s$ in a fit to all data  \cite{deFlorian:2009vb,sis} that leads to puzzling results. There is, in fact, an apparent tension between the 1st moments as determined by using the approximate $SU(3)$ symmetry and from fitting the data on SIDIS ($x\geq 0.001$) (in particular for the strange density). But the adequacy of the SIDIS data is questionable (in particular of the kaon data which fix $\Delta s$) and of their theoretical treatment (for example, the application of parton results at too low an energy and the ambiguities on the kaon fragmentation function). 

$\Delta \Sigma$ is conserved in perturbation theory at LO (i.e. it does not evolve in $Q^2$). For conserved quantities we would expect that they are the same for constituent and for parton quarks. But actually the conservation of $\Delta \Sigma$ is broken by the axial anomaly and, in fact, in perturbation theory beyond LO the conserved density is actually $\Delta \Sigma'=\Delta \Sigma+n_f/2\pi \alpha_s~\Delta g$  \cite{ross}. Note that also $\alpha_s \Delta g$ is conserved in LO, that is $\Delta g \sim \log{Q^2}$. This behaviour is not controversial but it will take long before the log growth of $\Delta g$ will be confirmed by experiment! But by establishing this behaviour  one would show that the extraction of $\Delta g$ from the data is correct and that the QCD evolution works as expected.  If $\Delta g$ was large enough it could account for the difference between partons ($\Delta \Sigma$) and constituents ($\Delta \Sigma'$). From the spin sum rule it is clear that the log increase should cancel between $\Delta g$ and  $L_z$. This cancelation is automatic as a consequence of helicity conservation in the basic QCD vertices.  $\Delta g$ can be measured indirectly by scaling violations and directly from asymmetries, e.g. in SIDIS. Existing measurements by HERMES, COMPASS, and at RHIC are still crude but show no hint of a large $\Delta g$ at accessible values of $x$ and $Q^2$. Present data, affected by large errors (see, in particular, ref. \cite{NNPol} for a discussion of this point)  are consistent \cite{NNPol,deFlorian:2009vb,Blumlein:2010rn,Leader:2010rb,Hirai:2008aj,sis} with a sizable contribution of $\Delta g$ to the spin sum rule in Eq.(\ref{spin}) but there is no indication that $\alpha_s \Delta g$ effects can explain the difference between constituents and parton quarks.

\subsection{Hadron Collider Processes and Factorization}
\label{sec:17}

There are three classes of hard processes: those with no hadronic particles in the initial state, like $e^+e^-$ annihilation, those initiated by a lepton and a hadron, like DIS, and those with two incoming hadrons. The parton densities, defined and measured in DIS, are instrumental to compute hard processes initiated by collisions of two hadrons, like $p \bar p$ (Tevatron) or $pp$ (LHC). Suppose you have a hadronic process of the form $h_1+h_2 \rightarrow X+all$ where $h_i$ are hadrons and  $X$ is some
triggering particle or pair of particles or one or more jets which specify the large scale $Q^2$ relevant for the process, in general somewhat,
but not much, smaller than s, the total c.o.m. squared mass. 
\begin{figure}[h]
\centering
\includegraphics[width=0.3\columnwidth]{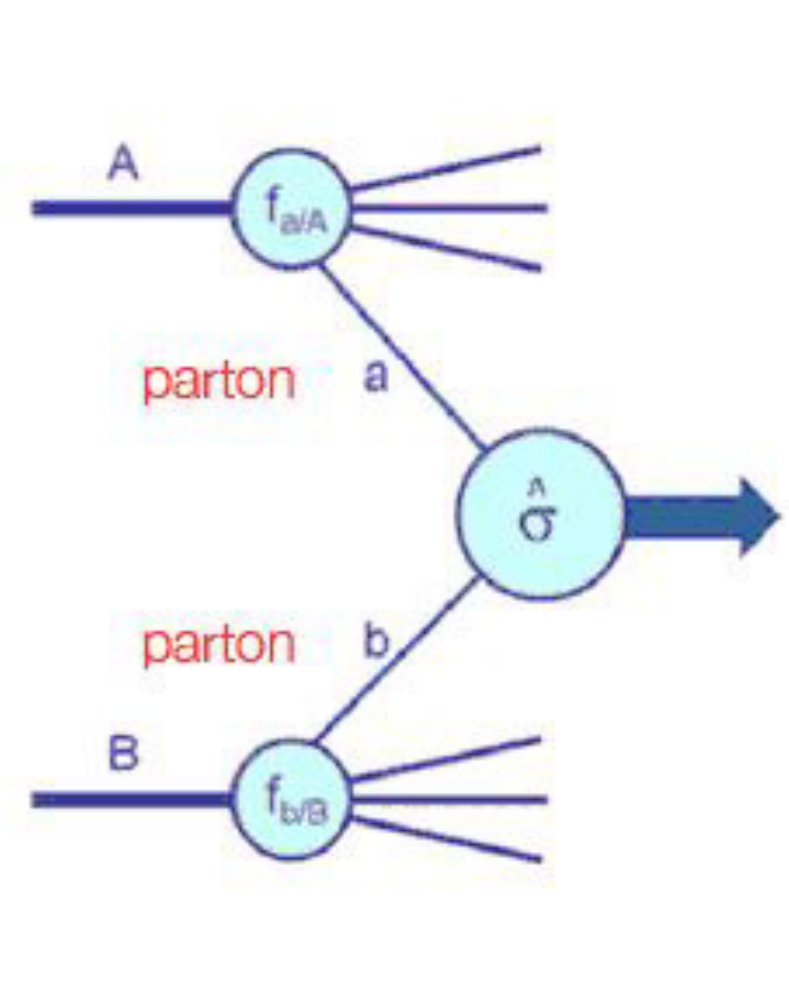} 
\caption[]{Diagram for the Factorization Theorem}
\label{figFT}
\end{figure}
For example, $X$ can be a $W^±$ or a Z or a virtual photon with large $Q^2$ (Drell-Yan processes), or a jet at large transverse momentum
$p_T$, or a pair of heavy ( of mass M) quark-antiquark. By "all" we mean a totally inclusive collection of hadronic particles. The Factorization Theorem (FT) states that for the total cross-section or some other sufficiently inclusive distribution we can write, apart from power suppressed corrections, the expression (see also Fig. \ref{figFT}):
\beq
\sigma(s,\tau)~=~\sum_{AB}\int dx_1dx_2 p_{1A}(x_1,Q^2)p_{2B}(x_2,Q^2)\sigma_{AB}(x_1x_2s,\tau)\label{FT}\\
\eeq
Here $\tau=Q^2/s$ is a scaling variable, $p_{iA}$ are the densities for a parton of type A inside the hadron $h_i$, $\sigma_{AB}$
is the partonic cross-section for parton-A + parton-B$ \rightarrow X + all'$. Here $all$' is the partonic version of $all$: a totally inclusive collection of quarks, antiquarks and gluons. This result is based on the fact that the mass
singularities that are associated with the initial legs are of universal nature, so that one can reproduce the same modified
parton densities, by absorbing these singularities into the bare parton densities, as in DIS. Once the
parton densities and $\alpha_s$ are known from other measurements, the prediction of the rate for a given hard process is
obtained with not much ambiguity (e.g from scale dependence or hadronization effects). At least a NLO calculation of the reduced partonic cross-section $\sigma_{AB}$ is needed in order to
correctly specify the scale and in general the definition of the parton densities and of the running coupling in the leading
term. The residual scale and scheme dependence is often the most important source of theoretical error.  An important question is: to what extension is the FT proven? In perturbation theory up to
NNLO it has been explicitly checked to hold for many processes: if corrections exist we already know that they must be small (we stress that we are only considering totally inclusive processes).  At all orders the most in depth discussions have been carried out in refs. \cite{collins}, in particular for  Drell-Yan processes. The LHC experiments offer a wonderful opportunity for testing the FT by comparing precise theoretical predictions with accurate data on a large variety of processes (for a recent review, see, for example, ref. \cite{bds}). 

A great effort has been and is being devoted to the theoretical preparation and interpretation of the LHC experiments. For this purpose very difficult calculations are needed at NLO and beyond because the strong coupling, even at the large $Q^2$ values involved,  is not that small. New powerful techniques for amplitude calculations have been developed.
An interesting development at the interface between string theory and QCD is twistor calculus. A precursor work was the Parke-Taylor result in 1986 \cite{pata} on the amplitudes for n incoming gluons with given ± helicities \cite{ber}. Inspired by dual models, they derived a compact formula for the maximum non vanishing helicity violating amplitude (with n-2  plus and 2 minus helicities) in terms of spinor products. Using the relation between strings and gauge theories in twistor space Witten developed in '03 \cite{Wit} a formalism in terms of effective vertices and propagators that allows to compute all helicity amplitudes. The method, alternative to other modern techniques for the evaluation of Feynman diagrams \cite{modFeyn}, leads to very compact results. Since then rapid progress followed (for reviews, see \cite{cacha}): the method was extended  to include massless external fermions \cite{Georgiou} and also external EW vector bosons \cite{Bern} and Higgs particles \cite{Dixon}. The level already attained is already important for multijet events at the LHC. The study of loop diagrams came next.
The basic idea is that loops can be fully reconstructed from their unitarity cuts. First proposed by Bern, Dixon and Kosower \cite{bdk} the technique was revived by Britto, Cachazo and Feng \cite{bcf}
and then perfected by Ossola, Papadopoulos and Pittau  \cite{opp} and further extended to massive particles in ref. \cite{egkm}. For a recent review of these new methods see ref. \cite{ekmz}. In parallel also the activity on event simulation has received a big boost from the LHC preparation (see, for example, the review \cite{revLHC}). Powerful techniques for the generation of numerical results at NLO for processes with complicated final states have been developed: the matching of matrix element calculation together with the modeling of parton showers has been realised in packages like Black Hat \cite{BH} (on-shell methods for loops) used in association with Sherpa \cite{Sherpa}(for real emission), or POWHEG BOX \cite{powheg}, or aMC@NLO \cite{aMC}, the automated version of the general framework MC@NLO \cite{frix}. In a complete simulation the matrix element calculation, improved by resummation of large logs, provides the hard skeleton (with large $p_T$ branchings) while the parton shower is constructed by a sequence of factorized collinear emissions fixed by the QCD splitting functions. In addition, at low scales a model of hadronization completes the simulation. The importance of all the components, matrix element, parton shower and hadronization can be appreciated in simulations of hard events compared with Tevatron and LHC data. One can say that the computation of NLO corrections in perturbative QCD has been by now completely automatized.

A partial list of examples of recent NLO calculations in pp collisions, obtained with these techniques is:  W + 3 jets \cite{W3j}, Z, $\gamma^*$  + 3 jets  \cite{Z3j}, W, Z + 4 jets \cite{W4j}, W + 5 jets \cite{W5j}, $t\bar t b \bar b$  \cite{ttbb}, $t \bar t$ + 2 jets \cite{ttjj}, $t\bar t$ W \cite{ttW}, WW+ 2 jets \cite{WWjj}, WW$b\bar b$ \cite{WWbb}, $b \bar b b \bar b$ \cite{bbbb} etc.
Here in the following we present in more detail a number of important, simplest, examples without any pretension to completeness.

\subsubsection{Vector Boson Production}
\label{sec:17.1}

\begin{figure}[h]
\centering
\includegraphics[width=0.8\columnwidth]{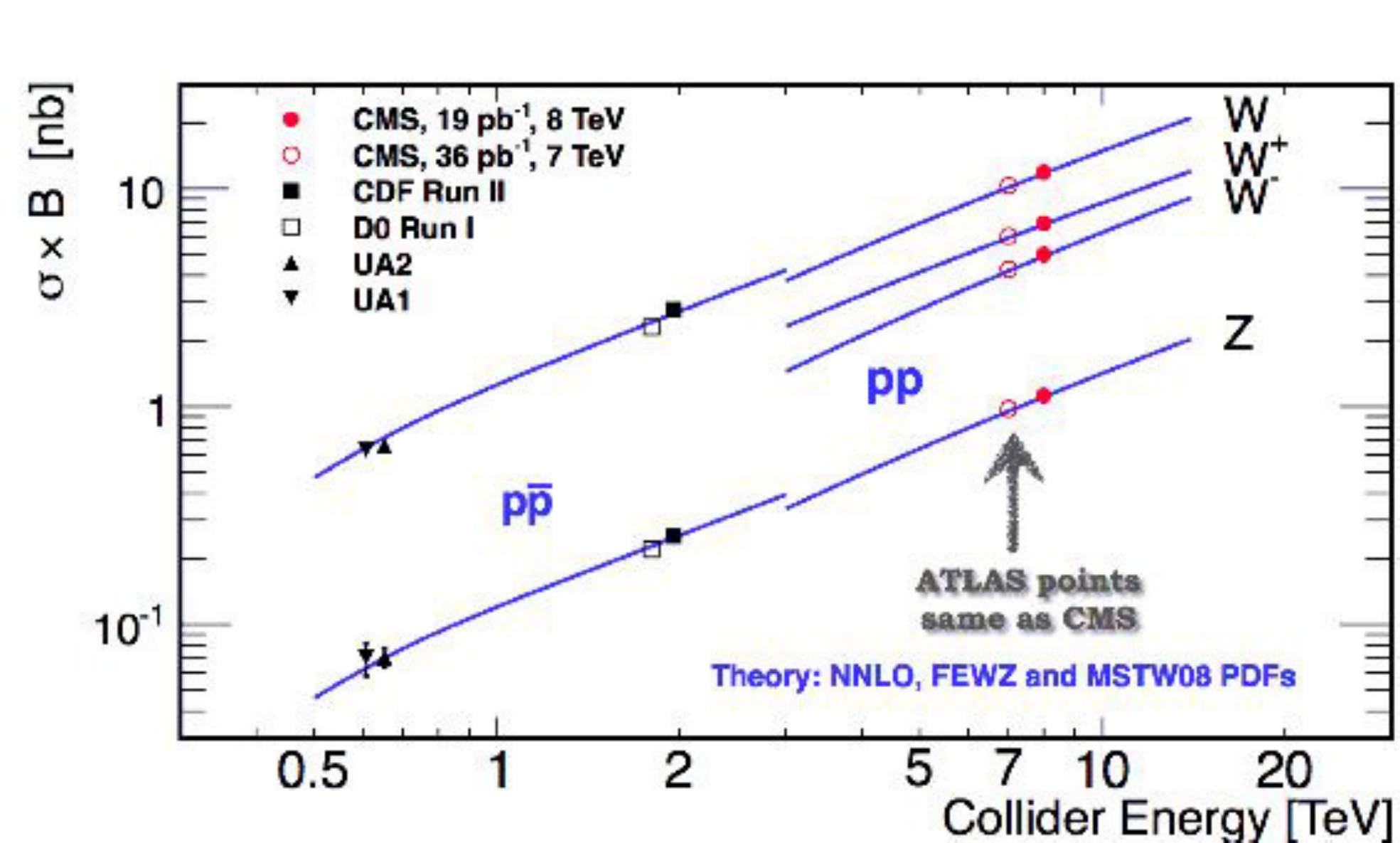}
\caption[]{Data vs. theory for $W$ and $Z$ production at hadron colliders \cite{WZ}.}
\label{fig19}
\end{figure}

Drell-Yan processes which include lepton pair production via virtual $\gamma$, W or Z exchange, offer a particularly good opportunity to
test QCD. This process, among those quadratic in parton densities with a totally inclusive final state, is perhaps the simplest one from a theoretical point of view. The large scale is
specified and measured by the invariant mass squared $Q^2$ of the lepton pair which itself is not strongly interacting (so there are no dangerous hadronization effects). The
QCD improved parton model leads directly to a prediction for the total rate as a function of $s$ and $\tau=Q^2/s$. The value of the LO
cross-section is inversely proportional to the number of colours $N_C$ because a quark of given colour can only annihilate
with an antiquark of the same colour to produce a colourless lepton pair. The order $\alpha_s(Q^2)$ NLO corrections to the
total rate were computed long ago  \cite{F2, kubar} and found to be particularly large, when the quark densities are defined from
the structure function $F_2$ measured in DIS at $q^2=-Q^2$. The ratio $\sigma_{corr}/\sigma_{LO}$ of the corrected and the
Born cross-sections, was called K-factor \cite{kfac}, because it is almost a constant in rapidity. More recently also the NNLO full
calculation of the K-factor was completed, a very remarkable calculation \cite{nee}. Over the years the QCD predictions for W and Z production, a better testing ground than the older fixed target Drell-Yan experiments,  have been compared with experiments at CERN $Sp \bar p S$ and Tevatron energies and now at the LHC. $Q\sim m_{W,Z}$ is large enough to make the prediction reliable (with a not too large K-factor) and the ratio $\sqrt{\tau}=Q/\sqrt{s}$ is not too small. Recall that in lowest order $x_1x_2s=Q^2$ so
that the parton densities are probed at $x$ values around $\sqrt{\tau}$. We have 
$\sqrt{\tau}=0.13-0.15$ (for W and Z production, respectively) at
$\sqrt{s}=630$ GeV (CERN
$Sp\bar pS$ Collider) and  $\sqrt{\tau}=0.04-0.05$ at the Tevatron. At the
LHC at 8 TeV or at 14 TeV one has $\sqrt{\tau}~\sim10^{-2}$  or $\sim$ 6  $\cdot 10^{-3}$, respectively (for both W and Z production). A comparison of the experimental total rates for $W$, $Z$ with the QCD predictions at hadron colliders \cite{WZ} is shown in Fig.~\ref{fig19}. It is also important to mention that the cross-sections for di-boson production (i.e. $WW, WZ, ZZ, W\gamma, Z\gamma$) have been measured at the Tevatron and the LHC and are in fair agreement with the SM prediction (see, for example, the summary in ref. \cite{Lomb} and refs. therein). The typical precision is comparable to or better than the size of NLO corrections.

\begin{figure}[h]
\centering
\includegraphics[width=10cm]{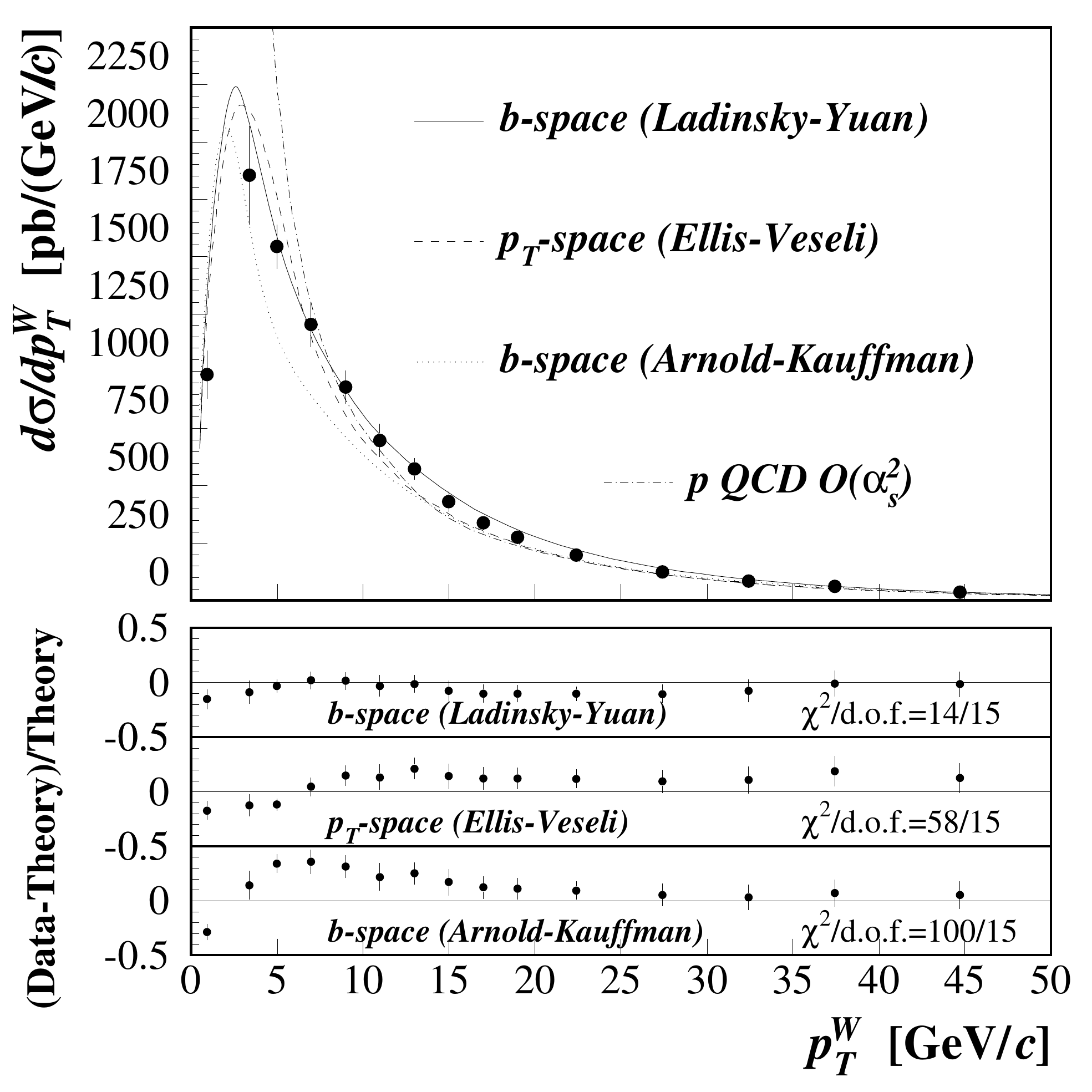} 
\caption[]{QCD predictions for the $W~p_T$ distribution compared with recent D0 data at the Tevatron ($\sqrt s$ = 1.8 TeV) \cite{D0} \cite{pTWfit}.}
\label{fig20}
\end{figure}

The calculation of the W/Z $p_T$ distribution is a classic challenge in QCD. For large $p_T$, for example $p_T\sim
O(m_W)$, the $p_T$ distribution can be reliably computed in perturbation theory, which was done up to NLO in the late '70's
and early '80's  \cite{pTpert}. A problem arises in the intermediate range $\Lambda_{QCD}<<p_T<<m_W$, where the bulk of the data is
concentrated, because terms of order
$\alpha_s(p_T^2)\log{m_W^2/p_T^2}$ become of order 1 and should included to all orders \cite{pTW}. At order $\alpha_s$ we have:
\beq
\frac{1}{\sigma_0}\frac{d\sigma_0}{dp_T^2}~=~(1+A)\delta(p_T^2)~+~\frac{B}{p_T^2}\log{\frac{m_W^2}{p_T^2}}_+
~+~\frac{C}{(p_T^2)_+}~+~ D(p_T^2)\label{pT}\\
\eeq
where A, B, C, D are coefficients of order $\alpha_s$. The "+" distribution is defined in complete analogy with
Eq.~(\ref{plus}):
\beq
\int_0^{p^2_{TMAX}}g(z)f(z)_+dz~=~\int_0^{p^2_{TMAX}}[g(z)-g(0)]f(z)dz\label{pluspT}\\
\eeq
The content of this, at first sight mysterious, definition is that the singular "+" terms do not contribute to the total
cross-section. In fact for the cross-section the weight function is $g(z)=1$ and we obtain:
\beq
\sigma~=~\sigma_0[(1+A)~+~\int_0^{p^2_{TMAX}}D(z)dz]\label{tot}\\
\eeq
The singular terms, of infrared origin, are present at the non completely inclusive level but disappear in the total
cross-section. Solid arguments have been given \cite{pTW} that these singularities exponentiate. Explicit calculations in low order support the exponentiation which leads to the following expression:
\beq
\frac{1}{\sigma_0}\frac{d\sigma_0}{dp_T^2}~=~\int \frac{d^2b}{4\pi}\exp{(-ib\cdot p_T)}(1+A)\exp{S(b)}\label{Sud}\\
\eeq
with:
\beq
S(b)~=~\int_0^{p_{TMAX}}\frac{d^2k_T}{2\pi}[\exp{ik_T\cdot b}-1][\frac{B}{k_T^2}\log{\frac{m_W^2}{k_T^2}}
~+~\frac{C}{k_T^2}]\label{s}\\
\eeq
At large $p_T$  the  perturbative expansion is recovered. At intermediate $p_T$ the infrared $p_T$ singularities are
resummed (the Sudakov log terms, which are typical of vector gluons, are related to the fact that for a charged particle in
acceleration it is impossible not to radiate, so that the amplitude for no soft gluon emission is exponentially suppressed). A delicate procedure for matching perturbative and resummed terms is needed \cite{match1}.
However this formula has problems at small $p_T$, for example, because of the presence of $\alpha_s$ under the integral for
$S(b)$: presumably the relevant scale is of order $k_T^2$. So it must be completed by some non perturbative ansatz or an
extrapolation into the soft region \cite{pTW}. All the formalism has been extended to NLO accuracy  \cite{pTWfit}, where one starts from the
perturbative expansion at order
$\alpha_s^2$, and generalises the resummation to also include NLO terms of order $\alpha_s(p_T^2)^2\log{m_W^2/p_T^2}$. The
comparison with the data is very impressive. In Fig.~\ref{fig20} we see the $p_T$ distribution as predicted in QCD (with a number of
variants that mainly differ in the approach to the soft region) compared with some recent data at
the Tevatron \cite{D0}. The W and Z $p_T$ distributions have also been measured at the LHC and are in fair agreement with the theoretical expectation \cite{pTLHC}.

\begin{figure}[h]
\centering
\includegraphics[width=10cm]{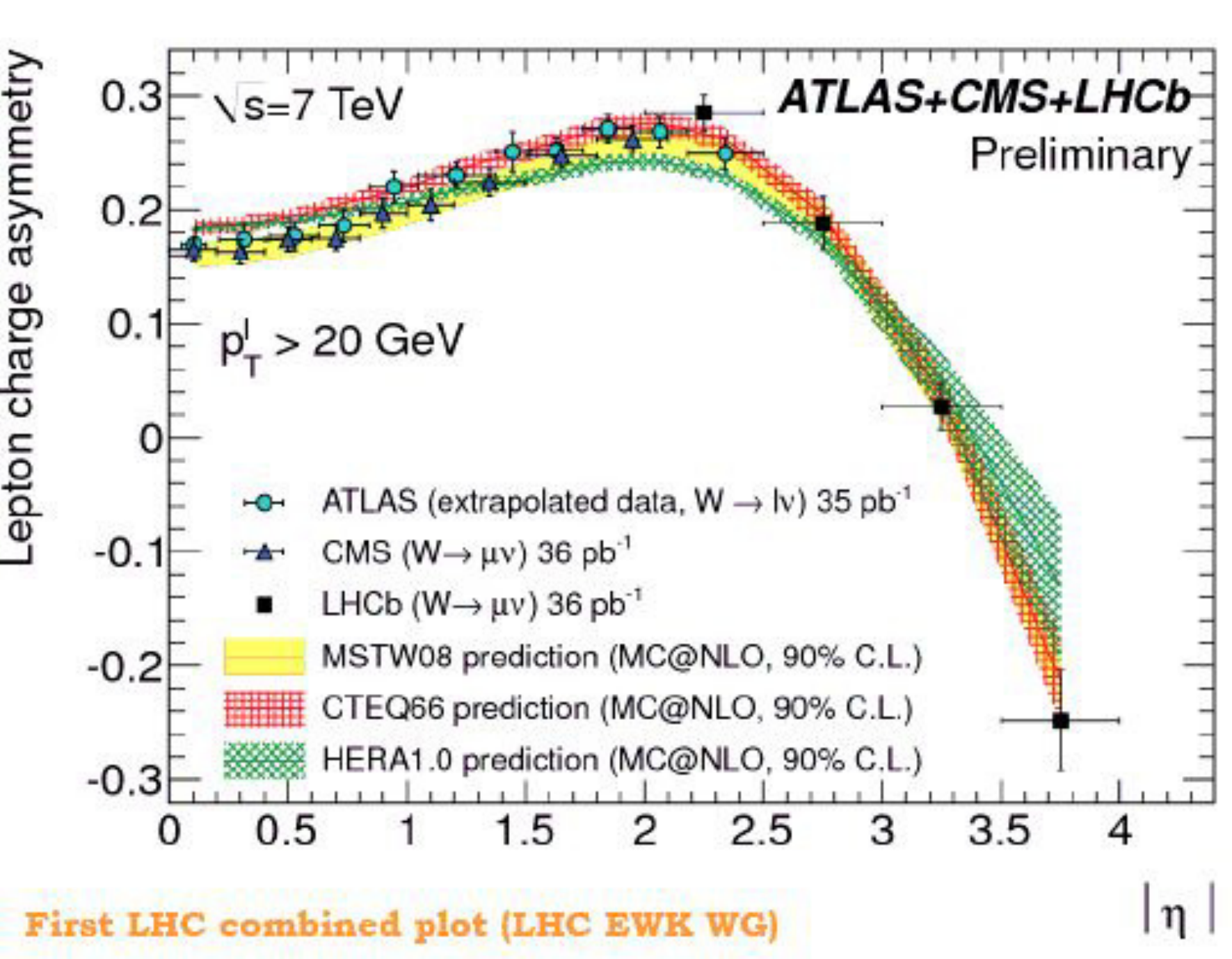} 
\caption[]{The combined LHC results for the W charge asymmetry (defined as $A~ \sim~ (W^+-W^-)/(W^++W^-)$ ) as a function of the pseudo rapidity $\eta$ \cite{etach}.}
\label{eta}
\end{figure}

The rapidity distributions of the produced W and Z have also been measured with fair accuracy at the Tevatron and at the LHC and predicted at NLO \cite{nlorap}. As a representative example of great significance we show in Fig. \ref{eta} the combined LHC results for the W charge asymmetry (defined as $A~ \sim~ (W^+-W^-)/(W^++W^-)$ ) as a function of the pseudo rapidity $\eta$ \cite{etach}. These data combine the ATLAS and CMS results at smaller values of $\eta$ with those of the LHCb experiments at larger $\eta$ (in the forward direction). This is a very important input for the disentangling of the different quark parton densities.

\subsubsection{Jets at Large Transverse Momentum}
\label{sec:17.2}

\begin{figure}[h]
\centering
\includegraphics[width=0.7\columnwidth]{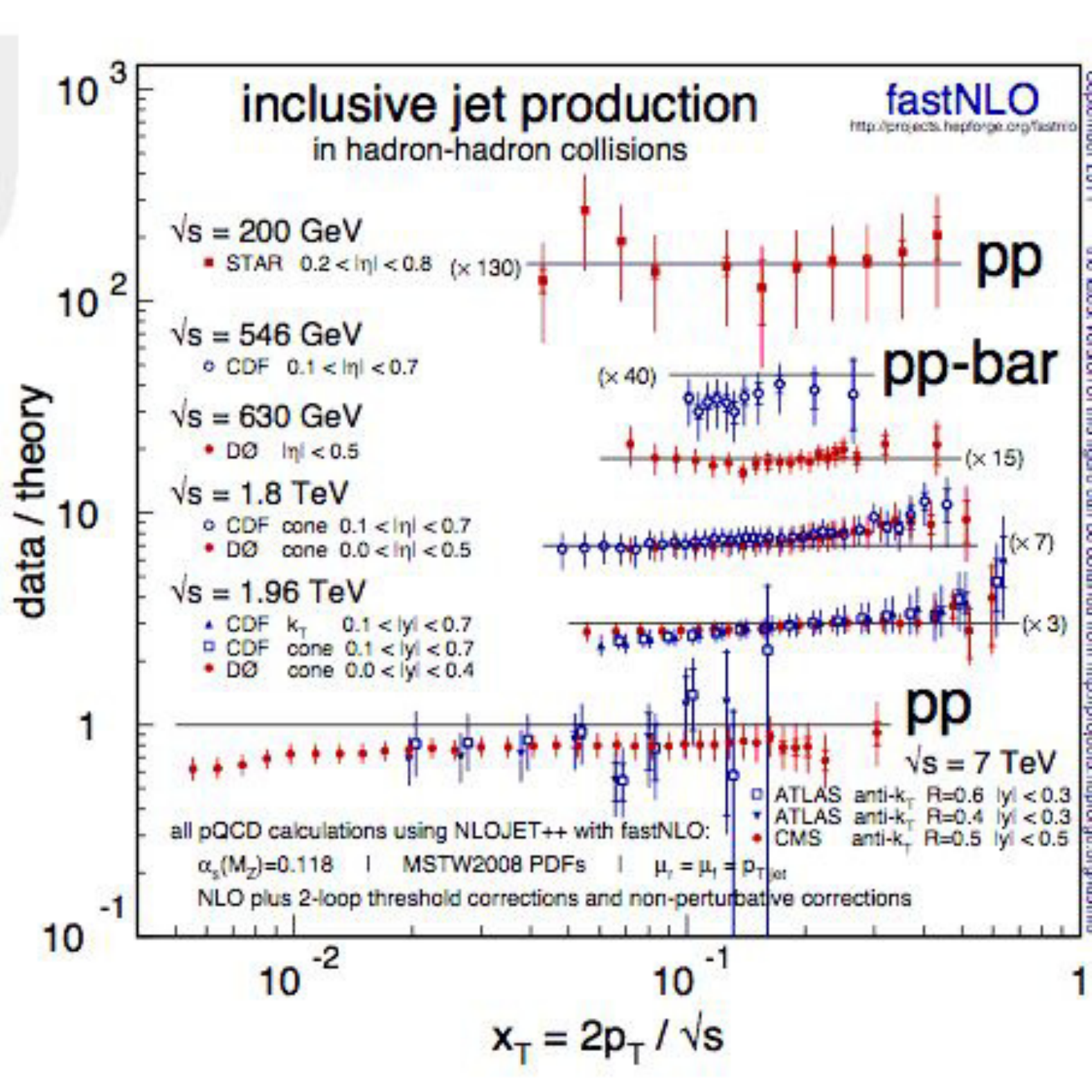} 
\caption[]{Jet production cross-section at $pp$ or $p\bar p$ colliders, as function of $p_T$ \cite{wobisch}. The theory predictions are from NLO perturbative calculations with state-of-the-art parton densities with the corresponding value of $\alpha_s$ plus a non perturbative correction factor due to hadronization and the underlying event, obtained using Monte Carlo event generators}
\label{fig15a}
\end{figure}

Another simple and important process at hadron colliders is the inclusive production of jets at large energy $\sqrt s$ and transverse momentum $p_T$. A comparison of the data with the QCD NLO predictions \cite{LOjets,NLOjets} in $pp$ or
$p\bar p$ collisions is shown in Fig.~\ref{fig15a} \cite{wobisch}. This is a particularly significant test because the rates at different
c.o.m. energies and, for each energy, at different values of $p_T$, span over many orders of magnitude. This steep behaviour is determined by the sharp falling of the parton densities with increasing $x$. Also, the
corresponding values of $\sqrt s$ and $p_T$ are large enough to be well inside the perturbative region. The overall
agreement of the data from ISR, UA1,2, STAR (at RHIC), CDF/D0,  and now ATLAS/CMS is indeed spectacular. In fact, the uncertainties on the resulting experiment/theory ratio, due to systematics and to ambiguities on parton densities, value of $\alpha_s$, scale choice and so on, which can reach a factor of 2-3, are much smaller than the spread of the cross-section values over many orders of magnitude.

\begin{figure}[h]
\centering
\includegraphics[width=0.8\columnwidth]{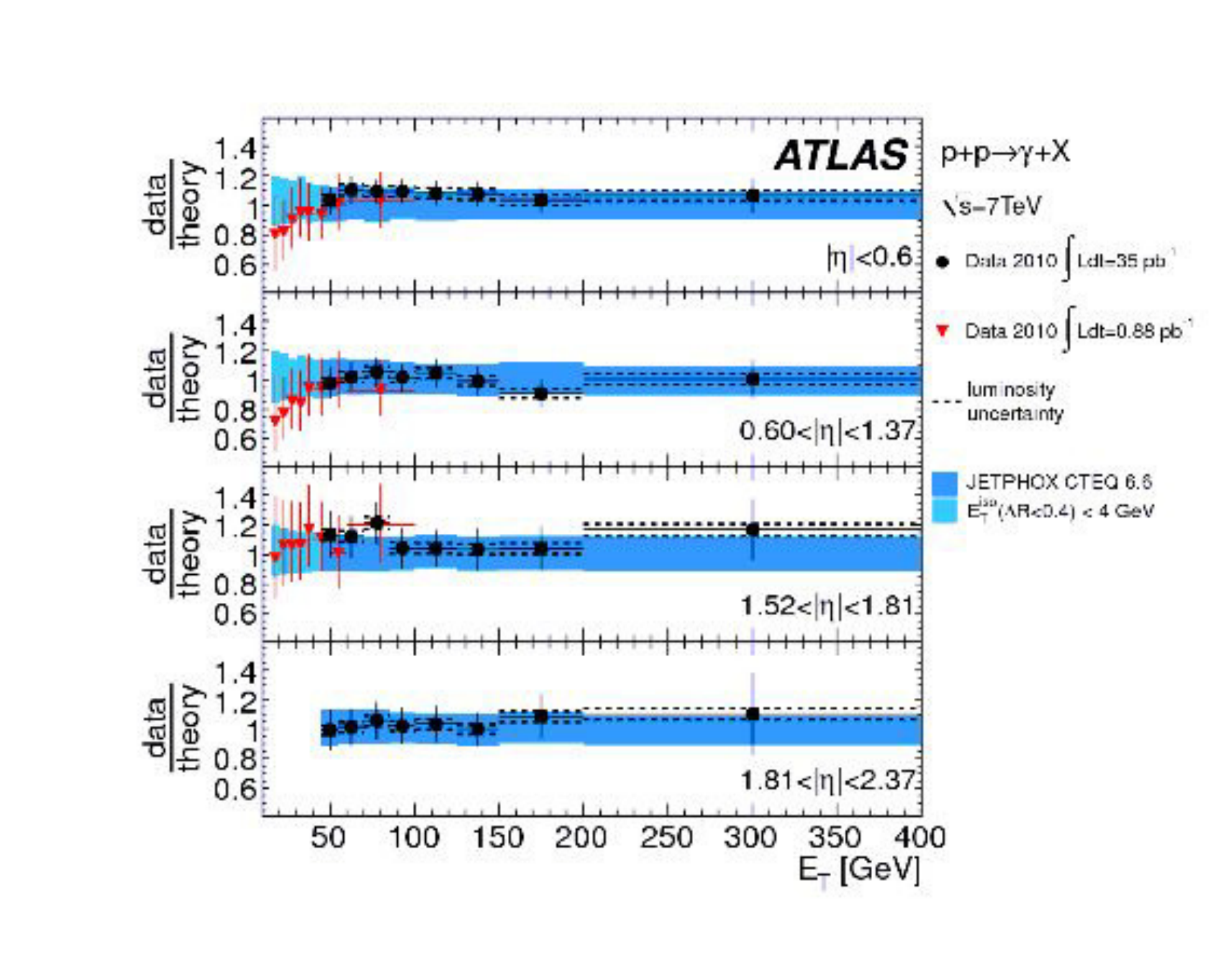} 
\caption[]{Single photon production in $p\bar p$ colliders as function of $p_T$ \cite{photatlas}}
\label{fig16a}
\end{figure}

Similar results also hold for the production of photons at large $p_T$. The ATLAS data \cite{photatlas}, shown in Fig.~\ref{fig16a}, are
in fair agreement with the theoretical predictions. For the same process less clear a situation was found with fixed target data. Here, first of
all, the experimental results show some internal discrepancies. Also, the $p_T$ accessible values being smaller, the
theoretical uncertainties are larger. 

\subsubsection{Heavy Quark Production}
\label{sec:17.3}

\begin{figure}[h]
\noindent
\includegraphics[width=15cm]{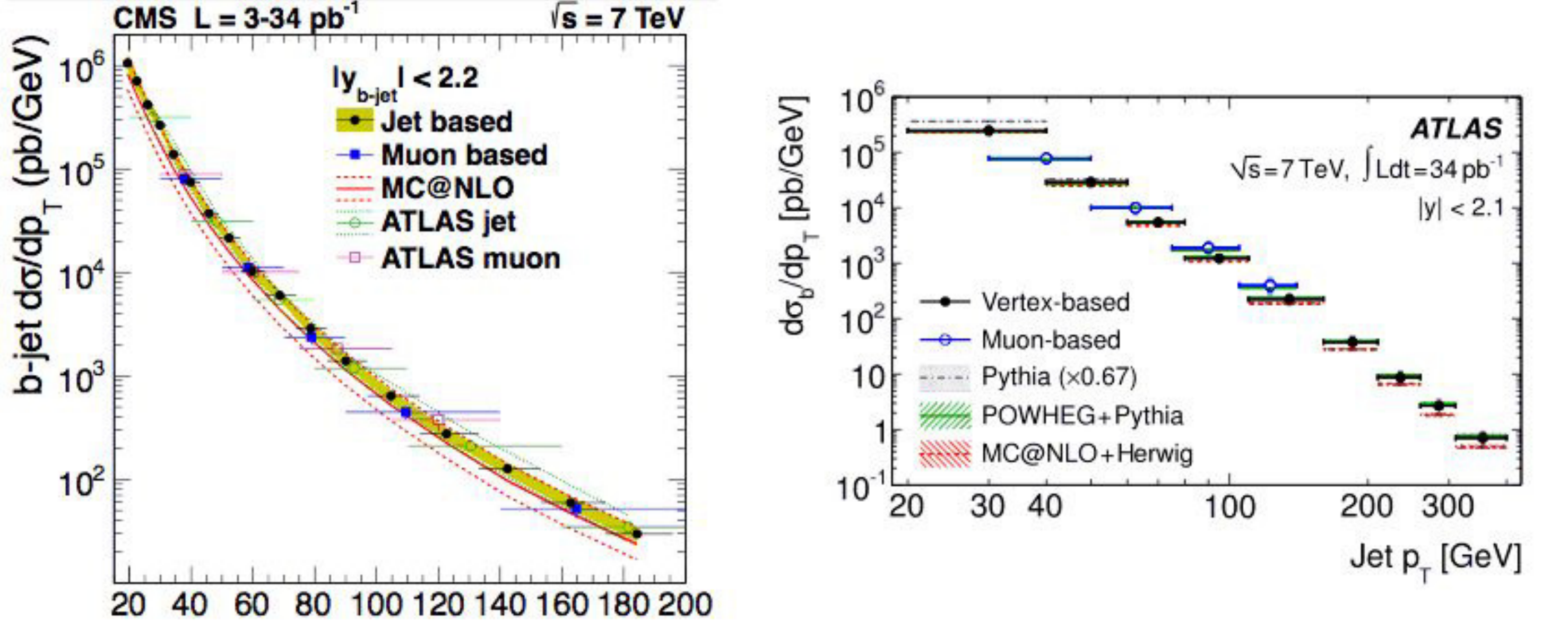} 
\caption[]{The $b$ production $p_T$ distribution at the LHC \cite{binclAtlCms}. }
\label{fig18a}
\end{figure}

We now discuss heavy quark production at colliders. The totally inclusive cross sections have been known at NLO since a long time  \cite{heavy}. The resummation of leading and next to leading  logarithmically enhanced effects in the vicinity of the threshold region have also been studied \cite{NLL}. The bottom production at the Tevatron has for some time represented a problem: the total rate and the $p_T$
distribution of b quarks observed at CDF and D0 appeared in excess of the prediction, up to the largest measured values of $p_T$ \cite{cacgre,bauer}.
But this is a complicated problem, with different scales being
present at the same time: $\sqrt{s}$, $p_T$, $m_b$.  Finally the discrepancy has been solved by better taking into account a number of small effects from resummation of large logarithms, the difference between b hadrons and b partons, the inclusion of better fragmentation functions etc. \cite{cacc}.
At present the LHC data on b production are in satisfactory agreement with the theoretical predictions (Fig.~\ref{fig18a} \cite{binclAtlCms}).

\begin{figure}[h]
\noindent
\includegraphics[width=16cm]{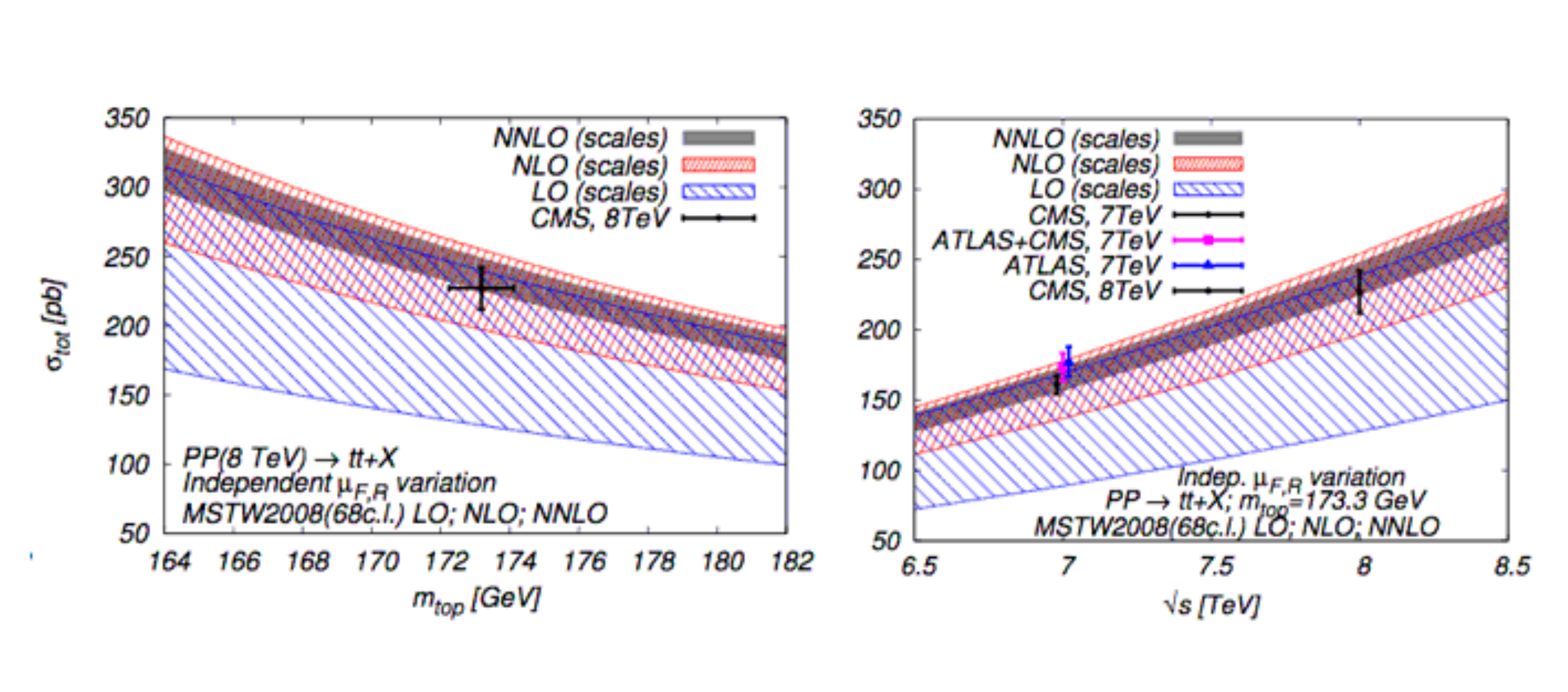} 
\caption[]{The $t \bar t$ production cross-section at the LHC collider. Scale dependence of the total cross-section at LO (blue), NLO (red) and NNLO (black) as a function of $m_{top}$ (left) or $\sqrt{s}$ (right) at the LHC 8 TeV \cite{tNNLO}}
\label{fig17-1}
\end{figure}

The top quark is really special: its mass is of the order of the Higgs VEV or its Yukawa coupling is of order 1 (in this sense it is the only "normal" case among all quarks and charged leptons).  Due to its heavy mass it decays so fast that it has no time to be bound in a hadron: thus it can be studied as a quark.  It is very important to determine its mass and couplings for different precision predictions of the SM. Perhaps the top quark could be particularly sensitive to new heavy states or have a connection to the Higgs sector in beyond the SM theories. Thus top physics has attracted much attention both from the experimental side, at hadron colliders, and from the theoretical point of view. In particular, the top-antitop inclusive cross-section has been measured in $p \bar p$ collisions at the Tevatron \cite{texp} and now in $pp$ collisions at the LHC \cite{tLHCATLAS,tLHCCMS}. The QCD prediction is at present completely known at NNLO \cite{tNNLO}. Soft gluon resummation has also been performed at NNLL \cite{nnll}. The agreement of theory and experiment is good for the best available parton density functions together with the values of $\alpha_s$ and of $m_t$ measured separately (the top mass is measured from the invariant mass of the decay products), as can be seen from Fig.~\ref{fig17-1} \cite{tNNLO}. The mass of the top (and the value of $\alpha_s$) can be determined from the cross section, assuming that QCD is correct, and compared with the more precise value from the decay final state.  The value of the pole top mass derived in ref.\cite{aldj} from the cross-section data, using the best available parton densities with the correlated value of $\alpha_s$, is: $ m_t^{pole}= 173.3\pm2.8$ GeV to be compared with the value measured at the Tevatron by the CDF and D0 collaborations $m_t^{exp}=173.2\pm0.9$ GeV. This quoted error is clearly too optimistic especially if one would identify this value with the pole mass which it resembles to. This error is only adequate within the specific procedure used by the experimental collaborations to define their mass (including which Montecarlo, which assumptions on higher order terms, non perturbative effects etc). The problem is how to export this value in other processes. Leaving aside the thorny issue of the precise relation of $m_t^{exp}$ with $m_t^{pole}$ it is clear that there is a good overall consistency.

The inclusive forward-backward asymmetry, $A_{FB}$,
in the $t \bar t$  rest frame has been measured by both the
CDF \cite{cdfFB} and  D0 \cite{d0FB} collaborations and found to be
in excess of  the SM prediction, by about  2 $\sigma$ \cite{AFBth}. For CDF the discrepancy increases at large $t \bar t$ invariant mass and reaches about 
2.5 $\sigma$ for $M_{t \bar t} \geq$  450 GeV.   Recently CDF has presented \cite{CDFcos} the first measurement
of the top-quark-pair production differential cross section as a
function of $\cos{\theta}$, with $\theta$ the production angle of the top quark. The coefficient of the $\cos{\theta}$ term in the dif-
ferential cross section, $a_1 = 0.40 \pm 0.12$ , is found in excess of
the NLO SM prediction, $0.15^{+0.07}_{-0.03}$, while all other terms are in good agreement with the NLO SM prediction and the $A_{FB}$ is dominated by this excess linear term. Is this a real discrepancy? The evidence is far from being compelling, but this effect has received much attention from theorists \cite{AFBRev}
A related observable at the LHC is the charge asymmetry
in $t \bar t$ production, $A_C$. In contrast to $A_{FB}$, the combined value of $A_C$ reported by ATLAS \cite{atlC} and CMS \cite{cmsC} agrees with the SM, within the still limited precision of the data.

\subsubsection{Higgs Boson Production}
\label{sec:17.4}

\begin{figure}[h]
\centering
\includegraphics[width=8cm]{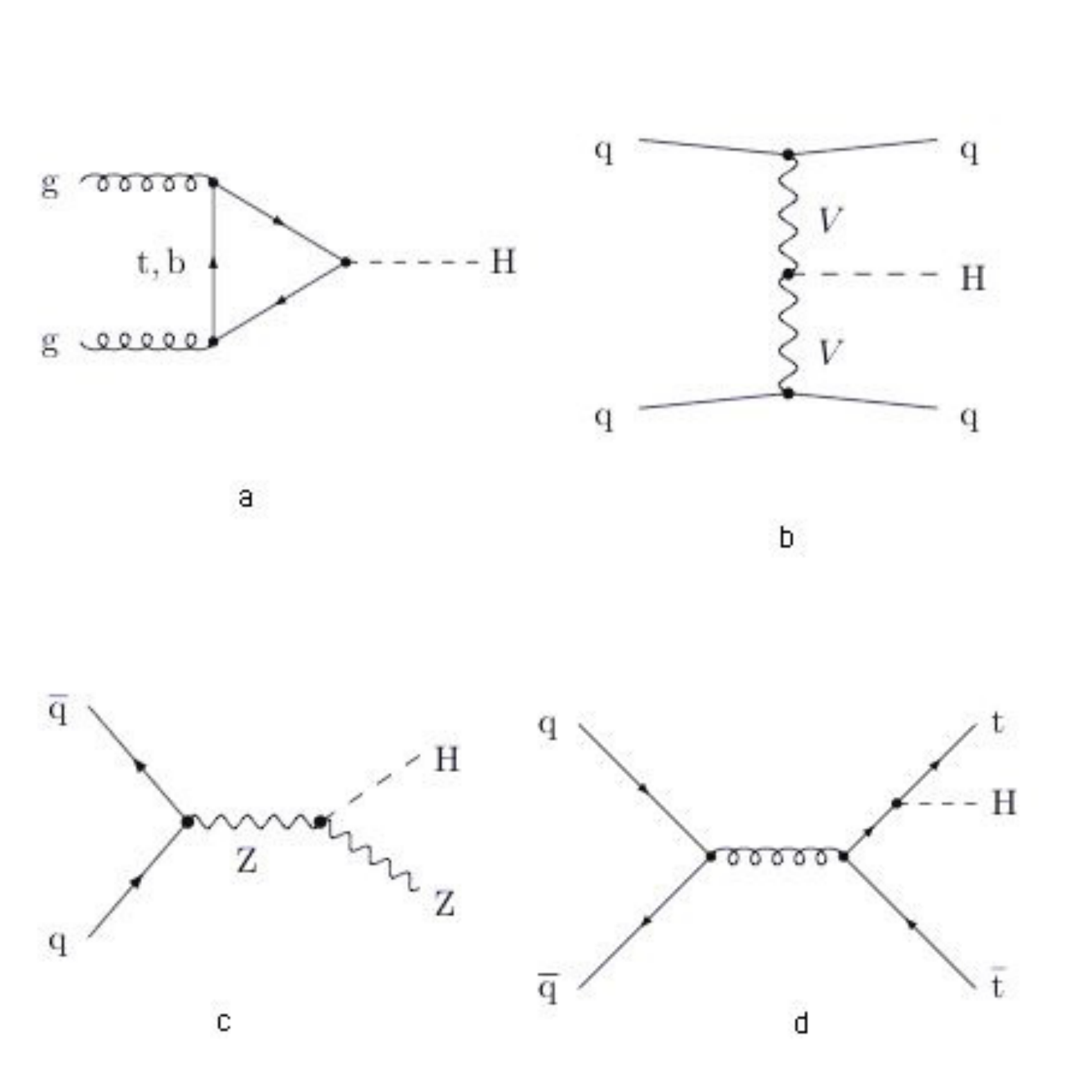} 
\caption[]{Representative Feynman diagrams for the Higgs production cross-section mechanisms: a) gluon fusion; b) Vector boson fusion (V=W, Z); c) Higgsstrahlung from a Z boson (an analogue diagram can be drawn for the W boson); d) $t \bar t$ associated production}
\label{fig21}
\end{figure}

\begin{figure}[h]
\centering
\includegraphics[width=9cm]{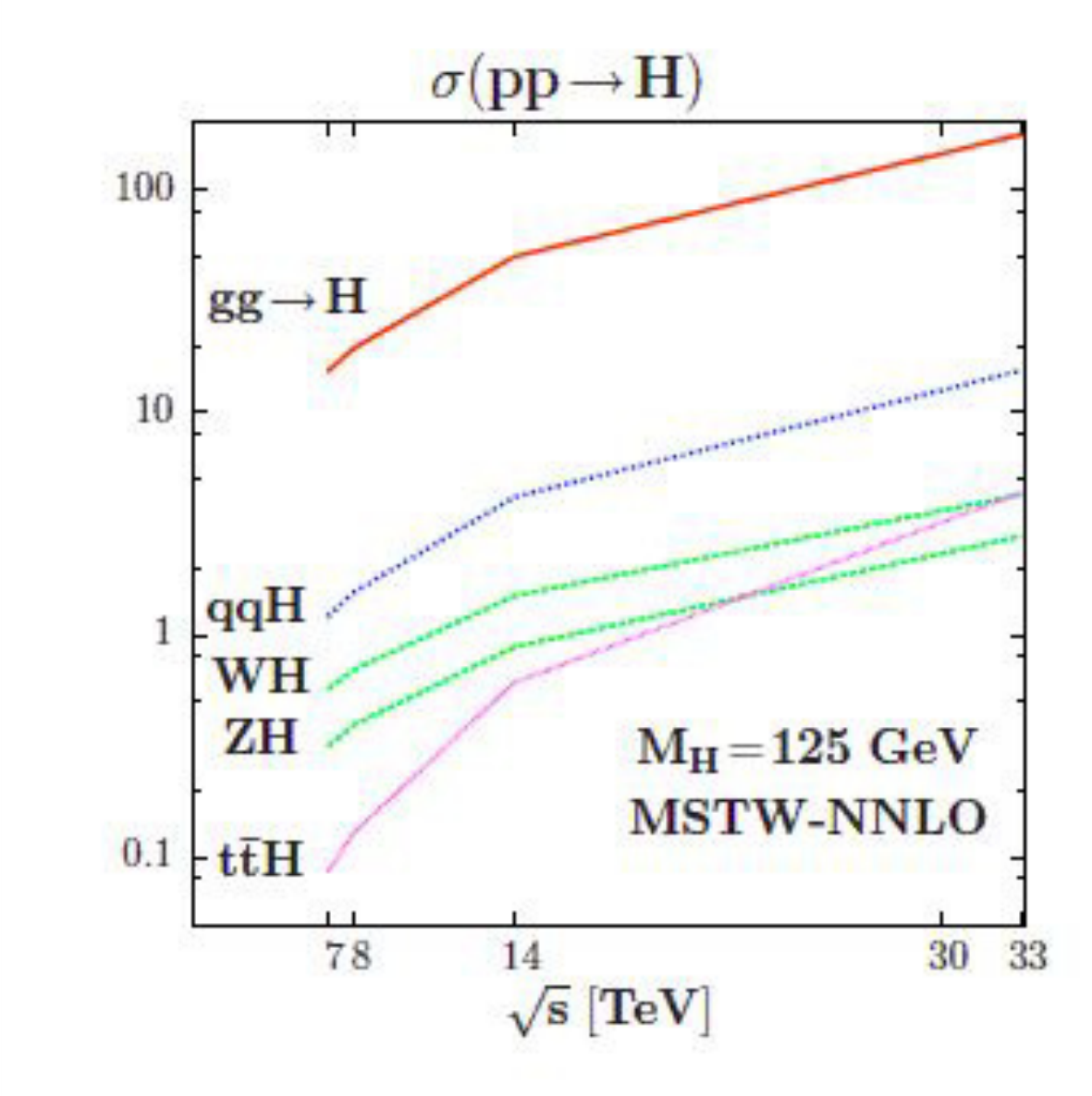} 
\caption[]{The production cross sections at the LHC for a Higgs with mass $M_H\sim 125$ GeV different c.m. energies \cite{djou12}.}
\label{Hrates}
\end{figure}

We now turn to the discussion of the SM Higgs inclusive production cross-section (for a review and a list of references see ref. \cite{handb}). The most important Higgs production modes are gluon fusion, vector boson fusion, Higgs strahlung, and associated production with top quark pairs. Some typical Feynman diagrams for those different modes are depicted in Fig.~\ref{fig21}. The predicted rates are shown in Fig. \ref{Hrates} \cite{djou12}. 

\begin{figure}[!h]
\centering
\includegraphics[width=9cm]{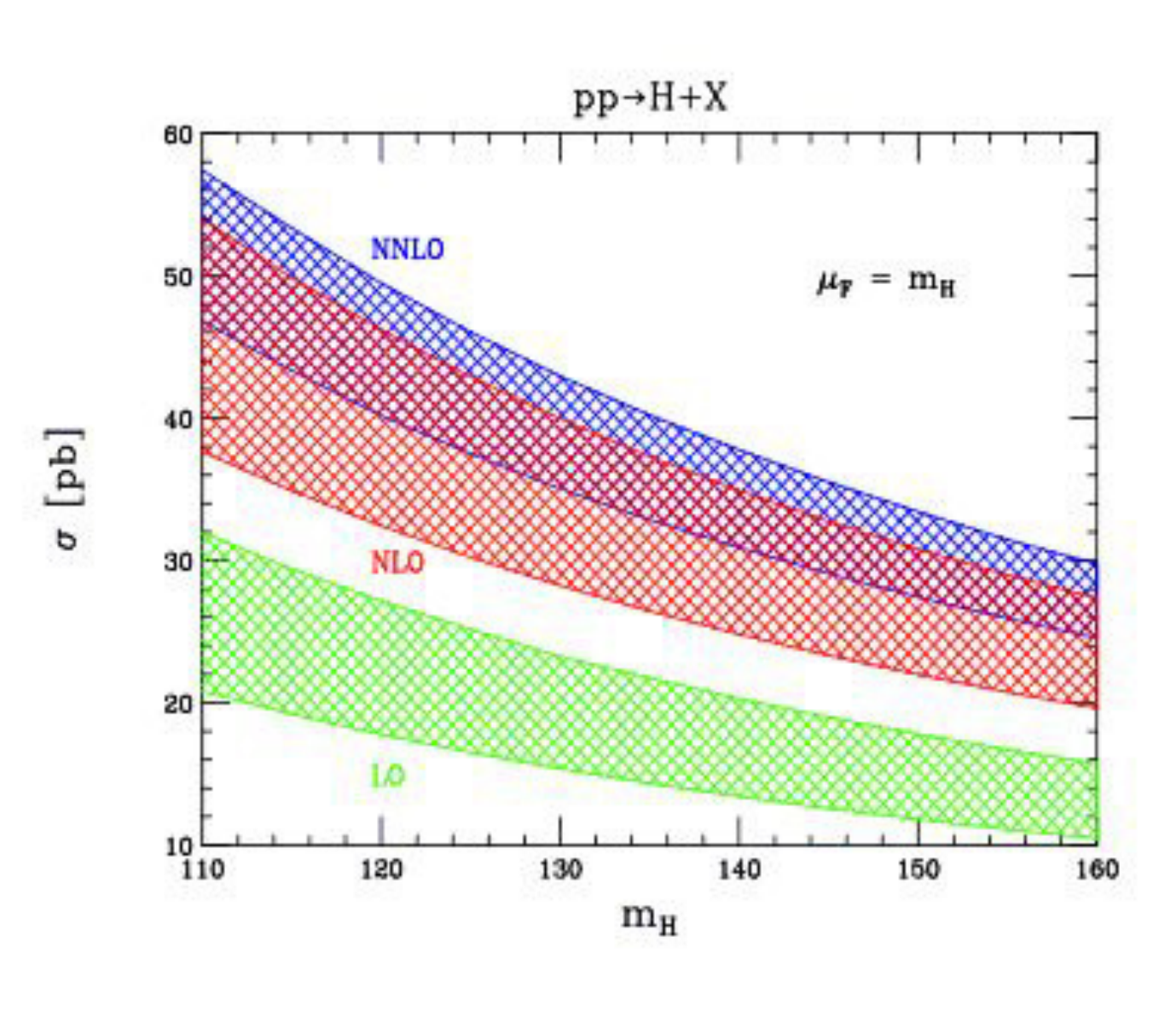}
\caption[]{The Higgs gluon fusion cross section in LO, NLO and NLLO \cite{Anasta1}.}
\label{fig21a}
\end{figure}

The most important channel at the LHC is Higgs production via $g ~+~ g \rightarrow H$. The amplitude is dominated by the top quark loop \cite{Georgi}. The NLO corrections turn out to be particularly large \cite{NLOHiggs}, as seen in Fig.~\ref{fig21a}. Higher order corrections can be computed either in the effective lagrangian approach, where the heavy top is integrated away and the loop is shrunk down to a point \cite{EllisGa} (the coefficient of the effective vertex is known to $\alpha_s^4$ accuracy \cite{Chety}), or in the full theory. At the NLO the two approaches agree very well for the rate as a function of $m_H$ \cite{Kramer}. The NNLO corrections have been computed in the effective vertex approximation \cite{NNLOHiggs} (see Fig.~\ref{fig21a}). Beyond fixed order, resummation of large logs were carried out \cite{Cat}. Also the NLO EW contributions have been computed \cite{Agli}. Rapidity (at  NNLO) \cite {Anasta} and $p_T$ distributions (at NLO) \cite{deFlo} have also been evaluated. At smaller $p_T$ the large logarithms [Log$(p_T/m_H)]^n$ have been resummed in analogy with what was done long ago for W and Z production \cite{Bozzi}. For additional recent works on Higgs physics at colliders see, for example, \cite{more}.

At different places in the previous pages we have seen examples of resummation of large logs. This is a very important chapter of modern QCD. The resummation of soft gluon logs enter in different problems and the related theory is subtle. I refer the reader here to some recent papers where additional references can be found \cite{soft}. A particularly interesting related development has to do with the so called non global logs (see, for example, \cite{nonglob}). If in the measurement of an observable some experimental cuts are introduced, which is a very frequent case, then a number of large logs can arise from the corresponding breaking of inclusiveness. It is also important to mention the development of software for the automated implementation of resummation (see, for example, \cite{numres}).

\begin{figure}[!h]
\noindent
\includegraphics[width=16cm]{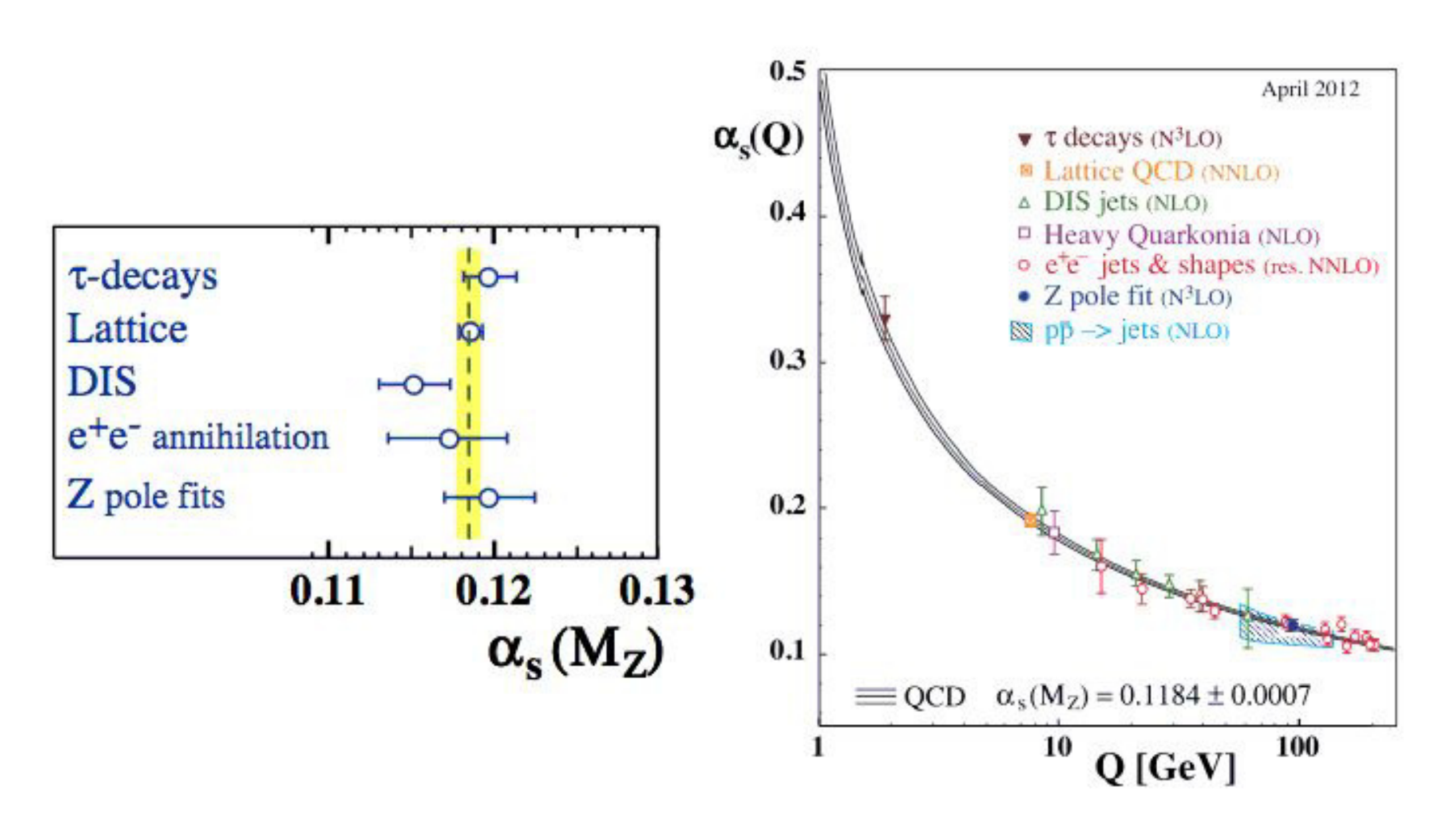}
\caption[]{Left: Summary of measurements of $\alpha_s(m_Z)$, used as input for the world average value of refs. \cite{pdg12,bethke}. The yellow band is the proposed average: $\alpha_s(m_Z)=0.1184\pm0.0007$. Right: Summary of measurements of $\alpha_s$ as a function of the respective energy scale Q}
\label{alfaspdg}
\end{figure}

\subsection{Measurements of $\alpha_s$}
\label{sec:18}

Very precise and reliable measurements of $\alpha_s(m_Z)$ are obtained from $e^+e^-$
colliders (in particular LEP), from deep inelastic scattering and from the hadron Colliders (Tevatron and LHC). The ''official''
compilation due to Bethke \cite{bethke,pich} and included in the 2012 edition of the PDG \cite{pdg12} is reproduced here in figs. \ref{alfaspdg}.
The agreement among so many different ways of measuring $\alpha_s$ is a strong quantitative test of QCD. However for some entries the stated error is taken directly from the original works and is not transparent enough as seen from outside (e.g. the lattice determination).
In my opinion one should select few theoretically cleanest processes for measuring $\alpha_s$ and consider all other ways as
tests of the theory. Note that in QED $\alpha$ is measured from one single very precise and theoretically clean observable (one possible calibration process is at present the electron g-2 \cite{ae1}).
The cleanest processes for measuring $\alpha_s$ are the totally inclusive ones (no hadronic corrections) with light cone dominance,
like Z decay, scaling violations in DIS and perhaps $\tau$ decay (but, for $\tau$, the energy scale is dangerously low). We will review these cleanest methods for measuring $\alpha_s$ in the following.

\subsubsection{$\alpha_s$ from $e^+e^-$ Colliders}
\label{sec:18.1}

The totally inclusive processes for measuring $\alpha_s$ at $e^+e^-$ colliders are hadronic Z decays ($R_l$, $\sigma_h$, $\sigma_l$, $\Gamma_Z$) and hadronic $\tau$ decays.
As we have seen in Sect. \ref{sec:15.1}, for a quantity like $R_l$ we can write a general expression of the form:
\beq
R_l~=~\frac{\Gamma(Z,\tau\rightarrow hadrons)}{\Gamma(Z,\tau\rightarrow
leptons)}~\sim~R^{EW}(1~+~\delta_{QCD}~+~\delta_{NP})~~\label{RR}\\
\eeq
where $R^{EW}$ is the electroweak-corrected Born approximation, $\delta_{QCD}$, $\delta_{NP}$ are the perturbative
(logarithmic) and non perturbative (power suppressed) QCD corrections. For a measurement of $\alpha_s$ (in the following we always refer to the $\overline{MS}$ definition of $\alpha_s$) at the Z resonance peak  one can use all the information from $R_l$, $\Gamma_Z=3\Gamma_l+\Gamma_h+\Gamma_{inv}$ and 
$\sigma_F=12\pi\Gamma_l\Gamma_F/(m_Z^2\Gamma_Z^2)$ (F=h or l). In the past the measurement from $R_l$ was preferred (by itself it leads to $\alpha_s(m_Z)=0.1226\pm0.0038$, a bit on the large side) but after LEP there is no reason for this preference. In all these quantities $\alpha_s$ enters through $\Gamma_h$, but the measurements of, say, $\Gamma_Z$, $R_l$ and $\sigma_l$ are really independent as they are affected by an entirely different systematics: $\Gamma_Z$ is extracted from the line shape, $R_l$ and $\sigma_l$ are measured at the peak but $R_l$ does not depend on the absolute luminosity while  $\sigma_l$ does. The most sensitive single quantity is $\sigma_l$. It gives $\alpha_s(m_Z)=0.1183\pm0.0030$. The combined value from the measurements at the Z (assuming the validity of the SM and the observed Higgs mass) is \cite{klu}:
\beq
\alpha_s(m_Z)=0.1187\pm0.0027\label{alZ}\\
\eeq
By adding all other electroweak precision electroweak tests (in particular $m_W$) one similarly finds \cite{ew}:
\beq
\alpha_s(m_Z)=0.1186\pm0.0026\label{alEW}\\
\eeq
These results have been obtained from the $\delta_{QCD}$ expansion up to and including the $c_3$ term of order $\alpha_s^3$. But by now the $c_4$ term (NNNLO!) has also been computed \cite{baik4} for inclusive hadronic $Z$ and $\tau$ decay. This remarkable calculation of about 20.000 diagrams, for the inclusive hadronic Z width, led to the result, for $n_f=5$ and $a_s=\alpha_s(m_Z)/\pi$:
\beq
\delta_{QCD}= [1~+~ a_s~+~0.76264~a_s^2  ~-~ 15.49~a_s^3~-~68.2~a_s^4~+~\dots]\\
\label{delc4}
\eeq
This result can be used to improve the value of  $\alpha_s(m_Z)$ from the EW fit given in Eq. (\ref{alEW}) that becomes:
\beq
\alpha_s(m_Z)=0.1190\pm0.0026 \label{alEW1}\\
\eeq
Note that the error shown is dominated by the experimental errors. Ambiguities from higher perturbative orders \cite{ste}, from power corrections and also from uncertainties on the Bhabha luminometer (which affect $\sigma_{h,l}$) \cite{debo} are very small. In particular, having now fixed $m_H$ does not decrease the error significantly \cite{gfitter,gru}. The main source of error is the assumption of no new physics, for example in the $Zb\bar b$ vertex that could affect the $\Gamma_h$ prediction. 

We now consider the measurement of $\alpha_s(m_Z)$ from $\tau$ decay. $R_\tau$ has a number of advantages that, at least in
part, tend to compensate for the smallness of $m_\tau=1.777$~GeV. First, $R_\tau$ is maximally inclusive, more than
$R_{e^+e^-}(s)$, because one also integrates over all values of the invariant hadronic squared mass:
\beq
R_\tau=\frac{1}{\pi}\int_0^{m_\tau^2}\frac{ds}{m_\tau^2}(1-\frac{s}{m_\tau^2})^2 Im\Pi_\tau(s)\label{incl}\\
\eeq
As we have seen, the perturbative contribution is now known at NNNLO \cite{baik4}. Analyticity can be used to transform the integral into one on the circle at $|s|=m_\tau^2$:
\beq
R_\tau=\frac{1}{2\pi i}\oint_{|s|=m_\tau^2}\frac{ds}{m_\tau^2}(1-\frac{s}{m_\tau^2})^2 \Pi_\tau(s)\label{incl1}\\
\eeq  
Also, the factor $(1-\frac{s}{m_\tau^2})^2$ is important to kill the sensitivity the region $Re[s]=m_\tau^2$ where the
physical cut and the associated thresholds are located. Still the sensitivity to hadronic effects in the vicinity of the cut is a non negligible source of theoretical error that the formulation of duality violation models try to decrease. But the main feature that has attracted attention to $\tau$ decays for the measurement of  $\alpha_s(m_Z)$ is that even a rough determination of $\Lambda_{QCD}$ at a low scale $Q \sim m_\tau$ leads to a very precise prediction of  $\alpha_s$ at the scale $m_Z$, just because in $\log{Q/\Lambda_{QCD}}$ the value of  $\Lambda_{QCD}$ counts less and less as $Q$ increases. The absolute error on $\alpha_s$ shrinks by a factor of about one order of magnitude going from  $\alpha_s(m_\tau)$ to  $\alpha_s(m_Z)$. Still I find a little suspicious that, in order to obtain a better measurement of  $\alpha_s(m_Z)$, you have to go down to lower and lower energy scales. And in fact, in general, one finds that the decreased control of higher order perturbative and of non perturbative corrections makes the apparent advantage totally illusory. For $\alpha_s$ from $R_\tau$ the quoted amazing precision is obtained by taking for granted that corrections suppressed by $1/m_\tau^2$ are negligible. The argument is that in the massless theory, the light cone expansion is given by:
\beq
\delta_{NP}=\frac{ZERO}{m_\tau^2}~+~c_4\cdot \frac{<O_4>}{m_\tau^4}~+~c_6\cdot \frac{<O_6>}{m_\tau^6}~+\cdots\label{OPEtau}\\
\eeq
In fact there are no dim-2 Lorentz and gauge invariant operators. For example, $Tr[{\bf g}_{\mu}{\bf g}^{\mu}]$ (recall Eq. \ref{10}) is not gauge invariant. In
the massive theory, the ZERO is replaced by the light quark mass-squared $m^2$. This is still negligible if $m$ is taken as a
lagrangian mass of a few MeV. If on the other hand the mass were taken to be the constituent mass 
of order $\Lambda_{QCD}$, this term would not be negligible at all and
would substantially affect the result (note that $\alpha_s(m_{\tau})/\pi\sim 0.1 \sim (0.6~{\rm GeV}/m_{\tau})^2$ and that $\Lambda_{QCD}$ for 3 flavours is large). The principle that coefficients in the operator expansion can be computed from the perturbative theory in terms of parton masses has never been really tested (due to ambiguities on the determination of condensates) and this particular case with a ZERO there is unique in making the issue crucial. Many distinguished theorists believe the optimistic version. I am not convinced that
the gap is not filled up by ambiguities of $0(\Lambda_{QCD}^2/m_\tau^2)$ from $\delta_{pert}$ \cite{alnari}. 

There is a vast and sophisticated literature on $\alpha_s$ from $\tau$ decay. Unbelievably small errors are obtained in one or the other of several different procedures and assumptions that have been adopted to end up with a specified result. With time there has been an increasing awareness on the problem of controlling higher orders and non perturbative effects. In particular fixed order perturbation theory (FOPT) has been compared to resummation of leading beta function effects in the so called contour improved perturbation theory (CIPT). The results are sizably different in the two cases and there have been many arguments in the literature on which method is best. One important progress comes from the experimental measurement of moments of the $\tau$ decay mass distributions, defined by modifying the weight function in the integral in Eq.(\ref{incl}). In principle one can measure  $\alpha_s$ from the sum rules obtained from different weight functions that emphasize different mass intervals and different operator dimensions in the light cone operator expansion. A thorough study of the dependence of the measured value of $\alpha_s$ on the choice of the weight function and in general of higher order and non perturbative corrections has appeared in ref.\cite{benbo} and I advise the interested reader to look at that paper and the references therein.

We consider here the recent evaluations of  $\alpha_s$ from $\tau$ decay based on the NNNLO perturbative calculations \cite{baik4} and different procedures for the estimate of all sorts of corrections. From the papers given in refs. \cite{taulist} we obtain an average value and error that agrees with the Erler and Langacker values given in the PDG'12 \cite{pdg12}:
\beq
\alpha_s(m_{\tau})=0.3285\pm0.018\label{altautau}\\
\eeq
or
\beq
\alpha_s(m_Z)=0.1194\pm0.0021\label{altau}\\
\eeq
In any case, one can discuss
the error, but what is true and remarkable, is that the central value of $\alpha_s$ from $\tau$ decay, obtained at very small $Q^2$, is in
good agreement with all other precise determinations of $\alpha_s$ at more typical LEP values of $Q^2$. 

\subsubsection{$\alpha_s$ from Deep Inelastic Scattering}
\label{sec:18.2}

In principle DIS is expected to be an ideal laboratory for the determination of $\alpha_s$ but in practice the outcome is still to some extent unsatisfactory. QCD predicts the $Q^2$ dependence of $F(x,Q^2)$ at each fixed $x$, not the $x$ shape. But the $Q^2$ dependence is related to the
$x$ shape by the QCD evolution equations. For each x-bin the data allow to extract the slope of an approximately straight line
in $dlogF(x,Q^2)/dlogQ^2$: the log slope. The $Q^2$ span and the precision of the data are not much sensitive to the
curvature, for most $x$ values. A single value of $\Lambda_{QCD}$ must be fitted to reproduce the collection of the log
slopes. For the determination of $\alpha_s$ the scaling violations of non-singlet structure functions would be ideal,
because of the minimal impact of the choice of input parton densities. We can write the non-singlet evolution equations in
the form:
\beq
\frac{d}{dt}logF(x,t)~=~\frac{\alpha_s(t)}{2\pi}\int_x^1\frac{dy}{y}\frac{F(y,t)}{F(x,t)}P_{qq}(\frac{x}{y},\alpha_s(t))
\label{NSEE}\\
\eeq
where $P_{qq}$ is the splitting function. At present NLO and NNLO corrections are  known. It is clear from this form that, for example, the normalization error on the input
density drops away, and the dependence on the input is reduced to a minimum (indeed, only a single density appears here, while in
general there are quark and gluon densities). Unfortunately the data on non-singlet structure functions are not very
accurate. If we take the difference of data on protons and neutrons, $F_p-F_n$, experimental errors add up in the difference
and finally are large. The $F_{3\nu N}$ data are directly non-singlet but are not very precise. 
Another possibility is to neglect sea and glue in $F_2$ at sufficiently large $x$. But by only taking data at $x > x_0$ one decreases
the sample, introduces a dependence on $x_0$ and an
error from residual singlet terms. A recent fit to non singlet structure functions in electro- or muon-production extracted from proton and deuterium data, neglecting sea
and gluons at $x > 0.3$ (error to be evaluated) has led to the results \cite{blum}: 
\bea
\alpha_s(m_Z)&=&0.1148\pm0.0019 (exp) ~+~?~~~~~(NLO) \\
\alpha_s(m_Z)&=&0.1134\pm0.0020(exp) ~+~?~~~~~(NNLO) 
\label{alblu}
\eea
The central values are rather low and there is not much difference between NLO and NNLO. The question marks refer to the uncertainties from the residual singlet component at $x > 0.3$ and also to the fact that the old BCDMS data, whose systematics has been questioned, are very important at $x > 0.3$ and push the fit towards small values of $\alpha_s$.
 
When one measures $\alpha_s$ from scaling violations in $F_2$, measured with e or $\mu$ beams, the data are abundant, the statistical errors are small, the ambiguities from the treatment of heavy quarks and the effects of the longitudinal structure function $F_L$ can be controlled,  but there is an increased dependence on input parton densities and especially a strong correlation between the result on $\alpha_s$ and the adopted parametrization of the gluon density. In the following we restrict our attention to recent determinations of $\alpha_s$ from scaling violations at NNLO accuracy, as, for example, those in refs. \cite{ABM,J-Dr} that report the results, in the order:
\bea
\alpha_s(m_Z)&=&0.1134\pm0.0011(exp) ~+~?~~~~~\\
\alpha_s(m_Z)&=&0.1158\pm0.0035~~~~~
\label{recdet}
\eea 
In the first line my question mark refers to the issue of the $\alpha_s$-gluon correlation. In fact $\alpha_s$ tends to slide towards low values ($\alpha_s\sim 0.113-0.116$) if the gluon input problem is not fixed. Indeed, in the second line, taken from ref. \cite{J-Dr}, the large error also includes an estimate of the ambiguity from the gluon density parametrization. One way to restrict the gluon density is to use the Tevatron and LHC high $p_T$ jet data to fix the gluon parton density at large $x$ that, via the momentum conservation sum rule, also constrain the small $x$ values of the same density. Of course in this way one has to go outside the pure domain of DIS. Also, the jet rates have been computed at NLO only. In a simultaneous fit of $\alpha_s$ and the parton densities from a set of data that, although dominated by DIS data, also contains Tevatron jets and Drell- Yan production, the result was \cite{MSTW}:
\beq
\alpha_s(m_Z)=0.1171\pm0.0014 ~+~?~~~~~\\
\label{mstw}
\eeq
The authors of ref. \cite{MSTW} attribute their larger value of $\alpha_s$ to a more flexible
parametrization of the gluon and the inclusion of Tevatron jet data that are important to fix the gluon at large $x$. An alternative way to cope with the gluon problem is to drastically suppress the gluon parametrization rigidity by adopting the neural network approach. With this method, in ref. \cite{NN}, from DIS data only, treated at NNLO accuracy, the following value was obtained:
\beq
\alpha_s(m_Z)=0.1166\pm0.0008(exp) \pm0.0009(th)~+~?~~~~\\
\label{nn1}
\eeq
where the stated theoretical error is that quoted by the authors within their framework, while the question mark has to do with possible additional systematics from the method adopted. Interestingly, in the same approach, by also including the Tevatron jets and the Drell-Yan data not much difference is found:
\beq
\alpha_s(m_Z)=0.1173\pm0.0007(exp) \pm0.0009(th)~+~?~~~~\\
\label{nn2}
\eeq
We see that when the gluon input problem is suitably addressed the fitted value of  $\alpha_s$ is increased.

As we have seen there is some spread of results, even among the most recent determinations based on NNLO splitting functions. We tend to favour determinations from the whole DIS set of data (i.e. beyond the pure non singlet case) and with attention paid to the gluon ambiguity problem (even if some non DIS data from Tevatron jets at NLO have to be included). A conservative proposal for the resulting value of $\alpha_s$ from DIS, that emerges from the above discussion is something like:  
\beq
\alpha_s(m_Z)=0.1165\pm0.0020~~~~~~\\
\label{myave}
\eeq
The central value is below those obtained from $Z$ and $\tau$ decays but perfectly compatible with those results.

\subsubsection{Recommended Value of $\alpha_s(m_Z)$}
\label{sec:18.3}

According to my proposal to calibrate $\alpha_s(m_Z)$ from the theoretically cleanest and most transparent methods, identified as the totally inclusive, light cone operator expansion dominated processes, I collect here my understanding of the results:
from $Z$ decays and EW precision tests, Eq.(\ref{alEW}):
\beq
\alpha_s(m_Z)=0.1190\pm0.0026; \label{alEW1pr}\\
\eeq
from scaling violations in DIS, Eq.(\ref{myave}):
\beq
\alpha_s(m_Z)=0.1165\pm0.0020;~~~~~~\\
\label{myavepr}
\eeq
from $R_\tau$, Eq.(\ref{altau}):
\beq
\alpha_s(m_Z)=0.1194\pm0.0021.\label{altaupr}\\
\eeq

If one wants to be on the safest side one can take the average of $Z$ decay and DIS:
\beq
\alpha_s(m_Z)=0.1174\pm0.0016.~~~~~~\\
\label{finave1}
\eeq
This is my recommended value. If one adds to the average the rather conservative $R_\tau$ value and error given above in Eq.\ref{altaupr}, that takes into account the dangerous low energy scale of the process, one obtains:
\beq
\alpha_s(m_Z)=0.1184\pm0.0011.~~~~~~\\
\label{finave2}
\eeq
Note that this is essentially coincident with the "official" average with a moderate increase of the error.

\subsubsection{Other $\alpha_s(m_Z)$ Measurements as QCD Tests}
\label{sec:18.4}

There are a number of other determinations of $\alpha_s$ that are important because they arise from qualitatively different observables and methods. Here I will give a few examples of the most interesting measurements.

A classic set of measurements is from a number of infrared safe observables related to event rates and jet shapes in $e^+e^-$ annihilation. One important feature of these measurements is that they can be repeated at different energies in the same detector, like the JADE detector in the energy range of PETRA (most of the intermediate energy points in the right panel of Fig. \ref{alfaspdg} are from this class of measurements) or the LEP detectors from LEP1 to LEP2 energies. As a result one obtains a striking direct confirmation of the running of the coupling according to the renormalization group prediction. The perturbative part is known at NNLO \cite{gher} and resummations of leading logs arising from the vicinity of cuts and/or boundaries have been performed in many cases using effective field theory methods. The main problem of these measurements is the possible large impact of non perturbative hadronization effects on the result and therefore on the theoretical error. 
According to ref.\cite{bethke} a summary result that takes into account the central values and the spread from the JADE measurements, in the range 14 to 46 GeV, at PETRA is given by: $\alpha_s(m_Z)=0.1172\pm 0.0051$, while from the ALEPH data at LEP, in the range 90 to 206 GeV, the reported value \cite{diss} is $\alpha_s(m_Z)=0.1224\pm 0.0039$.
It is amazing to note that among the related works there are a couple of papers by Abbate et al \cite{abb1,abb2} where an extremely sophisticated formalism is developed for the thrust distribution, based on NNLO perturbation theory with resummations at NNNLL plus a data/theory-based estimate of non perturbative corrections. The final quoted results are unbelievably precise: $\alpha_s(m_Z)=0.1135\pm 0.0011$ from the tail of the thrust distribution \cite{abb1} and  $\alpha_s(m_Z)=0.1140\pm 0.0015$ from the first moment of the thrust distribution \cite{abb2}. I think that this is a good example of an underestimated error which is obtained within a given machinery without considering the limits of the method itself. 
Another allegedly very precise determination of $\alpha_s(m_Z)$ is obtained from lattice QCD by several groups \cite{lat} with different methods and compatible results. A value that summarizes these different results is \cite{pdg12} $\alpha_s(m_Z)=0.1185\pm 0.0007$. With all due respect to lattice people I think this small error is totally unrealistic. 
But we have shown that a sufficiently precise measure of $\alpha_s(m_Z)$ can be obtained, Eqs. (\ref{finave1},\ref{finave2}), by only using the simplest processes where the control of theoretical errors is maximal. One is left free to judge whether a further restriction of theoretical errors is really on solid ground.

The value of $\Lambda$ (for $n_f=5$) which corresponds to Eq.~(\ref{finave1}) is:
\beq
\Lambda_5=202\pm 18~{\rm MeV}\label{lambda}\\
\eeq
while the value from Eq.~(\ref{finave2}) is:
\beq
\Lambda_5=213\pm 13~{\rm MeV}\label{lambda2}\\
\eeq

$\Lambda$ is the scale of mass that finally appears in massless QCD. It is the scale where $\alpha_s(\Lambda)$ is of order
1. Hadron masses are determined by $\Lambda$. Actually the $\rho$ mass or the nucleon mass receive little contribution from
the quark masses (the case of pseudoscalar mesons is special, as they are the pseudo Goldstone bosons of broken chiral
invariance). Hadron masses would be almost the same in massless QCD. 

\subsection{Conclusion}
\label{sec:19}

We have seen that perturbative QCD based on asymptotic freedom offers a rich variety of tests and we have described some
examples in detail. QCD tests are not as precise as for the electroweak sector. But the number and diversity of such tests
has established a very firm experimental foundation for QCD as a theory of strong interactions. The physics content of QCD is very large and our knowledge, especially in the non perturbative domain, is still very limited but progress both from experiment (Tevatron, RHIC, LHC......) and from theory is continuing at a healthy rate. And all the QCD predictions that we were able to formulate and to test appear to be in very good agreement with experiment.

The field of QCD appears as one of great maturity but also of robust vitality with many rich branches and plenty of new blossoms. I may mention the very exciting explorations of  Supersymmetric extensions of QCD and the connections with string theory (for a recent review and a list of references see ref.\cite{dix}). In particular N=4 SUSY QCD (that is, with 4 spinor charge generators) has a vanishing beta function and is loop finite. In the limit $N_C \rightarrow \infty$ with $\lambda=e_s^2 N_C$ fixed
planar diagrams are dominant. There is progress towards a solution of planar N=4 SUSY QCD. The large $\lambda$ limit corresponds by the AdS/CFT duality (Anti de Sitter/ Conformal Field Theory), a string theory concept, to the weakly coupled string (gravity) theory on $AdS_5~^.~S_5$ (the 10 dimensions are compactified in a 5-dimensional Anti de Sitter space times a 5-dimensional sphere. By moving along this very tentative route one can transfer some results (assumed to be of sufficiently universal nature) from the computable weak limit of the associated string theory to the non perturbative ordinary QCD domain. Further away on this line there are studies on the N = 8 Supergravity, related to N = 4 SUSY Yang-Mills, which has been proven finite up to 4 loops. It could possibly lead to a finite field theory of gravity in 4 dimensions.

\newpage

\section{\Large\bf{The Theory of Electroweak Interactions}}
\label{Chap:3}

\subsection{Introduction}
\label{sec:20}

In this Chapter, we summarize the structure of the standard EW theory \footnote {Some recent textbooks are listed in ref. \cite{BooksEW}; see also refs. \cite{spri3},\cite{qui}}  and specify the couplings of the intermediate vector bosons $W^{\pm}$, $Z$ and of the Higgs particle with the fermions and among themselves, as dictated by the gauge symmetry plus the observed matter content and the requirement of renormalizability. We discuss the realization of spontaneous symmetry breaking and of the Higgs mechanism. We then review the phenomenological implications of the EW theory for collider physics (that is we leave aside the classic low energy processes that are well described by the "old" weak interaction theory (see, for example, \cite{oku})). 
For this discussion we split the lagrangian into two parts by separating the terms with the Higgs field:
\begin{equation} {\cal L} = {\cal L}_{\rm gauge} + {\cal L}_{\rm Higgs}~.
\label{21}
\end{equation}
Both terms are written down as prescribed by the $SU(2)\bigotimes U(1)$ gauge symmetry and renormalizability, but the Higgs vacuum expectation value (VEV) induces the spontaneous symmetry breaking responsible for the non vanishing vector boson and fermion masses.

\subsection{The Gauge Sector}
\label{sec:21}

We start by specifying ${\cal L}_{\rm gauge}$, which involves only gauge bosons and fermions, according to the general formalism of gauge theories discussed in Chapter 1:
\begin{eqnarray} {\cal L}_{\rm gauge} &=& -\frac{1}{4}~\sum^3_{A=1}~F^A_{\mu\nu}F^{A\mu\nu} -
\frac{1}{4}B_{\mu\nu}B^{\mu\nu} +
\bar\psi_Li\gamma^{\mu}D_{\mu}\psi_L 
+  \bar\psi_Ri\gamma^{\mu}D_{\mu}\psi_R~.
\label{22}
\end{eqnarray} This is the Yang--Mills lagrangian for the gauge group $SU(2)\otimes U(1)$ with fermion matter fields.
Here
\begin{equation} B_{\mu\nu}  =  \partial_{\mu}B_{\nu} - \partial_{\nu}B_{\mu} \quad {\rm and} \quad F^A_{\mu\nu} =
\partial_{\mu}W^A_{\nu} - \partial_{\nu}W^A_{\mu}  - g \epsilon_{ABC}~W^B_{\mu}W^C_{\nu}
\label{23}
\end{equation} are the gauge antisymmetric tensors constructed out of the gauge field $B_{\mu}$ associated with $U(1)$,
and $W^A_{\mu}$ corresponding to the three $SU(2)$ generators; $\epsilon_{ABC}$ are the group structure constants [see
Eqs. (\ref{28},\ref{29})] which, for $SU(2)$, coincide with the totally antisymmetric Levi-Civita tensor (with $\epsilon_{123}=1$; recall the familiar
angular momentum commutators). The normalization of the $SU(2)$ gauge coupling $g$ is therefore specified by
Eqs.~(\ref{23}). As discussed in Sect. \ref{sec:5}  the standard EW theory is a chiral theory, in the sense that $\psi_L$ and $\psi_R$ behave
differently under the gauge group (so that parity and charge conjugation non conservation are made possible in principle). Thus, mass terms for fermions (of the form
$\bar\psi_L\psi_R$ + h.c.) are forbidden in the symmetric limit. In the following by
$\psi_{L,R}$ we mean column vectors, including all fermion types in the theory that span generic reducible representations of
$SU(2) \otimes U(1)$.  
In the absence of mass terms, there are only vector and axial vector interactions in
the lagrangian and those have the property of not mixing $\psi_L$ and $\psi_R$. Fermion masses will be introduced, together
with
$W^{\pm}$ and $Z$ masses, by the mechanism of symmetry breaking. The covariant derivatives $D_{\mu}\psi_{L,R}$ are
explicitly given by
\begin{equation} D_{\mu}\psi_{L,R} = 
\left[ \partial_{\mu} + ig \sum^3_{A=1}~t^A_{L,R}W^A_{\mu} + ig'\frac{1}{2}Y_{L,R}B_{\mu} \right] \psi_{L,R}~,
\label{27}
\end{equation}  where $t^A_{L,R}$ and $1/2Y_{L,R}$ are the $SU(2)$ and $U(1)$ generators, respectively, in the
reducible representations $\psi_{L,R}$. The commutation relations of the $SU(2)$ generators are given by
\begin{equation} [t^A_L,t^B_L] = i~\epsilon_{ABC}t^C_L \quad {\rm and} \quad [t^A_R,t^B_R] = i \epsilon_{ABC}t^C_R~.
\label{28}
\end{equation} We use the normalization as in Eq. (\ref{8}) [in the fundamental representation of
$SU(2)$]. The electric charge generator $Q$ (in units of $e$, the positron charge) is given by
\begin{equation} Q = t^3_L + 1/2~Y_L = t^3_R + 1/2~Y_R~.
\label{29}
\end{equation} Note that the normalization of the $U(1)$ gauge coupling $g'$ in (\ref{27}) is now specified as a
consequence of (\ref{29}). Note that $t^i_R \psi_R=0$, given that, for all known quark and leptons, $\psi_R$ is a singlet. But in the following, we keep $t^i_R \psi_R$ for generality, in case one day a non singlet right-handed fermion is discovered.

\subsection{Couplings of Gauge Bosons to Fermions}
\label{sec:22}

All fermion couplings of the gauge bosons can be derived directly from Eqs. (\ref{22}) and (\ref{27}). The charged $W_{\mu}$ fields are described by $W^{1,2}_{\mu}$, while the photon $A_{\mu}$ and weak neutral gauge boson $Z_\mu$ are obtained from combinations of $W^3_{\mu}$ and $B_{\mu}$. The
charged-current (CC) couplings are the simplest. One starts from the $W^{1,2}_{\mu}$ terms in Eqs. (\ref{22}) and (\ref{27}) which can be written as: 
\begin{eqnarray} g(t^1W^1_{\mu} + t^2W^2_{\mu}) &=& g \left\{ [(t^1 + it^2)/ \sqrt 2] (W^1_{\mu} - iW^2_{\mu})/\sqrt 2]
+ {\rm h.c.} \right\}\nonumber \\
 &= &g \left\{[(t^+W^-_{\mu})/\sqrt 2] + {\rm h.c.} \right\}~,
\label{30}
\end{eqnarray} where $t^{\pm}  = t^1 \pm it^2$ and $W^{\pm} = (W^1 \pm iW^2)/\sqrt 2$. By applying this generic relation to $L$ and $R$ fermions separately, we obtain the vertex
\begin{equation} V_{\bar \psi \psi W}  =  g \bar \psi \gamma_{\mu}\left[ (t^+_L/ \sqrt 2)(1 - \gamma_5)/2 + (t^+_R/
\sqrt 2)(1 + \gamma_5)/2 \right]
 \psi W^-_{\mu} + {\rm h.c.}
\label{31}
\end{equation}
Given that $t_R=0$ for all fermions in the SM, the charged current is pure $V-A$. 
In the neutral-current (NC) sector, the photon $A_{\mu}$ and the mediator
$Z_{\mu}$ of the weak NC are orthogonal and normalized linear combinations of
$B_{\mu}$ and $W^3_{\mu}$:
\begin{eqnarray} A_{\mu} &=& \cos \theta_WB_{\mu} + \sin \theta_WW^3_{\mu}~, \nonumber \\  Z_{\mu} &=& -\sin
\theta_WB_{\mu} + \cos \theta_WW^3_{\mu}~.
\label{32}
\end{eqnarray} and conversely: \begin{eqnarray} W^3_{\mu} &=& \sin \theta_WA_{\mu} + \cos \theta_WZ_{\mu}~, \nonumber \\  B_{\mu} &=& \cos
\theta_WA_{\mu} - \sin \theta_WZ_{\mu}~.
\label{32a}
\end{eqnarray}
Equations (\ref{32}) define the weak mixing angle $\theta_W$. We can rewrite the $W^3_{\mu}$ and $B_\mu$ terms in Eqs. (\ref{22}) and (\ref{27}) as follows:
\begin{eqnarray} gt^3W^3_{\mu} + g'Y/2B_\mu~&=&~[gt^3\sin \theta_W+g'(Q-t^3)\cos \theta_W]A_\mu~+\nonumber \\&+&~[gt^3\cos \theta_W-g'(Q-t^3)\sin \theta_W]Z_\mu~,
\label{32b}
\end{eqnarray} where Eq. (\ref{29}) for the
charge matrix $Q$ was also used. The photon is characterized by equal
couplings to left and right fermions with a strength equal to the electric charge. Thus we immediately obtain
\begin{equation} g~\sin \theta_W = g'\cos \theta_W = e~,
\label{33}
\end{equation} so that:
\begin{equation} {\rm tg}~\theta_W = g'/g
\label{34}
\end{equation} Once $\theta_W$ has been fixed by the photon couplings, it is a simple matter of algebra to derive the
$Z$ couplings, with the result
\begin{equation}
V_{\bar \psi \psi Z} = \frac{g}{2\cos \theta_W} \bar \psi \gamma_{\mu}
  [t^3_L(1-\gamma_5) + t^3_R(1+\gamma_5) - 2Q \sin^2\theta_W] \psi Z^{\mu}~,
\label{35}
\end{equation}  where $V_{\bar \psi \psi Z}$ is a notation for the vertex. Once again, recall that in the minimal SM, $t^3_R = 0$ and $t^3_L =
\pm 1/2$. 

In order to derive the effective four-fermion interactions that are equivalent, at low energies, to the CC and NC
couplings given in Eqs. (\ref{31}) and (\ref{35}), we anticipate that large masses, as experimentally observed, are
provided for $W^{\pm}$  and $Z$ by ${\cal L}_{\rm Higgs}$.  For left--left CC couplings, when the momentum transfer
squared can be neglected (with respect to
$m^2_W$) in the propagator of Born diagrams with single $W$ exchange (see, for example, the diagram for $\mu$ decay in  Fig.~3.1), from Eq.~(\ref{31}) we can write
 \begin{equation} {\cal L}^{\rm CC}_{\rm eff} \simeq \frac{g^2}{8m^2_W} [ \bar \psi \gamma_{\mu}(1 - \gamma_5)t^+_L\psi][
\bar \psi
\gamma^{\mu}(1 - \gamma_5) t^-_L\psi]~.
\label{36}
\end{equation} 

\begin{figure}[h]
\noindent
\includegraphics[width=6cm]{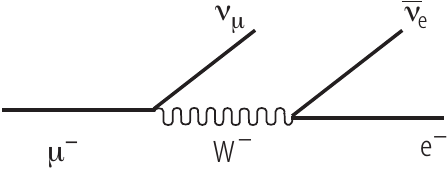} 
\caption[]{The Born diagram for $\mu$ decay.}
\label{fig:1}
\end{figure}

\begin{figure}[h]
\noindent
\includegraphics[width=6cm]{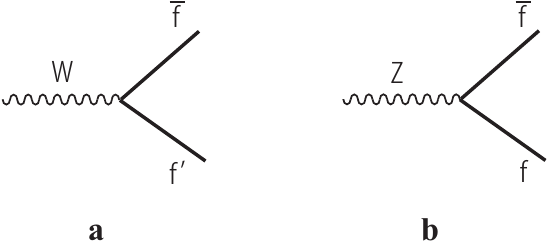} 
\caption[]{Diagrams for (a) the W and (b) the Z widths in Born approximation.}
\label{fig:2}
\end{figure}

By specializing further in the case of doublet fields such as $\nu_e-e^-$ or $
\nu_{\mu} - \mu^-$, we obtain the tree-level relation of $g$ with the Fermi coupling constant $G_F$  precisely measured from $\mu$
decay [see Chapter 1, Eqs. (2), (3)]:
\begin{equation}
 \frac{G_F}{\sqrt 2} = \frac{g^2}{8m^2_W}~.
\label{37}
\end{equation} By recalling that $g~\sin \theta_W = e$, we can also cast this relation in the form
\begin{equation} 
m_W = \frac{\mu_{\rm Born}}{\sin \theta_W}~,
\label{38}
\end{equation} with
\begin{equation}
\mu_{\rm Born} = (\frac{\pi \alpha}{\sqrt 2 G_F})^{1/2} \simeq 37.2802~{\rm GeV}~,
\label{39}
\end{equation} where $\alpha$ is the fine-structure constant of QED $(\alpha \equiv e^2/4\pi = 1/137.036)$. 

In the same way, for neutral currents we obtain in Born approximation from Eq.~(\ref{35}) the effective four-fermion
interaction given by
\begin{equation} {\cal L}^{\rm NC}_{\rm eff} \simeq \sqrt 2~G_F \rho_0\bar \psi \gamma_{\mu}[...]
\psi \bar \psi \gamma^{\mu}[...] \psi~,
\label{40}
\end{equation} where
\begin{equation} [...] \equiv t^3_L(1 - \gamma_5) + t^3_R (1 + \gamma_5) - 2Q \sin^2\theta_W
\label{41}
\end{equation} and
\begin{equation}
\rho_0 = \frac{m^2_W}{m^2_Z~\cos^2 \theta_W}~.
\label{42}
\end{equation}

All couplings given in this section are valid at tree level and are modified in higher orders of perturbation
theory. In particular, the relations between
$m_W$ and $\sin \theta_W$  [Eqs. (\ref{38}) and (\ref{39})] and the observed values of $\rho~(\rho = \rho_0$ at tree
level) in different NC processes, are altered by computable EW radiative corrections, as discussed in Sect. (\ref{sec:30}).

The partial width $\Gamma(W \rightarrow \bar f f')$ is given in Born approximation by the simplest diagram in Fig.~3.2 and one readily obtains from Eq.(\ref{31}) with $t_R=0$, in the limit of neglecting the fermion masses and summing over all possible $f'$ for a given $f$:
\begin{equation} \Gamma(W \rightarrow \bar f f') = N_C \frac{G_F m_W^3}{6\pi \sqrt{2}}= N_C \frac{\alpha m_W}{12 \sin^2{\theta_W}}, 
\label{31a}
\end{equation}
where $N_C=3~{\rm or}~1$ is the number of colours for quarks or leptons, respectively, and the relations Eqs.(\ref{33}, \ref{37}) have been used. Here and in the following expressions for the Z widths the one loop QCD corrections for the quark channels can be absorbed in a redefinition of $N_C$: $N_C \rightarrow 3[1+\alpha_s(m_Z)/\pi+...]$. Note that the widths are particularly large because the rate already occurs  at order $g^2$ or $G_F$.
The experimental values of the W total width and the leptonic branching ratio (the average of $e$, $\mu$ and $\tau$ modes) are \cite{pdg12}, \cite{ew} (see Sect. (\ref{sec:30})):
\beq
\Gamma_W=2.085\pm0.042~{\rm GeV},~~~~~~~B(W\rightarrow l\nu_l)=10.80\pm0.09.
\label{dataW}
\eeq
The branching ratio $B$ is in very good agreement with the simple approximate formula, derived from Eq.(\ref{31a}):
\beq
B(W\rightarrow l\nu_l)\sim \frac{1}{2^.3^.(1+\alpha_s(m_Z^2)/\pi)+3}\sim 10.8\%.
\label{apprBW}
\eeq
The denominator corresponds to the sum of the final states $d'\bar u$, $s' \bar c$, $e^-\bar \nu_e$, $\mu^-\bar \nu_\mu$, $\tau^-\bar \nu_\tau$ (for the definition of $d'$ and $s'$ see Eq. (\ref{km1})).

For $t_R=0$ the Z coupling to fermions in Eq.(\ref{35}) can be cast into the form:
\begin{equation}
V_{\bar \psi_f \psi_f Z} = \frac{g}{2~\cos \theta_W} \bar \psi_f \gamma_{\mu}
  [g_V^f-g_A^f \gamma_5] \psi_f Z^{\mu}~,
\label{35a}
\end{equation}
with:
\begin{equation}
g_A^f= t_L^{3f}~~~,~g_V^f/g_A^f~=~1-4|Q_f|\sin^2{\theta_W}~.
\label{35b}
\end{equation}
and $t_L^{3f}=\pm 1/2$ for up-type or down-type fermions. In terms of $g_{A,V}$ given in Eqs. (\ref{35b}) (the widths are proportional to $(g_V^2+g_A^2)$), the partial width $\Gamma(Z \rightarrow \bar f f)$ in Born approximation (see the diagram in Fig.~3.2), for negligible fermion masses, is given by:
\bea \Gamma(Z \rightarrow \bar f f)  
&=&N_C \frac{\alpha m_Z}{12 \sin^2{2\theta_W}}[1+(1-4|Q_f| \sin^2{\theta_W})^2]\nonumber \\
&=&N_C\rho_0\frac{G_Fm_Z^3}{24\pi \sqrt{2}}[1+(1-4|Q_f| \sin^2{\theta_W)^2}]. 
\label{31b}
\eea
where $\rho_0 =m_W^2/m_Z^2 \cos^2{\theta_W}$ is given in Eq. (\ref{60}). The experimental values of the Z total width and of the partial rates into charged leptons (average of $e$, $\mu$ and $\tau$), into hadrons and into invisible channels are \cite{pdg12}, \cite{ew}:
\bea
\Gamma_Z&=&2.4952\pm0.0023~{\rm GeV},\nonumber \\
\Gamma_{l^+l^-}&=&83.984\pm0.086~{\rm MeV},\nonumber \\
\Gamma_h&=&1744.4\pm2.0~{\rm MeV},\nonumber \\
\Gamma_{inv}&=&499.0\pm1.5~{\rm MeV}.
\label{gamZ}
\eea
The measured value of the Z invisible width, taking radiative corrections into account, leads to the determination of the number of light active neutrinos \cite{pdg12}, \cite{ew}:
\beq
N_\nu=2.9840\pm0.0082,
\label{nnu}
\eeq
well compatible with the 3 known neutrinos $\nu_e$, $\nu_\mu$ and $\nu_\tau$; hence there exist only the
three known sequential generations of fermions (with light neutrinos),
a result with important consequences also in astrophysics and cosmology.

At the Z peak, besides total cross sections, various types of asymmetries have been
measured.  The results of all asymmetry measurements are quoted in
terms of the asymmetry parameter $A_f$, defined in terms of the effective coupling constants, $g_V^f$ and $g_A^f$, as:
\begin{eqnarray}
A_f & = & 2\frac{g_V^fg_A^f}{g_V^{f2}+g_A^{f2}} ~ = ~ 
           2\frac{g_V^f/g_A^f}{1+(g_V^f/g_A^f)^2}\,, \qquad
A_{FB}^f ~ = ~ \frac{3}{4}A_eA_f.
\end{eqnarray}
The measurements are: the forward-backward asymmetry ($A_{FB}^f=
(3/4)A_eA_f$), the tau polarization ($A_\tau$) and its forward
backward asymmetry ($A_e$) measured at LEP, as well as the left-right
and left-right forward-backward asymmetry measured at SLC ($A_e$ and
$A_f$, respectively).  Hence the set of partial width and asymmetry
results allows the extraction of the effective coupling constants: widths measure $(g_V^2+g_A^2)$ and asymmetries measure $g_V/g_A$.

The top quark is heavy enough that it can decay into a real bW pair, which is by far its dominant decay channel. The next mode, $t \rightarrow sW$, is suppressed in rate by a factor $|V_{ts}|^2\sim 1.7^.10^{-3}$, see Eqs. (\ref{km8}-\ref{km9}). The associated width, neglecting $m_b$ effects but including 1-loop QCD corrections in the limit $m_W=0$, is given by (we have omitted a factor $|V_{tb}|^2$ that we set equal to 1) \cite{jezku}:
\begin{equation} 
\Gamma(t \rightarrow b W^+) =  \frac{G_F m_t^3}{8\pi \sqrt{2}}(1-\frac{m_W^2}{m_t^2})^2(1+2\frac{m_W^2}{m_t^2}) [1-\frac{2\alpha_s(m_Z)}{3\pi}(\frac{2\pi^2}{3}-\frac{5}{2})+...].
\label{36t}
\end{equation}
The top quark lifetime is so short, about $0.5^.10^{-24}$s, that it decays before hadronizing or forming toponium bound states.

\subsection{Gauge Boson Self-interactions}
\label{sec:23}

The gauge boson self-interactions can be derived from the
$F_{\mu\nu}$ term in ${\cal L}_{\rm gauge}$, by using Eq. (\ref{32}) and
$W^{\pm} = (W^1 \pm iW^2)/\sqrt 2$. 
Defining the three-gauge-boson vertex as in Fig.~3.3 (with all incoming  lines), we obtain $(V \equiv \gamma,Z)$
\begin{equation}
V_{W^-W^+V} = ig_{W^-W^+V}[g_{\mu\nu}(p-q)_{\lambda} + g_{\mu\lambda}(r-p)_{\nu} + g_{\nu\lambda}(q-r)_{\mu}]~,
\label{43}
\end{equation} with
\begin{equation} g_{W^-W^+\gamma} = g~\sin \theta_W = e \quad {\rm and} \quad g_{W^-W^+Z} = g~\cos \theta_W~.
\label{44}
\end{equation} 
Note that the photon coupling to the $W$ is fixed by the electric charge, as imposed by QED gauge invariance. The $ZWW$ coupling is larger by a $\cot{\theta_W}$ factor. This form of the triple gauge vertex is very special: in general, there could be departures from the above SM
expression, even restricting us to Lorentz invariant, electromagnetic-gauge symmetric and C and P conserving couplings. In fact some small
corrections are already induced by the radiative corrections. But, in principle, more important could be the modifications
induced by some new physics effect. The experimental testing of the triple gauge vertices has been done in the past mainly at LEP2 and at the Tevatron \cite{grugur} and now also at the LHC \cite{web}.

\begin{figure}[t]
\noindent
\includegraphics[width=6cm]{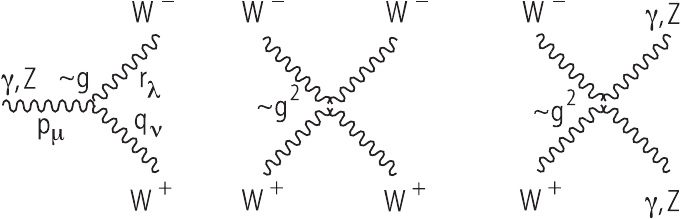} 
\caption[]{The 3- and 4-gauge boson vertices. The cubic coupling is of order $g$, while the quartic one is of order $g^2$.}
\label{fig:3}
\end{figure}

\begin{figure}[h]
\noindent
\includegraphics[width=6cm]{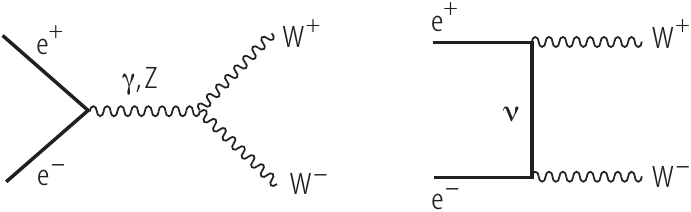} 
\caption[]{The lowest order diagrams for $e^+e^- \rightarrow W^+W^-$.}
\label{fig:4}
\end{figure}

\begin{figure}[!h]
\noindent
\includegraphics[width=10cm]{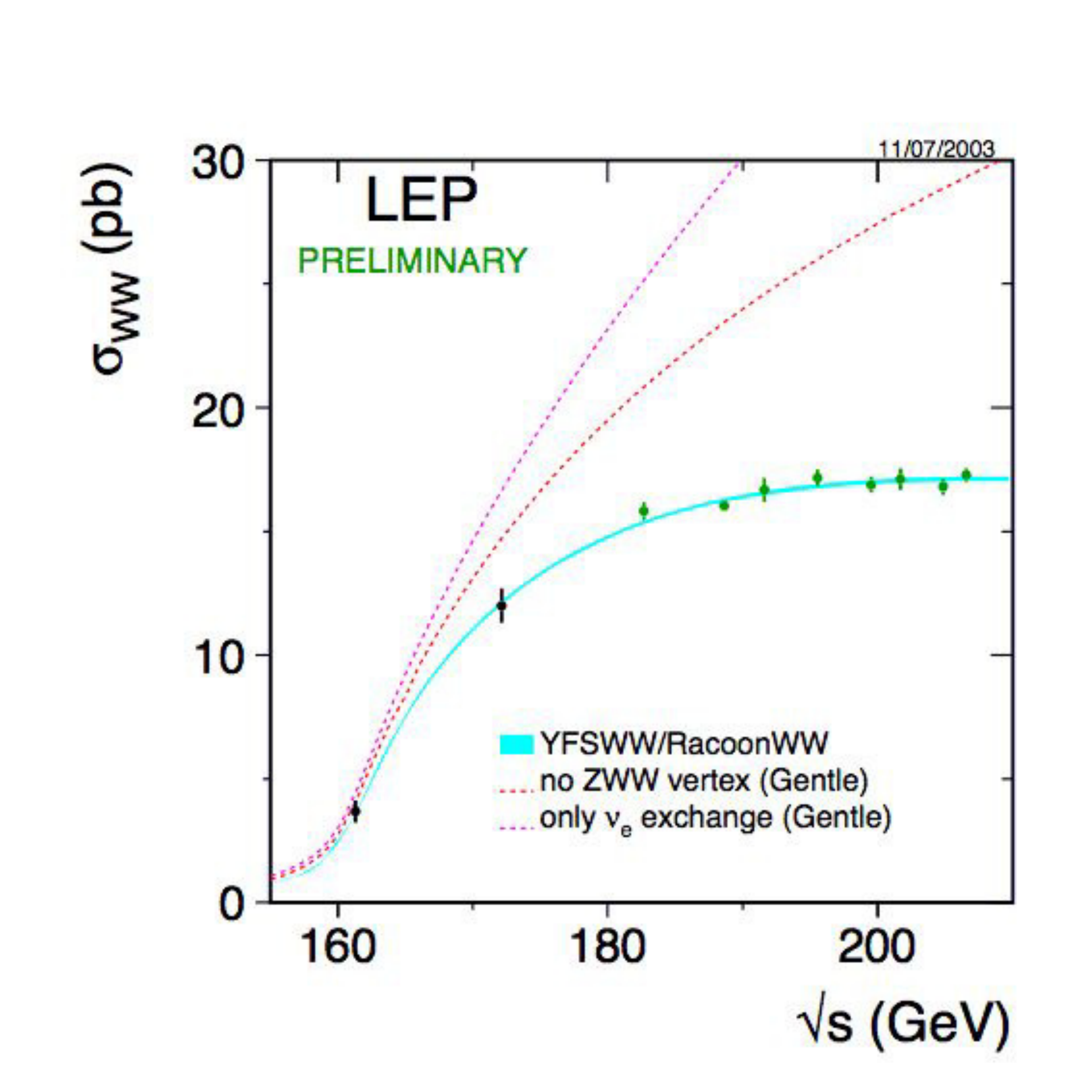} 
\caption[]{The measured production cross section for $e^+e^- \rightarrow W^+W^-$ compared to the SM and to fictitious theories not including trilinear gauge couplings, as indicated}
\label{figWW}
\end{figure}

 As a particularly important example, the cross-section and angular distributions for the process $e^+e^-\rightarrow W^+W^-$ have been studied at LEP2. In Born approximation the Feynman diagrams for the LEP2 process are shown in Fig.~3.4 \cite{LEP2}. Besides neutrino exchange which only involves the well established charged current vertex, the triple weak gauge vertices $V_{W^-W^+V}$ appear in the $\gamma$ and Z exchange diagrams. The Higgs exchange is negligible because the electron mass is very small. The analytic cross section formula in Born approximation can be found, for example, in ref. \cite{pdg12} (in the section ''Cross-section formulae for specific processes''). The experimental data are compared with the SM prediction in Fig.~\ref{figWW}. The agreement, within the present accuracy, is good. Note that the sum of all three exchange amplitudes has a better  high energy behaviour than its individual components. This is due to cancellations among the amplitudes implied by gauge invariance, connected to the fact that the theory is renormalizable (the cross-section can be seen as a contribution to the imaginary part of the $e^+e^-\rightarrow
e^+e^-$ amplitude).

The quartic gauge coupling is proportional to $g^2\epsilon_{ABC} W^BW^C \epsilon_{ADE} W^DW^E$. Thus in the term with A=3 we have 4 charged W's. For A=1 or 2 we have 2 charged W's and 2 $W^3$'s, each $W_3$ being a combination of $\gamma$ and $Z$ according to Eq. (\ref{32a}).
With a little algebra the quartic vertex can be cast in the form:
\beq
V_{WWVV}= i g_{WWVV} [2g_{\mu \nu}g_{\lambda \rho}-g_{\mu \lambda}g_{\nu \rho}-g_{\mu \rho}g_{\nu \lambda}]~,
\label{4V}
\eeq
where, $\mu$ and $\nu$ refer to $W^+W^+$ in the $4W$ vertex and to $VV$ in the $WWVV$ case and:
\beq
g_{WWWW}= g^2,~~~g_{WW\gamma \gamma} = -e^2,~~~g_{WW\gamma Z} = -eg\cos{\theta_W},~~~g_{WWZZ} = -g^2\cos^2{\theta_W}~.
\label{g4V}
\eeq
In order to obtain these result for the vertex the reader must duly take into account the factor of -1/4 in front of $F_{\mu \nu}^2$ in the lagrangian and the statistical factors which are equal to 2 for each pair of identical particles (like $W^+W^+$ or $\gamma \gamma$, for example).
The quartic coupling, being quadratic in g, hence small,  could not be directly tested so far.

\subsection{The Higgs Sector}
\label{sec:24}
We now turn to the Higgs sector of the EW lagrangian \cite{hig}. Until recently this simplest realization of  the EW symmetry breaking was a pure conjecture. But on July '12 the ATLAS and CMS Collaborations at the CERN LHC have announced \cite{atl,cms} the discovery of a particle with mass $m_H \sim 126~GeV$ that very much looks like the long sought Higgs particle. More precise measurements of its couplings and the proof that its spin is zero are necessary before the identification with the SM Higgs boson can be completely established. But the following description of the Higgs sector of the SM can now be read with this striking development in mind.

The Higgs lagrangian is specified by the gauge
principle and the requirement of renormalizability to be
\begin{equation} {\cal L}_{\rm Higgs} = (D_{\mu}\phi)^{\dag}(D^{\mu}\phi) - V(\phi^{\dag}\phi) -
\bar \psi_L \Gamma \psi_R \phi - \bar \psi_R \Gamma^{\dag} \psi_L \phi^{\dag}~,
\label{45}
\end{equation} where $\phi$ is a column vector including all Higgs fields; in general it transforms as a reducible representation
of the gauge group $SU(2)_L \otimes U(1)$. In the Minimal SM it is just a complex doublet. The quantities $\Gamma$ (which include all coupling constants) are matrices that make the Yukawa
couplings invariant under the Lorentz and gauge groups. The potential $V(\phi^{\dag}\phi)$, symmetric under $SU(2)_L
\otimes  U(1)$, contains, at most, quartic terms in $\phi$ so that the theory is renormalizable:
\beq
V(\phi^{\dag}\phi)=-\mu^2\phi^{\dag}\phi+\frac{1}{2}\lambda(\phi^{\dag}\phi)^2\label{44a}
\eeq

As discussed in Chapter 1, spontaneous symmetry
breaking is induced if the minimum of  V, which is the classical analogue of the quantum mechanical vacuum state, is not a single point but a whole orbit obtained for non-vanishing $\phi$ values. Precisely, we denote the vacuum
expectation value (VEV) of $\phi$, i.e. the position of the minimum, by $v$ (which is a doublet):
\begin{equation}
\langle 0 |\phi (x)|0 \rangle = v =\pmatrix {0 \cr v} \not= 0~.
\label{46}
\end{equation}
The reader should be careful that, for economy of notation,  the same symbol is used for the doublet and for the only non zero component of the same doublet.
The fermion mass matrix is obtained from the Yukawa couplings by replacing $\phi (x)$ by $v$:
\begin{equation} M = \bar \psi_L~{\cal M} \psi_R + \bar \psi_R {\cal M}^{\dag}\psi_L~,
\label{47}
\end{equation} with
\begin{equation} {\cal M} = \Gamma \cdot v~.
\label{48}
\end{equation} In the MSM, where all left fermions $\psi_L$ are doublets and all right fermions $\psi_R$ are singlets,
only Higgs doublets can contribute to fermion masses. There are enough free couplings in $\Gamma$ so that one single
complex Higgs doublet is indeed sufficient to generate the most general fermion mass matrix. It is important to observe
that by a suitable change of basis we can always make the matrix ${\cal M}$ Hermitian (so that the mass matrix is $\gamma_5$-free) and diagonal. In
fact, we can make separate unitary transformations on $\psi_L$ and $\psi_R$ according to
\begin{equation}
\psi'_L = U\psi_L, \quad \psi'_R = W\psi_R
\label{49}
\end{equation} and consequently
\begin{equation} {\cal M} \rightarrow {\cal M}' = U^{\dag}{\cal M}W~.
\label{50}
\end{equation} This transformation produces different effects on mass terms and on the structure of the fermion couplings in ${\cal L}_{\rm symm}$, because both the kinetic terms and the couplings to gauge bosons do not mix L and R spinors.  The combined effect of these unitary rotations leads to the phenomenon of mixing and, generically, to flavour changing neutral currents (FCNC), as we shall see in Sect. \ref{sec:25}.
If only one Higgs doublet is present, the change of basis that makes ${\cal M}$ diagonal will at the same time
diagonalize the fermion--Higgs Yukawa couplings. Thus, in this case, no flavour-changing neutral Higgs vertices
are present. This is not true, in general, when there are several Higgs doublets. But one Higgs doublet for each
electric charge sector i.e. one doublet coupled only to $u$-type quarks, one doublet to $d$-type quarks, one doublet to
charged leptons and possibly one for neutrino Dirac masses, would also be all right, because the mass matrices of fermions with different charges are
diagonalized separately. For several Higgs doublets in a given charge sector it is also possible to generate CP
violation by complex phases in the Higgs couplings. In the presence of six quark flavours, this CP-violation mechanism is
not necessary. In fact, at the moment, the simplest model with only one Higgs doublet could be adequate for describing all
observed phenomena.

We now consider the gauge-boson masses and their couplings to the Higgs. These effects are induced by the
$(D_{\mu}\phi)^{\dag}(D^{\mu}\phi)$ term in
${\cal L}_{\rm Higgs}$ [Eq. (\ref{45})], where
\begin{equation} D_{\mu}\phi = \left[ \partial_{\mu} + ig \sum^3_{A=1} t^AW^A_{\mu} + ig'(Y/2)B_{\mu} \right] \phi~.
\label{51}
\end{equation} Here $t^A$ and $Y/2$ are the $SU(2) \otimes U(1)$ generators in the reducible representation spanned by
$\phi$. Not only doublets but all non-singlet Higgs representations can contribute to gauge-boson masses. The condition
that the photon remains massless is equivalent to the condition that the vacuum is electrically neutral:
\begin{equation} Q|v\rangle = (t^3 + \frac{1}{2}Y)|v \rangle = 0~.
\label{52}
\end{equation} We now explicitlly consider the case of a single Higgs doublet:
\begin{equation}
\phi = \pmatrix { \phi^+ \cr
\phi^0}, \quad v = \pmatrix{ 0 \cr v}~, 
\label{53}
\end{equation}The charged $W$ mass is given by the quadratic terms in the $W$ field arising from
${\cal L}_{\rm Higgs}$, when $\phi (x)$ is replaced by $v$ in Eq.(\ref{46}). By recalling Eq.(\ref{30}), we obtain
\begin{equation} m^2_WW^+_{\mu}W^{- \mu} = g^2|(t^+v/ \sqrt 2)|^2 W^+_{\mu}W^{- \mu}~,
\label{54}
\end{equation} whilst for the $Z$ mass we get [recalling Eqs. (\ref{32}-\ref{32b})]
\begin{equation}
\frac{1}{2}m^2_ZZ_{\mu}Z^{\mu} = |[g \cos \theta_Wt^3 - g' \sin
\theta_W(Y/2)]v|^2Z_{\mu}Z^{\mu}~,
\label{55}
\end{equation} where the factor of 1/2 on the left-hand side is the correct normalization for the definition of the
mass of a neutral field. By using Eq. (\ref{52}), relating the action of $t^3$ and $Y/2$ on the vacuum $v$, and Eqs.
(\ref{34}), we obtain
\begin{equation}
\frac{1}{2}m^2_Z = (g \cos \theta_W + g' \sin \theta_W)^2 |t^3v|^2 = (g^2/ \cos^2 \theta_W)|t^3v|^2~.
\label{56}
\end{equation} For a Higgs doublet, as in Eq.(\ref{53}),
we have
\begin{equation} |t^+v|^2 = v^2, \quad |t^3v|^2 = 1/4v^2~,
\label{57}
\end{equation} so that
\begin{equation} m^2_W = \frac{1}{2}g^2v^2, \quad m^2_Z = \frac{g^2v^2}{2\cos^2\theta_W}~.
\label{58}
\end{equation} Note that by using Eq. (\ref{37}) we obtain
\begin{equation} v = 2^{-3/4}G^{-1/2}_F = 174.1~{\rm GeV}~.
\label{59}
\end{equation} It is also evident that for Higgs doublets
\begin{equation}
\rho_0 = \frac{m^2_W}{m^2_Z \cos^2\theta_W} = 1~.
\label{60}
\end{equation}
This relation is typical of one or more Higgs doublets and would be spoiled by the existence of Higgs triplets etc. In
general,
\begin{equation}
\rho_0 = \frac{\sum_i((t_i)^2 - (t^3_i)^2 + t_i ) v^2_i}{\sum _i2(t^3_i)^2v^2_i}
\label{61}
\end{equation} for several Higgs bosons with VEVs $v_i$, weak isospin $t_i$, and $z$-component $t^3_i$. These results are
valid at the tree level and are modified by calculable EW radiative corrections, as discussed in Sect. \ref{sec:30}.

The measured values of the W (combined from the LEP and Tevatron experiments)  and Z masses (from LEP) are \cite{pdg12}, \cite{ew}:
\beq
m_W=80.385\pm0.015~{\rm GeV},~~~~~~m_Z=91.1876\pm0.0021~{\rm GeV}.
\label{mwz}
\eeq

In the minimal version of the SM only one Higgs doublet is present. Then the fermion--Higgs couplings are in proportion to
the fermion masses. In fact, from the fermion $f$ Yukawa couplings $g_{\phi
\bar f f}(\bar f_L \phi f_R + h.c.)$, the mass $m_f$ is obtained by replacing
$\phi$ by $v$, so that $ m_f = g_{\phi \bar f f} v $. In the minimal SM
three out of the four Hermitian fields are removed from the physical spectrum by
the Higgs mechanism and become the longitudinal modes of $W^+, W^-$, and $Z$. The fourth neutral Higgs is physical and
should presumably be identified with the newly discovered particle at $\sim 126$ GeV. If more doublets are present, two more charged and two more neutral Higgs scalars should be around for
each additional doublet.

The couplings of the physical Higgs $H$ can be simply obtained from ${\cal L}_{\rm
Higgs}$, by the replacement (the remaining three hermitian fields correspond to the would-be Goldstone bosons that become the longitudinal modes of $W^\pm$ and $Z$):
\begin{equation}
\phi(x) = \pmatrix{ \phi^+(x) \cr
\phi^0(x)} \rightarrow 
\pmatrix {0 \cr v + (H/\sqrt2)}~,
\label{62}
\end{equation} [so that $(D_{\mu}\phi)^{\dag}(D^{\mu}\phi) = 1/2(\partial_{\mu}H)^2 + ...]$, with the results
\begin{eqnarray} {\cal L} [H,W,Z] &=& g^2\frac{v}{\sqrt 2}W^+_{\mu}W^{-\mu} H + \frac{g^2}{ 4} W^+_{\mu}W^{-\mu}H^2+ \nonumber \\
 &+& g^2\frac{v}{2 \sqrt 2 \cos^2\theta_W} Z_{\mu}Z^{\mu}H + \frac{g^2}{8
\cos^2\theta_W} Z_{\mu}Z^{\mu}H^2~.
\label{63}
\end{eqnarray}
Note that the trilinear couplings are nominally of order $g^2$, but the adimensional coupling constant is actually of order $g$ if we express the couplings in terms of the masses according to Eqs.(\ref{58}):
\begin{eqnarray} {\cal L} [H,W,Z] &=& g m_W W^+_{\mu}W^{-\mu} H + \frac{g^2}{ 4}W^+_{\mu}W^{-\mu}H^2 +\nonumber \\
 &+& \frac{g m_Z}{2 \cos^2\theta_W} Z_{\mu}Z^{\mu} H +  \frac{g^2}{8
\cos^2\theta_W} Z_{\mu}Z^{\mu}H^2~.
\label{63a}
\end{eqnarray} Thus the trilinear couplings of the Higgs to the gauge bosons are also proportional to the masses (at fixed g: if instead $G_F$ is kept fixed then, by Eq. \ref{37}, g is proportional to $m_W$, and the Higgs couplings are quadratic in $m_W$). The quadrilinear couplings are of order $g^2$. Recall that to go from the lagrangian  to the Feynman rules for the vertices the statistical factors must be taken into account: for example, the
Feynman rule for the $ZZHH$ vertex is $ig_{\mu \nu}g^2/2\cos^2\theta_W$.

The generic coupling of H to a fermion of type f is given by (after diagonalization):
\begin{equation}
{\cal L} [H,\bar{\psi}, \psi] = \frac{g_f}{\sqrt{2}}\bar{\psi}\psi H,
\label{63f}
\end{equation}
with
\begin{equation}
\frac{g_f}{\sqrt{2}}=\frac{m_f}{\sqrt{2}v}=2^{1/4}G_F^{1/2} m_f~.
\label{63ff}
\end{equation}

The Higgs self couplings are obtained from the potential in Eq.(\ref{44a}) by the replacement in
Eq.(\ref{62}). Given that, from the minimum condition:
\beq
v=\sqrt{\frac{\mu^2}{\lambda}}
\label{63b}
\eeq
one obtains:
\bea
V=
-\mu^2(v+\frac{H}{\sqrt{2}})^2 + \frac{\mu^2}{2v^2}(v+\frac{H}{\sqrt{2}})^4=
-\frac{\mu^2v^2}{2}+\mu^2 H^2+\frac{\mu^2}{\sqrt{2}v}H^3+\frac{\mu^2}{8v^2}H^4
\label{63c}
\eea
The constant term can be omitted in our context. We see that the Higgs mass is positive (compare with Eq.(\ref{44a})) and is given by:
\beq
m_H^2=2\mu^2=2\lambda v^2
\label{63d}
\eeq
By recalling the value of $v$ in Eq.(\ref{59}), we see that for $m_H \sim 126$ GeV $\lambda$ is small, $\lambda/2 \sim 0.13$ (note that $\lambda/2$ is the coefficient of $\phi^4$ in Eq.(\ref{44a}), and the Higgs self interaction is in the perturbative domain. 

The difficulty of the Higgs search is due to the fact that it is heavy and  coupled in proportion to mass: it is a heavy particle that must be radiated by another heavy particle. So a lot of phase space and of luminosity are needed. At LEP2 the main process for Higgs production was the Higgs-strahlung process $e^+e^-\rightarrow ZH$ shown in Fig.~3.5 \cite{Hstra}. The alternative process $e^+e^-\rightarrow H \nu \bar \nu$, via WW fusion, also shown in Fig.~3.5 \cite{87al1}, has a smaller cross-section at LEP2 energies but would become important, even dominant at higher energy $e^+e^-$ colliders, like the ILC or CLIC (the corresponding ZZ fusion process has a much smaller cross-section). The analytic formulae for the cross-sections of both processes can be found, for example, in \cite{LEP2}. The direct experimental limit on $m_H$ from LEP2 was $m_H\gtrsim 114~GeV$ at $95\%$ c.l.. The phenomenology of the SM Higgs particle and its production and detection at hadron colliders will be discussed in Sects. \ref{sec:32},  \ref{sec:33}.

\begin{figure}[h]
\noindent
\includegraphics[width=7cm]{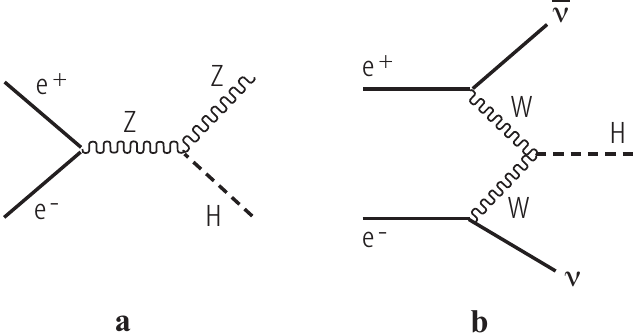} 
\caption[]{Higgs production diagrams in Born approximation for $e^+e^-$ annihilation: (a) The Higgs-strahlung process $e^+e^-\rightarrow ZH$, (b) the WW fusion process $e^+e^-\rightarrow H \nu \bar \nu$.}
\label{fig:5}
\end{figure}

\subsection{The CKM Matrix and Flavour Physics}
\label{sec:25}

Weak charged current vertices are the only tree level interactions in the SM that change flavour: for example, by emission of a $W^+$  an
up-type quark is turned into a  down-type quark, or a $\nu_l$ neutrino is turned into a $l^-$
charged lepton (all fermions are letf-handed). If we start from an up quark that is a mass
eigenstate, emission of a $W^+$ turns it into a down-type quark state d' (the weak isospin partner of
u) that in general is not a mass eigenstate. The mass eigenstates and the weak
eigenstates do not coincide and a unitary transformation connects the two sets:
\beq
D'=\left(\matrix{d^\prime\cr s^\prime\cr b^\prime}\right)=V\left(\matrix{d\cr s\cr b}\right)=VD
\label{km1}
\eeq
where V is the Cabibbo-Kobayashi-Maskawa (CKM) matrix \cite{CKM}. In analogy with $D$ we can denote by $U$ the column vector of the 3 up quark mass eigenstates.
Thus in terms of mass eigenstates the charged weak current of quarks is of the form:
\beq
J^+_{\mu}\propto\bar U \gamma_{\mu}(1-\gamma_5)t^+ VD 
\label{km2}
\eeq
where
\beq
V=U^\dagger_uU_d
\label{km3}
\eeq 
Here $U_u$ and $U_d$ are the unitary matrices that operate on left-handed doublets in the diagonalization of the $u$ and $d$ quarks, respectively (see Eq.(\ref{49})).
Since V is unitary (i.e. $VV^\dagger=V^\dagger V=1$) and commutes with $T^2$, $T_3$ and Q (because all d-type quarks
have the same isospin and charge), the neutral current couplings are diagonal both in the primed and unprimed basis (if
the down-type quark terms in the $Z$ current are written in terms of weak isospin eigenvectors as $\bar D^\prime \Gamma D^\prime$, then by changing basis we get $\bar D
V^\dagger \Gamma V D$ and V and $\Gamma$ commute because, as seen from Eq.(\ref{41}), $\Gamma$ is made of Dirac
matrices and of $T_3$ and Q generator matrices). It follows that $\bar D^\prime \Gamma D^\prime =\bar D \Gamma D$. This is
the GIM mechanism \cite{GIM} that ensures natural flavour conservation of the neutral current couplings at the tree level. 

For N generations of quarks, V is a NxN unitary matrix that depends on $N^2$ real numbers ($N^2$ complex entries with
$N^2$ unitarity constraints). However, the $2N$ phases of up- and down-type quarks are not observable. Note that an
overall phase drops away from the expression of the current in Eq.(\ref{km2}), so that only $2N-1$ phases can affect V.
In total, V depends on $N^2-2N+1=(N-1)^2$ real physical parameters. A similar counting gives $N(N-1)/2$ as the number of
independent parameters in an orthogonal NxN matrix. This implies that in V we have $N(N-1)/2$ mixing angles and
$(N-1)^2-N(N-1)/2=(N-1)(N-2)/2$ phases: for $N=2$ one mixing angle (the Cabibbo angle $\theta_C$) and no phases, for $N=3$ three angles ($\theta_{12}$, $\theta_{13}$ and $\theta_{23}$) and one
phase $\varphi$ etc. 

Given the experimental near diagonal structure of V a convenient parametrization is the one proposed by
Maiani \cite{mai}.  It can be cast in the form of a product of three independent 2x2 block matrices ($s_{ij}$ and $c_{ij}$ are shorthands for $\sin{\theta_{ij}}$ and $\cos{\theta_{ij}}$):
\beq  V~=~ 
\left(\matrix{1&0&0 \cr 0&c_{23}&s_{23}\cr0&-s_{23}&c_{23}     } 
\right)
\left(\matrix{c_{13}&0&s_{13}e^{i\varphi} \cr 0&1&0\cr -s_{13}e^{-i\varphi}&0&c_{13}     } 
\right)
\left(\matrix{c_{12}&s_{12}&0 \cr -s_{12}&c_{12}&0\cr 0&0&1     } 
\right)~.
\label{ufi}
\eeq
The advantage of this parametrization is that the 3 mixing angles are of different orders of magnitude. In fact, from experiment we know that $s_{12}\equiv\lambda$, $s_{23}\sim
O(\lambda^2)$ and 
$s_{13}\sim O(\lambda^3)$, where $\lambda = \sin{\theta_C}$ is the sine of the Cabibbo angle, and, as order of magnitude, $s_{ij}$ can be expressed in terms of small powers of $\lambda$. More precisely, following Wolfenstein \cite{wol}
one can set:
\beq
s_{12}\equiv\lambda,~~~~~~~~s_{23}=A\lambda^2,~~~~~~~~s_{13}e^{-i\phi}=A\lambda^3(\rho-i\eta)
\label{km7}
\eeq
As a result, by neglecting terms of higher order in $\lambda$ one can write down:
\beq
V= 
\left[\matrix{
V_{ud}&V_{us}&V_{ub} \cr
V_{cd}&V_{cs}&V_{cb}\cr
V_{td}&V_{ts}&V_{tb}     } 
\right ]~\sim~\left[\matrix{
1-\frac{\lambda^2}{2}&\lambda&A\lambda^3(\rho-i\eta) \cr
-\lambda&1-\frac{\lambda^2}{2}&A\lambda^2\cr
A\lambda^3(1-\rho-i\eta)&-A\lambda^2&1     } 
\right ]+O(\lambda^4).
\label{km8}
\eeq 
It has become customary to make the replacement $\rho,\eta \rightarrow \bar \rho, \bar \eta$ with:
\beq
\rho - i \eta = \frac{\bar \rho - i \bar \eta}{\sqrt{1-\lambda^2}}~\sim~(\bar \rho - i \bar \eta) (1+\lambda^2/2+\dots).
\label{km8a}
\eeq

The best values of the CKM parameters as obtained from experiment are continuously updated in refs. \cite{UTfit,CKMfitter} (a survey of the current status of the CKM
parameters can also be found in ref.\cite{pdg12}). A Summer 2013 fit \cite{UTfit} led to the values (compatible values, within stated errors, are given in ref. \cite{CKMfitter}):
\bea
\lambda&=&0.22535\pm0.00065\nonumber\\
A&=&0.822\pm0.012\nonumber\\
\bar \rho&=&0.127\pm0.023;~~~~~ \bar \eta = 0.353\pm0.014\label{km9}
\eea

\begin{figure}[h]
\noindent
\includegraphics[width=7cm]{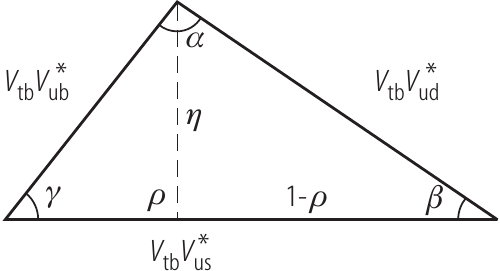} 
\caption[]{The unitarity triangle corresponding to Eq.(\ref{km10}).}
\label{fig:6}
\end{figure}

\begin{figure}[t]
\centerline{\includegraphics[height=4in]{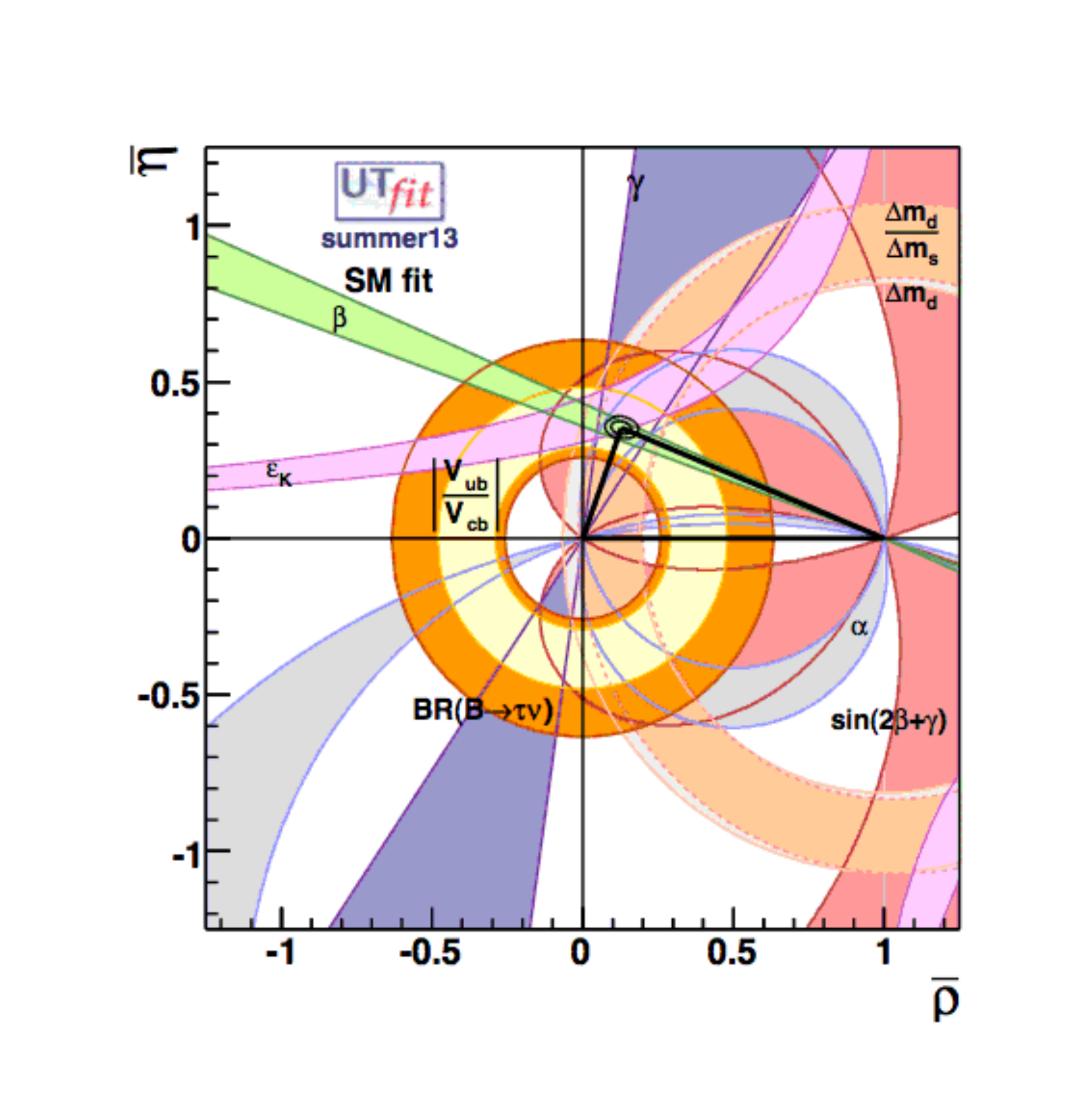}}     
\caption{Constraints in the $\bar \rho,\bar \eta$ plane including  the most recent data inputs (summer 2013) in the global CKM fit.}
\label{UT}
\end{figure}

In the SM the non vanishing of the $\bar \eta$ parameter (related to the phase $\varphi$ in Eqs.~\ref{ufi} and \ref{km7}) is the only source of CP violation in the quark sector (we shall see that new sources of CP violation very likely arise from the neutrino sector). Unitarity of the CKM matrix V implies
relations of the form
$\sum_a V_{ba}V^*_{ca}=\delta_{bc}$. In most cases these relations do not imply particularly instructive constraints on the
Wolfenstein parameters. But when the three terms in the sum are of comparable magnitude we get interesting information. The
three numbers which must add to zero form a closed triangle in the complex plane (unitarity triangle), with sides of comparable length. This is the
case for the t-u triangle shown in Fig.~\ref {fig:6} (or, what is equivalent in first approximation, for the d-b triangle):
\beq
V_{td}V^*_{ud}+V_{ts}V^*_{us}+V_{tb}V^*_{ub}=0\label{km10}
\eeq
All terms are of order $\lambda^3$. For $\eta$=0 the triangle would flatten down to vanishing area. In fact
the area of the triangle, J of order $J\sim \eta A^2 \lambda^6$, is the Jarlskog invariant \cite{85Ja1} (its value is independent of the
parametrization). In the SM, in the quark sector, all CP violating observables must be proportional to J, hence to the area of the triangle
or to $\eta$. Its experimental value is J $\sim (3.12 \pm 0.09)~10^{-5}$ \cite{UTfit}.  A direct and by now very solid evidence for
J non vanishing has been first obtained from the measurements of $\epsilon$ and $\epsilon'$ in K decay. Additional direct evidence has more recently been collected from experiments on B decays at beauty factories, at the Tevatron and at the LHC (in particular by the LHCb experiment). Very recently searches for CP violation in D decays (negative so far) have been reported by the LHCb experiment \cite{CPD}. The angles $\beta$ (the most precisely measured), $\alpha$ and $\gamma$ have been determined with fair precision. The angle measurements and the available information on the magnitude of the sides, taken together, are in good agreement with the predictions from the SM unitary triangle (see  Fig.~\ref {UT}) \cite{UTfit} \cite{CKMfitter}. Some alleged tensions are not convincing either because of their poor statistical significance or because of lack of confirmation from different potentially sensitive experiments or because the associated theoretical error estimates can be questioned. 

\begin{figure}[b]
\noindent
\includegraphics[width=8cm]{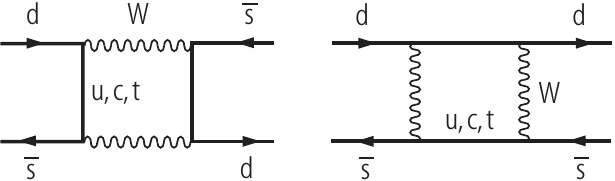} 
\caption[]{Box diagrams describing $K^0-\bar K^0$ mixing at the quark level at 1-loop.}
\label{fig:8}
\end{figure}

As we have discussed, due to the GIM mechanism, there are no flavour changing neutral current (FCNC) transitions at the tree level in the SM. Transitions with $|\Delta F|=1, 2$ are induced at one loop level. In particular, meson mixing, i.e. $M \rightarrow \bar M$ off diagonal $|\Delta F|=2$ mass matrix elements  (with $M=K, D$ or $B$ neutral mesons), are obtained from box diagrams. For example, in the case of $K^0-\bar K^0$ mixing the relevant transition is $ \bar s d \rightarrow s \bar d$ (see Fig.~\ref{fig:8}). In the internal quark lines all up-type quarks are exchanged. In the amplitude, two vertices and the connecting propagator (with virtual four momentum $p_\mu$) at one side contribute a factor ($u_i=u,c,t$):
\beq
F_{GIM}=\sum_i V^*_{u_i s} \frac{1}{p\llap{$/$}-m_{ui}}V_{u_i d}~,
\label{FGIM}
\eeq
which, in the limit of equal $m_{ui}$, is clearly vanishing due to the unitarity of the CKM matrix $V$. Thus the result is proportional to mass differences. 

For $K^0-\bar K^0$ mixing the contribution of virtual u quarks is negligible due to the small value of $m_u$ and the contribution of the t quark is also small due to the mixing factors $V^*_{t s} V_{t d} \sim O(A^2 \lambda^5)$. The dominant c quark contribution to the real part of the box diagram quark-level amplitude is approximately of the form (see, for example, \cite{92do1}):
\beq
Re H_{box}=\frac{G_F^2}{16\pi^2}m_c^2 Re(V^*_{cs} V_{c d})^2 \eta_1 O^{\Delta s=2}~,
\label{HGIM}
\eeq
where $\eta_1\sim 0.85$ is a QCD correction factor and $O^{\Delta s=2}=\bar d_L\gamma_\mu s_L~\bar s_L\gamma_\mu d_L$ is the relevant 4-quark, dimension-6,  operator. The $\eta_1$ factor arises from gluon exchanges among the quark legs of the 4-quark operator. Indeed the coefficients of the operator expansion, which arises when the heavy particles exchanged are integrated away, obey renormalization group equations and the associated logarithms can be resummed (the first calculation of resummed QCD corrections to weak non leptonic amplitudes was performed in refs. \cite{GLAM}; for a pedagogical introduction see, for example, ref. \cite{bur99}).
To obtain the $K^0-\bar K^0$ mixing amplitude the matrix element of $O^{\Delta s=2}$ between meson states must be taken which is parametrized in terms of a "$B_K$ parameter", defined in such a way that $B_K=1$ for vacuum state insertion between the two currents:
\beq
\langle K^0|O^{\Delta s=2}|\bar K^0\rangle =\frac{16}{3}f_K m_K^2 B_K~,
\label{BK}
\eeq
where $B_K \sim 0.75$ (this is the renormalization group independent definition usually denoted as $\hat B_K$) and $f_K\sim 113$ MeV, the kaon pseudoscalar constant, are best evaluated by QCD lattice simulations \cite{FLAG}. Clearly to the charm parton contribution in Eq.(\ref{HGIM}) additional non perturbative  terms must be added, some of them of $O(m_K^2/m_c^2)$, because the smallness of $m_c$ makes a completely partonic dominance inadequate. In Eq.(\ref{HGIM}) the factor $O(m_c^2/m_W^2)$ is the "GIM suppression" factor ($1/m^2_W$ is hidden in $G_F$ according to Eq. (\ref{37})). 

For B mixing the dominant contribution is from the t quark.  In this case, the partonic dominance is more realistic and the GIM factor  $O(m_t^2/m_W^2)$ is actually larger than 1. More recently also D mixing has been observed \cite{dmix}. In the corresponding box diagrams down-type quarks are involved. But starting from $D \sim c \bar u$ the b-quark contribution is strongly suppressed by the CKM angles (given that $V_{cb} V_{ub}^* \sim O(\lambda_C^5)$. The masses of the d and s quarks are too small for a partonic evaluation of the box diagram and non perturbative terms cannot be neglected. This makes a theoretical evaluation of  mixing and CP violation effects for D mesons problematic.

\begin{figure}[h]
\noindent
\includegraphics[width=8cm]{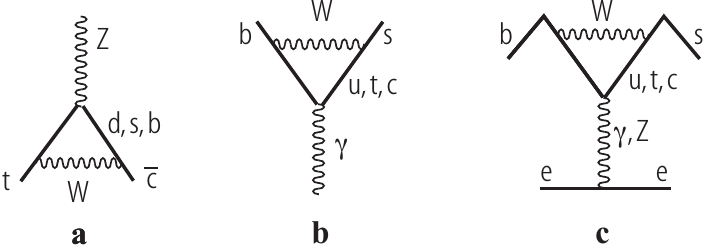} 
\caption[]{Examples of $|\Delta F|=1$ transitions at the quark level at 1-loop: (a) Diagram for a $Z\rightarrow t ~\bar c$ vertex, (b) $b \rightarrow s~ \gamma$, (c) a "penguin" diagram for $b \rightarrow s~ e^+e^-$.}
\label{fig:pen}
\end{figure}

All sorts of transitions with $|\Delta F|=1$ are also induced at loop level.  For example, an effective vertex $Z\rightarrow t \bar c$, which does not exist at tree level, is generated at 1-loop (see Fig.~\ref {fig:pen}). Similarly, transitions involving photons or gluons are also possible, like $t \rightarrow c~g$ or $b\rightarrow s ~\gamma$ (Fig.~\ref {fig:pen}) or $b\rightarrow s~g$. 

For light fermion exchange in the loop the GIM suppression is also effective in $|\Delta F|=1$ amplitudes. For example, analogous leptonic transitions like $\mu \rightarrow e~\gamma$ or $\tau \rightarrow \mu~\gamma$ also exist but in the SM are extremely small and out of reach for experiments, because the tiny neutrino masses enter in the GIM suppression factor. But new physics effects could well make these rare lepton flavour violating processes accessible to experiment. In fact, the present limits already pose stringent constraints on models of new physics. Of particular importance is the recent bound \cite{meg} obtained by the MEG Collaboration at SIN, near Zurich, Switzerland, on  the branching ratio for $\mu \rightarrow e~\gamma$: $B(\mu \rightarrow e~\gamma) \lesssim 5.7~10^{-13}$ at 90$\%$. 

The external $Z$, photon or gluon can be attached to a pair of light fermions, giving rise to an effective four fermion operator, as in  "penguin diagrams"  like the one shown in Fig.~\ref {fig:pen} for $b\rightarrow s ~l^+l^-$.  The inclusive rate $B \rightarrow X_s~\gamma$ (here $B$ stands for $B_d$) with $X_s$ a hadronic state containing a unit of strangeness corresponding to an s-quark, has been precisely measured. The world average result for the branching ratio with $E_\gamma >1.6~\rm{GeV}$ is \cite{dmix}:
$B(B \rightarrow X_s~\gamma)_{exp}= (3.55\pm0.26)^.10^{-4}~$.
The theoretical prediction for this inclusive process is to a large extent free of uncertainties from hadronization effects and is accessible to perturbation theory as the b-quark is heavy enough. The most complete result at order $\alpha_s^2$ is at present from ref. \cite{07Be1} (and refs. therein):
$B(B \rightarrow X_s~\gamma)_{th}= (2.98\pm0.26)^.10^{-4}~$.
Note that the theoretical value has recently become smaller than the experimental value. The fair agreement between theory and experiment imposes stringent constraints on possible new physics effects.

\begin{figure}[h]
\noindent
\includegraphics[width=15cm]{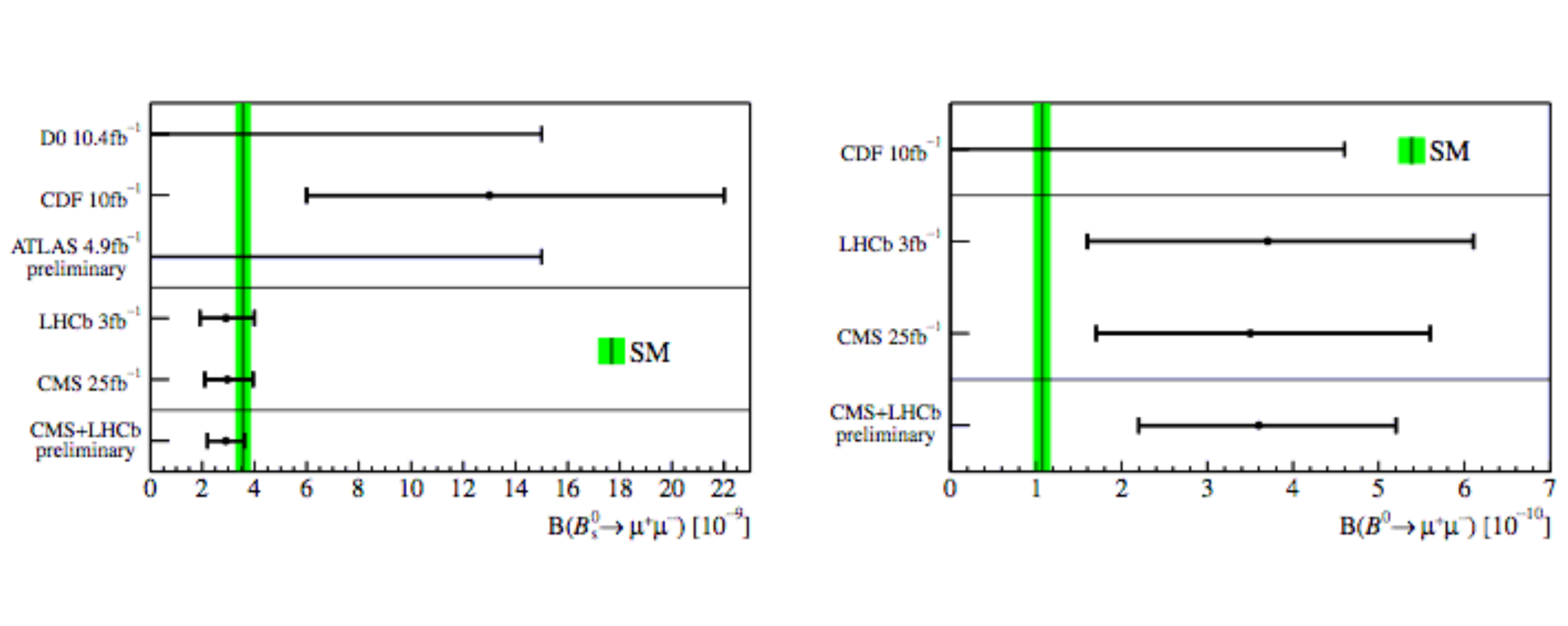} 
\caption[]{The experimental results on the $B_{s,d} \rightarrow \mu^+ \mu^-$ decays \cite{comb}.}
\label{bmm}
\end{figure}

Related processes are $B_{s,d} \rightarrow \mu^+ \mu^-$.  These decay are very rare in the SM, their predicted branching ratio being $B(B_s \rightarrow \mu^+ \mu^-) \sim (3.35\pm0.28)~10^{-9}$,  $B(B_d \rightarrow \mu^+ \mu^-) \sim (1.07\pm0.10)~10^{-10}$\cite{buras}. These very small expected branching ratios result because these decays are FCNC processes with helicity suppression in the purely leptonic final state (the decaying meson has spin zero and the muon pair is produced by vector exchange in the SM). Many models of new physics beyond the SM predict large deviations. Thus these processes pose very stringent tests to the SM. Recently the LHCb and CMS experiments have reached the sensitivity to observe the $B_s$ mode. The LHCb result is $B(B_s \rightarrow \mu^+ \mu^-) = 2.9^{+1.1}_{-1.0}~10^{-9}$ \cite{bmumu} (in the same paper the bound $B(B_d \rightarrow \mu^+ \mu^-) \leq 7.4~10^{-10}$ at 95$\%$ c.l. is set). On the same decays CMS has obtained $B(B_s \rightarrow \mu^+ \mu^-) = 3.0^{+1.0}_{-0.9}~10^{-9}$ \cite{bmumucms} and $B(B_d \rightarrow \mu^+ \mu^-) \leq 11~10^{-10}$ at 95$\%$ c.l. The LHCb and CMS results have been combined \cite{comb} and give $B(B_s \rightarrow \mu^+ \mu^-) = (2.9\pm~0.7)~10^{-9}$, in good agreement with the SM, and $B(B_d \rightarrow \mu^+ \mu^-) = 3.6^{+1.6}_{-1.4}~10^{-10}$ with the central value 1.7$\sigma$ above the SM (see Fig. \ref{bmm}). Another very demanding test of the SM has been passed!

Among the exclusive processes of the $b \rightarrow s$ type much interest is at present devoted to the channel $B \rightarrow K^*\mu^+\mu^-$ \cite{expKmumu,thKmumu}. The differential decay distribution depends on three angles and on the $\mu^+\mu^-$ invariant mass squared $q^2$. In general 12+12 form factors enter in the decay distribution (12 in $B$ decay and 12 in the CP conjugated $\bar B$ decay) and many observables can be defined. By suitable angular foldings and CP averages the number of form factors is reduced. A sophisticated theoretical analysis allows to identify and to study a number of quantities that can be measured and are "clean", i.e. largely independent of hadronic form factor ambiguities \cite{thKmumu}.  For those observables most of the results agree with the SM predictions (based on a Wilson operator expansion in powers of  $1/m_W$ and of $1/m_b$ with coefficients depending on $\alpha_s$) but a few discrepancies are observed. The significance, taking into account the number of observables studied and the theoretical ambiguities (especially on the estimate of $1/m_b$ corrections), is not compelling but a substantial activity, both on the experimental and the theoretical side, is under way (see, for example, \cite{th2}). To be followed!

In conclusion, the CKM theory of quark mixing and CP violation has been precisely tested in the last decade and turns out to be very successful. The expected deviations from new physics at the EW scale did not appear so far. The constraints on new physics from flavour phenomenology are extremely demanding:
when adding higher dimension effective operators to the SM, the flavour constraints generically lead  to powers of very large suppression scales $\Lambda$ in the denominators of the corresponding coefficients. In fact in the SM, as we have discussed in this section,  there are very powerful protections against flavour changing neutral currents and CP violation effects, in particular through the smallness of quark mixing angles. In this respect the SM is very special and, as a consequence, if there is new physics, it must
be highly non generic in order to satisfy the present flavour constraints. Only by imposing that the new physics shares the SM set of protections one can reduce the scale $\Lambda$ down to O(1) TeV. For example, the class of models with minimal flavour violation (MFV) \cite{MFV}, where the SM Yukawa couplings are the only flavour symmetry breaking terms also beyond the SM, have been much studied and represent a sort of extreme baseline. Alternative less minimal models that are currently under study are based on a suitably broken $U(3)^3$ or $U(2)^3$ flavour symmetry (the cube refers to the $Q_L=u_L,~d_L$ doublet and the two $u_R$ and $d_R$ singlets, while $U(3)$ or $U(2)$ mix the 3 or the first 2 generations) \cite{butta}.

\subsection{Neutrino Mass and Mixing}
\label{sec:26}

In the minimal version of the SM the right handed neutrinos $\nu_{iR}$, which have no gauge interactions, are not present at all. With no $\nu_R$ no Dirac mass is possible for neutrinos. If lepton number conservation is also imposed, then no Majorana mass is allowed either and, as a consequence, all neutrinos are massless. But, at present, from neutrino oscillation experiments, we know that at least 2 out of the 3 known neutrinos have non vanishing masses (for reviews, see, for example, refs. \cite{revnu}): the two mass squared differences measured from solar ($\Delta m^2_{12}$) and atmospheric oscillations ($\Delta m^2_{23}$) are given by $\Delta m^2_{12} \sim 8~10^{-5}~eV^2$ and $\Delta m^2_{23}\sim 2.5~10^{-3}~eV^2$ \cite{numassF,numassG,numassV}. 

Neutrino oscillations only measure $|m_i^2|$ differences. On the absolute values of each $m_i$ we know that they are very small, with an upper limit of a fraction of $eV$, obtained from 1) laboratory experiments (tritium $\beta$ decay near the end point: $m_\nu \lesssim 2 ~\rm{eV}$ \cite{pdg12}; 2) absence of visible neutrinoless double $\beta$ decay ($0\nu\beta\beta$): from $Ge^{76}$ one has obtained (the range is from nuclear matrix elements ambiguities) $|m_{ee}| \lesssim 0.2-0.4~ \rm{eV}$ \cite{gerda} ($m_{ee}$ is a combination of neutrino masses; for a review, see, for example  \cite{beta}). This result strongly disfavours, in a model independent way,  the claimed observation of $0\nu\beta\beta$ decay in $Ge^{76}$ decays \cite{klap}. From $Xe^{136}$ one obtains the combined result $|m_{ee}| \lesssim 0.12-0.25~ \rm{eV}$ \cite{exokam}. 3) from cosmological observations \cite{cosmo}: after the recent release of Planck data the quoted bounds for $\Sigma m_\nu$, the sum of (quasi)-stable neutrino masses, span a range, depending on the data set included and the cosmological priors, like  $\Sigma m_\nu \lesssim ~0.98$ or $\lesssim~0.32$ or $\lesssim ~0.23$ \cite{planck} (assuming 3 degenerate neutrinos these numbers have to be divided by 3 in order to obtain the limit on individual neutrino masses). 

If $\nu_{iR}$ are added to the minimal model and lepton number is imposed by hand, then  neutrino masses would in general appear as Dirac masses, generated by the Higgs mechanism, like for any other fermion.  But, for Dirac neutrinos, to explain the extreme smallness of neutrino masses, one should allow for very small Yukawa couplings. However, we stress that, in the SM, baryon B and lepton L number conservation, which are not guaranteed by gauge symmetries (which is instead the case for the electric charge $Q$), are understood as "accidental" symmetries. In fact the SM lagrangian should contain all terms allowed by gauge symmetry and renormalizability, but the most general renormalizable lagrangian (i.e. consisting of operator dimension $d \leq 4$), built from the SM fields, compatible with the SM gauge symmetry, in the absence of $\nu_{iR}$, is automatically B and L conserving (however, non perturbative instanton effects break the conservation of B+L while preserving B-L, as discussed in Sect. \ref{sec:27}). 
In the presence of 
$\nu_{iR}$, this is no more true and the right handed Majorana mass term is allowed:
\beq
M_{RR}=\bar{\nu}_{iR}^cM_{ij}\nu_{jR}=\nu_{iR}^TCM_{ij}\nu_{jR}~,
\label{mnu1}
\eeq
where $\nu_{iR}^c= C \bar \nu_{iR}^T$ is the charge conjugated neutrino field and C is the charge conjugation matrix in Dirac spinor space.  The Majorana mass term is an operator of dimension $d=3$ with $\Delta L =2$. Since the $\nu_{iR}$ are gauge singlets the Majorana mass $M_{RR}$ is fully allowed by the gauge symmetry and a coupling with the Higgs is not needed to generate this type of mass. As a consequence, the entries of the mass matrix $M_{ij}$ do not need to be of the order of the EW symmetry breaking scale $v$ and could be much larger. If one starts from the Dirac and RR Majorana mass terms for neutrinos, the resulting mass matrix, in the $L,R$ space, has the form:
\beq
m_\nu= 
\left[\matrix{
0&m_D\cr
m_D&M    } 
\right ]
\label{mnu3}
\eeq
where $m_D$ and $M$ are the Dirac and Majorana mass matrices ($M$ is the matrix $M_{ij}$ in Eq.(\ref{mnu1})). The corresponding eigenvalues are 3 very heavy neutrinos with masses of order $M$ and 3 light neutrinos with masses 
\beq
 m_\nu=-m_D^TM^{-1}m_D~,
\label{mnu4} 
\eeq
which are possibly very small if $M$ is large enough. This is the see-saw mechanism for neutrino masses \cite{ss}. Note that if no $\nu_{iR}$ existed a Majorana mass term could still be built out of $\nu_{jL}$. But $\nu_{jL}$ have weak isospin 1/2, being part of the left handed lepton doublet $l$. Thus, the left handed Majorana mass term has total weak isospin equal to 1 and needs 2 Higgs fields to make a gauge invariant term. The resulting mass term:
\beq 
O_5=\frac{(H l)^T_i \lambda_{ij} (H l)_j}{M}+~h.c.~~~,
\label{O5}
\eeq
with $M$ a large scale (apriori comparable to the scale of $M_{RR}$) and $\lambda$ a dimensionless coupling generically of O(1), is a non renormalizable operator of dimension 5, first pointed out by S. Weinberg \cite{weidim5}. The corresponding mass terms are of the order $m_\nu \sim \lambda v^2/M$, where $v$ is the Higgs VEV, hence of the same generic order of the light neutrino masses from Eq.(\ref{mnu4}). Note that, in general, the neutrino mass matrix has the form:
\beq
{\bf m_\nu}=\nu^T m_{\nu} \nu~~~~,
\label{mnuT}
\eeq
as a consequence of the Majorana nature of neutrinos.

In conclusion, neutrino masses are believed to be small because neutrinos are Majorana particles with masses inversely proportional to the large scale $M$ of energy where L non conservation is induced. This corresponds to an important enlargement of the original minimal SM where no $\nu_R$ was included and L conservation was imposed by hand (but this ansatz would be totally unsatisfactory because  L conservation is true "accidentally" only at the renormalizable level, but  is violated by  non renormalizable terms like the Weinberg operator and by instanton effects). Actually L and B non conservation are necessary if we want to explain baryogenesis and we have Grand Unified Theories (GUTs) in mind. It is interesting that the observed magnitudes of the mass squared splittings of neutrinos are well compatible with a scale $M$ remarkably close to the GUT scale, where indeed L non conservation is naturally expected. In fact, for
$m_{\nu}\approx \sqrt{\Delta m^2_{atm}}\approx 0.05$ eV (see Table(\ref{tab:data})) and 
$m_{\nu}\approx m_D^2/M$ with $m_D\approx v
\approx 200~GeV$ we find $M\approx 10^{15}~GeV$ which indeed is an impressive indication for
$M_{GUT}$.

In the previous Section we have discussed flavour mixing for quarks. But, clearly, given that non vanishing neutrino masses have been established, a similar mixing matrix is also introduced in the leptonic sector. We assume in the following  that there are only two distinct
neutrino oscillation frequencies, the atmospheric and the solar frequencies (both of them now also confirmed by experiments where neutrinos are generated on the earth like K2K, KamLAND, MINOS). At present the bulk of neutrino oscillation data are well reproduced in terms of three light neutrino species. However, some (so far not compelling) evidence for additional "sterile" neutrino species (i.e. not coupled to the weak interactions, as demanded by the LEP limit on the number of "active" neutrinos) are present in some data. We discuss here 3-neutrino mixing, which is in any case a good approximate framework to discuss neutrino oscillations, while for possible sterile neutrinos we refer to the comprehensive review in ref. \cite{white}. 

Neutrino oscillations are due to a misalignment between the flavour basis, $\nu'\equiv(\nu_e,\nu_{\mu},\nu_{\tau})$, where
$\nu_e$ is the partner of the mass and flavour eigenstate $e^-$ in a left-handed (LH) weak isospin SU(2) doublet (similarly
for 
$\nu_{\mu}$ and $\nu_{\tau})$) and the mass eigenstates $\nu\equiv(\nu_1, \nu_2,\nu_3)$ \cite{pon,lee,revnu}: 
\beq
\nu' =U \nu~~~,
\label{U}
\eeq  where $U$ is the unitary 3 by 3 mixing matrix. Given the definition of $U$ and the transformation properties of the
effective light neutrino mass matrix ${\bf m_{\nu}}$ in Eq. \ref{mnuT}:
\bea 
\label{tr} {\nu'}^T m_{\nu} \nu'&= &\nu^T U^T m_\nu U \nu\\ \nonumber  U^T m_{\nu} U& = &{\rm
Diag}\left(m_1,m_2,m_3\right)\equiv m_{diag}~~~,
\eea  we obtain the general form of $m_{\nu}$ (i.e. of the light $\nu$ mass matrix in the basis where the charged lepton
mass is a diagonal matrix):
\beq  m_{\nu}=U^* m_{diag} U^\dagger~~~.
\label{gen}
\eeq  The matrix $U$ can be parameterized in terms of three mixing angles $\theta_{12}$,
$\theta_{23}$ and $\theta_{13}$ ($0\le\theta_{ij}\le \pi/2$)  and one phase $\varphi$ ($0\le\varphi\le 2\pi$) \cite{cabnu},
exactly as for the quark mixing matrix $V_{CKM}$. The following definition of mixing angles can be adopted:
\beq  U~=~ 
\left(\matrix{1&0&0 \cr 0&c_{23}&s_{23}\cr0&-s_{23}&c_{23}     } 
\right)
\left(\matrix{c_{13}&0&s_{13}e^{i\varphi} \cr 0&1&0\cr -s_{13}e^{-i\varphi}&0&c_{13}     } 
\right)
\left(\matrix{c_{12}&s_{12}&0 \cr -s_{12}&c_{12}&0\cr 0&0&1     } 
\right)
\label{ufinu}
\eeq  where $s_{ij}\equiv \sin\theta_{ij}$, $c_{ij}\equiv \cos\theta_{ij}$.  In addition, if $\nu$ are Majorana particles,
we have two more phases \cite{bil} given by the relative phases among the Majorana masses
$m_1$, $m_2$ and $m_3$. If we choose $m_3$ real and positive, these phases are carried by $m_{1,2}\equiv\vert m_{1,2}
\vert e^{i\phi_{1,2}}$.  Thus, in general, 9 parameters are added to the SM when non-vanishing neutrino masses are included: 3
eigenvalues, 3 mixing angles and 3 CP  violating phases.

In our notation the two frequencies, $\Delta m^2_{I}/4E$ $(I=sun,atm)$, are parametrized in terms of the $\nu$ mass
eigenvalues by 
\beq
\Delta m^2_{sun}\equiv \vert\Delta m^2_{12}\vert ,~~~~~~~
\Delta m^2_{atm}\equiv \vert\Delta m^2_{23}\vert~~~.
\label{fre}
\eeq   where $\Delta m^2_{12}=\vert m_2\vert^2-\vert m_1\vert^2 > 0$ and $\Delta m^2_{23}= m_3^2-\vert m_2\vert ^2$. The
numbering 1,2,3 corresponds to a definition of the frequencies and in principle may not coincide with the ordering from
the lightest to the heaviest state. "Normal hierarchy" is the case where $m_3$ is the largest mass in absolute value, otherwise one has an "inverse hierarchy". 

\begin{figure}[t]
\centerline{\includegraphics[height=13cm]{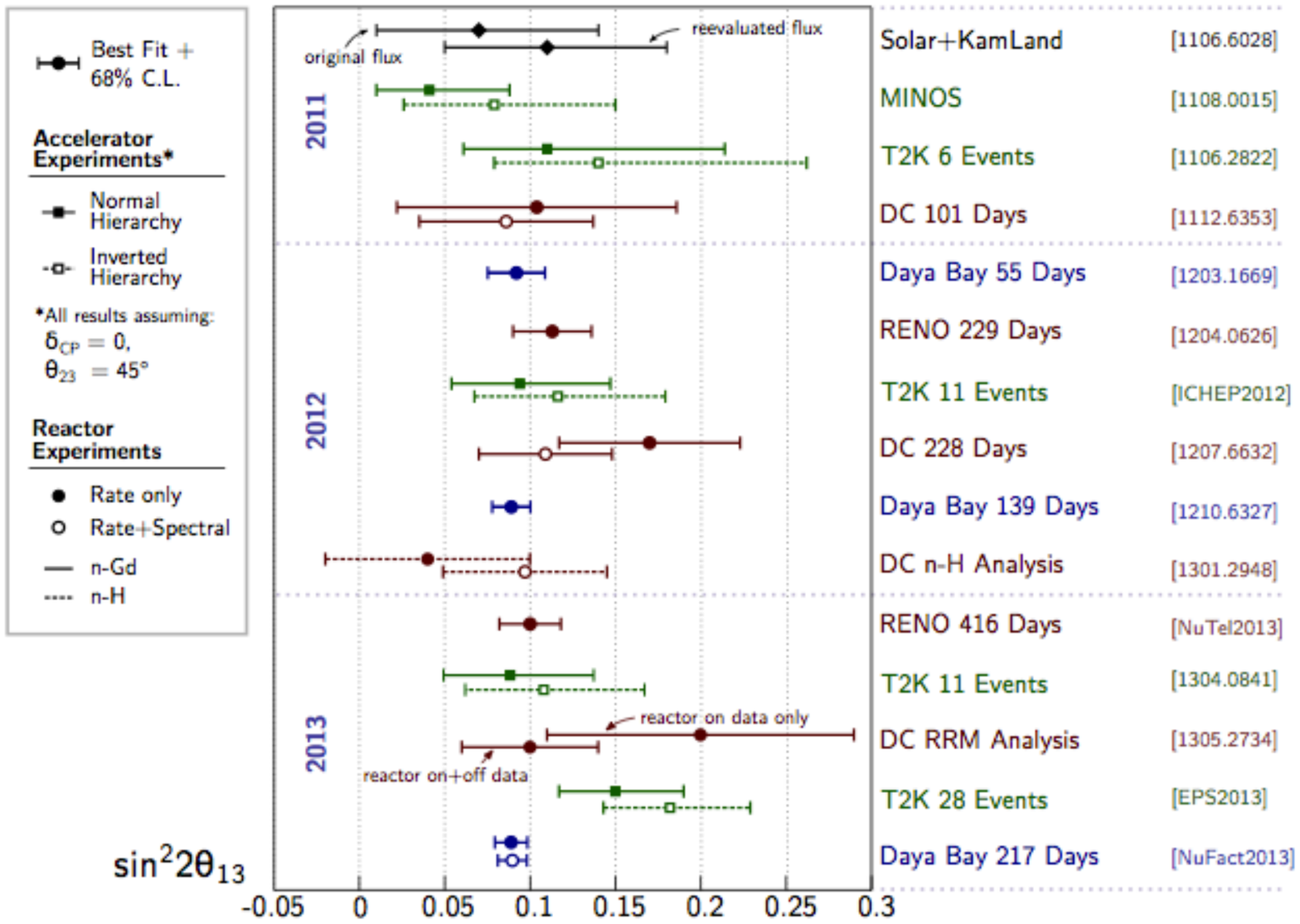}}     
\caption{The reactor angle measurements, updated to the NUFACT13 Conference, August 2013 \cite{kettel}, from the experiments T2K\cite{Abe:2011sj}, MINOS\cite{Adamson:2011qu}, DOUBLE CHOOZ\cite{Abe:2011fz}, Daya Bay \cite{An:2012eh} and RENO \cite{Ahn:2012nd}, for the normal (inverse) hierarchy.}
\label{teta13}
\end{figure}

Very important developments in the data have occurred in 2012. The value of the mixing angle $\theta_{13}$ has been proven to be non vanishing and its value is by now known with fair precision. Several experiments have been involved in the $\theta_{13}$ measurement and their results are reported in Fig. \ref{teta13}. 
The most precise result is from the Daya Bay reactor experiment in China:
\beq
\sin^2{2\theta_{13}}=~0.090\pm0.012~~~{\rm or}~~~	\sin^2{\theta_{13}}=0.023\pm0.003~~~{\rm or}~~~	\theta_{13}\sim 0.152\pm0.010
\label{DB}
\eeq
Note that $\theta_{13}$ is somewhat smaller but of the same order than the Cabibbo angle $\theta_C$.

The present data on the oscillation parameters are summarized in table \ref{tab:data} \cite{numassG}.
\vspace{0.1cm}
\begin{table}[h]
\begin{center}
\begin{tabular}{|c|c|}
   \hline
  &\\[-2mm]
  $\Delta m^2_{sun}~(10^{-5}~{\rm eV}^2)$ 			&$7.45^{+0.19}_{-0.16}$   \\[1mm]
  $\Delta m^2_{atm}~(10^{-3}~{\rm eV}^2)$ 			&$2.417\pm0.013$ ($-2.410\pm0.062$)   \\[1mm]
  $\sin^2\theta_{12}$ 											&$0.306\pm0.012$  \\[1mm]
  $\sin^2\theta_{23}$ 											&$0.446\pm0.007 \bigoplus  0.587^{+0.032}_{-0.037}$ \\[1mm]
  $\sin^2\theta_{13}$ 											&$0.0229^{+0.0020}_{-0.0019}$) 	  \\[1mm]
  $\delta_{CP}~(^o)$ 												&$265^{+56}_{-61}$	  \\[1mm]
  \hline
  \end{tabular}
\caption{\label{tab:data} Fits to neutrino oscillation data from Ref.~\cite{numassG} (free fluxes, including short baseline reactor data). The results for both the normal and the inverse (in the brackets) hierarchies are shown.}
\end{center}
\end{table}

Neutrino mixing is important because it could in principle provide new clues for
the understanding of the flavour problem. Even more so since neutrino mixing angles
show a pattern that is completely different than that of quark mixing: for quarks all
mixing angles are small, for neutrinos two angles are large (one is still compatible
with the maximal value) and only the third one is small. In reality it is frustrating that no real illumination was sparked on the problem of flavour. We can reproduce in models the data on neutrino mixing, in a wide range of dynamical setups that goes from anarchy to discrete flavour symmetries (for reviews and references see, for example refs. \cite{rmp,capri,afm,afms,afmm,kilu}) but we have not yet been able to single out a unique and convincing baseline for the understanding of fermion masses and mixings. In spite of many interesting ideas and the formulation of many elegant models the mysteries of the flavour structure of the three generations of fermions have not been much unveiled.

\subsection{Quantization and Renormalization of the Electroweak Theory}
\label{sec:27}

The Higgs mechanism gives masses to the Z, the $W^\pm$ and to fermions while the lagrangian density is still
symmetric. In particular the gauge Ward identities and the symmetric form of the gauge currents are preserved. The validity of
these relations is an essential ingredient for renormalizability. In the previous Sections we have specified the Feynman vertices in the "unitary" gauge where only physical particles appear. However, as discussed in Chapter 1, in this gauge
the massive gauge boson propagator would have
a bad ultraviolet behaviour:
\beq
W_{\mu\nu}=\frac{-g_{\mu\nu}+\frac{q_\mu q_\nu}{m^2_W}}{q^2-m^2_W}.\label{propW}
\eeq
A formulation of the standard EW theory with good apparent ultraviolet behaviour can be obtained by introducing the renormalizable or $R_\xi$ gauges \cite{AL}, in analogy with the abelian case discussed in detail in Chapter 1. One parametrizes the Higgs doublet as:
\begin{equation}
\phi = \pmatrix { \phi^+ \cr \phi^0}=\pmatrix { \phi_1+i \phi_2 \cr\phi_3+i \phi_4} =\pmatrix { -iw^+ \cr v+ \frac{H+iz}{\sqrt{2}}}~, 
\label{phixi}
\end{equation}
and similarly for $\phi ^\dagger$, where $w^-$ appears. The scalar fields $w^\pm$ and $z$ are the pseudo Goldstone bosons associated with the longitudinal modes of the physical vector bosons $W^\pm$ and $Z$. The $R_\xi$ gauge fixing lagrangian has the form: 
\beq
\Delta {\cal L}_{GF}=-\frac{1}{\xi}|\partial^\mu W_{\mu}-\xi m_W w|^2-\frac{1}{2\eta}(\partial^\mu Z_{\mu}-\eta m_Z z)^2 -\frac{1}{2\alpha}(\partial^\mu A_{\mu})^2~.
\label{xi5bis}
\eeq
The $W^\pm$ and $Z$ propagators, as well as those of the scalars $w^\pm$ and $z$, have exactly the same general forms as for the abelian case in Eqs. (67)-(69) of Chapter 1, with parameters $\xi$ and $\eta$, respectively (and the pseudo Goldstone bosons $w^\pm$ and $z$ have masses $\xi m_W$ and $\eta m_Z$). In general, a set of associated ghost fields must be added, again in direct analogy with the treatment of $R_\xi$ gauges in the abelian case of Chapter 1. The complete Feynman rules for the standard EW theory can be found in a number of textbooks (see, for example, \cite{txtb}). 

The pseudo Goldstone bosons $w^\pm$ and $z$ are directly related to the longitudinal helicity states of the corresponding massive vector bosons $W^\pm$ and $Z$. This correspondence materializes in a very interesting "equivalence theorem": at high energies of order $E$ the amplitude for the emission of one or more longitudinal gauge bosons $V_L$ (with $V=W,Z$) becomes equal (apart from terms down by powers of $m_V/E$) to the amplitude where each longitudinal gauge boson is replaced by the corresponding Goldstone field $w^\pm$ or $z$ \cite{eqth}. For example, consider top decay with a longitudinal $W$ in the final state: $t \rightarrow b W^+_L$. The equivalence theorem asserts that we can compute the dominant contribution to this rate from the simpler $t \rightarrow b w^+$ matrix element:
\beq
\Gamma(t \rightarrow b W^+_L)=\Gamma(t \rightarrow b w^+)[1+O(m_W^2/m_t^2)]~.
\label{eqtht}
\eeq
In fact one finds:
\beq
\Gamma(t \rightarrow b w^+)=\frac{h_t^2}{32\pi}m_t=\frac{G_Fm_t^3}{8\pi\sqrt{2}}~,
\label{eqtht1}
\eeq
where $h_t=m_t/v$ is the Yukawa coupling of the top quark (numerically very close to 1), and we used $1/v^2=2\sqrt{2}G_F$ (see Eq.(\ref{59})). If we compare with Eq.(\ref{36t}), we see that this expression coincides with the total top width (i.e. including all polarizations for the $W$ in the final state), computed at tree level, apart from terms down by powers of $O(m_W^2/m_t^2)$. In fact, the longitudinal $W$ is dominant in the final state because $h_t^2>>g^2$. Similarly the equivalence theorem can be applied to find the dominant terms at large $\sqrt{s}$ for the cross-section $e^+e^- \rightarrow W^+_L W^-_L$, or the leading contribution, in the limit $m_H>>m_V$, to the width for the decay $\Gamma(H \rightarrow VV)$.

The formalism of the $R_\xi$ gauges is also very useful in proving that spontaneously broken gauge theories are renormalizable. In fact, the non singular behaviour of propagators at large momenta is very suggestive of the result. Nevertheless it is by far not a simple matter to prove this statement.
The fundamental theorem that in general a gauge theory with spontaneous symmetry breaking and the Higgs
mechanism is renormalizable was proven by 't Hooft and Veltman \cite{HV}, \cite{LZJ}. 

For a chiral theory like the SM an additional complication arises from
the existence of chiral anomalies. But this problem is avoided in the SM because the quantum numbers of the quarks and
leptons in each generation imply a remarkable (and, from the point of view of the SM, mysterious) cancellation of the anomaly, as originally
observed in ref. \cite{BIM}. In quantum field theory one encounters an anomaly when a symmetry of the
classical lagrangian is broken by the process of quantization, regularization and renormalization of the theory.  Of direct relevance for the EW theory is the Adler-Bell-Jackiw (ABJ) chiral anomaly \cite{ABJ}. The classical lagrangian
of a theory with massless fermions is invariant under a U(1) chiral transformations
$\psi\prime=e^{i\gamma_5\theta}\psi$ (see also Sect. \ref{sec:10}). The associated axial Noether current is conserved at the classical level. But, at the
quantum level, chiral symmetry is broken due to the ABJ anomaly and the current is not conserved. The chiral breaking is
produced by a clash between chiral symmetry, gauge invariance and the regularization procedure. 

\begin{figure}[h]
\noindent
\includegraphics[width=4cm]{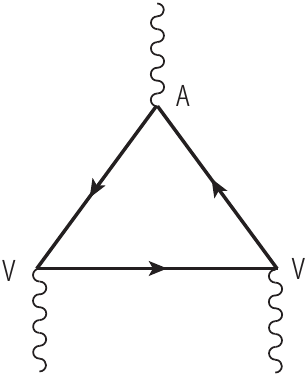} 
\caption[]{Triangle diagram that generates the ABJ anomaly \cite{ABJ}.}
\label{fig:9}
\end{figure}

The anomaly is generated
by triangular fermion loops with one axial and two vector vertices (Fig.~\ref{fig:9}).
For example, for the Z  the axial coupling is proportional to the 3rd component of weak isospin $t_3$, while the vector coupling is proportional to a linear combination of
$t_3$ and the electric charge Q. Thus in order for the chiral anomaly to vanish all traces of the form $tr\{t_3QQ\}$,
$tr\{t_3t_3Q\}$, $tr\{t_3t_3t_3\}$ (and also  $tr\{t_+t_-t_3\}$ when charged currents are also included) must vanish, where
the trace is extended over all fermions in the theory that can circulate in the loop. Now all of these
traces happen to vanish for each fermion family separately. For example, take $tr\{t_3QQ\}$. In one family there are, with
$t_3=+1/2$, three colours of up quarks with charge $Q=+2/3$ and one neutrino with $Q=0$ and, with $t_3=-1/2$, three colours
of down quarks with charge $Q=-1/3$ and one $l^-$ with $Q=-1$. Thus we obtain $tr\{t_3QQ\}=1/2^.3^.4/9-1/2^.3^.1/9-1/2^.1=0$.
This impressive cancellation suggests an interplay among weak isospin, charge and colour quantum numbers which appears as a
miracle from the point of view of the low energy theory but is in fact understandable from the point of view of the high energy
theory. For example, in Grand Unified Theories (GUTs) (for reviews, see, for example, \cite{GUTs}) there are similar relations where charge quantization and colour are related: in the 5 of SU(5)
we have the content $(d,d,d,e^+,\bar\nu)$ and the charge generator has a vanishing trace in each SU(5) representation (the
condition of unit determinant, represented by the letter S in the SU(5) group name, translates into zero trace for the
generators). Thus the charge of d quarks is -1/3 of the positron charge because there are three colours. A whole family fits perfectly in one 16 of SO(10) which is anomaly free. So GUTs can naturally explain the cancellation of the chiral anomaly.

An important implication of chiral anomalies together with the topological properties of the vacuum in non abelian gauge theories is that the conservation of the charges associated to baryon (B) and lepton (L) numbers is broken by the anomaly \cite{tHo}, so that B and L conservation is actually violated in the standard electroweak theory (but B-L remains conserved). B and L are conserved to all orders in the perturbative expansion but the violation occurs via non perturbative instanton effects \cite{Bela}
(the amplitude is proportional to the typical non perturbative factor $\exp{- c/g^2}$, with $c$ a constant and $g$ the $SU(2)$ gauge coupling). The corresponding effect is totally negligible at zero temperature $T$, but becomes relevant at temperatures close to the electroweak symmetry breaking scale, precisely at $T\sim O(TeV)$. The non conservation of B+L and the conservation of B-L near the weak scale plays a role in the theory of baryogenesis that quantitatively aims at explaining the observed matter antimatter asymmetry in the Universe (for reviews and references, see, for example, \cite{bupe}).

\subsection{QED Tests: Lepton Anomalous Magnetic Moments}
\label{sec:28}

\begin{figure}[[!htbp]
\noindent
\includegraphics[width=8cm]{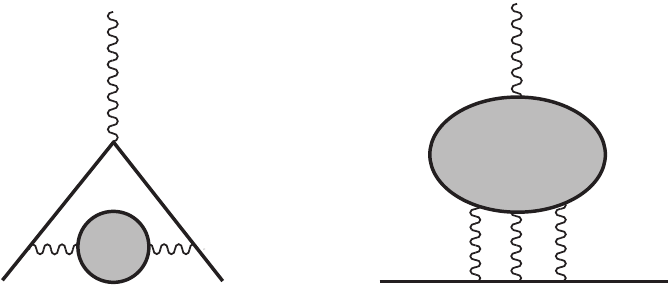} 
\caption[]{The hadronic contributions to the anomalous magnetic moment: vacuum polarization (left) and light by light scattering (right).}
\label{fig:10}
\end{figure}

The most precise tests of the electroweak theory apply to the QED sector. Here we discuss the anomalous magnetic moments of the electron and of the muon that are among the most precise measurements in the whole of physics. The magnetic moment $\vec \mu$ and the spin $\vec S$ are related by $\vec \mu = - g e\vec S/2m$, where $g$ is the gyromagnetic ratio ($g=2$ for a pointlike Dirac particle). The quantity $a = (g-2)/2$ measures the anomalous magnetic moment of the particle. Recently there have been new precise measurements of $a_e$ and $a_\mu$ for the electron \cite{ae1} and the muon \cite{amu}:
\beq
a_e^{exp} = 11596521807.3(2.8)~^.10^{-13},~~~~~~~~a_\mu^{exp} = 11659208.9(6.3)~^.10^{-10}.
\label{aemuexp}
\eeq
The theoretical calculations in general contain a pure QED part plus the sum of hadronic and weak 
contribution terms:
\beq
a = a^{QED} + a^{hadronic} +a^{weak} = \sum_i C_i (\frac{\alpha}{\pi})^{i}  + a^{hadronic} +a^{weak}.
\label{ath}
\eeq
The QED part has been computed analytically for $i=1,2,3$, while for $i=4$ there is a numerical calculation with an error (see, for example, \cite{kino} and refs therein). The complete numerical evaluation of $i=5$ for the muon case has been published in 2012 \cite{kino12} as a new impressive achievement of Kinoshita and his group. The hadronic contribution is from vacuum polarization insertions and from light by light scattering diagrams (see Fig.~\ref{fig:10}). The weak contribution is from $W$ or $Z$ exchange.
    
For the electron case the weak contribution is essentially negligible and the hadronic term ($a_e^{hadronic} \sim (16.82~\pm~0.19)^.10^{-13}$) does not introduce an important uncertainty.  As a result this measurement can be used to obtain the most precise determination of the fine structure constant \cite{kino12}: 
\beq
\alpha^{-1} \sim 137.0359991657(340)~,
\label{alfaqed}
\eeq

In the muon case the experimental precision is less by about 3 orders of magnitude, but the sensitivity to new physics effects is typically increased by a factor $(m_\mu/m_e)^2 \sim 4^.10^4$ (one mass factor arises because the effective operator needs a chirality flip and the second one is because, by definition, one must factor out the Bohr magneton $e/2m$). From the theory side, the QED term (using the value of $\alpha$ from $a_e$
in Eq.(\ref{alfaqed})), and the weak contribution \cite{amuw} are affected by small errors and are given by (all theory numbers given here are taken from ref.  \cite{kino12})
\beq
a_\mu^{QED} = (116584718.853\pm0.037)^.10^{-11},~~~~~~a_\mu^{weak} = (154\pm2.0)^.10^{-11}
\label{muth1}
\eeq

\begin{figure}
\centerline{\includegraphics[height=3in]{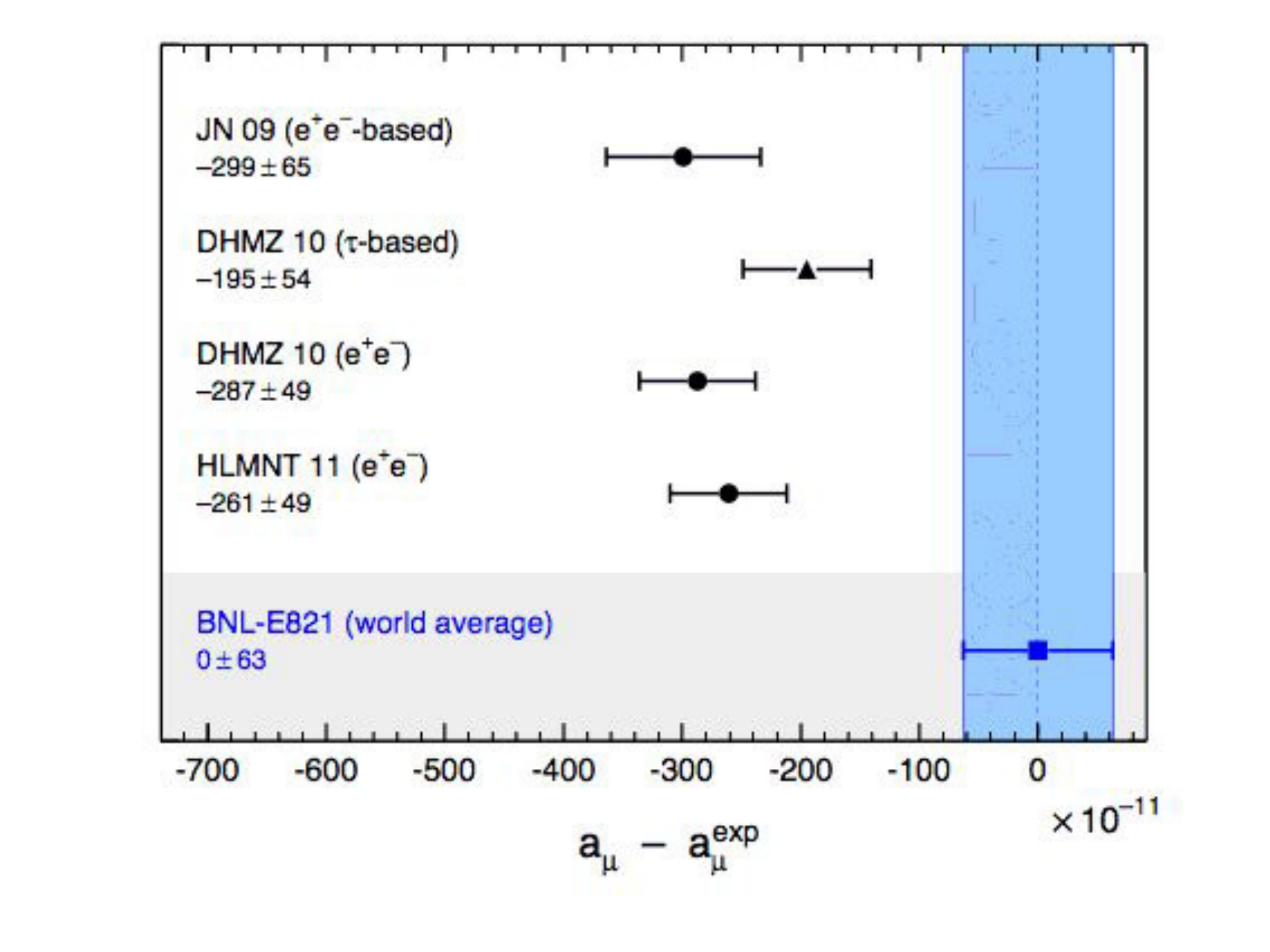}}     
\caption{Compilation of recently published results for $a_\mu$ (in units of $10^{-11}$) \cite{HMpdg}: JN \cite{JN}, DHMZ \cite{vacpol}, HLMNT \cite{hagi}.}
\label{anmu}
\end{figure}

The dominant ambiguities arise from the hadronic term. The lowest order (LO) vacuum polarization contribution can be evaluated from the measured cross sections in $e^+e^-\rightarrow \rm{hadrons}$ at low energy via dispersion relations (the largest contribution is from the $\pi \pi$ final state) \cite{vacpol,hagi}, with the result $a_\mu^{LO.}10^{-11} = 6949\pm43$. The higher order (HO) vacuum polarization contribution (from 2-loop diagrams containing an hadronic insertion) is given by: $a_\mu^{HO.}10^{-11} = -98.4\pm0.7$ \cite{hagi}. The contribution of the light by light (LbL) scattering diagrams is estimated to be: $a_\mu^{LbL.}10^{-11} = 116\pm40$ \cite{lbyl}. Adding the above contributions up the total hadronic result is reported as:
\beq
a_\mu^{hadronic} = (6967\pm 59)^.10^{-11}.
\label{muth2}
\eeq
At face value this would lead to a $2.9 \sigma$ deviation from the experimental value $a_\mu^{exp}$ in Eq.(\ref{aemuexp}):
\beq
a_\mu^{exp} - a_\mu^{th (e^+e^-)}= (249\pm 87)^.10^{-11}.
\label{delamu}
\eeq
For a recent exchange on the significance of the discrepancy see Refs. \cite{JegDav}. However, the error estimate on the LbL term, mainly a theoretical uncertainty, is not compelling, and it could well be somewhat larger (although probably not by as much as to make the discrepancy to completely disappear). A minor puzzle is the fact that, using the conservation of the vector current (CVC) and isospin invariance, which are well established tools at low energy, $a_\mu^{LO}$ can also be evaluated from $\tau$ decays. But the results on the hadronic contribution from $e^+e^-$ and from $\tau$ decay, nominally of comparable accuracy,  are still somewhat apart (although the two are now closer than in the past), and the (g-2) discrepancy would be attenuated if one takes the $\tau$ result (see Fig. \ref{anmu} which refers to the most recent results). Since it is difficult to find a theoretical reason for the $e^+e^-$ vs $\tau$ difference, one must conclude that perhaps there is something which is not understood either in the data or in the assessment of theoretical errors. The prevailing view is to take the $e^+e^-$ determination as the most directly reliable, which leads to Eq.(\ref{delamu}), but some doubts remain. Finally, we note that, given the great accuracy of the $a_\mu$ measurement and the relative importance of the non QED contributions, it is not unreasonable that a first signal of new physics would appear in this quantity.

\subsection{Large Radiative Corrections to Electroweak Processes}
\label{sec:29}
Since the SM theory is renormalizable higher order perturbative corrections can be reliably computed. Radiative corrections
are very important for precision EW tests. The SM inherits all successes of the old V-A theory of charged currents and of
QED. Modern tests have focussed on neutral current processes, the W mass and the measurement of triple gauge vertices.  For Z physics
and the W mass the state of the art computation of radiative corrections include the complete one loop diagrams and selected
dominant multi loop corrections. In addition some resummation techniques are also implemented, like Dyson resummation of vacuum
polarization functions and important renormalization group improvements for large QED and QCD logarithms. We now discuss in
more detail sets of large radiative corrections which are particularly significant (for reviews of radiative corrections for LEP1 physics, see, for example: \cite{radcorr}; for a more pedagogical description of LEP physics see \cite{verz}).

Even leaving aside QCD corrections, a set of important quantitative contributions to the radiative corrections arise from large logarithms [e.g. terms of the
form $(\alpha/\pi ~{\rm ln}~(m_Z/m_{f_ll}))^n$ where $f_{ll}$ is a light fermion]. The sequences of leading and
close-to-leading logarithms are fixed by well-known and consolidated techniques ($\beta$ functions, anomalous
dimensions, penguin-like diagrams, etc.). For example, large logarithms from pure QED effects dominate the running of
$\alpha$ from $m_e$, the electron mass, up to $m_Z$. Similarly large logarithms of the form $[\alpha/\pi~{\rm
ln}~(m_Z/\mu)]^n$ also enter, for example, in the relation between $\sin^2\theta_W$ at the scales $m_Z$ (LEP, SLC) and $\mu$
(e.g. the scale of low-energy neutral-current experiments). Also, large logs from initial state radiation dramatically distort
the line shape of the Z resonance as observed at LEP1 and SLC and this effect was accurately taken into account for the measurement of
the Z mass and total width.
The
experimental accuracy on $m_Z$ obtained at LEP1 is $\delta m_Z = \pm 2.1$~MeV. Similarly, a
measurement of the total width to an accuracy $\delta \Gamma = \pm 2.3$~MeV has been achieved. The prediction of the
Z line-shape in the SM to such an accuracy has posed a formidable challenge to theory, which has been
successfully met. For the inclusive process $e^+e^- \rightarrow f \bar fX$, with $f \not= e$ (for a concise discussion, we leave
Bhabha scattering aside) and $X$ including $\gamma$'s and gluons, the physical cross-section can be written in the form of a
convolution
\cite{radcorr}: 
\begin{equation} \sigma(s) = \int^1_{z_0} dz~\hat \sigma(zs)G(z,s)~,
 \label{130}
\end{equation}  where $\hat \sigma$ is the reduced cross-section, and $G(z,s)$ is the radiator function that describes the
effect of initial-state radiation; $\hat \sigma$ includes the purely weak corrections, the effect of final-state radiation
(of both $\gamma$'s and gluons), and also non-factorizable terms (initial- and final-state radiation interferences, boxes,
etc.) which, being small, can be treated in lowest order and effectively absorbed in a modified $\hat \sigma$. The radiator
$G(z,s)$ has an expansion of the form
\begin{eqnarray} G(z,s) & = &
\delta(1-z) + \alpha /\pi(a_{11}L + a_{10}) + (\alpha/\pi)^2 (a_{22}L^2 + a_{11}L + a_{20})~+~...~+ \nonumber \\
 &+& 
(\alpha/\pi)^n~\sum^n_{i=0} a_{ni}L^i~,
\label{131}
\end{eqnarray} where $L = {\rm ln}~s/m^2_e \simeq 24.2$ for $\sqrt s \simeq m_Z$. All first- and second-order terms are known
exactly. The sequence of leading and next-to-leading logs can be exponentiated (closely following the formalism
of structure functions in QCD). For $m_Z \approx 91$~GeV, the convolution displaces the peak by  +110~MeV, and reduces it by
a factor of about 0.74. The exponentiation is important in that it amounts to an additional shift of about 14~MeV in the peak position with respect to the 1 loop radiative correction.
 
Among the one loop EW radiative corrections, a very remarkable class of contributions are those terms that increase
quadratically with the top mass.  The sensitivity
of radiative corrections to $m_t$ arises from the existence of these terms. The quadratic dependence on
$m_t$ (and on other possible widely broken isospin multiplets from new physics) arises because, in spontaneously broken
gauge theories, heavy virtual particles do not decouple. On the contrary, in QED or QCD, the running of
$\alpha$ and $\alpha_s$ at a scale $Q$ is not affected by heavy quarks with mass
$M \gg Q$. According to an intuitive decoupling theorem \cite{deco}, diagrams with heavy virtual particles of mass $M$ can be
ignored at $Q \ll M$ provided that the couplings do not grow with $M$ and that the theory with no heavy particles is still
renormalizable. In the spontaneously broken EW gauge theories both requirements are violated. First, one important difference
with respect to unbroken gauge theories is in the longitudinal modes of weak gauge bosons. These modes are generated by the
Higgs mechanism, and their couplings grow with masses (as is also the case for the physical Higgs couplings). Second, the
theory without the top quark is no more renormalizable because the gauge symmetry is broken as the (t,b) doublet  would
not be complete (also the chiral anomaly would not be completely cancelled). With the observed value of
$m_t$ the quantitative importance of the terms of order $G_Fm^2_t/4\pi^2\sqrt{2}$ is substantial but not dominant (they are
enhanced by a factor $m^2_t/m^2_W\sim 5$ with respect to ordinary terms). Both the large logarithms and the
$G_Fm^2_t$ terms have a simple structure and are to a large extent universal, i.e. common to a wide class of processes. In
particular the $G_Fm^2_t$ terms appear in vacuum polarization diagrams which are universal (virtual loops inserted in gauge boson internal lines are independent of the nature of the vertices on each side of the propagator) and in the $Z\rightarrow b \bar b$
vertex which is not. This vertex is specifically sensitive to  the top quark which, being the partner of the b quark in a doublet, runs in the loop. Instead all types of heavy particles
could  in principle  contribute to vacuum polarization diagrams. The study of universal vacuum polarization contributions, also called "oblique" corrections, and of top enhanced terms is important for an understanding of the
pattern of radiative corrections. 
More in general, the important consequence of non
decoupling is that precision tests of the electroweak theory may apriori be sensitive to new physics even if the new particles are
too heavy for their direct production, but aposteriori no signal of deviation has clearly emerged.

While radiative corrections are quite sensitive to the top mass, they are unfortunately much less dependent on the Higgs
mass. In fact, the dependence of one
loop diagrams on
$m_H$ is only logarithmic:
$\sim G_Fm^2_W
\log(m^2_H/m^2_W)$. Quadratic terms $\sim G^2_Fm^2_H$ only appear at two loops \cite{bij} and are too small to be detectable. The
difference with the top case is that the splitting $m^2_t-m^2_b$ is a direct breaking of the gauge symmetry that
already affects the 1- loop corrections, while the Higgs couplings are "custodial" SU(2) symmetric in lowest order.

\subsection{Electroweak Precision Tests}
\label{sec:30}
\begin{figure}[t]
\centerline{\includegraphics[height=5in]{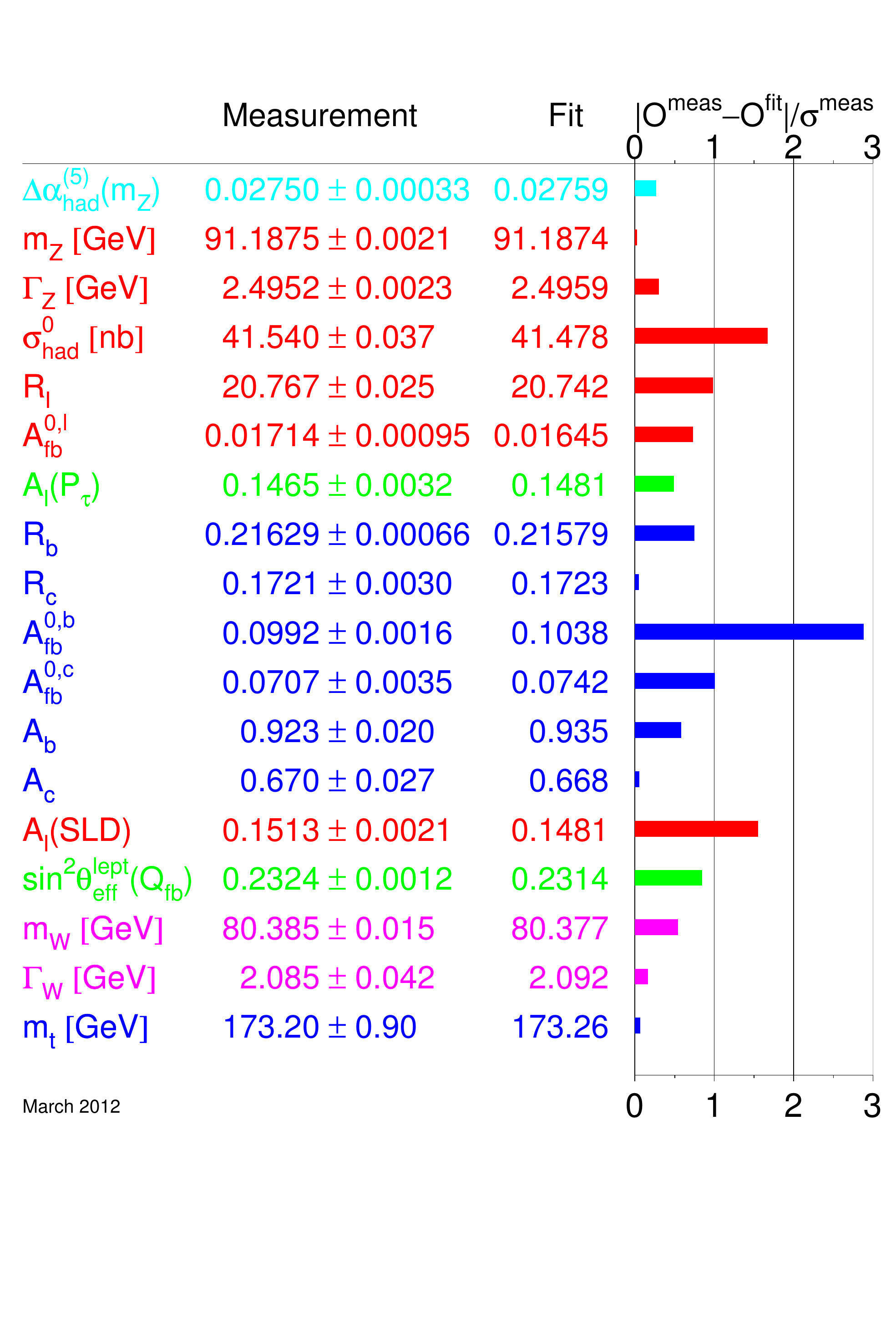}}     
\caption{Summary of electroweak precision
measurements at high $Q^2$~\cite{ew}. The first block shows
the Z-pole measurements.  The second block shows additional results
from other experiments: the mass and the width of the W boson measured
at the Tevatron and at LEP-2, the mass of the top quark measured at
the Tevatron, and the contribution to $\alpha$ of the hadronic
vacuum polarization. The SM
fit results are also shown with the corresponding pulls (differences data - fits in units of standard deviations).}
\label{w12_pull}
\end{figure}

For the analysis of electroweak data in the SM one starts from the
input parameters: as is the case in any renormalizable theory, masses and couplings
have to be specified from outside. One can trade one parameter for
another and this freedom is used to select the best measured ones as
input parameters. Some of them, $\alpha$, $G_F$ and
$m_Z$, are very precisely known, as we have seen, some other ones, $m_{f_{light}}$,
$m_t$ and $\alpha_s(m_Z)$ are less well determined while $m_H$, before the LHC, was
largely unknown. In this Section we discuss the EW fit without the new input on $m_H$ from the LHC, in order to compare the limits so derived on  $m_H$ with the LHC data. The discussion of the LHC results will follow in the next Sections. Among the light fermions, the quark masses are badly known, but
fortunately, for the calculation of radiative corrections, they can be
replaced by $\alpha(m_Z)$, the value of the QED running coupling at
the Z mass scale. The value of the hadronic contribution to the
running, embodied in the value of  $\Delta \alpha^{(5)}_{had}(m_Z^2)$ (see Fig.~\ref{w12_pull},~\cite{ew} ) is obtained
through dispersion relations from the data on $e^+e^-\rightarrow \rm{hadrons}$ at
moderate centre-of-mass energies. From the input parameters
one computes the radiative corrections to a sufficient precision to
match the experimental accuracy. Then one compares the theoretical
predictions with the data for the numerous observables which have been
measured \cite{prlep}, checks the consistency of the theory and derives constraints
on $m_t$, $\alpha_s(m_Z)$ and $m_H$. 

The basic tree level relations:
\beq
\frac{g^2}{8m^2_W}=\frac{G_F}{\sqrt{2}},~~~~~~g^2\sin^2\theta_W=e^2=4\pi\alpha\label{bb1}
\eeq
can be combined into
\beq
\sin^2\theta_W=\frac{\pi\alpha}{\sqrt{2}G_Fm^2_W}\label{bb2}
\eeq
Always at tree level, a different definition of $\sin^2\theta_W$ is from the gauge boson masses:
\beq
\frac{m^2_W}{m^2_Z\cos^2\theta_W}=\rho_0=1~~~\Longrightarrow~~~\sin^2\theta_W=1-\frac{m^2_W}{m^2_Z}\label{bb3}
\eeq
where $\rho_0=1$ assuming that there are only Higgs doublets. The last two relations can be put into the convenient form
\beq
(1-\frac{m^2_W}{m^2_Z})\frac{m^2_W}{m^2_Z}=\frac{\pi\alpha}{\sqrt{2}G_Fm^2_Z}\label{bb4}
\eeq
Beyond tree level, these relations are modified by radiative corrections:
\bea
(1-\frac{m^2_W}{m^2_Z})\frac{m^2_W}{m^2_Z}&=&\frac{\pi\alpha(m_Z)}{\sqrt{2}G_Fm^2_Z}\frac{1}{1-\Delta r_W}\nonumber\\
\frac{m^2_W}{m^2_Z\cos^2\theta_W}&=&1+\Delta \rho_m\label{bb5}
\eea
The Z and W masses are to be precisely defined, for example, in terms of the pole position in the respective propagators. Then, in the first relation the replacement of $\alpha$ with the running coupling at the Z mass $\alpha(m_Z)$ makes $\Delta r_W$
completely determined at 1-loop by purely weak corrections ($G_F$ is protected from logarithmic running as an indirect consequence of (V-A) current conservation in the massless theory). This relation defines $\Delta r_W$ unambiguously, once the meaning of $m_{W,Z}$ and of 
$\alpha(m_Z)$ is specified (for example, $\bar M\bar S$). On the contrary, in the second relation $\Delta \rho_m$ depends on the definition of
$\sin^2\theta_W$ beyond the tree level. For LEP physics $\sin^2\theta_W$ is usually defined from the
$Z\rightarrow\mu^+\mu^-$ effective vertex. At the tree level the vector and axial-vector couplings $g_V^\mu$ and $g_A^\mu$ are given in Eqs.(\ref{35b}). Beyond the tree level a corrected vertex can be written down
in terms of modified effective couplings. Then $\sin^2\theta_W\equiv\sin^2\theta_{eff}$
is in general defined through the muon vertex:
\bea
g^\mu_V/g^\mu_A&=&1-4\sin^2\theta_{eff}\nonumber\\
\sin^2\theta_{eff}&=&(1+\Delta k)s^2_0,~~~~~~~s^2_0 c^2_0=\frac{\pi\alpha(m_Z)}{\sqrt{2}G_Fm^2_Z}\nonumber\\
g^{\mu2}_A&=&\frac{1}{4}(1+\Delta\rho)\label{bb7}
\eea
We see that $s_0^2$ and $c_0^2$ are "improved" Born approximations (by including the running of $\alpha$) for $\sin^2\theta_{eff}$ and $\cos^2\theta_{eff}$. Actually, since in the SM lepton universality is only broken by masses and is in agreement with experiment within the
present accuracy, in practice the muon channel can be replaced with the average over charged leptons. 

We can write a symbolic equation that summarizes the status of what has been computed up to now for the radiative corrections (we list some recent work on each item from where older references can be retrieved)
$\Delta r_W$ \cite{rW}, $\Delta \rho$ \cite{ro} and $\Delta k$ \cite{kap}:
\beq
\Delta r_W, \Delta \rho, \Delta k=g^2(1+\alpha_s) +g^2 \frac{m^2_t}{m^2_W}(\alpha^2_s+\alpha^3_s) + g^4 + g^4
\frac{m^4_t}{m^4_W}\alpha_s + g^6
\frac{m^6_t}{m^6_W}...\label{bb8}
\eeq
The meaning of this relation is that the one loop terms of order $g^2$ are completely known, together with their first 
order QCD corrections; the second and third order QCD corrections are only known for the $g^2$ terms enhanced by
$m^2_t/m^2_W$; the two loop terms of order $g^4$ are completely known, while, only for $\Delta \rho$, the terms $g^4\alpha_s$ enhanced by the ratio $m^4_t/m^4_W$ and the terms $g^6
\frac{m^6_t}{m^6_W}$ are also computed.

In the SM the quantities $\Delta r_W$, $\Delta \rho$, $\Delta k$, for
sufficiently  large $m_t$, are all dominated by quadratic terms in $m_t$ of order $G_Fm^2_t$.   The quantity $\Delta \rho_m$ is not independent and can expressed in terms of them. As new physics
can  more easily be disentangled if not masked by large conventional $m_t$ effects, it is convenient to keep
$\Delta \rho$ while trading $\Delta r_W$
and 
$\Delta k$ for two quantities with no contributions of order $G_Fm^2_t$. One thus introduces the
following linear combinations (epsilon parameters) \cite{eps}:
\bea
\epsilon_1~&=&~\Delta \rho, \nonumber \\
\epsilon_2~&=&~c^2_0 \Delta \rho~+~\frac{s^2_0 \Delta r_W}{c^2_0-s^2_0}~-~2s^2_0 \Delta k, \nonumber\\
\epsilon_3~&=&~c^2_0 \Delta \rho~+~(c^2_0-s^2_0) \Delta k. \label{8n}
\eea
The quantities $\epsilon_2$ and $\epsilon_3$ no longer contain terms of order $G_Fm^2_t$ but only logarithmic
terms in $m_t$. The leading terms for large Higgs mass, which are logarithmic, are contained in
$\epsilon_1$ and $\epsilon_3$. To complete the set of top-enhanced radiative corrections one adds $\epsilon_b$ defined from the loop corrections to the $Zb \bar b$ vertex. One modifies $g_V^b$ and $g_A^b$ as follows:
\bea
g_A^b~=~-\frac{1}{2}(1+\frac{\Delta \rho}{2})(1+\epsilon_b), \nonumber\\
\frac{g_V^b}{g_A^b}~=~\frac{1-4/3\sin^2\theta_{eff}+\epsilon_b}{1+\epsilon_b}.
\label{7n}
\eea
$\epsilon_b$ can be measured from $R_b=\Gamma(Z\rightarrow b \bar b)/\Gamma(Z\rightarrow {\rm hadrons})$ (see Fig.~\ref{w12_pull}). This is clearly not the most general deviation from the SM in the $Z\rightarrow b \bar b$ vertex but $\epsilon_b$ is
the quantity where the large
$m_t$ corrections are located in the SM.
Thus, summarizing, in the SM one has the following "large"
asymptotic contributions:
\bea
\epsilon_1~&=&~\frac{3G_F m_t^2}{8 \pi^2 \sqrt{2}}~-~\frac{3G_F m_W^2}{4 \pi^2 \sqrt{2}} \tan^2{\theta_W}
\ln{\frac{m_H}{m_Z}}~+....,\nonumber \\
\epsilon_2~&=&~-\frac{G_F m_W^2}{2 \pi^2 \sqrt{2}}\ln{\frac{m_t}{m_Z}}~+....,\nonumber \\
\epsilon_ 3~&=&~\frac{G_F m_W^2}{12 \pi^2 \sqrt{2}}\ln{\frac{m_H}{m_Z}}~-~\frac{G_F m_W^2}{6 \pi^2
\sqrt{2}}\ln{\frac{m_t}{m_Z}}....,\nonumber \\
\epsilon_b~&=&~-\frac{G_F m_t^2}{4 \pi^2 \sqrt{2}}~+.... \label{9n}
\eea

The $\epsilon_i$ parameters vanish in the limit where only tree level SM effects are kept plus pure QED and/or QCD corrections. So they describe the effects of quantum corrections (i.e. loops) from weak interactions. A similar set of parameters are the S, T, U parameters \cite{pt}: the shifts induced by new physics on S, T and U are proportional to those induced on $\epsilon_3$, $\epsilon_1$ and $\epsilon_2$, respectively. In principle, with no model dependence, one can measure the four $\epsilon_i$ from the basic observables of LEP physics $\Gamma(Z\rightarrow\mu^+\mu^-)$, $A_{FB}^\mu$  and $R_b$ on the Z peak plus $m_W$. With increasing model dependence, one can include other measurements in the fit for the $\epsilon_i$. For example, use lepton universality to average the $\mu$ with the $e$ and $\tau$ final states, or include all lepton asymmetries and so on. The present experimental values of the $\epsilon_i$, obtained from a fit of all LEP1-SLD measurements plus $m_W$, are given by \cite{ciuch}:
\bea
\epsilon_1~^.10^3&=&5.6\pm1.0,~~~~~\epsilon_2~^.10^3=~ -7.8\pm0.9,\nonumber \\
\epsilon_3~^.10^3&=&5.6\pm0.9,~~~~~\epsilon_b~^.10^3=~ -5.8\pm1.3.
\label{epsexp}
\eea
Note that the $\epsilon$ parameters are of order a few in $10^{-3}$ and are known with an accuracy in the range $15-30\%$. These values are in agreement with the predictions of the SM with a 126 GeV Higgs \cite{ciuch}: 
\bea
\epsilon_1^{SM}~^.10^3&=&5.21\pm0.08,~~~~~\epsilon_2^{SM}~^.10^3=~ -7.37\pm0.03,\nonumber \\
\epsilon_3^{SM}~^.10^3&=&5.279\pm0.004,~~~~~\epsilon_b^{SM}~^.10^3=~ -6.94\pm0.15.
\label{epssm}
\eea
All models of new physics must be compared with these findings and pass this difficult test.

\subsection{Results of the SM Analysis of Precision Tests}
\label{sec:31}


The electroweak Z pole measurements, combining the results of all the
experiments, plus the $W$ mass and width and the top mass $m_t$, are summarised in Fig~\ref{w12_pull}, as of March 2012 \cite{ew}. The primary rates are given by the pole cross sections for the various
final states $\sigma^0$; ratios thereof correspond to ratios of
partial decay widths:
\begin{eqnarray}
\sigma^0_h & = & \frac{12\pi}{m_Z^2}\ \frac{\Gamma_{ee}\Gamma_h}{\Gamma_Z^2}\,, \qquad
R^0_l ~ = ~ \frac{\sigma^0_h}{\sigma^0_l} ~ = ~ \frac{\Gamma_h}{\Gamma_{ll}}\,, \qquad
R^0_q ~ = ~ \frac{\Gamma_{q\bar q}}{\Gamma_h} \,.
\end{eqnarray}
Here $\Gamma_{ll}$ is the partial decay width for a pair of massless charged
leptons.  The partial decay width for a given fermion species contains
information about the effective vector and axial-vector coupling
constants of the neutral weak current:
\begin{eqnarray}
\Gamma_{ff} & = & N_C^f \frac{G_Fm_Z^3}{6\sqrt{2}\pi} 
\left( g_{af}^2 C_{\mathrm{Af}} +g_{vf}^2 C_{\mathrm{Vf}} \right) 
          +  \Delta_{\rm ew/QCD}\,,
\end{eqnarray}
where $N_C^f$ is the QCD colour factor, $C_{\mathrm{\{A,V\}f}}$ are
final-state QCD/QED correction factors also absorbing imaginary
contributions to the effective coupling constants, $g_{af}$ and $g_{vf}$
are the real parts of the effective couplings, and $\Delta$ contains
non-factorisable mixed corrections.

Besides total cross sections, various types of asymmetries have been
measured.  The results of all asymmetry measurements are quoted in
terms of the asymmetry parameter $A_f$, defined in terms of the real
parts of the effective coupling constants, $g_{af}$ and $g_{vf}$, as:
\begin{eqnarray}
A_f & = & 2\frac{g_{vf}g_{af}}{g_{vf}^2+g_{af}^2} ~ = ~ 
           2\frac{g_{vf}/g_{af}}{1+(g_{vf}/g_{af})^2}\,, \qquad
A_{FB}^{0,f}~ = ~ \frac{3}{4}A_eA_f\,.
\end{eqnarray}
The measurements are: the forward-backward asymmetry ($A_{FB}^{0,f}$), the tau polarization ($A_\tau$) and its forward
backward asymmetry ($A_e$) measured at LEP, as well as the left-right
and left-right forward-backward asymmetry measured at SLC ($A_e$ and
$A_f$, respectively).  Hence the set of partial width and asymmetry
results allows the extraction of the effective coupling constants. 

\begin{figure}[h]
\centerline{\includegraphics[height=3in]{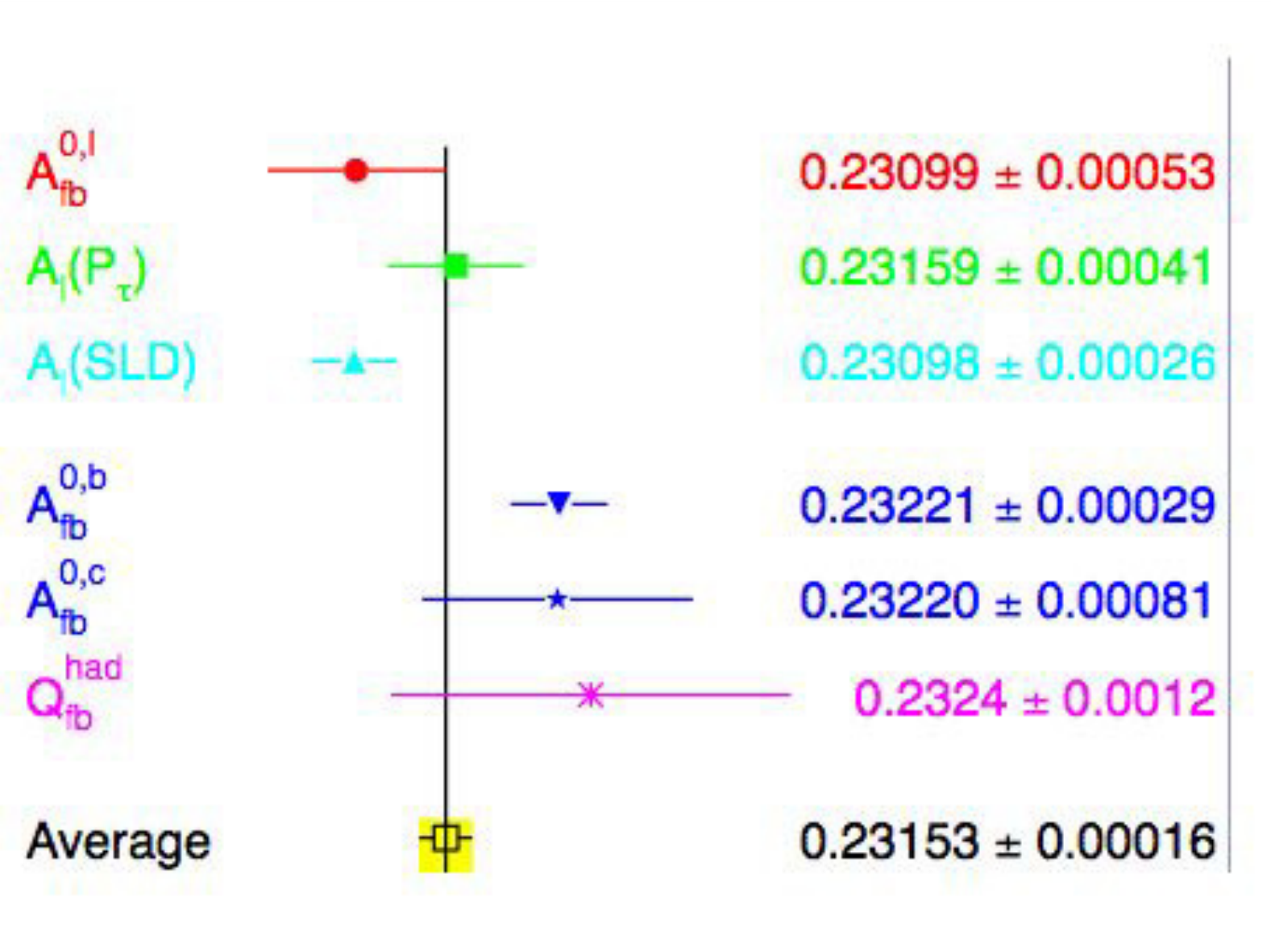}}    
\caption{Summary of $\sin^2\theta_{eff}$ precision
measurements at high $Q^2$~\cite{ew}. }
\label{sineff}
\end{figure}

The various asymmetries determine the effective electroweak mixing
angle for leptons with highest sensitivity (see Fig. \ref{sineff}). The weighted
average of these results, including small correlations, is:
\beq
\sin^2\theta_{eff}=  0.23153\pm0.00016 ,
\label{sin2teff}
\eeq
Note, however, that this average has a $\chi^2$ of 11.8 for 5 degrees
of freedom, corresponding to a probability of a few \%. The $\chi^2$ is
pushed up by the two most precise measurements of $\sin^2\theta_{eff}$, namely
those derived from the measurements of $A_l$ by SLD, dominated by the
left-right asymmetry $A_{LR}^0$, and of the forward-backward asymmetry
measured in $b \bar b$ production at LEP, $A_{FB}^{0,b}$, which differ by about
$3 \sigma$s.

We now expand the discussion on the  SM fit of the data. One can think of different
types of fit, depending on which experimental results are included or
which answers one wants to obtain. For example, in
Table~\ref{tab:ew2} we present in column~1 a fit of all Z pole
data plus $m_W$ and $\Gamma_W$ (this is interesting as it shows the value
of $m_t$ obtained indirectly from radiative corrections, to be
compared with the value of $m_t$ measured in production experiments),
in column~2 a fit of all Z pole data plus $m_t$ (here it is $m_W$
which is indirectly determined), and, finally, in column~3 a fit of
all the data listed in Fig.~\ref{w12_pull} (which is the most
relevant fit for constraining $m_H$).  From the fit in column~1  we see that the extracted value of $m_t$ is
in good agreement with the direct measurement (see Fig~\ref{w12_pull}).  Similarly we see that the experimental
measurement of $m_W$ is larger by about
one standard deviation with respect to the value from the fit in
column~2.  We have seen that quantum corrections depend only
logarithmically on $m_H$.  In spite of this small sensitivity, the
measurements are precise enough that one still obtains a quantitative
indication of the mass range. From the fit in column~3 we obtain:
$\log_{10}{m_H(\rm{GeV})}=1.97\pm 0.12$ (or $m_H=94^{+29}_{-24}~\rm{GeV}$).
This result on the Higgs mass is particularly remarkable. The value of
$\log_{10}{m_H(\rm{GeV})}$ is compatible with the small window between
$\sim 2$ and $\sim 3$ which is allowed, on the one side, by the direct
search limit ($m_H > 114~\rm{GeV}$ from LEP-2~\cite{ew}),
and, on the other side, by the theoretical upper limit on the Higgs
mass in the minimal SM, $m_H\lesssim 600-800~\rm{GeV}$ \cite{hr} to be discussed in Sect. \ref{sec:32}.

\begin{table}[h!]
\begin{center}
\begin{tabular}{|c|c|c|c|}
  \hline
  &&\\[-3mm]
Fit       & 1 & 2 & 3 \\
\hline
Measurements      &$m_W,~\Gamma_W$         &$m_t$            &$m_t,~m_W,~\Gamma_W$\\
\hline
$m_t~(\rm{GeV})$      &$178.1^{+10.9}_{-7.8}$&$173.2\pm0.9$    &$173.26\pm0.89$\\
$m_H~(\rm{GeV})$      &$148^{+237}_{-81}$    &$122^{+59}_{-41}$&$94^{+29}_{-24}$\\
$\log~[m_H(\rm{GeV})]$&$2.17\pm{+0.38}$&$2.09\pm0.17$    &$1.97\pm0.12$ \\
$\alpha_s(m_Z)$   &$0.1190\pm0.0028$     &$0.1191\pm0.0027$&$0.1185\pm0.0026$\\
\hline
$m_W~(\rm{MeV})$      &$80381 \pm 13$    &$80363 \pm 20$   &$80377 \pm 12$  \\
\hline
  \end{tabular}
\caption{\label{tab:ew2} Standard Model fits of electroweak data \cite{ew}. All fits use the
Z pole results and $\Delta \alpha^{(5)}_{had}(m_Z^2)$ as listed in Fig.~\ref{w12_pull}. In
addition, the measurements listed on top of each column are included in that case. The fitted W mass is also shown \cite{ew} (the directly measured value is
$m_W=80385\pm 15~\rm{MeV} $).}
\end{center}
\end{table}

Thus the whole picture of a perturbative theory with a fundamental
Higgs is well supported by the data on radiative corrections. It is
important that there is a clear indication for a particularly light
Higgs: at $95\%$ c.l. $m_H\lesssim 152~\rm{GeV}$ (which becomes $m_H\lesssim 171~\rm{GeV}$ including the input from the LEP2 direct search result).  This was quite
encouraging for the LHC search for the Higgs particle.  More in
general, if the Higgs couplings are removed from the Lagrangian the
resulting theory is non renormalizable. A cutoff $\Lambda$ must be
introduced. In the quantum corrections $\log{m_H}$ is then replaced by
$\log{\Lambda}$ plus a constant. The precise determination of the
associated finite terms would be lost (that is, the value of the mass
in the denominator in the argument of the logarithm).  A heavy Higgs
would need some unfortunate accident: the finite terms, different in
the new theory from those of the SM, should by chance compensate
for the heavy Higgs in a few key parameters of the radiative
corrections (mainly $\epsilon_1$ and $\epsilon_3$, see, for example,
\cite{eps}).  Alternatively, additional new physics, for example in
the form of effective contact terms added to the minimal SM
lagrangian, should accidentally do the compensation, which again needs
some sort of conspiracy.

To the list of precision tests of the SM one should add the results on low energy
tests obtained from neutrino and antineutrino deep
inelastic scattering (NuTeV~\cite{NuTeV}), parity violation
in Cs atoms (APV~\cite{QWC}) and the recent measurement of
the parity-violating asymmetry in Moller scattering \cite{E158}.  When these
experimental results are compared with the SM predictions the agreement is good
except for the NuTeV result that shows a deviation by three standard
deviations.  The NuTeV measurement is quoted as a measurement of
$\sin^2\theta_W=1-m_W^2/m_Z^2$ from the ratio of neutral to charged current
deep inelastic cross-sections from $\nu_{\mu}$ and $\bar{\nu}_{\mu}$
using the Fermilab beams. But it has been argued and it is now generally accepted that the NuTeV
anomaly probably simply arises from an underestimation of the theoretical
uncertainty in the QCD analysis needed to extract $\sin^2\theta_W$.  In fact,
the lowest order QCD parton formalism on which the analysis has been
based is too crude to match the experimental accuracy.  

\begin{figure}
\centerline{\includegraphics[height=4in]{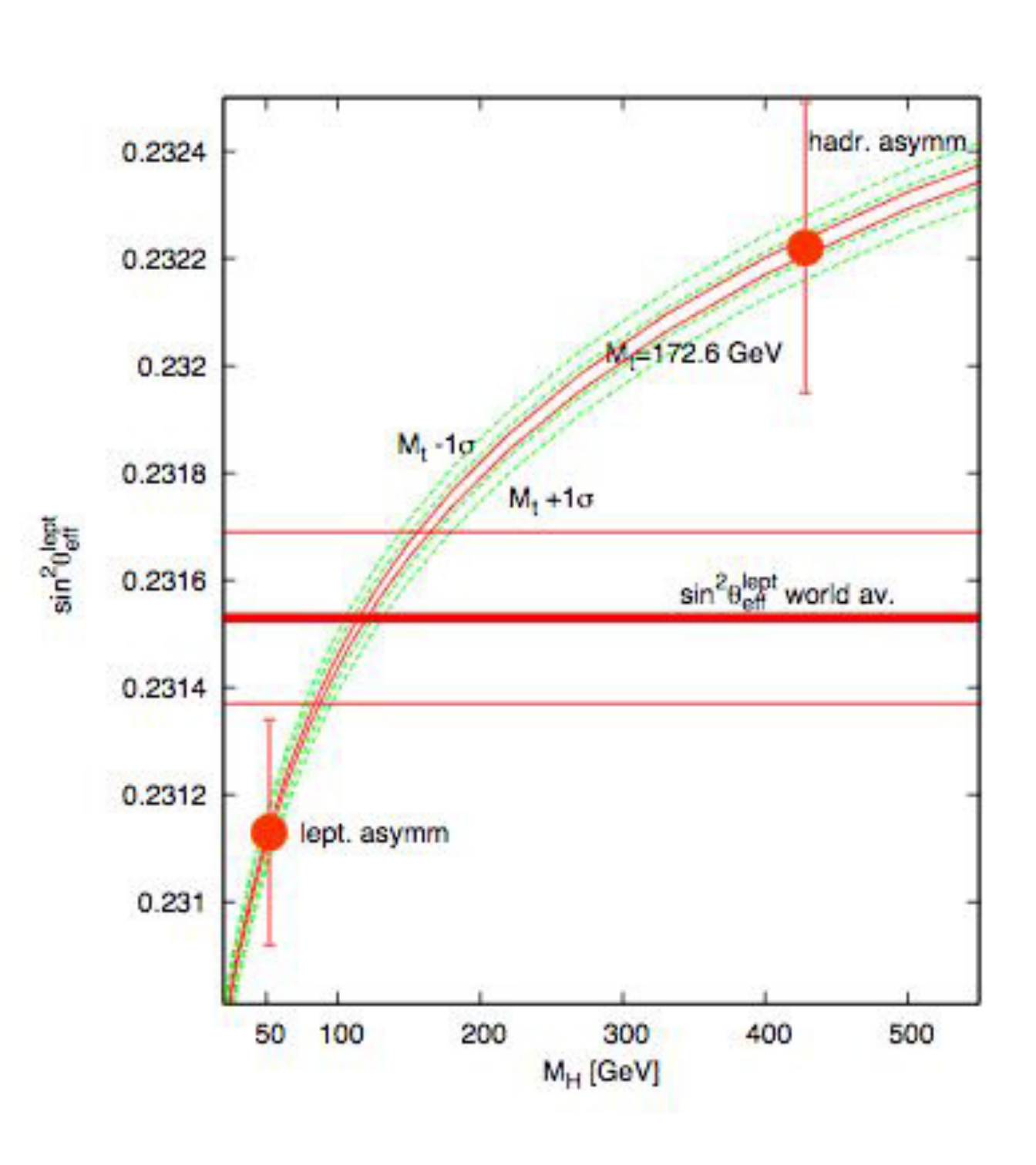}}     
\caption{The data for $\sin^2\theta_{\rm eff}^{\rm lept}$ are
plotted vs $m_H$. The theoretical prediction for the measured value of $m_t$ is also shown. For presentation purposes the measured points are
shown each at the $m_H$ value that would ideally correspond to it,
given the central value of $m_t$ (updated from \cite{P-Gambino})}
\label{fig:11}  
\end{figure}

\vspace*{12pt}
When confronted with these results, on the whole the SM performs
rather well, so that it is fair to say that no clear indication for
new physics emerges from the data. However, as already mentioned, one
problem is that the two most precise measurements of $\sin^2\theta_{\rm eff}$ from
$A_{LR}$ and $A_{FB}^b$ differ by about $3\sigma$'s.  In
general, there appears to be a discrepancy between $\sin^2\theta_{\rm eff}$
measured from leptonic asymmetries ($(\sin^2\theta_{\rm eff})_l$) and
from hadronic asymmetries ($(\sin^2\theta_{\rm eff})_h$). In fact, the result from $A_{LR}$ is in good
agreement with the leptonic asymmetries measured at LEP, while all
hadronic asymmetries, though their errors are large, are better
compatible with the result of $A_{FB}^b$.
These two results for $\sin^2\theta_{\rm eff}$ are shown in Fig.~\ref{fig:11} \cite{P-Gambino}. Each of them is plotted at the $m_H$
value that would correspond to it given the central value of $m_t$.
Of course, the value for $m_H$ indicated by each $\sin^2\theta_{\rm eff}$ has an
horizontal ambiguity determined by the measurement error and the width
of the $\pm1\sigma$ band for $m_t$.  Even taking this spread into
account it is clear that the implications on $m_H$ are sizably
different.  One might imagine that some new physics effect could be
hidden in the $\mathrm{Z b \bar b}$ vertex.  Like for the top quark
mass there could be other non decoupling effects from new heavy states
or a mixing of the b quark with some other heavy quark.  However, it
is well known that this discrepancy is not easily explained in terms
of some new physics effect in the $\mathrm{Z b \bar b}$ vertex. A rather large
change with respect to the SM of the b-quark right handed coupling to the Z  is needed in
order to reproduce the measured discrepancy (precisely a $\sim 30\%$
change in the right-handed coupling), an effect too large to be a loop
effect but which could be produced at the tree level, e.g., by mixing
of the b quark with a new heavy vectorlike quark \cite{CTW}), or some mixing of the Z with ad hoc heavy states \cite{DMR}.  But
then this effect should normally also appear in the direct measurement
of $A_b$ performed at SLD using the left-right polarized b asymmetry,
even within the moderate precision of this result.  The measurements of neither
$A_b$ at SLD nor $R_b$ confirm the need of such a large effect (recently a numerical calculation of NLO corrections to $R_b$ \cite{freit} at first appeared to indicate a rather large result but finally the full correction turned out to be rather small).  Alternatively, the observed
discrepancy could be simply due to a large statistical fluctuation or an
unknown experimental problem. As a consequence of this problem, the ambiguity in the measured value of
$\sin^2\theta_{\rm eff}$ is in practice larger than the nominal error, reported in
Eq.~\ref{sin2teff}, obtained from averaging all the existing
determinations, and the interpretation of precision tests is less sharp than it would otherwise be.

\begin{figure}
\centerline{\includegraphics[height=4in]{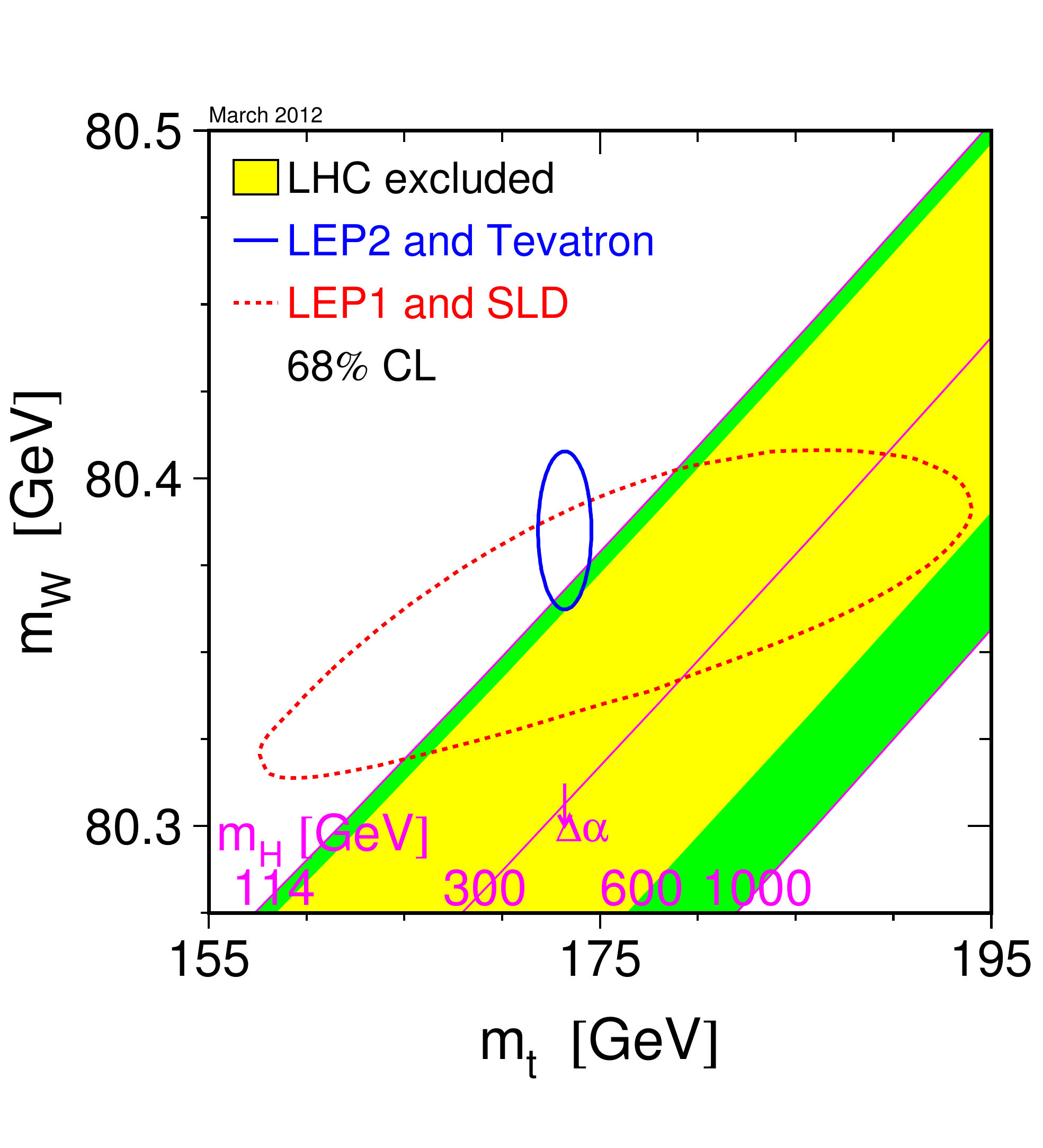}}     
\caption{The data for $m_W$ are
plotted vs $m_t$ \cite{ew}.}
\label{mWmt}  
\end{figure}

We have already observed that the experimental value of $m_W$ (with
good agreement between LEP and the Tevatron) is a bit high compared to
the SM prediction (see Fig.~\ref{mWmt}). The value of $m_H$
indicated by $m_W$ is on the low side, just in the same interval as
for $\sin^2\theta_{\rm eff}^{\rm lept}$ measured from leptonic
asymmetries. 

\vspace*{12pt}

In conclusion, the experimental information on the Higgs sector, obtained from EW precision tests at LEP-1,2 and the Tevatron can be summarized as follows. First, the relation $M_W^2=M_Z^2\cos^2{\theta_W}$, Eq.(\ref{60}), modified by small, computable
radiative corrections, has been
experimentally proven. This relation means that the effective Higgs
(be it fundamental or composite) is indeed a weak isospin doublet.
The direct lower limit $m_H \gtrsim 114.5$~GeV (at $95\%$ c.l.) was
obtained from  searches at LEP-2.  The radiative corrections
computed in the SM when compared to the data on precision EW
tests lead to a clear indication for a light Higgs, not too far from
the direct LEP-2 lower bound. The upper limit for $m_H$ in the SM from the EW tests depends on the value of the top quark mass $m_t$.  The CDF and D0 combined value after Run II is at present \cite{ew} $m_t= 173.2\pm0.9~GeV$. As a consequence the limit on $m_H$ from the LEP and Tevatron measurements is rather stringent \cite{ew}: $m_H < 171~GeV$ (at $95\%$ c.l., after including the information from the 114.5 GeV direct bound). 
 
\subsection{The Search for the SM Higgs}
\label{sec:32}
The Higgs problem is really central in particle physics today. On the one hand, the experimental verification of the Standard Model (SM) cannot be considered complete until the structure of the  Higgs sector is not established by experiment. On the other hand, the Higgs is also related to most of the major problems of particle physics, like the flavour problem and the hierarchy problem, the latter strongly suggesting the need for new physics near the weak scale (that so far was not found). In turn the discovery of new physics could clarify  the dark matter identity. It was already clear before the LHC that some sort of Higgs mechanism is at work. The W or the Z with longitudinal polarization that we observe are not present in an unbroken gauge theory (massless spin-1 particles, like the photon, are transversely polarized): the longitudinal degrees of freedom for the W or the Z are borrowed from the Higgs sector and are an evidence for it. Also, at LEP it has been precisely
established that the gauge symmetry is unbroken in the vertices of the
theory: all currents and charges are indeed symmetric. Yet there is obvious
evidence that the symmetry is instead badly broken in the
masses. Not only the W and the Z have large masses, but the large splitting of, for example,  the t-b doublet shows that even a global weak SU(2) is not at all respected by the fermion spectrum. This is a clear signal of spontaneous
symmetry breaking and the implementation of spontaneous
symmetry breaking in a gauge theory is via the Higgs mechanism. The big questions are about
the nature and the properties of the Higgs particle(s). The search for the Higgs boson and for possible new physics that could accompany it has been the main goal of the LHC from the start. On the Higgs the LHC should answer the following questions: do some Higgs particles exist? Which ones: a single doublet, more doublets, additional singlets? SM Higgs or SUSY Higgses? Fundamental or composite (of fermions, of WW...)? Pseudo-Goldstone boson of an enlarged symmetry? A manifestation of large extra dimensions (5th component of a gauge boson, an effect of orbifolding or of boundary conditions...)? Or some combination of the above or something so far unthought of? By now we have a candidate Higgs boson that really looks like the simplest realization of the Higgs mechanism, as described by the minimal SM Higgs. In the following we first consider the apriori expectations for the Higgs sector and then the profile of the Higgs candidate discovered at the LHC.

\subsection{Theoretical Bounds on the SM Higgs Mass}

A strong argument indicating that the solution of the Higgs problem could not be too far away (that is, either discovering the Higgs or finding the new physics that complicates the picture)  is the fact that, in the absence of a Higgs particle or of an alternative mechanism, violations of unitarity appear in some scattering amplitudes at energies in the few TeV range \cite{unit}.  In particular, amplitudes involving longitudinal gauge bosons (those most directly related to the Higgs sector) are affected. For example, at tree level in the absence of Higgs exchange, for $s>>m_Z^2$ one obtains:
\beq
A(W^+_LW^-_L \rightarrow Z_L Z_L)_{no~Higgs} \sim  i \frac{s}{v^2}
\label{viol1}
\eeq
In the SM this unacceptable large energy behaviour is quenched by the Higgs exchange diagram contribution:
\beq
A(W^+_LW^-_L \rightarrow Z_L Z_L)_{Higgs} \sim  -i \frac{s^2}{v^2(s-m_H^2)}
\label{viol2}
\eeq
Thus the total result in the SM is:
\beq
A(W^+_LW^-_L \rightarrow Z_L Z_L)_{SM} \sim  -i \frac{s m_H^2}{v^2(s-m_H^2)}
\label{viol3}
\eeq
which at large energies saturates at a constant value. To be compatible with unitarity bounds one needs $m_H^2 < ~4\pi \sqrt{2} /G_F$ or $m_H < ~1.5 ~TeV$. This is an important theorem that guarantees that either the Higgs boson(s) or new physics or both must be present in the few TeV energy range.

It is well known that, as described in \cite{hri} and references therein, in the SM with only one Higgs doublet an upper bound on $m_H$ (with mild dependence on $m_t$ and the QCD coupling $\alpha_s$) is obtained from the requirement that the perturbative description of the theory remains valid up to  a large energy scale $\Lambda$ where the SM model breaks down and new physics appears.  Similarly a lower limit on
$m_H$ can be derived from the requirement of vacuum stability \cite{zzi}, \cite{zzii}, \cite{aaiiii} (or, in milder form, of a moderate instability, compatible with the lifetime of the Universe  \cite{isi,degr}). The Higgs mass enters because it fixes the initial value of the quartic Higgs coupling $\lambda$ in its running up to the large scale $\Lambda$. We now briefly recall the derivation of these limits.

The upper limit on
the Higgs mass in the SM is clearly important for an apriori assessment of the chances of success for the LHC as an accelerator designed
to solve the Higgs problem. One way to estimate the upper limit  \cite{hri} is to require that the Landau pole associated with the non
asymptotically free behaviour of the $\lambda \phi^4$ theory does not occur below the scale $\Lambda$. The running of
$\lambda(\Lambda)$ at one loop is given by: 
\begin{equation}
\frac{d\lambda}{dt} = \frac{3}{4\pi^2} [ \lambda^2 + 3\lambda h^2_t - 9h^4_t + {\rm small~gauge~and~Yukawa~terms }]~,
\label{131h}
\end{equation} with the normalization such that at $t=0, \lambda = \lambda_0 = m^2_H/2v^2$ (from the minimum condition in Eq.(\ref{63b})) and the top Yukawa coupling is given by
$h_t^0 = m_t/v$. The initial value of
$\lambda$ at the weak scale increases with $m_H$ and the derivative is positive at large $\lambda$ (because of the positive $\lambda^2$ term - the $\lambda \varphi^4$ theory is not asymptotically free - which overwhelms the negative top-Yukawa term). Thus, if $m_H$ is too large, the point where $\lambda$ computed from the perturbative beta function becomes infinite (the Landau pole) occurs at too low an energy.  Of course in the vicinity of the Landau pole the 2-loop evaluation of the beta function is not reliable. Indeed the limit indicates the frontier of the domain where the theory is well described by the perturbative  expansion. Thus the quantitative evaluation of the limit is only indicative, although it has been to some extent supported by simulations of the Higgs sector of the EW theory on the lattice. For the upper limit on $m_H$ one finds \cite{hri}
\bea
m_H&\lesssim& 180~GeV~{\rm for}~\Lambda\sim M_{GUT}-M_{Planck} \nonumber \\
m_H&\lesssim&0.5-0.8~TeV~{\rm for}~ \Lambda \sim 1~TeV.
\label{25hi}
\eea

As for a lower limit on the SM Higgs mass, a possible instability of the Higgs potential $V[\phi]$ is generated by the quantum loop corrections to the
classical expression of $V[\phi]$. At large $\phi$ the derivative $V'[\phi]$ could become negative and the potential
would become unbound from below. The one-loop corrections to $V[\phi]$ in the SM are well known and change the dominant
term at large $\phi$ according to $\lambda \phi^4 \rightarrow (\lambda +
\gamma~{\rm log}~\phi^2/\Lambda^2)\phi^4$. This one-loop approximation is not enough in this case, because it fails
at large enough $\phi$, when
$\gamma~{\rm log}~\phi^2/\Lambda^2$ becomes of order 1. The renormalization group improved version of the corrected
potential leads to the replacement $\lambda\phi^4 \rightarrow
\lambda(\Lambda)\phi'^4(\Lambda)$ where $\lambda(\Lambda)$ is the running coupling and
$\phi'(\mu) =\phi~{\rm exp}\int^t \gamma(t')dt'$, with $\gamma(t)$ being an anomalous dimension function and $t = {\rm
log}\Lambda/v$ ($v$ is the vacuum expectation value
$v = (2\sqrt 2 G_F)^{-1/2}$). As a result, the positivity condition for the potential amounts to the requirement that
the running coupling $\lambda(\Lambda)$ never becomes negative. A more precise calculation, which also takes into
account the quadratic term in the potential, confirms that the requirements of positive
$\lambda(\Lambda)$ leads to the correct bound down to scales $\Lambda$ as low as $\sim$~1~TeV.  We see that, for $m_H$ small and $m_t$ fixed at its measured value,
$\lambda$ decreases with $t$ and can become negative.  If one requires that
$\lambda$ remains positive up to $\Lambda = 10^{16}$--$10^{19}$~GeV, then the resulting bound on $m_H$ in the SM with
only one Higgs doublet, obtained from a recent state of the art calculation  \cite{degr,buttaz} is given by: 
\begin{equation} m_H(\rm{GeV}) > 129.6 + 2.0 \left[ \frac{m_t(GeV) - 173.35}{0.7} \right] - 0.5~ \frac{\alpha_s(m_Z) - 0.1184}{0.0007} \pm~0.3~.
\label{25h}
\end{equation}
The estimate of the ambiguity associated with $m_t$ can be questioned: is the definition of mass as measured at the Tevatron relevant for this calculation \cite{aledjou}?
Note that this limit is evaded in models with more Higgs doublets. In this case the limit applies to some average mass but the lightest Higgs particle can well be below, as it is the case in the minimal SUSY extension of the SM (MSSM).

In conclusion, for
$m_t \sim$ 173~GeV, only a small range of values for $m_H$ is allowed, $130 < m_H <\sim 180$~GeV, if the SM holds and the vacuum is absolutely stable up
to an energy scale $\Lambda \sim M_{GUT}$ or $M_{Planck}$. For Higgs masses below this range one can still have a domain where the SM is viable because the vacuum can be unstable but with a lifetime longer than the age of the Universe \cite{degr,buttaz,branch}. We shall come back on that later (see Fig. \ref{fig:degr}).

\subsection{SM Higgs Decays}

\begin{figure}[h]
\noindent
\includegraphics[width=16cm]{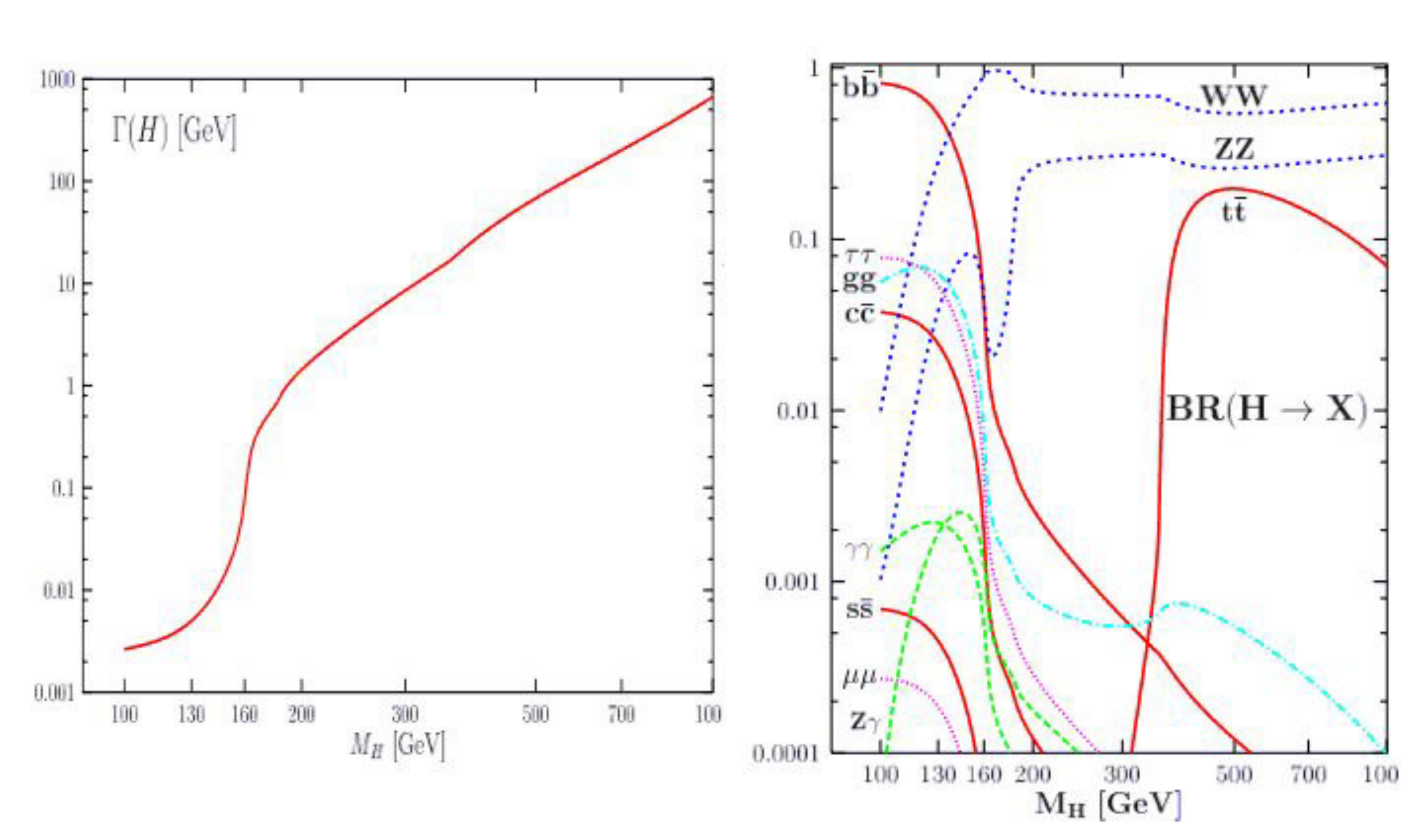} 
\caption[]{Left: The total width of the SM Higgs boson as function of the mass. Right: The branching ratios of the SM Higgs boson  as function of the mass (solid: fermions, dashed: bosons) \cite{djou}}
\label{HwidBrR}
\end{figure}

The total width and the branching ratios for the SM Higgs as function of $m_H$ are given in Fig.\ref{HwidBrR} \cite{djou}. 
Since the couplings of the Higgs particle are in proportion to masses, when $m_H$ increases the Higgs particle becomes strongly coupled. This is reflected in the sharp rise of the total width with $m_H$. For $m_H$ in the range 114-130 GeV, the width is below 5 MeV, much less than the widths of the W or the Z which have a comparable mass. The dominant channel for such a Higgs is $H\rightarrow b \bar b$. In Born approximation the partial width into a fermion pair is given by \cite{djou}, \cite{hunt}:
\beq
\Gamma(H\rightarrow f \bar f)=N_C \frac{G_F}{4\pi \sqrt{2}} m_H m_f^2\beta_f^3
\label{gaH1}
\eeq
where $\beta_f=(1-4m_f^2/m_H^2)^{1/2}$. The factor of $\beta^3$ appears because parity requires that the fermion pair must be in a p-state of orbital angular momentum for a  scalar Higgs (with parity P=+1), (this factor would be $\beta$ for a pseudoscalar Higgs boson). We see that the width is suppressed by a factor $m_f^2/m_H^2$ (the Higgs coupling is proportional to the fermion mass) with respect to the natural size $G_Fm_H^3$ for the width of a particle of mass $m_H$ decaying through a diagram with only one weak vertex. 

A glance to the branching ratios shows that the branching ratio into $\tau$ pairs is larger by more than a factor of 2 with respect to the $c \bar c$ channel. This is at first sight surprising because the colour factor $N_C$  favours the quark channels and the masses of $\tau$'s and of $D$ mesons are quite similar. This is due to the fact that the QCD corrections replace the charm mass at the scale of charm with the charm mass at the scale $m_H$, which is lower by about a factor of 2.5. The masses run logarithmically in QCD, similar to the coupling constant. The corresponding logs are already present in the 1-loop QCD correction that amounts to the replacement $m_q^2 \rightarrow m_q^2[1+2\alpha_s/\pi(\log{m_q^2/m_H^2}+3/2)]\sim m_q^2(m_H^2)$.

The Higgs width sharply increases as the WW threshold is approached. For decay into a real pair of $V$'s, with $V=W,Z$, one obtains in Born approximation \cite{djou}, \cite{hunt}:
\beq
\Gamma(H\rightarrow VV)= \frac{G_Fm_H^3}{16\pi \sqrt{2}} \delta_V\beta_V(1-4x+12x^2)
\label{gaH2}
\eeq
where $\beta_V=\sqrt{1-4x}$ with $x=m_V^2/m_H^2$ and $\delta_W=2$, $\delta_Z=1$. Much above threshold the $VV$ channels are dominant and the total width, given approximately by:
\beq
\Gamma_H \sim 0.5~{\rm TeV}(\frac{m_H}{1~{\rm TeV}})^3
\label{gaH3}
\eeq
becomes very large, signalling that the Higgs sector is becoming strongly interacting (recall the upper limit on the SM Higgs mass in Eq.(\ref{25hi})). The $VV$ dominates over the $t \bar t$ because of the $\beta$ threshold factors 
that disfavour the fermion channel and, at large $m_H$, by the cubic versus linear behaviour with $m_H$ of the partial widths for $VV$ versus $t \bar t$. Below the $VV$ threshold the decays into virtual $V$ particles is important: $VV^*$ and $V^*V^*$. Note in particular the dip of the $ZZ$ branching ratio just below the $ZZ$ threshold: this is due to the fact that the $W$ is lighter than the Z and the opening of its threshold depletes all other branching ratios. When the $ZZ$ threshold is also passed then the $ZZ$ branching fraction comes back to the ratio of approximately 1:2 with the $WW$ channel (just the number of degrees of freedom: two hermitian fields for the $W$, one for the $Z$).

\begin{figure}[h]
\includegraphics[height=0.7in]{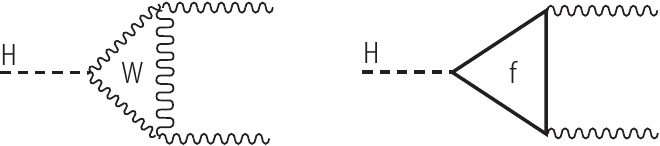}     
\caption{ Typical one-loop diagrams for Higgs decay into $\gamma \gamma$, $Z \gamma$ and, for only the quark loop, to $gg$.}
\label{fig:15}  
\end{figure}

The decay channels into $\gamma \gamma$, $Z \gamma$ and $gg$ proceed through loop diagrams, with the contributions from W (only for $\gamma \gamma$ and $Z \gamma$ ) and from fermion loops (for all) (Fig.~\ref{fig:15}).

We reproduce here the results for $\Gamma(H\rightarrow \gamma \gamma)$ and  $\Gamma(H\rightarrow gg)$ \cite{djou}, \cite{hunt}:
\beq
\Gamma(H\rightarrow \gamma \gamma)=\frac{G_F\alpha^2m_H^3}{128\pi^3\sqrt{2}}\vert A_W(\tau_W)+\sum_f N_C Q_f^2 A_f(\tau_f)\vert^2 
\label{gaH4}
\eeq
\beq
\Gamma(H\rightarrow gg)=\frac{G_F\alpha_s^2m_H^3}{64\pi^3\sqrt{2}}\vert \sum_{f=Q}  A_f(\tau_f)\vert^2 
\label{gaH5}
\eeq
where $\tau_i=m_H^2/4m_i^2$ and:
\bea
A_f(\tau)&=&\frac{2}{\tau^2}[\tau+(\tau-1)f(\tau)]\nonumber \\
A_W(\tau)&=&-\frac{1}{\tau^2}[2\tau^2+3\tau+3(2\tau-1)f(\tau)]
\label{gaH6}
\eea
with:
\bea
f(\tau)&=&\arcsin^2{\sqrt \tau}~~~~~{\rm for}~~\tau \le 1\nonumber \\
f(\tau)&=&-\frac{1}{4}[\log{\frac{1+\sqrt{1-\tau^{-1}}}{1-\sqrt{1-\tau^{-1}}}}-i\pi]^2~~~~~{\rm for}~~\tau > 1
\label{gaH7}
\eea
For $H\rightarrow \gamma \gamma$ (as well as for $H\rightarrow Z \gamma$) the W loop is the dominant contribution at small and moderate $m_H$. We recall that the $\gamma \gamma$ mode is a possible channel for Higgs discovery only for $m_H$ near its lower bound (i.e for $114<m_H<150$ GeV). In this domain of $m_H$ we have $\Gamma(H\rightarrow \gamma \gamma)\sim 6-23$ KeV. For example, in the limit $m_H<<2m_i$, or $\tau \rightarrow 0$, we have $A_W(0)=-7$ and  $A_f(0)=4/3$. The two contributions become comparable only for $m_H\sim$ 650 GeV where the two amplitudes, still of opposite sign, nearly cancel. The top loop is dominant among fermions 
(lighter fermions are suppressed by $m_f^2/m_H^2$ modulo logs ) and, as we have seen, it approaches a constant for large $m_t$. Thus the fermion loop amplitude for the Higgs would be sensitive to effects from very heavy fermions, in particular the $H\rightarrow gg$ effective vertex would be sensitive to all possible very heavy coloured quarks (of course there is no W-loop in this case and the top quark gives the dominant contribution in the loop). As discussed in the QCD Chapter, the $gg \rightarrow H$ vertex provides one of the main production channels for the Higgs boson at hadron colliders (another important channel at present is the WH associate production).

\subsection{The Higgs Discovery at the LHC}
\label{sec:33}

On July 4th, 2012 at CERN the ATLAS and CMS Collaborations \cite{ATLH,CMSH} announced the observation of a particle with mass around 126 GeV that, within the present accuracy, indeed looks like the SM Higgs boson. This is a great breakthrough that, by itself, already makes an adequate return for the LHC investment. With the Higgs discovery the main missing block for the experimental validation of the SM is now in place. The Higgs discovery is the last milestone in the long history (some 130 years) of the development of  a field theory of fundamental interactions (apart from quantum gravity),
starting with the Maxwell equations of classical electrodynamics, going through the great revolutions of Relativity
and Quantum Mechanics, then the formulation of Quantum Electro Dynamics (QED) and the gradual build up of the gauge part of the Standard Model
and finally completed with the tentative description of the  Electro-Weak (EW) symmetry breaking sector of the SM in terms of a simple formulation of the Englert- Brout- Higgs mechanism \cite{ebh}. 
The other extremely important result from the LHC at 7 and 8 TeV center of mass energy is that no new physics signals have been seen so far. This negative result is certainly less exciting than a positive discovery, but it is a crucial new input that, if confirmed in the future LHC runs at 13 and 14 TeV, will be instrumental in re-directing our perspective of the field.  In this Section we summarize the relevant data on the Higgs signal as they are known at present while the analysis of the data from the 2012 LHC run is still in progress. 

The Higgs particle has been observed by ATLAS and CMS in five channels $\gamma \gamma$, $ZZ^*$, $WW^*$, $b \bar b$ and $\tau^+ \tau^-$. Also including the Tevatron experiments, especially important for the $b \bar b$ channel, the combined evidence is by now totally convincing.  The ATLAS (CMS) combined values for the mass, in GeV$/c^2$, are $m_H = 125.5 \pm 0.6$ ($m_H = 125.7 \pm 0.4$). This light Higgs is what one expects from a direct interpretation
of EW precision tests \cite{ew,ciuch,smfit}. The possibility of a "conspiracy" (the Higgs is heavy but it falsely appears as light because of confusing new physics effects) has been discarded: the EW precision tests of the SM tell the truth and in fact, consistently, no "conspirators", namely no new particles, have been seen around. 

\begin{figure}
\centerline{\includegraphics[height=2.5in]{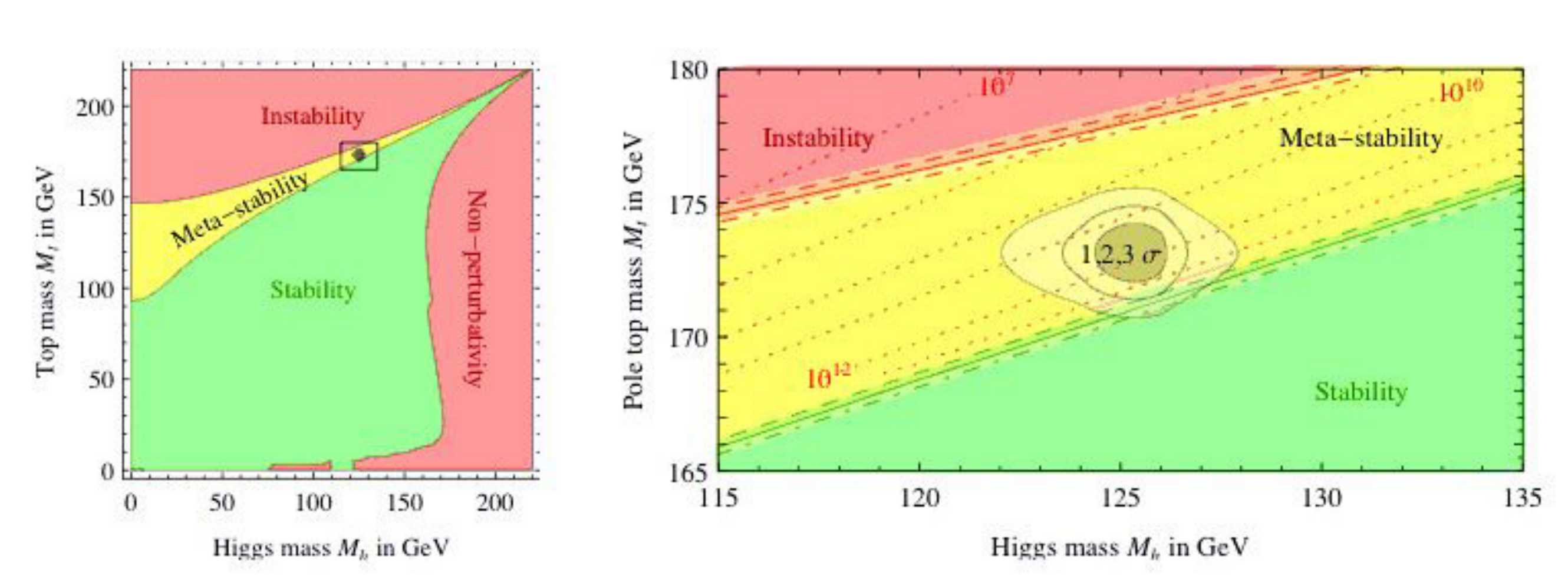}}  
\caption{Vacuum stability domains in the SM for the observed values of $m_t$ and $m_H$ \cite{degr,buttaz}. On the right an expanded view of the most relevant domain in the $m_t$-$m_H$ plane. The dotted contour-lines show the scale $\Lambda$ in GeV where the instability sets in, for $\alpha_s(m_Z)$ = 0.1184.}
\label{fig:degr}  
\end{figure}

As shown in the previous Section the observed value of $m_H$ is a bit too low for the SM to be valid up to the Planck mass with an absolutely stable vacuum (see Eq. \ref{25h}) but it corresponds to a metastable value with a lifetime longer than the age of the universe, so that the SM can well be valid up to the Planck mass (if one is ready to accept the immense fine tuning that this option implies, as discussed in Sect. \ref{sec:34}). This is shown in Fig. \ref{fig:degr} where the stability domains as functions of $m_t$ and $m_H$ are shown, as obtained from a recent state-of-the-art evaluation of the relevant boundaries \cite{degr,buttaz}. It is puzzling to find that, with the measured values of the top and Higgs masses and of the strong coupling constant, the evolution of the Higgs quartic coupling ends up into a narrow metastability wedge at very large energies. This criticality looks intriguing and perhaps it should tell us something.

In order to be sure that this is the SM Higgs boson one must confirm that the spin-parity is $0^+$ and that the couplings are as predicted by the theory. Also it is essential to search for possible additional Higgs states as, for example, predicted in supersymmetric extensions of the SM.

\vspace*{12pt}

As for the spin (see, for example, \cite{ellis}), the existence of the $H \rightarrow \gamma \gamma$ mode proves that spin cannot be 1 and must be either 0 or 2, in the assumption of an s-wave decay. The $b \bar b$ and $\tau^+ \tau^-$ modes are compatible with both possibilities. With large enough statistics the spin-parity can be determined from the  distributions of $H \rightarrow ZZ^*\rightarrow 4~leptons$, or $WW^* \rightarrow 4~leptons$. Information can also be obtained from the HZ invariant mass distributions in the associated production \cite{ellis}. The existing data already appear to strongly favour a $J^P=0^+$ state against $0^-,~1^{+/-},~2^+$ \cite{Land}. We do not expect surprises on the spin-parity assignment because, if different, then all the lagrangian vertices would be changed and the profile of the SM Higgs particle would be completely altered. 
 
\begin{figure}
\centerline{\includegraphics[height=3in]{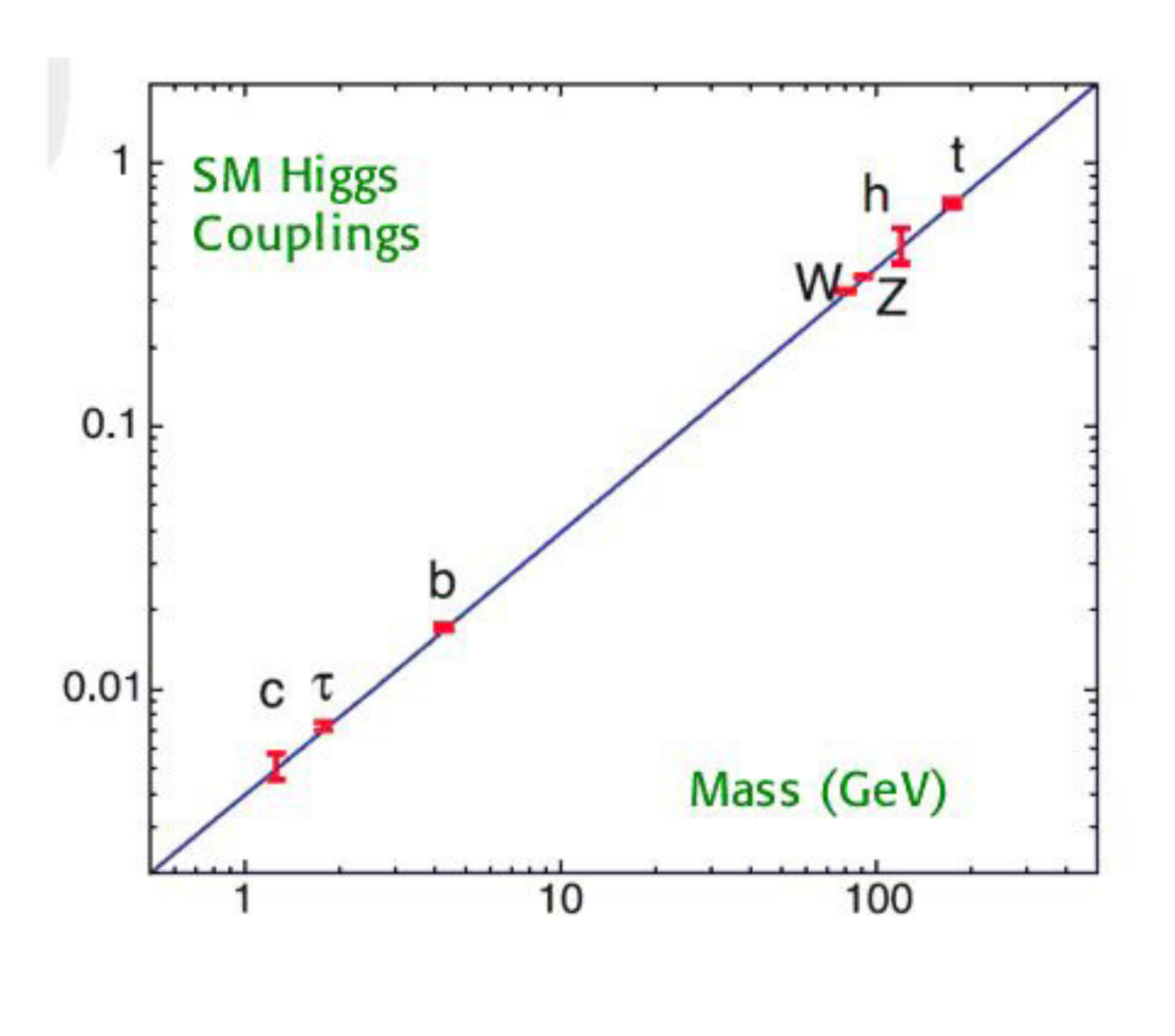}}  
\caption{The predicted couplings of the SM Higgs.}
\label{Hcoupl}  
\end{figure}

\begin{figure}
\centerline{\includegraphics[height=3in]{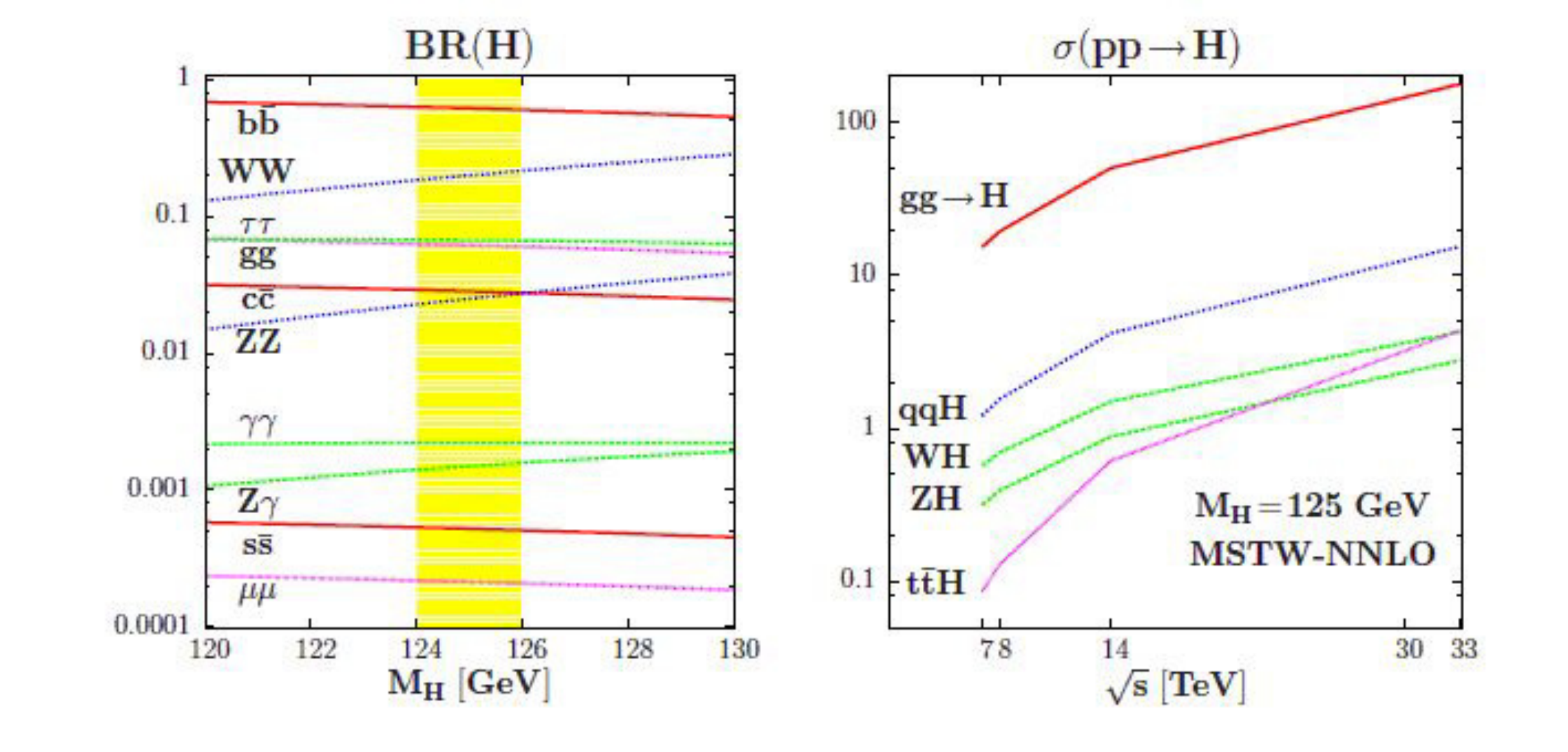}}  
\caption{The branching ratios of the SM Higgs boson in the mass range $m_H = 120-130 $GeV
(left) and its production cross sections at the LHC for various c.m. energies (right) \cite{djou'12}.}
\label{fig:15a}  
\end{figure}

 The tree level couplings of the Higgs are in proportion to masses and, as a consequence, are very hierarchical. The loop effective vertices to $\gamma  \gamma$ and $g g$, $g$ being the gluon, are also completely specified in the SM, where no states heavier than the top quark exist and contribute in the loop. As a consequence the SM Higgs couplings are predicted to exhibit a very special and very pronounced pattern (see Fig. \ref{Hcoupl}) which would be extremely difficult to fake by a random particle (only a dilaton, particle coupled to the energy-momentum tensor, could come close to simulate a Higgs particle, at least for the $H$ tree level couplings, although in general there would be a common proportionality factor in the couplings). The hierarchy of couplings is reflected in the branching ratios and the rates of production channels, as seen in Figs. \ref{fig:15a}. The combined signal strengths (which, modulo acceptance and selection cuts deformations, correspond to $\mu=\sigma Br/(\sigma Br)_{SM}$) are obtained as $\mu=0.8\pm0.14$ by CMS and $\mu=1.30\pm0.20$ by ATLAS. Taken together these numbers make a triumph for the SM!

\vspace*{12pt}

\vspace*{12pt}

Within the present (October '13) limited, accuracy the measured Higgs couplings are in reasonable agreement (at about a 20$\%$ accuracy) with the sharp predictions of the SM. Great interest was excited by a hint of an enhanced Higgs signal in $\gamma \gamma$ but, if we put the ATLAS and CMS data together, the evidence appears now to have evaporated. All included, if the CERN particle is not the SM Higgs it must be a very close relative! Still it would be really astonishing if the H couplings would exactly be those of the minimal SM, meaning that no new physics distortions reach an appreciable contribution level. Thus, it becomes a firm priority to establish a roadmap for measuring the H couplings as precisely as possible. The planning of new machines beyond the LHC has already started. Meanwhile the strategies for analyzing the already available and the forthcoming data in terms of suitable effective lagrangians have been formulated (see, for example,\cite{eff} and refs. therein). A very simple test is to introduce a universal factor multiplying all $H \bar \psi \psi$ couplings to fermions, denoted by $c$ and another factor $a$ multiplying the $HWW$ and $HZZ$ vertices. Both $a$ and $c$ are 1 in the SM limit. All existing data on production times branching ratios are compared with the $a$- and $c$-distorted formulae to obtain the best fit values of these parameters (see \cite{giastru,azat,falk} and refs. therein). At present this fit is performed routinely by the experimental Collaborations \cite{atlcoup,cmscoup} each using its own data. But theorists have not refrained from abusively combining the data from both experiments and the result is well in agreement with the SM, as shown in Fig. \ref{a-cFit} \cite{giastru,falk}. Actually, a more ambitious fit in terms of 7 parameters has also been performed \cite{falk} with a common factor like $a$ for couplings to $WW$ and $ZZ$, 3 separate $c$-factors, $c_t$, $c_b$ and $c_\tau$ for up-type and d-type quarks and for charged leptons, and 3 parameters,  $c_{gg}$,   $c_{\gamma\gamma}$ and $c_{Z\gamma}$ for additional gluon-gluon, $\gamma-\gamma$ and $Z-\gamma$ terms, respectively. In the SM $a=c_t=c_b=c_\tau=1$ and $c_{gg}=c_{\gamma\gamma}=c_{Z\gamma}=0$.
The present data allow a meaningful determination of all 7 parameters which turns out to be in agreement with the SM \cite{falk}.
 For example, in the MSSM, at the tree level, $a=\sin{(\beta - \alpha})$, for fermions the u- and d-type quark couplings are different: $c_t=\cos{\alpha}/\sin{\beta}$ and $c_b= - \sin{\alpha}/\cos{\beta}=c_\tau$. At the tree-level (but radiative corrections are in many cases necessary for a realistic description), the $\alpha$ angle is related to the $A$, $Z$ masses and to $\beta$ by $\tan{2\alpha}=\tan{2\beta} (m_A^2-m_Z^2)/(m_A^2+m_Z^2)$.  If $c_t$ is enhanced, $c_b$ is suppressed. In the limit of large $m_A$, $a=\sin{(\beta - \alpha)} \rightarrow 1$. 

In conclusion it really appears that the Higgs sector of the minimal SM, with good approximation,  is realized in nature. Apparently, what was considered just as a toy
model, a temporary addendum to the gauge part of the SM, presumably to be replaced by a more complex reality and likely to be accompanied by new physics,
has now been experimentally established as the actual realization of the EW symmetry breaking (at least to a very good approximation).
If the role of the newly discovered particle in the EW symmetry breaking will be confirmed it would be the only known example in physics of a fundamental,
weakly coupled, scalar particle with vacuum expectation value (VEV). We know many composite types of Higgs-like particles, like the Cooper pairs of superconductivity or the quark condensates that break the chiral symmetry of massless QCD, but the Higgs found at the LHC is the only possibly elementary one.
This is a death blow not only to Higgsless models, to straightforward technicolor
models and other unsophisticated strongly interacting Higgs sector models but actually a threat to all models without
fast enough decoupling (in that, if new physics comes in a model with decoupling, the
absence of new particles at the LHC helps in explaining why large corrections
to the H couplings are not observed).

\begin{figure}
\centerline{\includegraphics[height=3.2in]{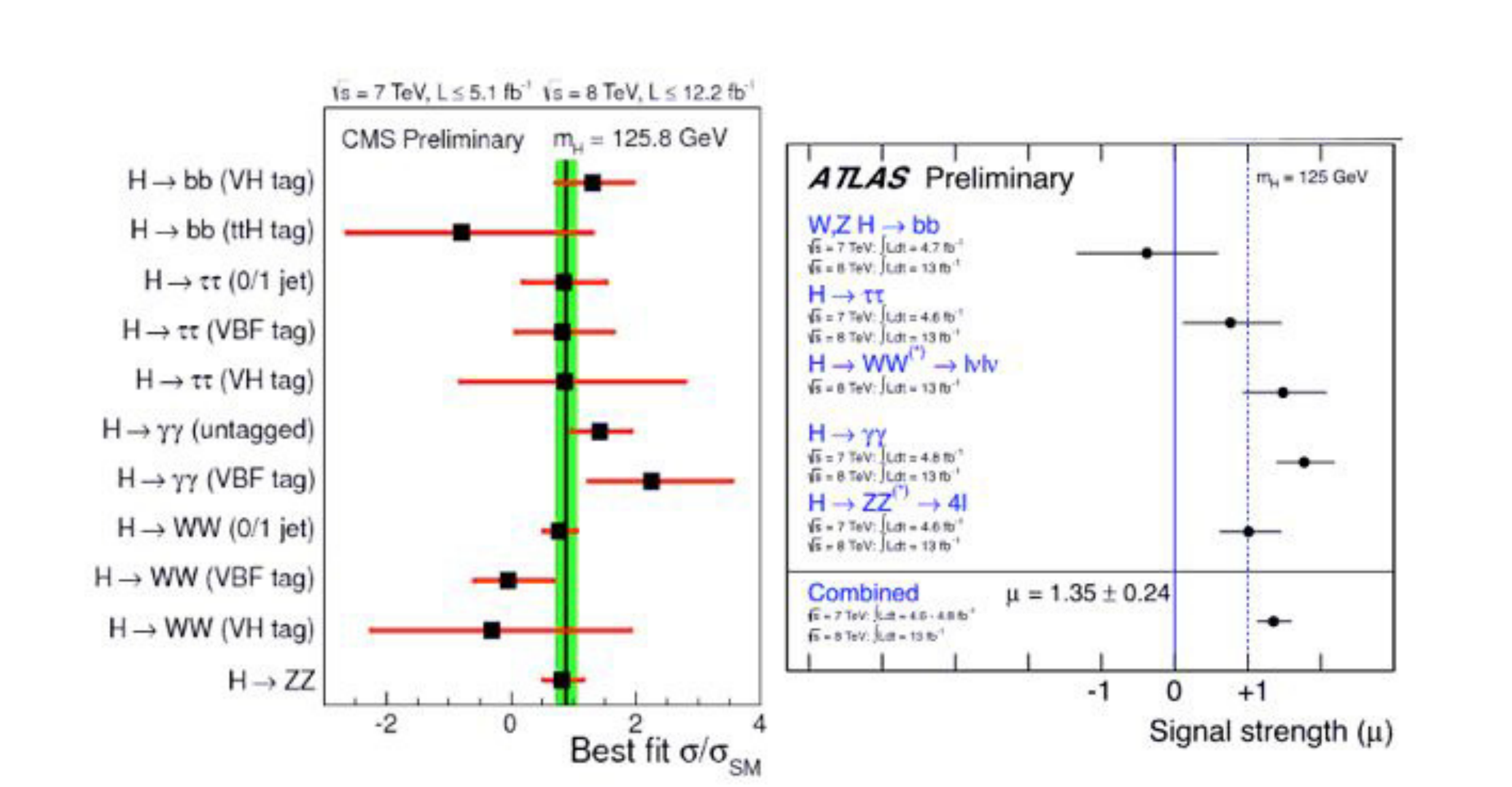}}  
\caption{The measured H couplings compared with the SM predictions by the CMS and ATLAS Collaborations.}
\label{fig:16}  
\end{figure}
\vspace*{12pt}

\begin{figure}
\centerline{\includegraphics[height=12cm]{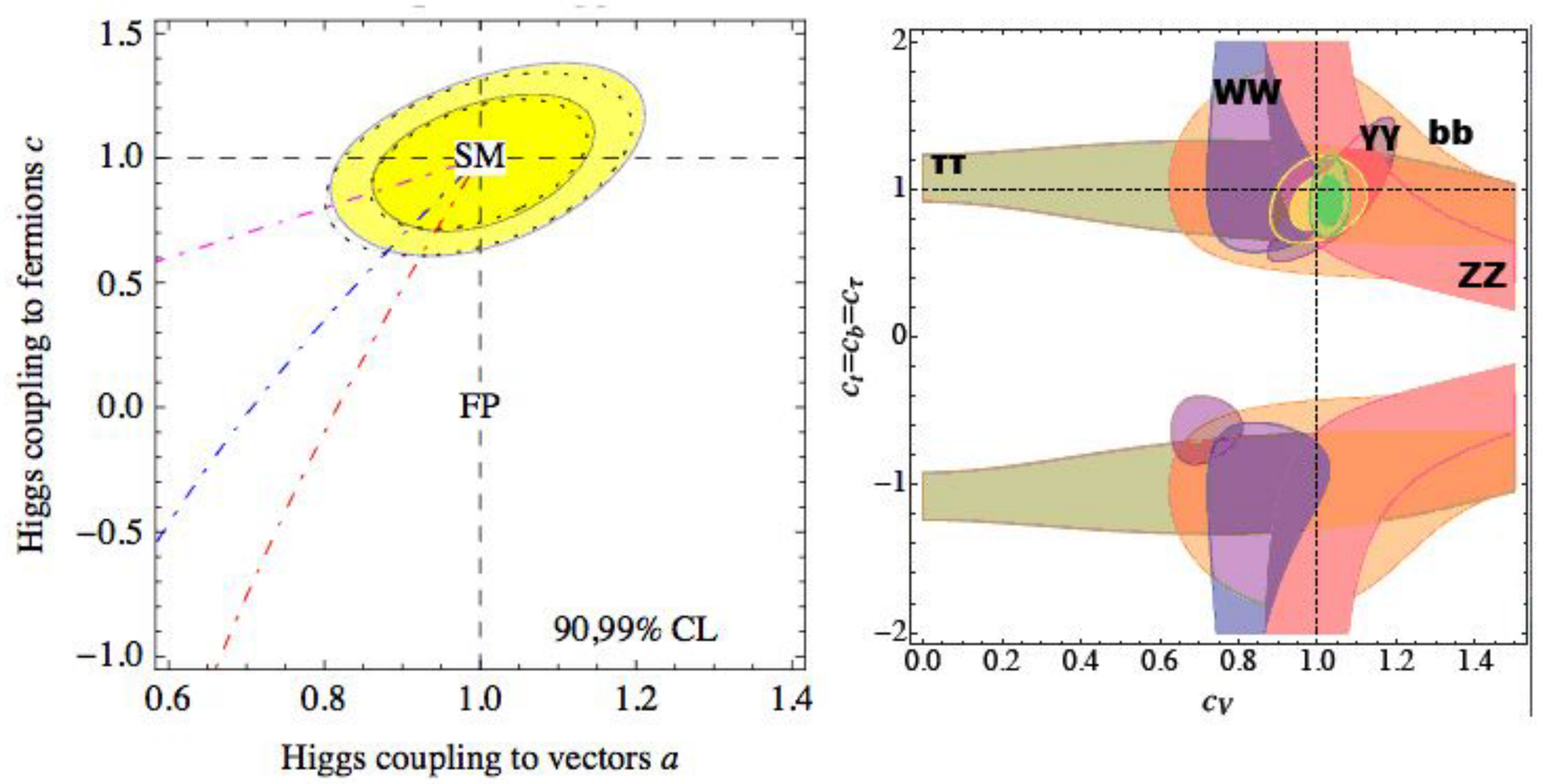}}  
\caption{Fit of the Higgs boson couplings obtained from the (unofficially) combined ATLAS and CMS data assuming common rescaling factors a and c with respect to the SM prediction for couplings to vector bosons and fermions, respectively. Left: from ref. \cite{giastru}: the dashed lines correspond to different versions of composite Higgs models. The dashed vertical line, denoted by FP (Fermio-Phobic) corresponds to a=1 and c=$1-\xi$. Then from bottom to top: c=$(1-3\xi)/a$,  c=$(1-2\xi)/a$, a=c=$\sqrt{1-\xi}$, with $\xi$ defined in sect. 5. Right:  taken from ref. \cite{falk} with $c_t=c_b=c_{\tau}=c$ and $c_V=a$ .}
\label{a-cFit}  
\end{figure}
\vspace*{12pt}

\subsection{Limitations of the Standard Model}
\label{sec:34}

No signal of new physics has been
found neither by direct production of new particles at the LHC nor in the electroweak precision tests nor in flavour physics. Given the success of the SM why are we not satisfied with this theory? Once the Higgs particle has been found,
why don't we declare particle physics closed? The reason is that there are
both conceptual problems and phenomenological indications for physics beyond the SM. On the conceptual side the most
obvious problems are that quantum gravity is not included in the SM and that the famous hierarchy (or naturalness or fine-tuning) problem remains open. Among the main
phenomenological hints for new physics we can list coupling unification, dark matter, neutrino masses (discussed in Sect. (\ref{sec:26})), 
baryogenesis and the cosmological vacuum energy.  At accelerator experiments the most plausible departure from the SM is the muon anomalous magnetic moment that, as discussed in Sect. \ref{sec:28} shows a deviation by about 3 $\sigma$, but some caution should be applied as a large fraction of the uncertainty is of theoretical origin, in particular that due to the hadronic contribution to light-light scattering \cite{HMpdg}.

\begin{figure}
\centerline{\includegraphics[height=3in]{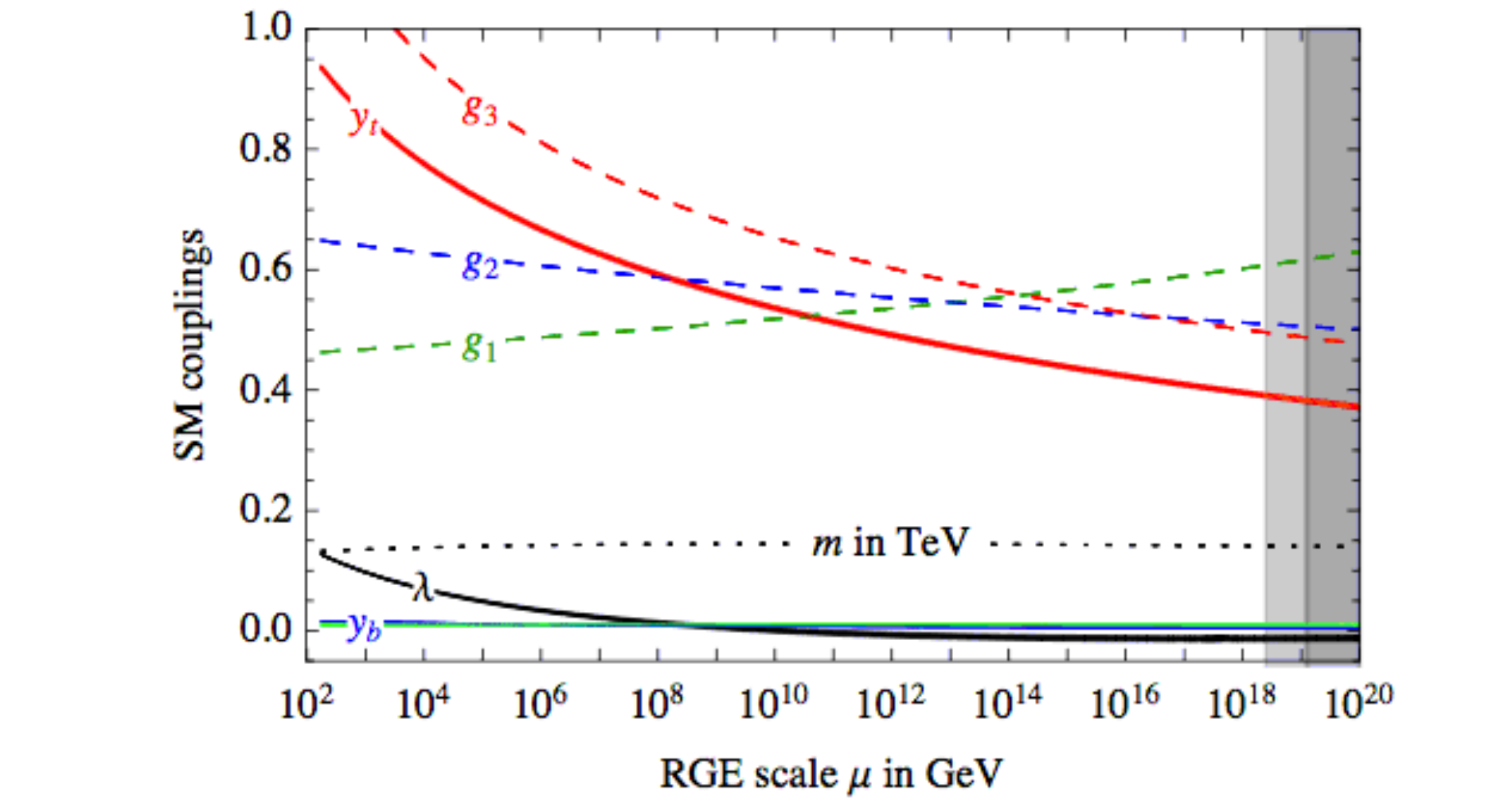}}  
\caption{Renormalisation of the SM gauge couplings $g_1 = \sqrt{5/3}g_Y$ , $g_2$, $g_3$, of the top, bottom and $\tau$ couplings ($y_t, ~y_b, ~y_\tau $), of the Higgs quartic coupling $\lambda$ and of the Higgs mass parameter $m$. In the figure $y_b$ and $y_\tau$ are not easily distinguished. All parameters are defined in the $\bar{M}\bar{S}$  scheme \cite{buttaz}.}
\label{evol}  
\end{figure}
\vspace*{12pt}

The computed evolution with energy
of the effective SM gauge couplings clearly points towards the unification of the electro-weak and strong forces (GUTs) at scales of energy
$M_{GUT}\sim  10^{15}-10^{16}~ $GeV \cite{GUTs} which are close to the scale of quantum gravity, $M_{Planck}\sim 10^{19}~ $GeV.  The crossing of the 3 gauge couplings at a single point is not perfect in the SM and is much better in the supersymmetric extensions of the SM. But still the matching is sufficiently close in the SM (see Fig. \ref{evol}, \cite{buttaz}) that one can imagine some atypical threshold effect at the GUT scale to fix the apparent residual mismatch. One is led to
imagine  a unified theory of all interactions also including gravity (at present superstrings \cite{strings} provide the best attempt at such
a theory). 

Thus GUTs and the realm of quantum gravity set a very distant energy horizon that modern particle theory cannot
ignore. Can the SM without new physics be valid up to such large energies? One can imagine that some obvious problems of the SM could be postponed to the more fundamental theory at the Planck mass. For example, the explanation of the three generations of fermions and the understanding of fermion masses and mixing angles can be postponed. But other problems must find their solution in the low energy theory. In particular, the structure of the
SM could not naturally explain the relative smallness of the weak scale of mass, set by the Higgs mechanism at $v\sim
1/\sqrt{G_F}\sim  250~ $GeV  with $G_F$ being the Fermi coupling constant. This so-called hierarchy problem \cite{hiera} is due to the instability of the SM with respect to quantum corrections. 
In fact, nobody can believe that the SM is the definitive, complete theory but, rather, we all believe it is only an effective low energy theory. The dominant terms at low energy correspond to the SM renormalizable lagrangian
but additional non renormalizable terms should be added which are suppressed by powers (modulo logs) of the large scale $\Lambda$ where physics beyond the SM becomes relevant (for simplicity we write down only one such scale of new physics, but there could be different levels). The complete Lagrangian takes the general form:
\bea
{\cal L}& = &O(\Lambda^4)+O(\Lambda^2){\cal L}_2+O(\Lambda){\cal L}_3+O(1){\cal L}_4+\nonumber\\
&+& O(\frac{1}{\Lambda}){\cal L}_5+ O(\frac{1}{\Lambda^2}){\cal L}_6+\dots
\label{efflag}
\eea
Here ${\cal L}_D$ are lagrangian vertices of operator dimension $D$. In particular ${\cal L}_2=\Phi^\dagger\Phi$ is a scalar mass term, ${\cal L}_3= \bar{\Psi}\Psi$ is a fermion mass term (that in the SM only appears after EW symmetry breaking), ${\cal L}_4$ describes all dimension-4 gauge and Higgs interactions, ${\cal L}_5$ is the Weinberg operator \cite{weidim5}  (with two lepton doublets and two Higgs fields) that leads to neutrino masses (see sect. \ref{sec:26}) and ${\cal L}_6$ includes 4-fermion operators (among others). The first line in Eq. \ref{efflag} corresponds to the renormalizable part (that is, what we usually call the SM). The baseline power of the large scale $\Lambda$ in the coefficient of each ${\cal L}_D$ vertex is fixed by dimensions.  A deviation from the baseline power can only be naturally expected if some symmetry or some dynamical principle justifies a suppression. For example, for the fermion mass terms, we know that all Dirac masses vanish in the limit of gauge invariance and only arise when the Higgs VEV $v$ breaks the EW symmetry. The fermion masses also break chiral symmetry. Thus the fermion mass coefficient is not linear in $\Lambda$ modulo logs but actually behaves as $v\log{\Lambda}$. An exceptional case is the Majorana mass term of right-handed neutrinos $\nu_R$, $M_{RR}\bar{\nu_R^c}\nu_R$ , which is lepton number non conserving but gauge invariant (because $\nu_R$ is a gauge singlet). In fact,  in this case,  one expects that $M_{RR} \sim \Lambda$. As another example, proton decay arises from a 4-fermion operator in ${\cal L}_6$ suppressed by $1/\Lambda^2$, where, in this case, $\Lambda$ could be identified with the large mass of lepto-quark gauge bosons that appear in GUTs. 

The hierarchy problem arises because the coefficient of ${\cal L}_2$ is not suppressed by any symmetry. This term, which appears in the Higgs potential, fixes the scale of the Higgs VEV and of all related masses. Since empirically the Higgs mass is light (and, by naturalness, it should be of $O(\Lambda)$) we would expect that $\Lambda$, i.e. some form of new physics, should appear near the TeV scale. The hierarchy problem can be put in very practical terms (the "little hierarchy problem"): loop corrections to the Higgs mass squared are
quadratic in the cut off  $\Lambda$, which can be interpreted as the scale of new physics. The most pressing problem is from the top loop.
 With $m_h^2=m^2_{bare}+\delta m_h^2$ the top loop gives 
 \begin{eqnarray}
\delta m_{h|top}^2\sim -\frac{3G_F}{2\sqrt{2} \pi^2} m_t^2 \Lambda^2\sim -(0.2\Lambda)^2 \label{top}
\end{eqnarray}
If we demand that the correction does not exceed the light Higgs mass observed by experiment (that is, we exclude an unexplained fine-tuning) $\Lambda$ must be
close, $\Lambda\sim O(1~$TeV$)$. Similar constraints also arise from the quadratic $\Lambda$ dependence of loops with exchanges of gauge bosons and
scalars, which, however, lead to less pressing bounds. So the hierarchy problem strongly indicates that new physics must be very close (in
particular the mechanism that quenches or compensates the top loop). The restoration of naturalness would occur if new physics implemented an approximate symmetry implying the cancellation of the $\Lambda^2$ coefficient.
Actually, this new physics must be rather special, because it must be
very close, yet its effects are not already clearly visible neither in precision electroweak tests (the "LEP Paradox" \cite{BS}) nor in flavour changing processes and CP violation.

It is important to note that although the hierarchy problem is directly related to the quadratic divergences in the scalar sector of the SM, actually the problem can be formulated without any reference to divergences, directly in terms of renormalized quantities. After renormalization the hierarchy problem is manifested by the quadratic sensitivity of $\mu^2$ to the physics at large energy scales. If there is a threshold at large energy, where some particles of mass $M$ coupled to the Higgs sector can be produced and contribute in loops, then the renormalized running mass $\mu$ would evolve slowly (i.e. logarithmically according to the relevant beta functions \cite{strubeta}), up to $M$ and there, as an effect of the matching conditions at the threshold, rapidly jump to become of order $M$ (see, for example, \cite{barThr}). In fact in Fig. \ref{evol} we see  that, in the assumption of no thresholds, the running Higgs mass $m$ slowly evolves, starting from the observed low energy value, up to very high energies. In the presence of a threshold at $M$ one needs a fine tuning of order $\mu^2/M^2$ in order to fix the running mass at low energy to the observed value. Thus for naturalness either new thresholds appear endowed with a mechanism for the cancellation of the sensitivity or they would better not appear at all. But certainly there is the Planck mass, connected to the onsetting of quantum gravity, that sets an unavoidable threshold. A possible point of view is that there are no new thresholds up to $M_{Planck}$ (at the price of giving up GUTs, among other things) but, miraculously, there is a hidden mechanism in quantum gravity that solves the fine tuning problem related to the Planck mass \cite{shapo,gian}. For this one would need to solve all phenomenological problems, like dark matter, baryogenesis and so on, with physics below the EW scale. Possible ways to do so are discussed in ref. \cite{shapo}.  This point of view is extreme but allegedly not yet ruled out.

The main classes of orthodox solutions to the hierarchy problem are:

1) Supersymmetry \cite{susy}. In the limit of exact boson-fermion symmetry the quadratic bosonic divergences cancel so that
only log divergences remain. However, exact SUSY is clearly unrealistic. For approximate SUSY (with soft breaking terms and R-parity conervation),
which is the basis for most practical models, $\Lambda^2$ is essentially replaced by the splitting of SUSY multiplets, $\Lambda^2 \sim
m_{SUSY}^2-m_{ord}^2$ (with $m_{ord}$ being the SM particle masses). In particular, the top loop is quenched by partial cancellation with s-top exchange, so the s-top cannot be too heavy.  After the the bounds from the LHC, the present emphasis is to build SUSY models
where naturalness is
restored not too far from the weak scale but the related
new physics is arranged in such a way that it was not visible so far. The simplest ingredients introduced in order to decrease the fine tuning are either the assumption of a split spectrum with heavy first two generations of squarks (for some recent work along this line see, for example,  ref. \cite{natsusy}) or the enlargement of  the Higgs sector of the MSSM by adding a singlet Higgs field (see, for example, ref. \cite{nmssm} (Next-to minimal SUSY SM: NMSSM) or both.

2) A strongly interacting EW symmetry breaking sector. The archetypal model of this class is Technicolor where the Higgs is a condensate of new fermions \cite{tech1}. In these theories there is no fundamental scalar Higgs field, hence no
quadratic divergences associated to the $\mu^2$ mass in the scalar potential. But this mechanism needs a very strong binding force,
$\Lambda_{TC}\sim 10^3~\Lambda_{QCD}$. It is  difficult to arrange for such a nearby strong force not to show up in
precision tests. Hence this class of models has been abandoned after LEP, although some special classes of models have been devised a posteriori, like walking TC, top-color assisted TC etc \cite{tech2} (for reviews see, for example, ref. \cite{tech3}). But the simplest Higgs observed at the LHC has now eliminated another score of these models. Modern strongly interacting models, like little Higgs models \cite{little} (in these models extra symmetries allow $m_h\not= 0$ only at two-loop level, so that $\Lambda$
can be as large as
$O(10~$TeV$)$), or composite Higgs models  \cite{compoold,compo} (where a non perturbative dynamics modifies the linear realization of the gauge symmetry and the Higgs has both elementary and composite components) are more sophisticated. All models in this class share the idea that the Higgs is light because it is the pseudo-Goldstone boson of an enlarged global symmetry of the theory, for example $SO(5)$ broken down to $SO(4)$. There is a gap between the mass of the Higgs (similar to a pion) and the scale $f$ where new physics appears in the form of resonances (similar to the $\rho$ etc). The ratio $\xi = v^2/f^2$ defines a degree of compositeness that interpolates between the SM at $\xi=0$ up to technicolor at $\xi=1$. Precision EW tests impose that $\xi <0.05-0.2$. In these models the bad quadratic behaviour from the top loop is softened by the exchange of new vector-like fermions with charge 2/3 or even with exotic charges like 5/3, see, for example, refs. \cite{extop,CMST5/3}.

3) Extra~dimensions \cite{ed1,rs} (for pedagogical introductions, see, for example, ref. \cite{ed2}). The idea is that $M_{Planck}$ appears very large, or equivalently that gravity appears very weak,
because we are fooled by hidden extra dimensions so that either the real gravity scale is reduced down to a lower scale, even possibly down to
$O(1~TeV)$ or the intensity of gravity is red shifted away by an exponential warping factor \cite{rs}. This possibility is very exciting in itself and it is really remarkable that it is compatible with experiment. It provides a very rich framework with many different scenarios.

4) The anthropic evasion of the problem. The observed value of the cosmological constant $\Lambda$ also poses a tremendous, unsolved naturalness problem \cite{tu}. Yet the value of $\Lambda$ is close to the Weinberg upper bound for galaxy formation \cite{We}. Possibly our Universe is just one of infinitely many bubbles (Multiverse) continuously created from the vacuum by quantum fluctuations. A different physics takes place in different Universes according to the multitude of string theory solutions \cite{doug} ($\sim 10^{500}$). Perhaps we live in a very unlikely Universe but the only one that allows our existence \cite{anto,giu,sche}. Personally, I find the application of the anthropic principle to the SM hierarchy problem somewhat excessive. After all one can find plenty of models that easily reduce the fine tuning from $10^{14}$ to $10^2$: why make our Universe so terribly unlikely? If  we add, say, supersymmetry to the SM, does the Universe become less fit for our existence? In the Multiverse there should be plenty of less fine tuned Universes where more natural solutions are realized and yet are suitable for our living. By comparison the case of the cosmological constant is a lot different: the context is not as fully specified as that for the SM (quantum gravity, string cosmology, branes in extra dimensions, wormholes through different Universes....). Also, while there are many natural extensions of the SM,  so far there is no natural theory of the cosmological constant.

It is true that the data impose a substantial amount of apparent fine tuning and certainly our criterion of naturalness has failed so far, so that we are now lacking a reliable argument on where precisely the new physics threshold is located, but still many of us remain confident that some new physics will appear not too far from the weak scale. 

While I remain skeptical I would like to sketch here one possibility of how the SM can be extended in agreement with the anthropic idea.  If we ignore completely the fine tuning problem and only want to reproduce, in a way compatible with GUTs,  the most compelling data that demand new physics beyond the SM, a possible scenario is the following one. The SM spectrum is completed by the just discovered light Higgs and no other new physics is in the LHC range (how sad!). In particular there is no SUSY in this model. At the GUT scale of $M_{GUT} \ge 10^{16}$ GeV the unifying group is $SO(10)$, broken at an intermediate scale, typically $M_{int} \sim  10^{10}-10^{12}$ down to a subgroup like the Pati-Salam group $SU(4)\bigotimes SU(2)_L \bigotimes SU(2)_R$ or $SU(3)\bigotimes U(1) \bigotimes
SU(2)_L \bigotimes SU(2)_R$ \cite{mal}. Note that, in general, unification in $SU(5)$ would not work because we need a group of rank larger than 4 to allow for a two step (at least) breaking: this is needed, in the absence of SUSY,  to restore coupling unification and to avoid a too fast proton decay. An alternative is to assume some ad hoc intermediate threshold to modify the evolution towards unification \cite{uniax}. The Dark Matter problem is one of the strongest evidences for new physics. In this model it should be solved by axions \cite{pequi,kimcp,kim}. It must be said that axions have the problem that their mass should be fixed ad hoc to reproduce the observed amount of Dark Matter. In this respect the WIMP (Weakly Interacting Massive Particle) solution, like the neutralinos in SUSY models,  is much more attractive.  Lepton number violation, Majorana neutrinos and the see-saw mechanism give rise to neutrino mass and mixing. Baryogenesis occurs through leptogenesis \cite{bupe}. One should one day observe proton decay and neutrino-less beta decay. None of the alleged indications for new physics at colliders would survive (in particular even the claimed muon (g-2) \cite{amu} discrepancy should be attributed, if not to an experimental problem, to an underestimate of the theoretical uncertainties or, otherwise, to some specific addition to the above model \cite{stru3}). This model is in line with the non observation of the decay $\mu \rightarrow  e \gamma$ at MEG \cite{meg}, of the electric dipole moment of the neutron \cite{nedm}
etc. It is a very important challenge to experiment to falsify such a  scenario by establishing a firm evidence of new physics at the LHC or at another "low energy" experiment. 

In 2015 the LHC will restart at 13-14 TeV and, in the following years, should collect a much larger statistical sample than available at present at 7-8 TeV. From the above discussion it is clear that it is extremely important for the future of particle physics to know whether the extraordinary and unexpected success of the SM, including the Higgs sector, will continue or if  clear signals of new physics will finally appear as we very much hope.

\vskip 4.0cm

Acknowledgement

I am very grateful to Giuseppe Degrassi, Ferruccio Feruglio, Paolo Gambino, Mario Greco, Martin Grunewald, Vittorio Lubicz, Richard Ball, Keith Ellis,  Stefano Forte,  Ashutosh Kotwal, Lorenzo Magnea, Michelangelo Mangano, Luca Merlo, Silvano Simula, Graham Watt for their help and advice.

This work has been partly supported by the Italian Ministero dell'Uni\-ver\-si\-t\`a e della Ricerca Scientifica, under the COFIN program (PRIN 2008), by the European Commission, under the networks ``LHCPHENONET'' and ``Invisibles''.

\end{document}